%% file: moxc3.tex
\documentclass[preprint,11pt,trackchanges]{aastex62}

\usepackage{grffile}  %grffile allows .'s in filenames
\usepackage{url}                                                                                  

\graphicspath{{AEAur/}{AFGL4029/}{Berk87/}{DR15/}{G305/}{hPer/}{NGC2264/}{NGC281/}{NGC6193/}{NGC6231/}{NGC6357/}{RCW49/}}

\newcommand{\anchorfoot}[2] {\href{#1}{#2}\footnote{\url{#1}}}
\newcommand{\anchorparen}[2]{\href{#1}{#2} (\url{#1})}

\newcommand{\tablenotetextbox}[3]{\vspace{\baselineskip}\hspace{-0.6in}#1\hspace{1ex}\parbox[t]{#2}{#3}}
\renewcommand{\S}{Section }
\newcommand{\tnm}[1]{\tablenotemark{#1}}
\renewcommand{\dataset}[2][]{{#1}}

\newcommand{\hii}{H{\scriptsize II} }

\newcommand{\Gaia} {{\em Gaia~}}
\newcommand{\Spitzer} {{\em Spitzer~}}
\newcommand{\WISE} {{\em WISE~}}
\newcommand{\Herschel} {{\em Herschel~}}
\newcommand{\XMM}      {{\em XMM~}}
\newcommand{\Chandra} {{\em Chandra~}}
\newcommand{\ACIS}    {{ACIS}}
\newcommand{\CIAO}    {{\em CIAO}}

\newcommand{\DSnine}  {{\em DS9}}
\newcommand{\MARX}    {{\em MARX}}
\newcommand{\AEacro}  {{\em AE}}

\newcommand{\TOPCAT}    {{\em TOPCAT}}
\newcommand{\XSPEC}   {{\em XSPEC}}

\shorttitle{MOXC3}
\shortauthors{Townsley et al.} 

\begin{document}

\title{The Massive Star-Forming Regions Omnibus X-Ray Catalog, \\ Third Installment }

\email{townsley@astro.psu.edu, patrick.broos@icloud.com}

\author{Leisa K. Townsley}
\affil{Department of Astronomy \& Astrophysics, 525 Davey Laboratory, 
Pennsylvania State University, University Park, PA 16802, USA}

\author{Patrick S. Broos}
\affil{Department of Astronomy \& Astrophysics, 525 Davey Laboratory, 
Pennsylvania State University, University Park, PA 16802, USA}

\author{Gordon P. Garmire}
\affil{Huntingdon Institute for X-ray Astronomy, LLC, 10677 Franks Road, Huntingdon, PA 16652, USA}

\author{Matthew S. Povich}
\affil{Department of Physics and Astronomy, California State Polytechnic University, 3801 West Temple Avenue, Pomona, CA 91768, USA}

\begin{abstract}
We offer to the star formation community the third installment of the Massive Star-forming Regions (MSFRs) Omnibus X-ray Catalog (MOXC3), a compilation of X-ray point sources detected in 50 archival {\em Chandra}/ACIS observations of 14 Galactic MSFRs and surrounding fields.  The MOXC3 MSFRs are NGC~2264, NGC~6193, RCW~108-IR, Aur~OB1, DR15, NGC~6231, Berkeley~87, NGC~6357, AFGL~4029, h~Per (NGC~869), NGC~281, Onsala~2S, G305, and RCW~49 (Wd~2); they have distances of 0.7~kpc to 4.2~kpc.  Most exhibit clumped or clustered young stellar populations; several contain at least two distinct massive young stellar clusters.  The total MOXC3 catalog includes 27,923 X-ray point sources.  
% This counts AE Aur's 945 srcs.
We take great care to identify even the faintest X-ray point sources across these fields.  This allows us to remove this point source light, revealing diffuse X-ray structures that pervade and surround MSFRs, often generated by hot plasmas from massive star feedback.  As we found in MOXC1 and MOXC2, diffuse X-ray emission is traceable in all MOXC3 MSFRs; here we perform spectral fitting to investigate the origins of selected diffuse regions.  Once again, MOXC3 shows the value of high spatial resolution X-ray studies of MSFRs enabled by {\em Chandra}.
\end{abstract}

% Six keywords are allowed: 
% The list of available keywords is at:
%   http://journals.aas.org/authors/keywords2013.html
\keywords{\hii regions --- stars: early-type --- stars: formation --- X-Rays: stars} 

% Enclose first mention of each named astronomical object in an \object{} macro; see:
%   http://journals.aas.org/authors/aastex.html
% NOTE -- as of 7 Dec 2017, Greg Schwarz says nobody is using this \object{} information.

% =============================================================================
\section{Introduction \label{sec:intro}}

With this, the third Massive Star-forming Regions (MSFRs) Omnibus X-ray Catalog (MOXC3), we continue our efforts to identify the young stars in MSFRs by analyzing archival data from the {\em Chandra X-ray Observatory} using its primary camera, the Advanced CCD Imaging Spectrometer (ACIS) \citep{Garmire03}.  Our original study, MOXC1 \citep{Townsley14}, amassed a catalog of X-ray properties for $>$20,000 X-ray point sources in 12 MSFRs and images of diffuse X-ray emission for those 12 MSFRs plus 11 others.  This was followed by MOXC2 \citep{Townsley18}, a similar catalog paper with $>$18,000 X-ray point sources and diffuse X-ray images for 16 MSFRs.  Our \Chandra data analysis methodologies evolved and improved in the years between MOXC1 and MOXC2, as described in MOXC2.  

MOXC3 presents the {\em Chandra}/ACIS observations of 14 more MSFRs (plus the massive runaway star AE~Aurigae) with nearly 28,000 X-ray point sources (Table~\ref{targets.tbl}).  This table includes a rough limiting luminosity ($L_{tc}$) and the corresponding limiting mass ($M_{50\%}$) where the brighter half of the pre-main sequence (pre-MS) X-ray population is sampled, based on results from the \Chandra Orion Ultradeep Project \citep[COUP,][]{Preibisch05}.  This and other MOXC3 tables and figures are constructed to be comparable to their counterparts in MOXC1 and MOXC2.  Please see those papers for details of our science motivations and analysis methods.

\input{target_table.tex}
\section{{\em Chandra} Observations and Data Analysis \label{sec:data}}

The analysis methods for MOXC3 are nearly identical to those employed in MOXC2.  The one major extension is that MOXC3 introduces the analysis of zeroth-order data from ACIS observations that included the \Chandra High Energy Trasmission Gratings \citep[HETG,][]{Canizares05}.  Including HETG data allows us to extend the \Chandra spatial and temporal coverage of several important MSFRs.  MOXC3 also incorporates newly-available {\em Gaia}~DR2 \citep{Gaia16,Gaia18} distances from the literature \citep{Bailer18,Binder18,Cantat18,Kuhn19} for most targets.  Lastly, we revive our workflow for spectral fitting of diffuse X-ray emission regions from the \Chandra Carina Complex Project \citep[CCCP,][]{Townsley11a,Townsley11b,Townsley11c} to ascertain the origins of prominent patches of diffuse X-ray emission in MOXC3 MSFRs.

%-----------------------------------------------------------------------------
\subsection{Observations}

Table~\ref{tbl:obslog} enumerates the 50 archival {\em Chandra}/ACIS observations used in MOXC3 by listing their unique Observation Identification (ObsID) numbers.  Details are given in MOXC2.  Observations listing the Principal Investigator (PI) as S.~Murray or G.~Garmire were part of the \Chandra Guaranteed Time Observations (GTO) program; GTO data have been a mainstay of \Chandra MSFR studies since the beginning of the mission and continue as such today.

%\clearpage
\input{observing_log.tex}
% \label{tbl:obslog}
%There are NO ACIS-S aimpoint observations in MOXC3.

Most MOXC3 data were obtained with the ACIS Imaging Array (ACIS-I) configuration; MOXC2 demonstrated that our analysis methods also support the ACIS-S imaging configuration, but no MOXC3 observations used that configuration.  HETG observations employ ACIS spectroscopy CCDs S0--S5 with the telescope aimpoint on the S3 device.  Not all X-ray photons are dispersed by the gratings; we make use of the undispersed ``zeroth-order'' HETG image.
%CCDs S1 and S3 are back-side illuminated CCDs, while all other ACIS CCDs are front-side illuminated \citep{Garmire03}.  We analyze data from these two types of detectors separately because they have very different backgrounds. 
Although only a small (energy-dependent) fraction of the light ends up in the zeroth-order image, it is worth including HETG data on MSFRs when available.  In some cases, no other ACIS data cover the HETG field.  Even when other ACIS data exist, the HETG data provide a different time sample; this is useful for MSFRs, where most sources exhibit variable X-ray emission when monitored for sufficiently long timescales.

Complications do arise when including HETG data in our analysis workflow, however.  Dispersed spectra from bright sources in the field must be masked by hand to avoid spurious source detections.  Backgrounds are higher for HETG source extractions.  HETG data cannot be used for our diffuse analysis, since the transmission gratings disperse most of the diffuse photons irreversibly to different positions on the ACIS CCDs, scrambling the spatial distribution of diffuse emission.  Thus, for targets that employed HETG data, the fields of view for the point source catalog and for the diffuse X-ray mosaics (described below) differ.  
% Additionally, we find that ACIS CCDs S0 and S5, which are usually included in HETG observations, are too far off-axis to be useful for source-finding; we plan to exclude them from our analysis of future MOXC targets.

%\clearpage
%-----------------------------------------------------------------------------
\subsection{Data Analysis}

% I can find no significant algorithmic changes since November 2017 (end of MOXC2 processing).  PSB

The algorithms, procedures, and tools we used for data reduction, astrometric alignment of observations, point source detection and validation, point source extraction, point source masking, construction of smoothed diffuse images, diffuse region extraction, and point source spectral fitting were the same as those used in MOXC2 \citep[][Section~2.2]{Townsley18}.  Please see that paper (and its references) for explanations of our data analysis methods.

We employ a variety of standard and custom X-ray analysis software packages in our workflows.
\ACIS\ data are manipulated and calibrated using the \Chandra data analysis system, \anchorfoot{http://cxc.harvard.edu/ciao/}{\CIAO} \citep{Fruscione06}, and the \anchorfoot{http://space.mit.edu/ASC/marx}{\MARX} observatory simulator \citep{Davis12}.
Throughout our workflow, data products are visually reviewed using the \anchorfoot{http://ds9.si.edu}{{\it SAOImage DS9}} tool \citep{Joye03} and \anchorfoot{http://www.star.bris.ac.uk/~mbt/topcat/}  {\TOPCAT} \citep{Taylor05}.
Models are fit to extracted spectra using the \anchorfoot{https://heasarc.gsfc.nasa.gov/xanadu/xspec/}{\XSPEC} tool \citep{Arnaud96}.
The algorithms and scripts driving these standard tools are implemented in the  \anchor{http://personal.psu.edu/psb6/TARA/ae_users_guide.html}{{\em ACIS Extract}} \citep[\AEacro,][]{Broos10,AE12,AE16} software package\footnote{
The {\em ACIS Extract} software package and User's Guide are publicly available from the Astrophysics Source Code Library, from Zenodo, and at \url{http://personal.psu.edu/psb6/TARA/ae_users_guide.html}. 
}
and in a set of detailed procedures we developed.
\AEacro\ and most other custom software we use are written in the \anchorfoot{www.harrisgeospatial.com/ProductsandTechnology/Software/IDL.aspx}{{\it Interactive Data Language}} (IDL).

Exposure maps and Ancillary Response Files (ARFs) for HETG ObsIDs in MOXC3 represent the zeroth-order effective area and exposure.  In \AEacro, HETG data get no special treatment.  When a source has both HETG and ACIS-only extractions, they compete for inclusion in optimized ObsID merges (to assess source position and photometry) in the usual way, based on signal-to-noise ratio.  In blind ObsID merges (used to calculate source validity and for timing), HETG and ACIS-only extractions are treated identically.

%-------------------
%\subsubsection{Diffuse Emission \label{sec:diffuse}} 

MOXC1 and MOXC2 both showed that a wide range of massive star-forming environments exhibit extensive diffuse X-ray emission.  This emission can be quite faint but we are able to image it for most MSFRs.  For MOXC3, we extract spectra from a few select patches of diffuse X-ray emission in and around our MSFRs.  We fit those spectra using a complicated model that allows for three {\it pshock} diffuse plasma components (with different absorptions, plasma temperatures, and ionization timescales), unresolved pre-MS stars, background from the Galactic Ridge Emission and unresolved active galactic nuclei, and (rarely) a power law component for non-thermal sources.  Similar models were used in our CCCP studies of diffuse X-ray emission in the Carina Nebula \citep{Townsley11b} and to compare other MSFRs to Carina \citep{Townsley11c}; model details can be found in those papers.

\clearpage
%=============================================================================
\section{MOXC3 Data Products \label{sec:products}}

The primary purpose of this paper is to announce the availability of high-level custom-processed {\em Chandra}/ACIS data products for a large number of MSFRs.  MOXC3 builds on our group's long history of creating and disseminating such data products, including COUP \citep[PI E.~Feigelson,][]{Getman05}, CCCP \citep[PI L.~Townsley,][]{Broos11}, the Massive Young Star-forming Complex Study in Infrared and X-ray \citep[MYStIX; PIs E.~Feigelson and L.~Townsley,][]{Kuhn13,Townsley14,Feigelson18}, MOXC1 \citep{Townsley14}, the Star Formation in Nearby Clouds project \citep[SFiNCs; PI K.~Getman,][]{Getman17}, and MOXC2 \citep{Townsley18}.  With every new project, we endeavor to improve and enhance our analysis methodologies and data products; MOXC3 follows this tradition.

%-----------------------------------------------------------------------------
\subsection{The MOXC3 Chandra Point Source Catalog \label{sec:catalog}}

The primary product of our MOXC3 efforts is the point source catalog.  As noted above, the total MOXC3 catalog includes 27,923 X-ray point sources from 12 ACIS MSFR mosaics.  Several of these mosaics include more than one distinct young stellar cluster.  As in the past, we amass all of these X-ray point sources into a single electronic table.  {\bf Table~\ref{xray_properties.tbl} defines the columns of that table.}
This catalog contains a large number of properties for each source, including details on its ACIS observations, validity, astrometry, variability, and photometry.  It is available in FITS format from the electronic edition of this article.

\input{xray_column_labels.tex}

% \label{xray_properties.tbl}

The MOXC3 point source catalog contains the same columns as MOXC2, with one addition:  the new boolean column ``IsOccasional'' provides additional information on the validation of each point source.  As described in \citet[][Section~2.2.3]{Townsley18}, when a source is extracted from multiple observations, we test the validity of the source using a set of pre-defined combinations of those extractions.  The ``IsOccasional'' flag is set when source validation failed in all attempted multi-ObsID extraction combinations, but succeeded in one or more single-ObsID extractions.  

Note that this flag is not a claim of flux variability; we specifically coined the term ``occasional'' to make this distinction.  Many occasional sources are too faint to assess their variability.  Occasional sources could be true detections of variable objects, true detections of weak constant objects that are not validated in the ObsID merges due to Poisson variations (in the source or the background), or spurious background fluctuations that managed to pass all of our validation tests.  Of course the next step in assessing the validity of occasional sources is to find multiwavelength counterparts for them (or to acquire more ACIS data).  Conversely, multiwavelength sources coincident with an X-ray source---even a faint occasional X-ray source---are much more likely to be cluster members, so occasional sources serve a valuable role in establishing the cluster population.  As demonstrated below, identifying occasional sources is also an important step in our diffuse X-ray analysis, because identifying and removing even faint point sources improves our ability to recognize and characterize diffuse plasma emission.

%Our description of the source coordinate columns (RAdeg, DEdeg) in MOXC2 (and in other published catalogs) contains an inconsequential mistake.  We wrote that our \Chandra\ source coordinates are in the J2000 equatorial coordinate system, but we have recently realized that our astrometric reference (the Two Micron All Sky Survey \citep[][]{Skrutskie06}) is technically in the ICRS coordinate system.  Thus, our \Chandra\ source coordinates are also in the ICRS coordinate system.

%\clearpage
%-----------------------------------------------------------------------------
\subsection{Sources Suffering Photon Pile-up \label{sec:pileup}}

Table~\ref{pile-up_risk.tbl} lists the point source extractions that are significantly impacted by an instrumental non-linearity known as \anchorfoot{http://cxc.harvard.edu/ciao/why/pileup_intro.html}{{\em photon pile-up}}, which is described in MOXC1, MOXC2, and references therein.  When possible, we model the piled-up spectra in these extractions \citep{Broos11} and use that model to estimate a ``reconstructed'' ACIS spectrum free from pile-up effects.

Our pile-up corrector works on a per-ObsID basis, assuming that the lightcurve and spectrum of a source are constant within a given ObsID (which of course is not always the case).  The level of pile-up is characterized by the ratio of the pile-up-free count rate to the observed (piled-up) count rate in the total (0.5--8~keV) energy band.  Table~\ref{pile-up_risk.tbl} also notes which piled-up sources are variable, either within a single ObsID or between multiple ObsIDs, because variable sources often are only piled up in some of their ObsIDs or in only part of a single ObsID, so they probably require more detailed pile-up correction than what we have done here.  For all sources in Table~\ref{pile-up_risk.tbl}, several quantities in the MOXC3 point source catalog are expected to be biased by pile-up effects, where no attempt has been made to correct for pile-up distortions.

In order to get approximate luminosities for sources corrected for pile-up, the reconstructed spectra are fit in \XSPEC\ with thermal plasma models; fit parameters and luminosities are shown in Table~\ref{pile-up_risk.tbl}.  Uncertainties on fit parameters are not reported because the bins in the reconstructed spectra have correlated errors that are incompatible with the standard parameter uncertainty estimation tools in \XSPEC.  Fitting with more sophisticated models (e.g., using more specialized abundances or time-resolved pile-up correction and spectroscopy for variable sources) would certainly be appropriate, but such work is beyond the scope of MOXC3.  We include pile-up corrected data products in the MOXC3 data archive, described below, to facilitate further analysis.

%\clearpage

\input{piled_table.tex}
% \label{pile-up_risk.tbl}

Lightcurves made from piled-up extractions also suffer from non-linear distortion, because the fraction of events lost to pile-up is a non-linear function of the incident photon rate.  Thus, a piled-up lightcurve under-estimates the true variability of the source.  Under the assumption that the source has a constant spectral shape throughout the ObsID under investigation, we can vary the photon rate in our model of the reconstructed spectrum (derived from the observed spectrum obtained from the entire ObsID) to reproduce the observed event rate in each lightcurve time bin, and then estimate the pile-up corrected event rate in each time bin.  Examples are shown in Section~\ref{sec:targets}.

%\clearpage
%-----------------------------------------------------------------------------
\subsection{MOXC3 Diffuse Spectral Fits \label{sec:diffuse_fits}}

As described in detail in Section~\ref{sec:targets} below, we perform X-ray spectral fitting on example diffuse X-ray emission regions for each MOXC3 target.  This is accomplished by first excising all detected point sources to minimize point source contamination of the diffuse spectra; this is one reason we push our source detection to the faintest possible limits.  Next, smoothed images of the remaining emission are made, using our adaptive-kernel smoothing code \citep{Broos10,Townsley03}.  Subtle diffuse structures are apparent in these smoothed images because we take great care to remove instrumental background.  

We then define sample diffuse extraction regions by contouring the apparent surface brightness in the total-band (0.5--7~keV) adaptively smoothed diffuse images, using the contour analysis capability of {\it SAOImage DS9} \citep{Joye03}.  Our \AEacro\ software accepts diffuse extraction regions as well as point source extraction regions; it generates appropriate calibration files and performs automated \XSPEC\ spectral fitting, error estimation, and flux calculations, then tabulates the results \citep{Broos10}.

Diffuse spectral fit parameters are gathered together for all MOXC3 targets in Table~\ref{tbl:diffuse_spectroscopy_style2} to facilitate comparison between targets.  Distances are needed to calculate luminosities; the assumed distances are shown in the table.  For sample diffuse regions located far from the centers of MSFRs, our distance assumptions could be incorrect.  While this does not affect the surface brightnesses shown in the table, it does impact the luminosities.  

The generic {\em XSPEC} model form we use is shown in the table notes.  It is nearly identical to that used to study brighter diffuse emission in Carina \citep{Townsley11b} and other famous MSFRs \citep{Townsley11c}.  It allows for three different {\it pshock} plasma components with solar abundances; variable abundance plasmas might provide better fits for some spectra, but generally our spectra have few counts so such complicated models are not justified.  Short ionization timescales in the {\it pshock} model suggest non-equilibrium ionization (NEI); this often comes from low-density, recently shocked plasmas \citep{Smith10}.  Long ionization timescales suggest collisional ionization equilibrium (CIE). 

Unresolved pre-MS stars in our diffuse model are represented by a two-temperature {\it apec} component.  A component for hard emission from the celestial background is included, but its parameters are frozen so they are not shown in the table.  An optional power law component is also available; this is useful for fitting pulsar wind nebulae (PWN) or other non-thermal emission.

The images of diffuse X-ray emission described above are generated for ACIS-only observations of all MOXC3 targets; diffuse analysis cannot be performed on HETG data because the gratings disperse the photons, scrambling their positions in ACIS images.  The diffuse extraction regions named in Table~\ref{tbl:diffuse_spectroscopy_style2} are defined in the figures in Section~\ref{sec:targets}.  Interpretations of these fits are also given in Section~\ref{sec:targets}, in the context of the MSFRs from which the spectra were extracted.

\input{diffuse_spectroscopy_style2.tex}

% \label{tbl:diffuse_spectroscopy_style2}

Fitting spectra extracted from regions of faint diffuse X-ray emission is always a complicated enterprise.  Large extraction regions are needed to gather enough X-ray events to avoid a noisy spectrum, but large extraction regions also capture a wide diversity of physical processes that then require a more complicated spectral model to describe.  We attempt to balance these competing requirements, but the results are seldom ideal; in particular, our spectral fits sometimes have imperfect reduced $\chi^{2}$, unmodeled spectral features, and are not necessarily unique.  Nevertheless, this is the first time that diffuse X-ray plasma emission has been detected for many MOXC3 targets, so even a basic, qualitative characterization of that emission is worthwhile.  As always in the MOXC paper series, we will refer to unresolved X-ray emission as ``diffuse,'' even though in many cases this may actually mean ``a mix of unresolved point source emission and truly diffuse X-ray structures.''  One of the main goals of our diffuse spectral fitting is to determine how much unresolved emission from low-mass pre-MS stellar populations remains in these spectra.

%\clearpage
%-----------------------------------------------------------------------------
\subsection{Archive of Reduced Data Products \label{sec:repository}}

%Limited resources prevent us from performing extensive science analysis in MOXC3.  Instead, 
We make all of our ACIS data analysis products available on Zenodo, for use by the star formation community.  We hope that both X-ray astronomers and those with longer-wavelength expertise will find applications for MOXC3 data products that further our understanding of these important MSFRs.

The MOXC2 analysis products are posted on Zenodo \citep{Townsley17}.  An introduction to available data products can be found in the README file there.  Our MOXC3 analysis products are similarly available on Zenodo \citep{Townsley19}, with an updated README file describing spectra and other data products new to MOXC3.

% MOXC3 DOI is 10.5281/zenodo.3238562

%\clearpage
%=============================================================================
\section{X-Ray Characterizations of MOXC3 Targets \label{sec:targets}}

Here we give some cursory details on MOXC3 results for each target in Table~\ref{targets.tbl}, emphasizing images and spectra of diffuse X-ray emission.  Please see the much more extensive references in the small set of papers we cite for multiwavelength overviews and science analysis of these famous targets.  Three standard figures are given for each target; details on the construction of the first two of these figures can be found in MOXC2.  

The first figure shows the 1~keV exposure map of the ACIS mosaic, with the brighter ($\geq$5 net counts) X-ray point sources superposed, represented as dots and color-coded by their median energy.  A legend gives the number of plotted sources in each median energy range.  The total number of ACIS point sources found for the target (shown in Table~\ref{targets.tbl}) is also noted on the figure.  Very faint sources are not depicted in this image because the median energy statistic becomes highly uncertain for faint sources.  Additionally, the spatial distribution of very faint sources is strongly governed by the telescope sensitivity; they are always centrally concentrated and their distribution falls off strongly with off-axis angle.  The purpose of this figure is to show the breadth and depth of ACIS coverage of each MSFR and to give a qualitative indication of the clumps and clusters of X-ray sources found across the field.  In particular, many targets show obscured young clusters as clumps of green and blue dots; revealed clusters are seen as groups of red dots.

The second figure displays the observed surface brightness of the total-band (0.5--7~keV) diffuse X-ray emission as an image in celestial J2000 coordinates, superposed on \Spitzer or \WISE images to give a cold interstellar medium (ISM) context to the hot plasmas imaged by ACIS.  Our diffuse images are constructed by masking out point sources from the observed event list, the particle background event list, and the 1~keV exposure map for each ObsID.  These masked images are combined to make ``target-level'' images; a surface brightness image is computed from these constituents and smoothed with an adaptive kernel to achieve a pre-defined signal-to-noise ratio \citep[][Section~9.1]{Townsley03,Broos10}.

We add a new standard figure to MOXC3:  diffuse X-ray emission shown as two separate smoothed images, in soft (0.5--2~keV) and hard (2--7~keV) bands, again superposed on a long-wavelength image for context.  This is displayed in Galactic coordinates, for ease of comparison with the literature.  Many MOXC3 targets show different spatial distributions for hard and soft diffuse emission, caused by changes in absorption across the field or changes in the temperatures of emitting plasmas.  

Additional figures include zoomed versions of the standard figures to detail certain features in the MSFR; these are often overlaid with polygons showing the point source extraction regions to indicate where point sources were located (before they were excised from the data to create the smoothed images of diffuse emission).  Other figures show these point source extraction regions superposed on images of the ACIS event data, with soft events in red, hard events in green.  We distinguish ``occasional'' sources (Section~\ref{sec:catalog}) by dark blue or yellow extraction regions, while ``regular'' sources are depicted as medium blue or white extraction regions.

In Section~\ref{sec:diffuse_fits}, we attempted basic spectral fitting on sample diffuse X-ray structures in all MOXC3 targets.  The extraction polygons for those diffuse regions are shown in a variety of figures below, as black or cyan outlines that are much larger than X-ray point source extraction regions.  The extracted spectra and our spectral fits are displayed next to the hard+soft diffuse X-ray images.  Model components are color-coded:  red, green, and blue represent soft, medium, and hard {\it pshock} components; the two {\it apec} plasmas that model unresolved pre-MS stars are shown in cyan; celestial background emission is magenta; the one example of nonthermal emission (from the PWN in RCW~49) is shown in purple.

For a few important massive stars, we give spectral fit parameters below.  The simple \XSPEC\ model used for the fit was {\em TBabs*apec} or {\em TBabs(apec$^1$ + apec$^2$)} with solar abundances.  The luminosity $L_{X}$ is total-band (0.5--8~keV), corrected for absorption.

%\clearpage
%-----------------------------------------------------------------------------
\subsection{AE~Aurigae and Aur OB1 \label{sec:aeaur}}
% l,b = 172.08, -2.26 (near ONC, in front of Aur OB1)
% Simbad coords for AE Aur:  05 16 18.15 +34 18 44.3
% AE Aur has A_V = 1.82, from Maiz18, so NH=0.29e22.  It is ACIS src 051618.17+341845.3, c1412.
% Aur OB1:  Avg A_V is 1.5 mag, so NH=0.24e22. D = 1.63 kpc, so 4*pi*D^2 = 3.180e+44 cm^2.

We begin our examination of the X-ray properties of MOXC3 MSFRs with the only \Chandra target that is not a MSFR, the O9.5V star \citep{Sota11} AE~Aurigae (HD~34078).  This massive star ionizes the Flaming Star Nebula \citep[IC~405;][]{Herbig58}.  It is thought to be a runaway from the cluster NGC~1980 in the Orion Nebula region \citep{Bally08}, so it is not expected to be accompanied by the usual population of lower-mass pre-MS stars found in a MSFR.  It was observed by \Chandra to search for an X-ray signature from its prominent IR bow shock \citep{France07}; no such diffuse X-ray emission was detected, either by an earlier \XMM observation \citep{Toala17} or by \Chandra \citep{Rangelov18,Binder19}.  

We analyzed the deep (141~ks) ACIS-I observations of AE~Aur to search once more for faint diffuse X-ray emission from its bow shock or from its wind interactions with IC~405.  
%We also wanted to use this long \Chandra observation of a nearby Galactic Plane region to sample the X-ray point source population outside of a known MSFR.  Such sources normally constitute the ``contaminants'' in our \Chandra observations of MSFRs, so we thought this would be a good opportunity to characterize those contaminants.  Based on previous experience, we expected to detect a few dozen foreground stars, perhaps a few background Galactic sources, and maybe up to $\sim$200 AGN.
We find that AE~Aur is a slightly piled-up soft X-ray source (Table~\ref{pile-up_risk.tbl}) with no variability.  We confirm the X-ray non-detection of its IR bow shock. 

\begin{figure}[htb]
\centering
\includegraphics[width=0.49\textwidth]{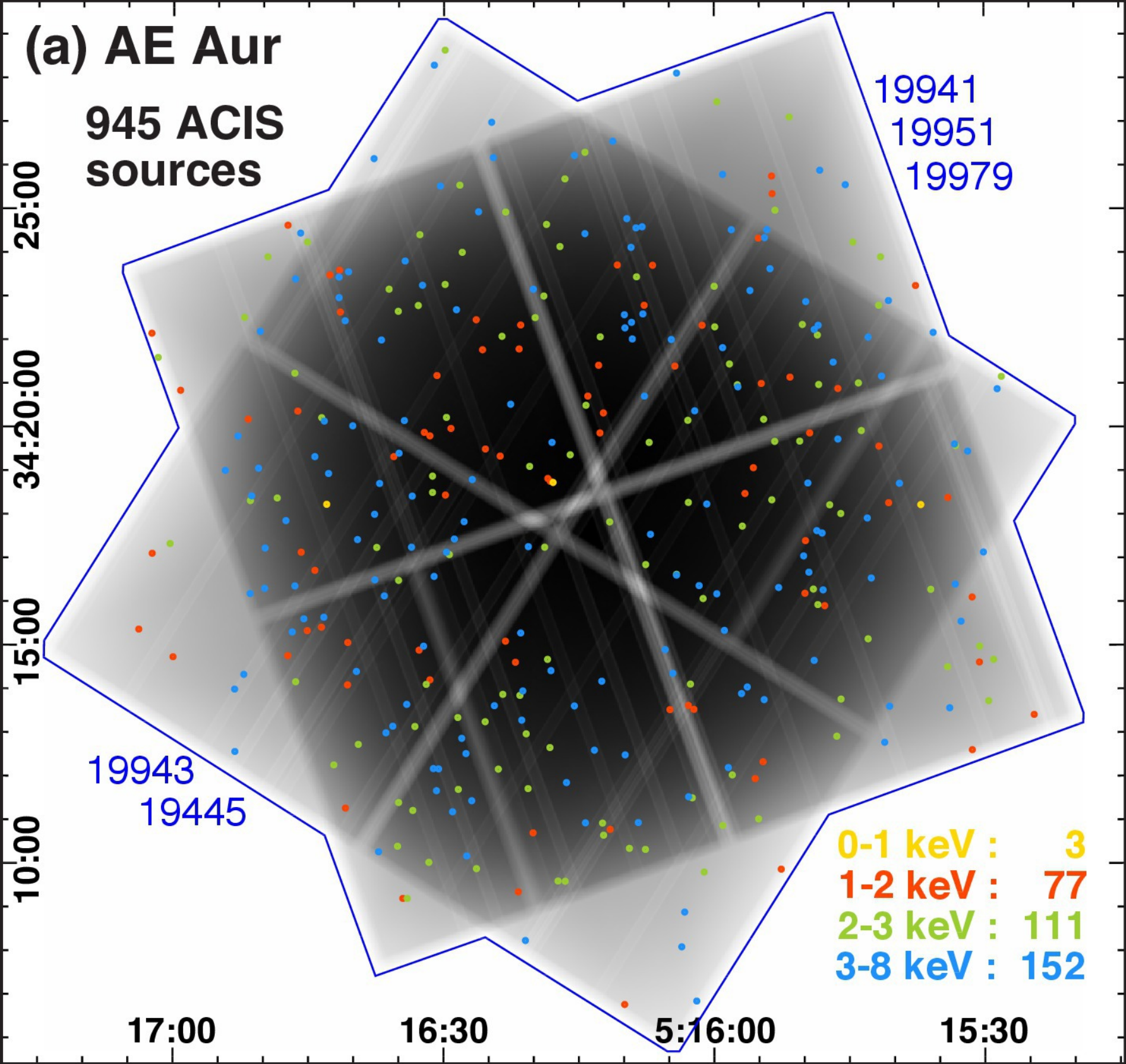}
\includegraphics[width=0.49\textwidth]{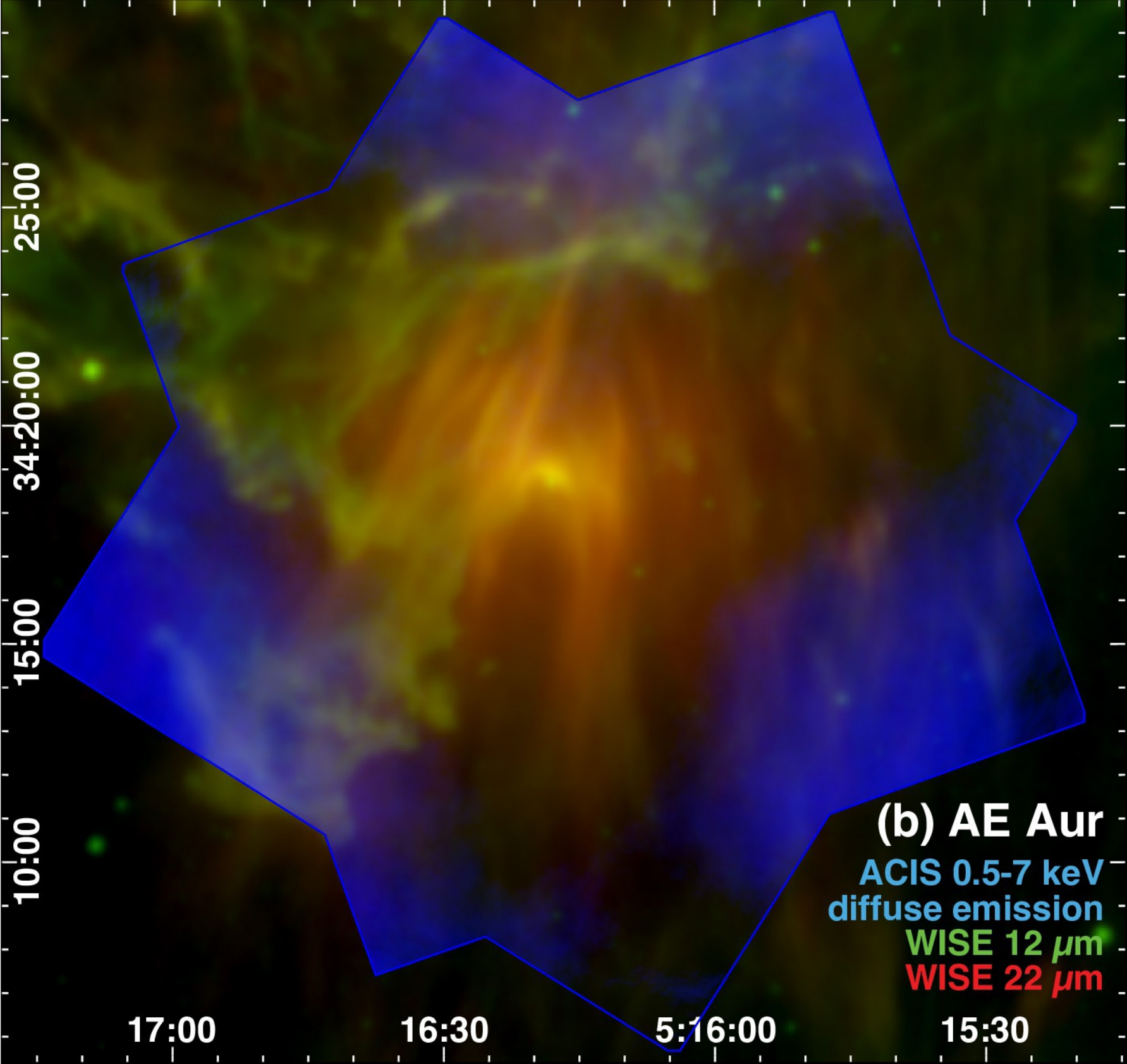}
\caption{AE~Aurigae.
(a) ACIS exposure map with 343 brighter ($\geq$5 net counts) ACIS point sources overlaid; colors denote median energy for each source.  ObsID numbers are shown in blue.
(b) ACIS diffuse emission in the \WISE context.  
\label{aeaur.fig}}
\end{figure}

We were surprised to find 945 X-ray point sources (Figure~\ref{aeaur.fig}(a)) in this field, far more than we expected.  No \Chandra Source Catalog \citep[CSC,][]{Evans10} results are currently available for the AE Aur field, so we could not use that for comparison.
% Pat looked -- see e-mail 29 April 2019.
Instead, we studied \citet{Ayres18}, who recently investigated 10 of the 20 serendipitous X-ray sources tabulated by the CSC in the central $16\arcmin \times 16\arcmin$ of the combined data from 19 10-ks {\em Chandra}/HRC snapshots of $\alpha$~Centauri.  This dataset is comparable to ours:  it is a long multi-ObsID \Chandra integration on a target near the Galactic Plane, with a similar field of view.  \citet{Ayres18} demonstrated that the Plane harbors a variety of mostly variable X-ray sources with a wide range of variability timescales, including many M dwarf stars.  But again, the $\alpha$~Cen field yielded only a handful of X-ray sources, whereas the AE~Aur field has hundreds---almost 50 times as many X-ray sources as the $\alpha$~Cen study.
% Alpha Cen l,b = 315.73, -00.68.  So we are looking through the Plane there.

%Although many of these sources are too faint to assess their variability, we do find 67 ``definitely variable'' and an additional 154 ``possibly variable'' sources, distributed uniformly across the field.  X-ray variability can be an indicator of stellar youth, so this is some evidence that the extra sources in the AE~Aur field could be young stars.

% Pat notes an interesting lightcurve for 051646.41+341409.6, c2274.  This is 2MASS 05164639+3414094, also an XMM source and OIR source; VizieR entries imply that it's a galaxy.
% Another bright source is 051545.93+341345.8 ; 'c341', with pronounced inter-ObsID variability.  This is a ROSAT and XMM source; VizieR entries imply that it's a galaxy.

Further investigation of these Galactic Plane sightlines yields an explanation:  AE~Aur lies in the foreground of the Aur~OB1 association and has a tight grouping of {\em IRAS} sources projected in its vicinity \citep{Kawamura98}, so many of our ACIS sources are likely young stars in Aur~OB1.  Thus the ACIS observation of AE~Aur is, by accident of alignment, also an observation of a more distant MSFR.  In contrast, $\alpha$~Cen does not fall along such a serendipitous sightline \citep{Reipurth+Bally08}.  This exercise demonstrates that any future high spatial resolution, higher-sensitivity X-ray mission should be prepared to find diverse populations of X-ray sources in its Galactic Plane observations, from just a few field M dwarfs along some sightlines to rich young stellar populations along others.

Prior to the \Gaia era, Aur~OB1 was estimated to be at a distance of $1.32 \pm 0.1$~kpc \citep{Humphreys78,Reipurth+Yan08}.  We registered the ACIS point source catalog astrometry for this target to \Gaia DR2 data \citep{Gaia16,Gaia18} then matched the ACIS and \Gaia catalogs; we used 49 matches (assumed to be association members) to estimate a median \Gaia distance to Aur~OB1 of $1.63^{+0.41}_{-0.27}$~kpc.  The methodology used in this distance calculation is identical to that presented by \citet{Binder18} and \citet{Povich19} and similar to that of \citet{Kuhn19}.  This distance is reported in Table~\ref{targets.tbl}.

\begin{figure}[htb]
\centering
\includegraphics[width=0.99\textwidth]{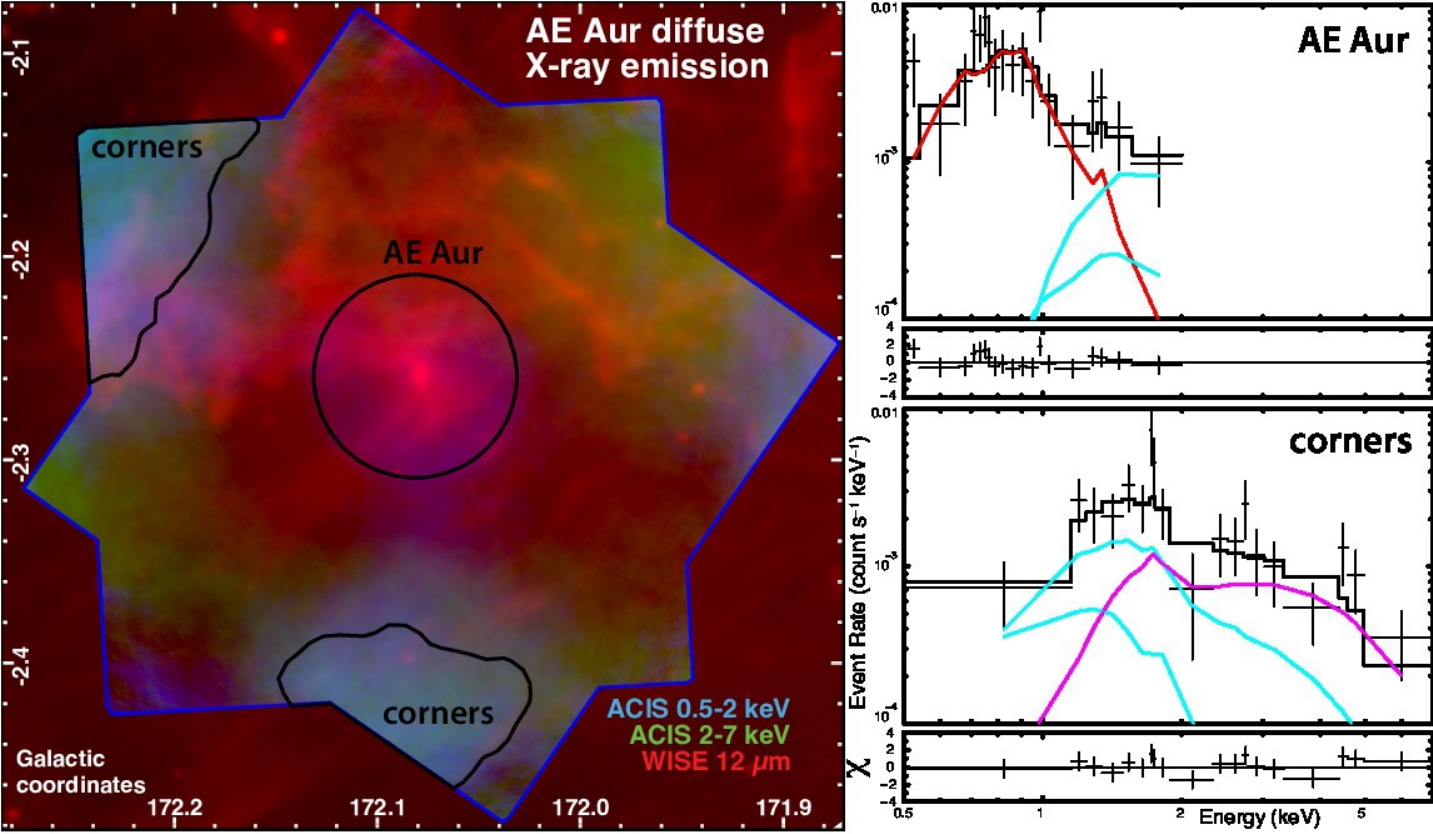}
\caption{Characterizing diffuse X-ray emission around AE~Aurigae.  
This image shows ACIS soft-band (0.5--2~keV, in blue) and hard-band (2--7~keV, in green) diffuse emission in the \WISE context, now in Galactic coordinates.  Extraction regions for diffuse spectral fitting are shown in black.  Corresponding spectra are on the right; axis ranges are the same for both spectra.  Table~\ref{tbl:diffuse_spectroscopy_style2} gives fit parameters for the spectral models.
\label{aeaurgal+spectra.fig}}
\end{figure}

In Figure~\ref{aeaur.fig}(b), we find bright total-band diffuse X-ray emission at the edges of the ACIS field; its absence across much of the field interior probably indicates that IC~405 is shadowing the X-ray background (and any diffuse X-ray emission from the more distant Aur~OB1).  Rescaling the diffuse X-ray emission to bring out fainter structures, Figure~\ref{aeaurgal+spectra.fig} shows faint, soft diffuse X-ray emission in a wide region around AE~Aur at field center although, again, the IR bow shock is not detected in X-rays.  

We usually extract diffuse X-ray spectra using the total-band energy range of 0.5--7~keV.  For the large circular diffuse region around AE~Aur, however, the emission is faint; this spectrum has the fewest counts in all of MOXC3.  To avoid excessively noisy channels at low and high energies, we had to restrict our fitting to the narrower range of 0.52--2.0~keV.  Our spectral fitting (Figure~\ref{aeaurgal+spectra.fig} and Table~\ref{tbl:diffuse_spectroscopy_style2}) indicates that a soft, unobscured plasma surrounds AE~Aur; unresolved pre-MS stars seen behind a large obscuring column account for the harder part of the spectrum.  This large column implies that these unresolved stars lie behind IC~405 and are obscured by it.

Unmodeled line emission at $\sim$0.74~keV is apparent in the fit residuals.  Similar line features were ubiquitous in CCCP \citep{Townsley11b} and in ACIS spectra of many other MSFRS \citep{Townsley11c}; speculation on the origin of these spectral features can be found there.  We will see these unmodeled lines again in many MOXC3 diffuse spectra discussed below. 

The morphology of the soft plasma surrounding AE~Aur (Figure~\ref{aeaur.fig}(b) and Figure~\ref{aeaurgal+spectra.fig}) may also indicate some shadowing by the IC~405 nebula.  This plasma could be the faint signature of feedback from AE~Aur's wind; it is distictly softer than the X-ray emission from AE~Aur itself (Table~\ref{pile-up_risk.tbl}).  We have used the distance to AE~Aur \citep[0.402~kpc,][]{Bailer18} to calculate the X-ray luminosity of this diffuse emission in Table~\ref{tbl:diffuse_spectroscopy_style2}.
%The X-ray luminosity of this soft diffuse emission, enclosed by the large central circular extraction region shown in Figure~\ref{aeaur.fig}(c), is $3 \times 10^{30}$~erg~s$^{-1}$, nine times fainter than the lowest ACIS estimates of AE~Aur's luminosity (Table~\ref{pile-up_risk.tbl}).

Spectral fitting shows that the emission in the ``corners'' regions is mainly attributable to unresolved pre-MS stars, plus celestial background to fit the hard tail of the spectrum (Figure~\ref{aeaurgal+spectra.fig} and Table~\ref{tbl:diffuse_spectroscopy_style2}).  The absorbing column is much lower in these regions than it was for the unresolved pre-MS component in the ``AE Aur'' diffuse region.  Notably, the surface brightness is the same, suggesting that there is a uniform spatial distribution of unresolved pre-MS stars across the entire ACIS field.  This is probably (at least in part) the distributed population of young stars in Aur~OB1, so we have used our \Gaia distance estimate to Aur~OB1 (1.63~kpc) to calculate the area enclosed by the extraction regions (hence the X-ray luminosity) for the unresolved pre-MS components of both diffuse spectra in the AE~Aur field.  This necessitated dividing the ``AE Aur'' diffuse region in Table~\ref{tbl:diffuse_spectroscopy_style2} into two rows.

In summary, although the ACIS observations of AE~Aur did not find X-ray emission from its bow shock, they did reveal hot plasma emission from its winds interacting with the Flaming Star Nebula.  Through great serendipity, these observations also allowed us to characterize part of the pre-MS population of the more distant MSFR Aur~OB1, both via a large number of resolved stars and a ubiquitous distribution of unresolved pre-MS star emission.  Due to the great penetrating power of X-rays and the precision of high spatial resolution imaging, \Chandra observations of Galactic Plane sightlines are often more complicated (and interesting) than expected.

%\clearpage
%-----------------------------------------------------------------------------
\subsection{NGC~2264 \label{sec:n2264}}
% NGC 2264 -- 3373 point sources in a 4-pointing mosaic.
% SF Handbook review:  Dahm08, Volume 1 (north)
% At D = 0.738 kpc, 4*pi*D^2 = 6.5180011e+43 cm^2.
% Avg A_V is 0.5 mag, so NH=0.08e22.
% W90 is Dahm05 src 125, at 06 40 44.61 +09 48 02.4 -- not detected in X-rays.
% V642~Mon is at 06 40 46.08 +09 49 17.3
% HETG observations were of 15 Mon (aka S Mon, HD 47839), at 06:40:58.69 +09:53:44.6.  Simbad calls this ``* 15 Mon -- Be Star'' with K=5.3 mag and sptype O7V+B1.5/2V from Skiff13 ``General Catalogue of Stellar Spectral Classifications'' from VizieR.  Maiz18 says A_V = 0.237 mag for 15 Mon.
% Simbad coords for 15 Mon are 06 40 58.66 +09 53 44.7.
%Examining the central cluster in our mosaic ("IRS2 region"), it looks like we failed to find a pretty obvious hard X-ray source, at roughly  6:41:03.28 +9:35:44.0.  Overlaying removed_sources.reg, it looks like we never even nominated that position.  Sigh. 

NGC~2264 \citep{Dahm08} is a nearby ``cluster of clusters'' \citep[e.g.,][]{Bastian07,Elmegreen08} complex, hosting as its most massive component the O7V+B2:V hierarchical triple system 15~Mon (S~Mon, HD~47839) \citep{Maiz18b}.  \Chandra has performed a variety of NGC~2264 observations over the years, including a long ACIS-I observation centered on the main cluster IRS2 \citep{Flaccomio06} and an HETG study of 15~Mon \citep{Waldron07}.  MYStIX included the four ACIS-I datasets available at the time \citep{Feigelson13,Kuhn13}, finding 1328 X-ray point sources; we presented an image of the diffuse X-ray emission in the MYStIX field in MOXC1 \citep[][Figure~15.4]{Townsley14}.  Several MYStIX science analysis papers include results for NGC~2264; \citet{Feigelson18} provides an overview of that large body of work.

A 2011 December ACIS-I Large Project centered on the IRS2 region (PI G.~Micela) was part of an extensive simultaneous multiwavelength campaign called the ``Coordinated Synoptic Investigation of NGC 2264'' (CSI~2264), designed to study variability in pre-MS stars across the electromagnetic spectrum \citep{Cody14}.  Many CSI~2264 papers resulted; a recent one \citep{Flaccomio18} explores the X-ray flaring behavior of NGC~2264's pre-MS stars in detail.
% Did any of the CSI 2264 papers give a complete X-ray source list?

\begin{figure}[htb]
\centering
\includegraphics[width=0.415\textwidth]{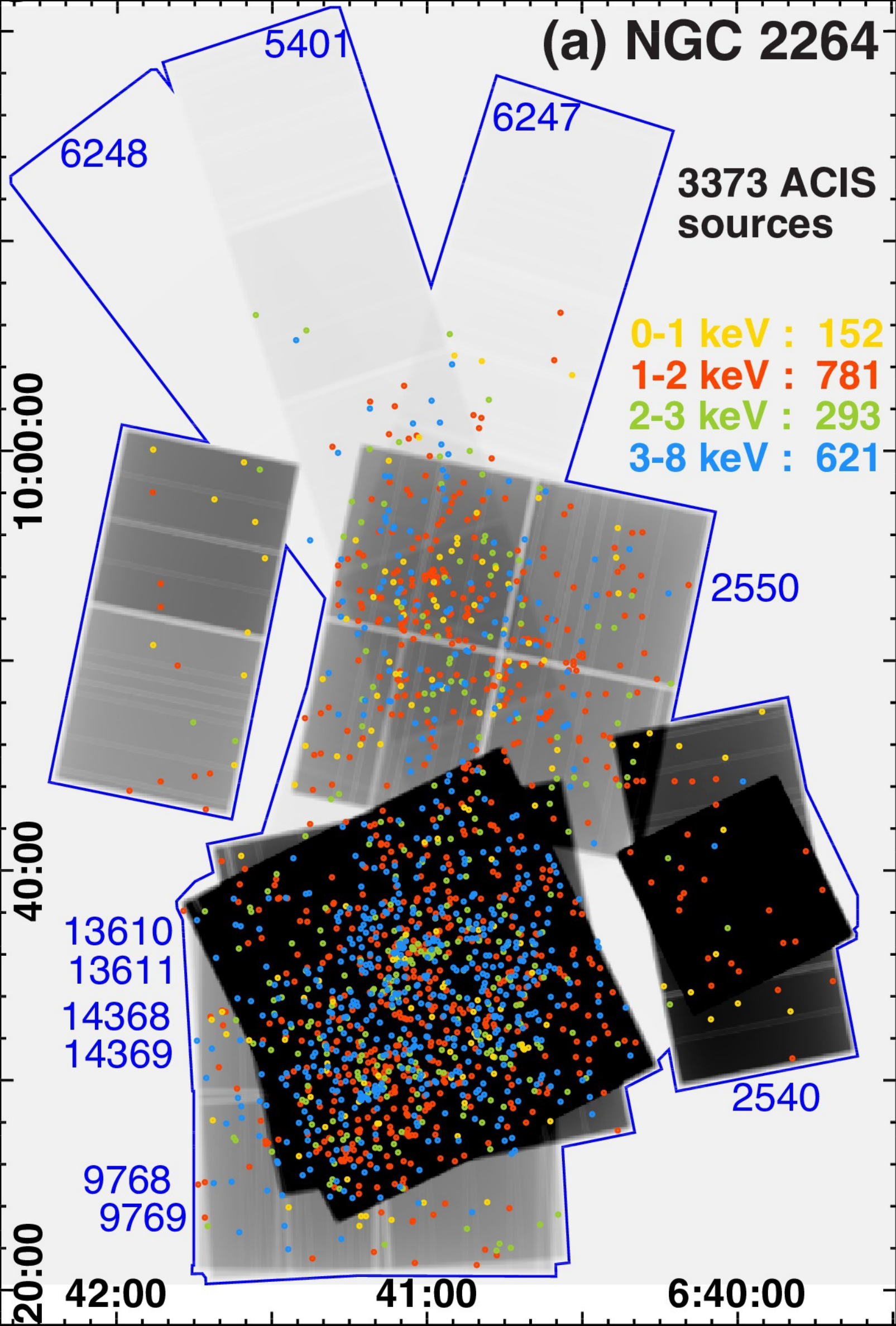}
\includegraphics[width=0.575\textwidth]{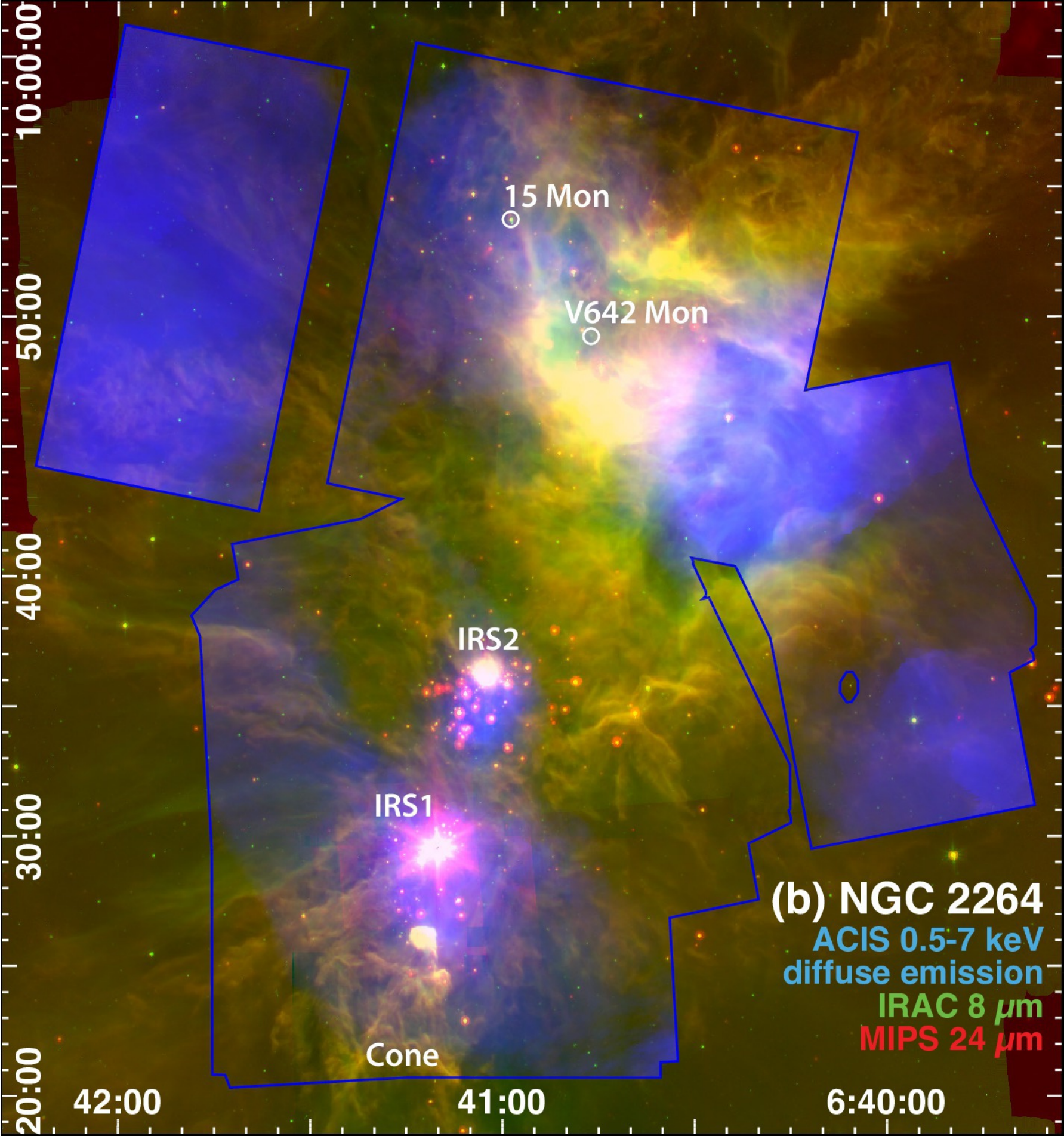}
\caption{NGC~2264.
(a) ACIS exposure map with 1847 brighter ($\geq$5 net counts) ACIS point sources overlaid; colors denote median energy for each source.  ObsID numbers are shown in blue.
(b) ACIS total-band (0.5--7~keV) diffuse emission in the \Spitzer context.  The field of view is smaller here than in (a) because HETG data cannot be used to map diffuse X-ray emission.  Objects mentioned in the text are marked.
\label{ngc2264.fig}}
\end{figure}

We include NGC~2264 in MOXC3 because it represents an incremental step in our analysis procedures, combining all available \Chandra data (including the HETG data on 15~Mon and the far off-axis ACIS-S CCDs from the many ACIS-I observations) for a final wide-field census of 3373 X-ray point sources and a portrait of extensive, spatially complex diffuse X-ray emission (Figure~\ref{ngc2264.fig}).  The zeroth-order HETG data document the X-ray point source population north of 15~Mon; the off-axis ACIS-S CCDs show that diffuse X-ray emission extends farther east and west of the main star-forming ridge than the area captured by the ACIS-I CCDs.

Due to its small distance, NGC~2264 has 19 piled-up X-ray sources (Table~\ref{pile-up_risk.tbl}), many of which are lower-mass pre-MS stars that pile up when they flare.  The O7V binary 15~MonAaAb \citep{Maiz18b} is piled up even in the zeroth-order HETG data.  (The B2:V component 15~MonB \citep{Maiz18b} is not detected by ACIS.)
% Maiz18 says A_V = 0.237 mag (NH = 0.04e22) for 15 Mon.  Pat finds that simultaneous XSPEC fitting shows no variability in 15 Mon!
The spectroscopic binary V642~Mon \citep{Merle17}, located near the center of ObsID~2550, suffers the worst pile-up in this target.  V642~Mon (and many other piled-up sources in NGC~2264) are variable X-ray sources; pile-up correction is important for determining the amplitude of such variability.  
% V642 Mon is at 06 40 46.08 +09 49 17.3.

\begin{figure}[htb]
\centering
\includegraphics[width=0.49\textwidth]{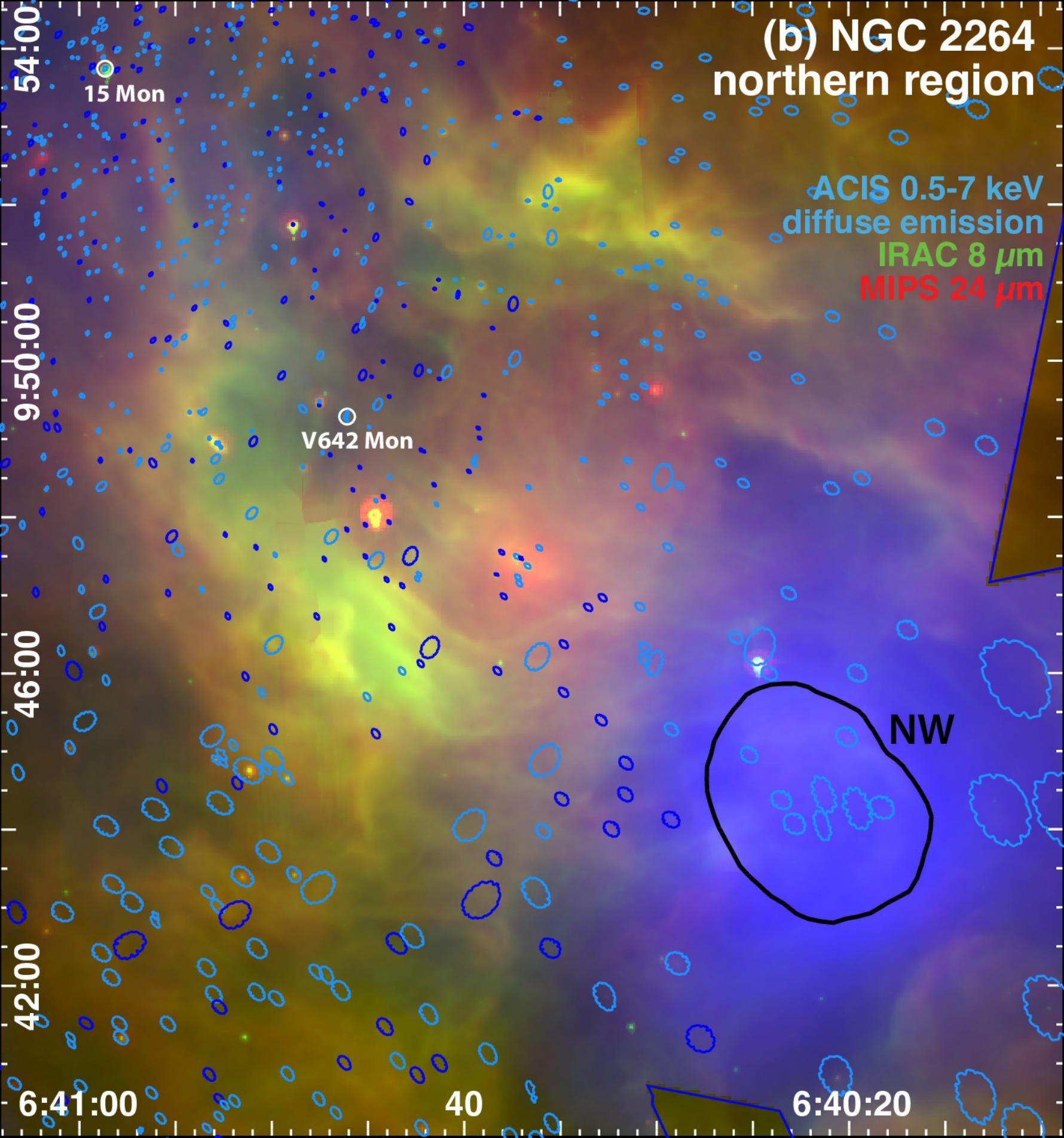}
\includegraphics[width=0.49\textwidth]{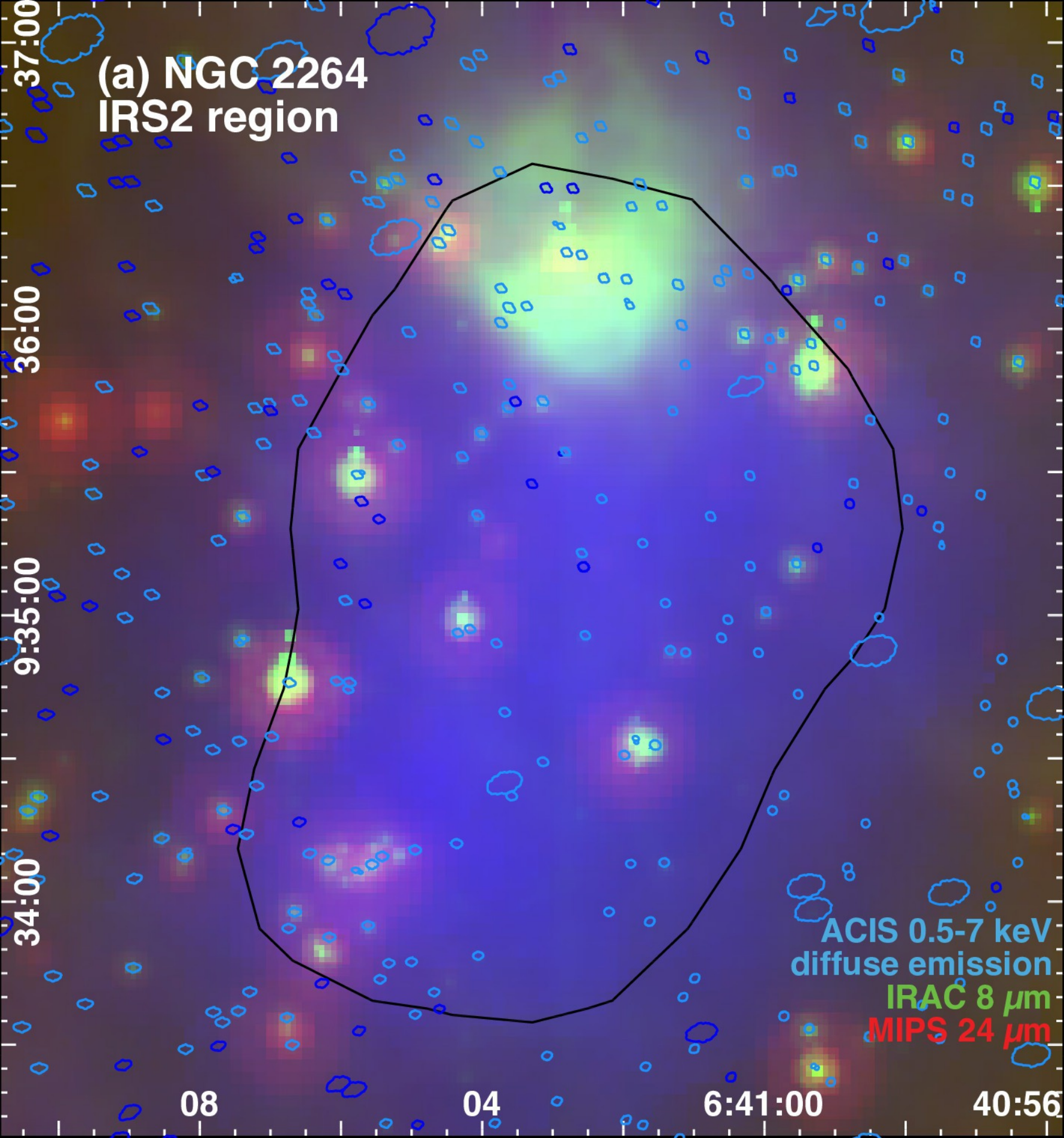}
\caption{Areas of interest in NGC~2264, now including X-ray point source extraction regions.  Dark blue apertures denote ``occasional'' sources.  The large black polygons outline extraction regions for diffuse X-ray spectra shown in Figure~\ref{ngc2264gal+spectra.fig}.
(a) A zoom of the northern part of Figure~\ref{ngc2264.fig}(b).  
(b) The IRS2 region, hosting the central cluster of NGC~2264.
\label{ngc2264zooms.fig}}
\end{figure}

More detailed images for the northern part of NGC~2264 (just above the center of Figure~\ref{ngc2264.fig}(a)) and the IRS2 region are given in Figure~\ref{ngc2264zooms.fig}.  Along with many X-ray point source extraction regions, Figure~\ref{ngc2264zooms.fig}(a) shows the northwest patch of diffuse X-ray emission that extends west onto the S-array CCDs.  This is the brightest diffuse X-ray structure in the ACIS mosaic of NGC~2264, but spectral fitting (described below) suggests that it is probably not related to the MSFR.  Figure~\ref{ngc2264zooms.fig}(b) shows the main NGC~2264 cluster near IRS2 (the IR-bright nebulosity at the top center of this image); it hosts a large number of X-ray sources but is not particularly crowded or confused in the ACIS data.  Diffuse X-ray emission is prominent to the south of IRS2.
% I didn't add a tight zoom of IRS2 because there's not a whole lot to see there.
% From Simbad, IRS2 is at 06 41 02.7 +09 36 10.  There is no X-ray source at the Simbad coordinates for IRS2. 

\begin{figure}[htb]
\centering
\includegraphics[width=0.49\textwidth]{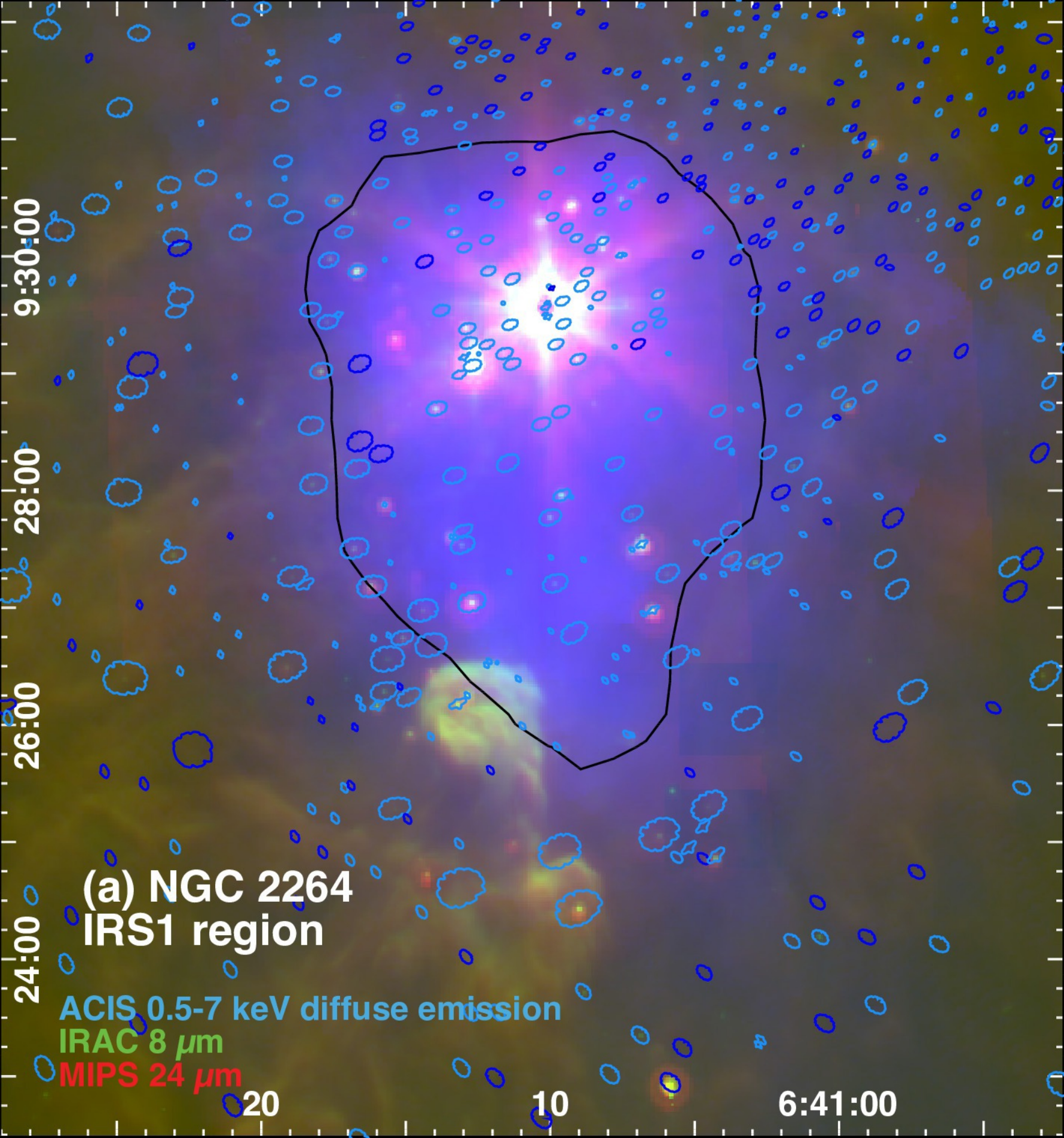}
\includegraphics[width=0.49\textwidth]{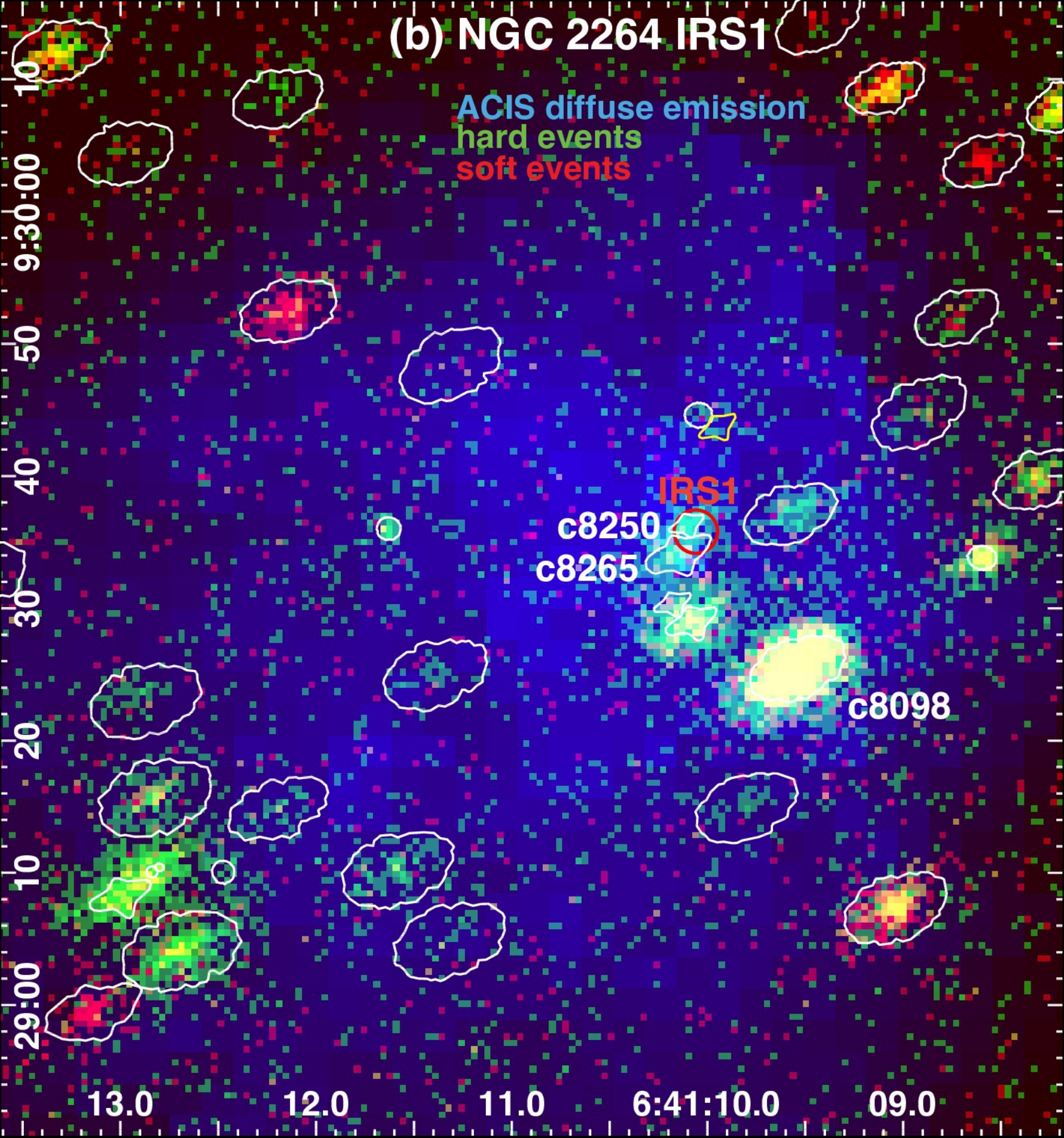}
\caption{The IRS1 region.
(a)  A zoom of the southern part of Figure~\ref{ngc2264.fig}(b), showing ACIS total-band diffuse emission in the \Spitzer context.  The top of the Cone Nebula is seen just below the center of the image.  The large black polygon outlines the extraction region for the diffuse X-ray spectrum shown in Figure~\ref{ngc2264gal+spectra.fig}.
(b) ACIS event data and total-band diffuse emission around IRS1.  ACIS point source extraction regions are now shown in white, for better contrast with the diffuse X-ray emission.
\label{ngc2264irs1.fig}}
\end{figure}

In the south, a strong concentration of X-ray sources surrounds NGC~2264 IRS1 (``Allen's Source'', Figure~\ref{ngc2264irs1.fig}), an early-B massive young stellar object (MYSO) with a disk \citep{Grellmann11}.  The piled-up source c8098 (the YSO 2MASS~J06410954+0929250) is nearby; its high X-ray luminosity and flare-like variability suggests that it is an intermediate-mass pre-MS star \citep[IMPS,][]{Povich11,Gregory16,Povich19} that will become X-ray-quiet on the main sequence, but for now is an exhuberant X-ray emitter.  Many ACIS sources around IRS1 are similar obscured, hard X-ray emitters (Figure~\ref{ngc2264irs1.fig}(b)).

Two X-ray sources (c8265 and c8250) are crowded together in the vicinity of IRS1; they have extraction apertures reduced to $\sim$50\% to minimize cross-contamination.  Source CXOU~J064110.15+092933.9 (c8265) has a flaring lightcurve; a spectral fit (including data from all ObsIDs) yields $N_{H} = 4 \times 10^{22}$~cm$^{-2}$, $kT = 6$~keV, and $L_{X} = 2 \times 10^{30}$~erg~s$^{-1}$.  The most likely counterpart to IRS1 itself is CXOU~J064110.11+092936.1 (c8250), with 122 net counts and a very high median energy of 4.6~keV.  Its lightcurve shows the decaying tail of a bright X-ray flare in ObsID~13611.  Its spectral fit gives $N_{H} = 23 \times 10^{22}$~cm$^{-2}$, $kT > 10$~keV, and $L_{X} = 7 \times 10^{30}$~erg~s$^{-1}$.  We have seen such hard, variable X-ray emission behind $>$100~mag of extinction from other MYSOs in MOXC1 and MOXC2, but never with so many counts or with such a clearly decaying lightcurve reminiscent of magnetic reconnection flaring in low-mass pre-MS stars.  The X-ray emission from this object very much mimics its lower-mass pre-MS cousins, lending further evidence that the formation mechanism for MYSOs is a scaled-up version of that for lower-mass stars.
% Searching Simbad for ``NGC 2264 IRS 1'' yields ``NAME Allen's Source -- Young Stellar Object'' at 06 41 10.06 +09 29 35.8, sptype B2.
%Searched Simbad for the southern IR bright source at around 6:41:10.130 +9:29:33.98 and got this:  ``Cl* NGC 2264 LBM 6076 -- Star in Cluster'' at 06 41 10.16 +09 29 33.7.  This is 2MASS J06411015+0929336, with K = 4.9 mag.  See Flaccomio06, ACIS data ObsID 2540; this is [FMS2006] 305.  

\begin{figure}[htb]
\centering
\includegraphics[width=0.99\textwidth]{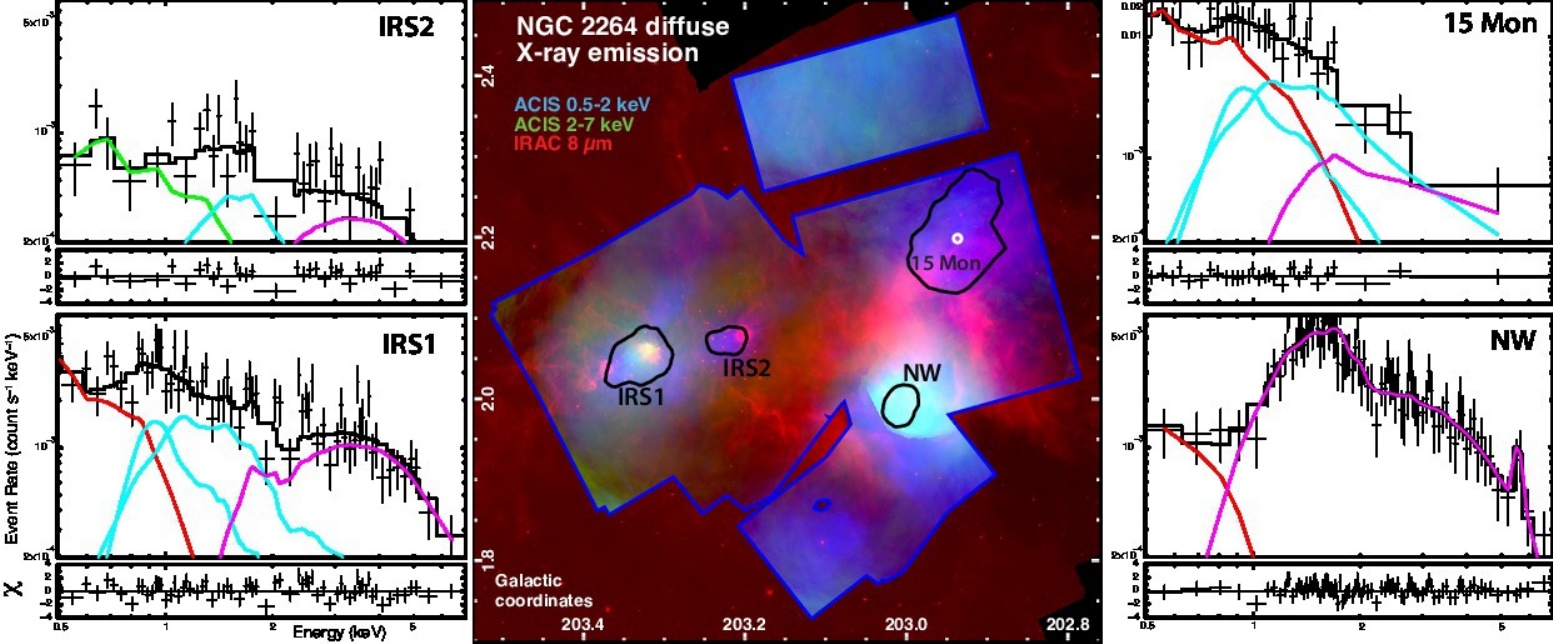}
\caption{Characterizing the NGC~2264 diffuse X-ray emission.
This image shows ACIS soft-band and hard-band diffuse emission in the \Spitzer context.  The position of 15~Mon is shown with a small white circle; extraction regions for diffuse spectral fitting are shown in black.  Corresponding spectra flank the image; axis ranges are the same for all spectra except the ``15~Mon'' diffuse region.  Table~\ref{tbl:diffuse_spectroscopy_style2} gives fit parameters for the spectral models.
\label{ngc2264gal+spectra.fig}}
\end{figure}

Diffuse extraction regions and spectra are shown in Figure~\ref{ngc2264gal+spectra.fig}.  15~Mon is surrounded by soft diffuse X-ray emission.  A good fit is obtained from a soft unobscured {\it pshock} plasma ($kT = 0.3$~keV) with a short ionization timescale, plus an unresolved pre-MS component with absorption consistent with our X-ray spectral fitting results for 15~Mon itself (Table~\ref{pile-up_risk.tbl}).  The soft, shocked plasma is consistent with feedback from 15~Mon and the population of resolved X-ray sources around it (shown in Figure~\ref{ngc2264zooms.fig}(a) above) demonstrates the plausibility of an unresolved population of pre-MS stars contributing to this diffuse emission.  

The spectral fit of the northwest region shows a conspicuous emission line at $\sim$5.7~keV.  Thawing the redshift in the background {\it apec} component of our standard model, we arrived at a good fit, with a hard thermal plasma behind a large absorbing column; the emission line appears to be an iron line at a redshift of 0.20 (see Table~\ref{tbl:diffuse_spectroscopy_style2}).  Thus it seems that we have discovered another cluster of galaxies situated behind a Galactic MSFR; we might have been surprised, except that we saw a similar alignment between the Carina South Pillars region and a $z = 0.10$ galaxy cluster in CCCP \citep{Townsley11a}.

Diffuse X-ray emission is concentrated south of IRS2; we extracted a spectrum of this diffuse emission from the large polygon shown in black in Figure~\ref{ngc2264zooms.fig}(b).  Spectral fitting indicates that it is composed of a lightly obscured NEI plasma with $kT \sim 0.5$~keV and heavily obscured unresolved pre-MS stars; the diffuse emission surface brightness is 50\% larger than that of the unresolved stars.  Its comparatively short ionization timescale implies a shocked plasma.  
% *** What IRS2 source(s) could be shocking the plasma? ***

Diffuse X-ray emission is also apparent in the southern part of the field, centered on IRS1 but extending south, to the top of the Cone Nebula (Figure~\ref{ngc2264irs1.fig}(a)).  Its spectrum is modeled with a soft ($kT \sim 0.14$~keV), unobscured NEI plasma and unresolved pre-MS stars with light obscuration.  Here, the surface brightness of unresolved stars is 60\% brighter than that of the diffuse plasma.  The apparent surface brightness and extent of diffuse X-ray emission in the IRS1 region exceeds that of the IRS2 region, but we find the IRS2 intrinsic surface brightness to be much higher than that of IRS1 (by a factor of 6 for the diffuse plasma and a factor of 2.6 for the unresolved young stars) and its plasma temperature to be substantially higher.

The candidate galaxy cluster in our NGC~2264 mosaic reminds us that even nearby Galactic Plane targets can fall along sightlines that harbor prominent extragalactic X-ray sources.  The early-B MYSO IRS1 has striking X-ray emission properties that mimic those of pre-MS stars and a disk \citep{Grellmann11}, strongly suggesting that it formed in a manner similar to lower-mass stars.  Soft diffuse emission traces a variety of hot plasmas associated with the massive binary 15~Mon and sites of recent star formation (IRS1 and IRS2) in NGC~2264.  Harder diffuse emission suggests that many young stars across NGC~2264 have yet to be resolved in X-rays, despite this long \Chandra observation.

%\clearpage
%-----------------------------------------------------------------------------
\subsection{NGC~6193 and RCW~108-IR \label{sec:n6193}}
% RCW 108-IR + NGC 6193 -- 2596 point sources in a 2-pointing mosaic, with HETG data for NGC 6193.
% SF Handbook review:  Wolk, Comeron, & Bourke 2008, Volume 2 (south); \citep{Wolk08a}
% At D = 1.19 kpc, 4*pi*D^2 = 1.6947109e+44 cm^2.
% NGC 6193:  Avg A_V is 2.6 mag, so NH=0.4e22.
% RCW 108-IR:  Avg A_V is 14 mag, so NH=2.2e22.
% HD 150136 is at 16 41 20.42 -48 45 46.7.

We consider these two MSFRs together in this section because they are astrophysically related:  they are at the same distance and probably formed in the same large star-forming complex Ara OB1a \citep{Wolk08a}.  They are listed separately in Table~\ref{targets.tbl} because they had very different ACIS observing configurations and integration times.

\begin{figure}[htb]
\centering
\includegraphics[width=0.478\textwidth]{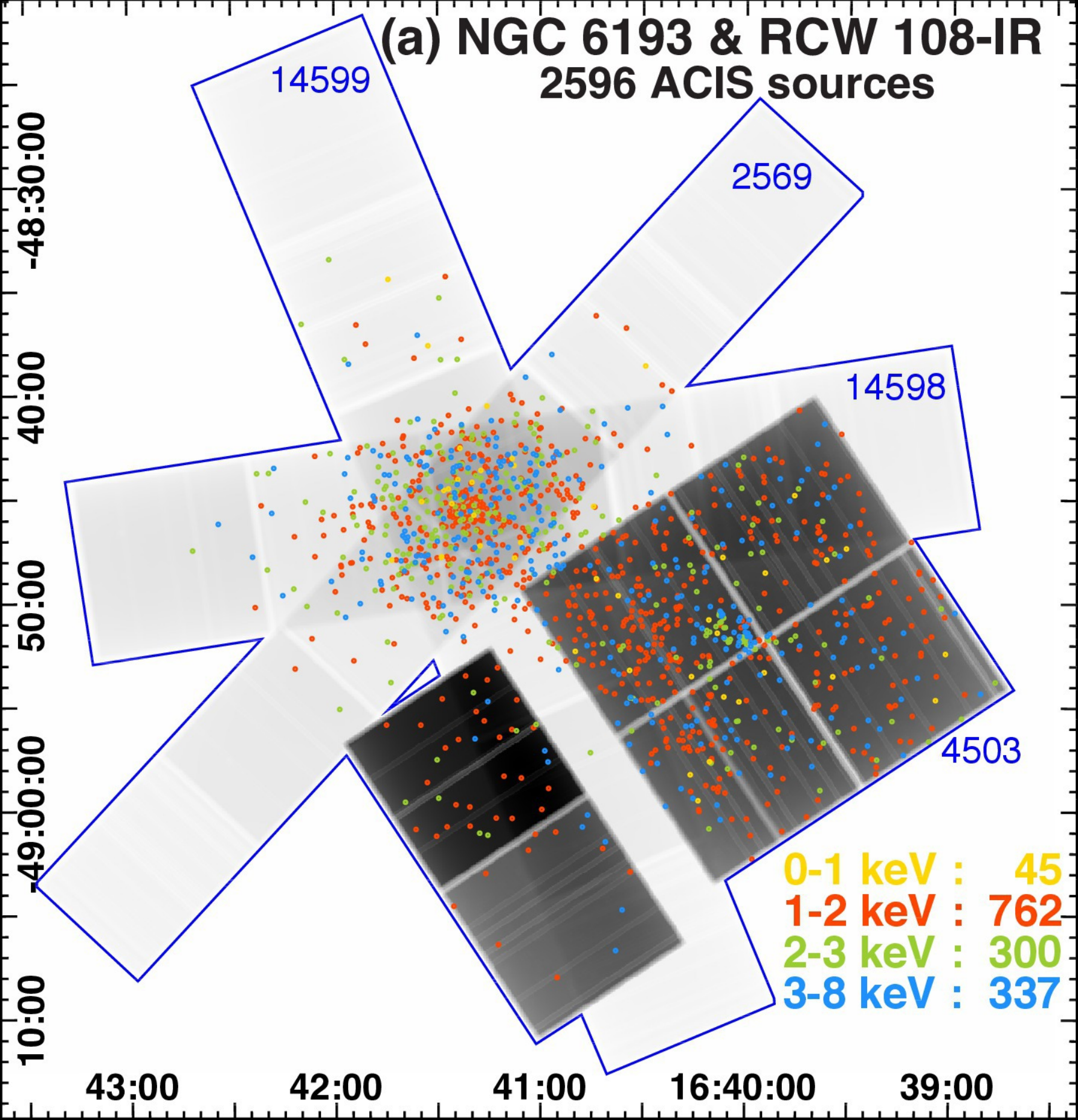}
\includegraphics[width=0.515\textwidth]{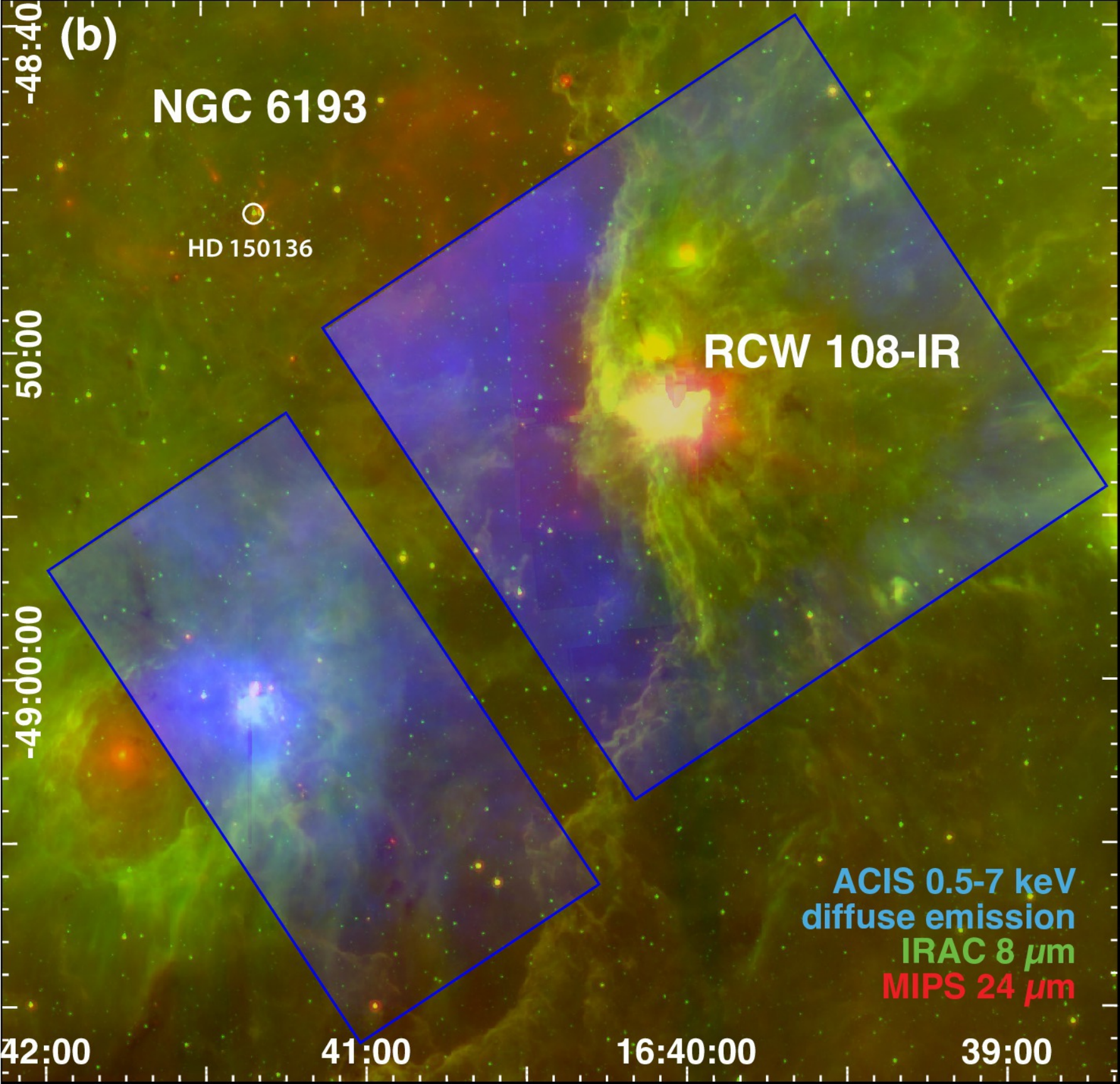}
\caption{NGC~6193 and RCW~108-IR.
(a) ACIS exposure map with 1444 brighter ($\geq$5 net counts) ACIS point sources overlaid; colors denote median energy for each source.  ObsID numbers are shown in blue.
(b) ACIS diffuse emission in the \Spitzer context.  The field of view is smaller here than in (a) because HETG data cannot be used to map diffuse X-ray emission.
\label{ngc6193.fig}}
\end{figure}

The revealed monolithic MSFR NGC~6193 provided strong motivation for expanding our analysis methods to include HETG observations, because these are the only \Chandra data that exist for this nearby massive cluster (Figure~\ref{ngc6193.fig}(a)).  The HETG data were obtained primarily to study the O6.5V star HD~150135 and the O3V+O5.5V+O6.5V hierarchical triple system HD~150136 \citep{Skinner05,Mahy12}.  \citet{Skinner05} examined 43 X-ray sources in the central 2.1$\arcmin$ of the cluster by analyzing the zeroth-order data from ObsID~2569.  They found near-IR counterparts to all of these sources and suggest that they all might be cluster members.

\begin{figure}[htb]
\centering
\includegraphics[width=0.99\textwidth]{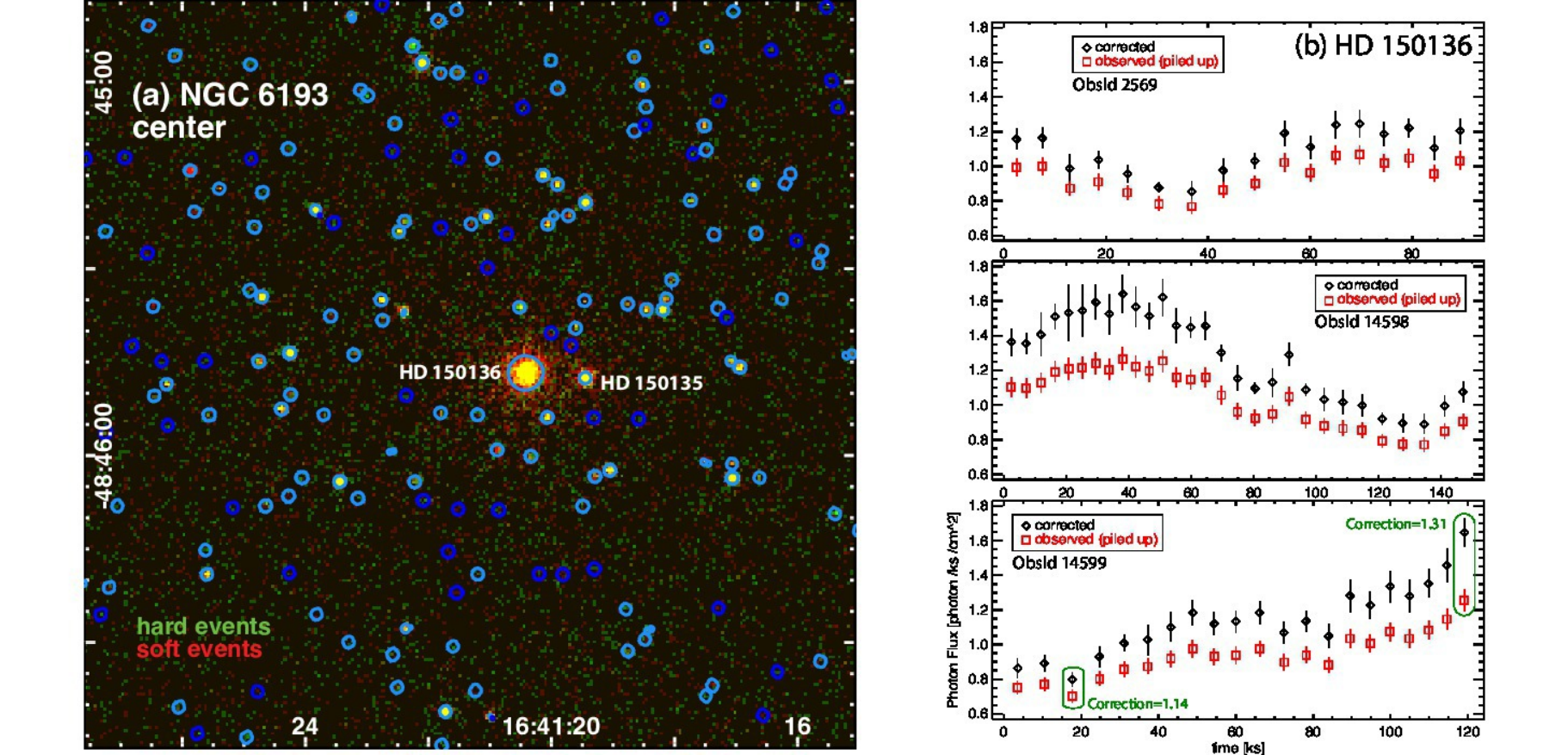}
\caption{
(a) ACIS event data at the center of NGC~6193, from HETG observations.  Dark blue apertures denote ``occasional'' sources.  
(b) Pile-up corrected lightcurves for the three HETG observations of HD~150136.  In the last plot, pile-up correction ranging from a factor of 1.14 to a factor of 1.31 is demonstrated.  
\label{ngc6193details.fig}}
\end{figure}

Subsequently, another 268~ks of HETG data were obtained to monitor variability in HD~150136 (PI J.-C.~Leyder).  By combining the 358~ks of HETG data now available for NGC~6193, we reach an effective ACIS-I exposure of $\sim$42~ks in the zeroth-order, at the center of the cluster.  For comparison with \citet{Skinner05}, we find 139 sources in a 2.1$\arcmin$ box centered on HD~150136 (Fig~\ref{ngc6193details.fig}(a)).

HD~150136 still suffers substantial pile-up in the zeroth-order HETG data; all three ObsIDs show highly variable X-ray lightcurves.  Pile-up correction on those lightcurves restores the amplitude of the variations (Figure~\ref{ngc6193details.fig}(b)), which pile-up suppressed.  
As described in Section~\ref{sec:pileup}, our pile-up correction works on a per-ObsID basis, acting on the time-averaged spectrum for the whole ObsID; for a bright, variable source such as HD~150136, breaking each ObsID into shorter intervals and performing pile-up correction and spectral fitting on such ``time-resolved'' spectra would be a useful future exercise.  Our spectral fits on the pile-up corrected spectra for HD~150136 (Table~\ref{pile-up_risk.tbl}) give higher absorptions and softer thermal plasma temperatures than those found by \citet{Skinner05}.  These result in high X-ray luminosities of $L_{X} = 1.2-2.3 \times 10^{34}$~erg~s$^{-1}$.  Comparing to other MOXC3 massive stars, these luminosities are similar to the (pile-up corrected) X-ray luminosities for WR~20a and WR~21a in RCW~49 and are exceeded only by WR~48a in G305.

\begin{figure}[htb]
\centering
\includegraphics[width=0.9\textwidth]{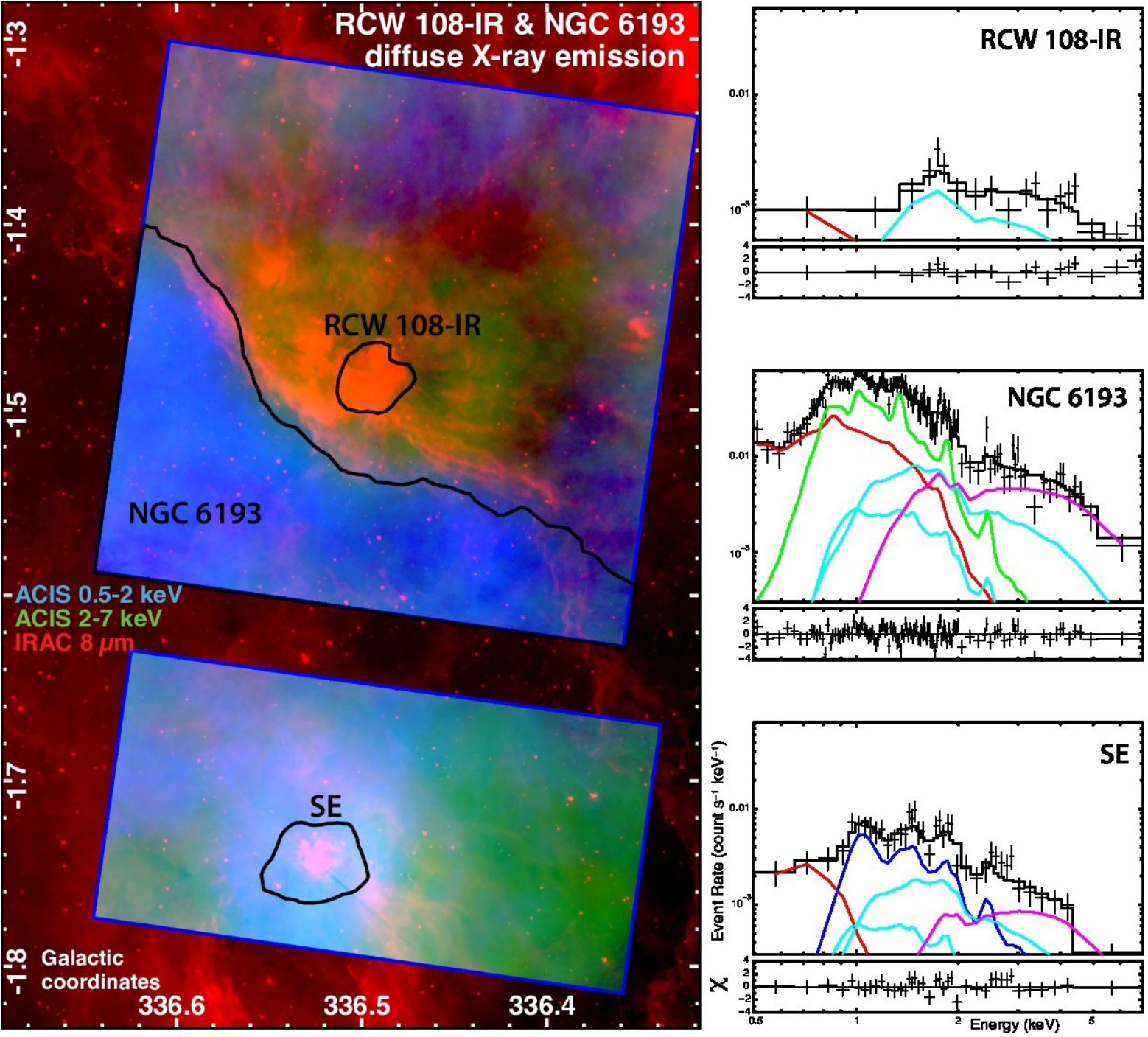}
\caption{Characterizing diffuse X-ray emission in Ara~OB1a.
This image shows ACIS soft-band and hard-band diffuse emission in the \Spitzer context; extraction regions for diffuse spectral fitting are shown in black.  Corresponding spectra are on the right; axis ranges are the same for all spectra.  Table~\ref{tbl:diffuse_spectroscopy_style2} gives fit parameters for the spectral models.
\label{ngc6193gal+spectra.fig}}
\end{figure}

Massive star feedback in NGC~6193 might have influenced the formation of the dense, obscured cluster RCW~108-IR in the molecular cloud west of NGC~6193 \citep{Wolk08a,Wolk08b}.  We find prominent diffuse X-ray emission east of the ionization front that forms the molecular cloud boundary (Figure~\ref{ngc6193.fig}(b), Figure~\ref{ngc6193gal+spectra.fig}), clear evidence for the feedback in NGC~6193 that may have compressed the molecular cloud to form RCW~108-IR.  

% Wolk did NOT find the same diffuse emission from NGC 6193 that we are reporting here.

We extracted a diffuse spectrum from the large region of soft diffuse emission labeled ``NGC~6193'' in Figure~\ref{ngc6193gal+spectra.fig}.  This spectrum has $>$7000 net counts; future work might include subdividing this region into smaller segments to try to simplify the physics captured by each diffuse spectrum.  Our fit shows an interesting phenomenon, described before in CCCP \citep{Townsley11b}:  a single $kT = 0.5$~keV plasma is seen with two very different ionization timescales, as both an unobscured NEI plasma and an obscured plasma closer to CIE.  Faint unresolved pre-MS star emission also contributes to the spectrum, but at a low surface brightness; its absorption is not well-constrained, so we set it equal to that of the obscured diffuse plasma component.  

This recombining plasma suggests strong shocks or even supernova activity around NGC~6193.  An ACIS-only observation of the center of NGC~6193 is highly warranted, to explore the extent of the diffuse X-ray emission around this cluster and the details of its physical state closer to the cluster's massive stars.

\begin{figure}[htb]
\centering
\includegraphics[width=0.32\textwidth]{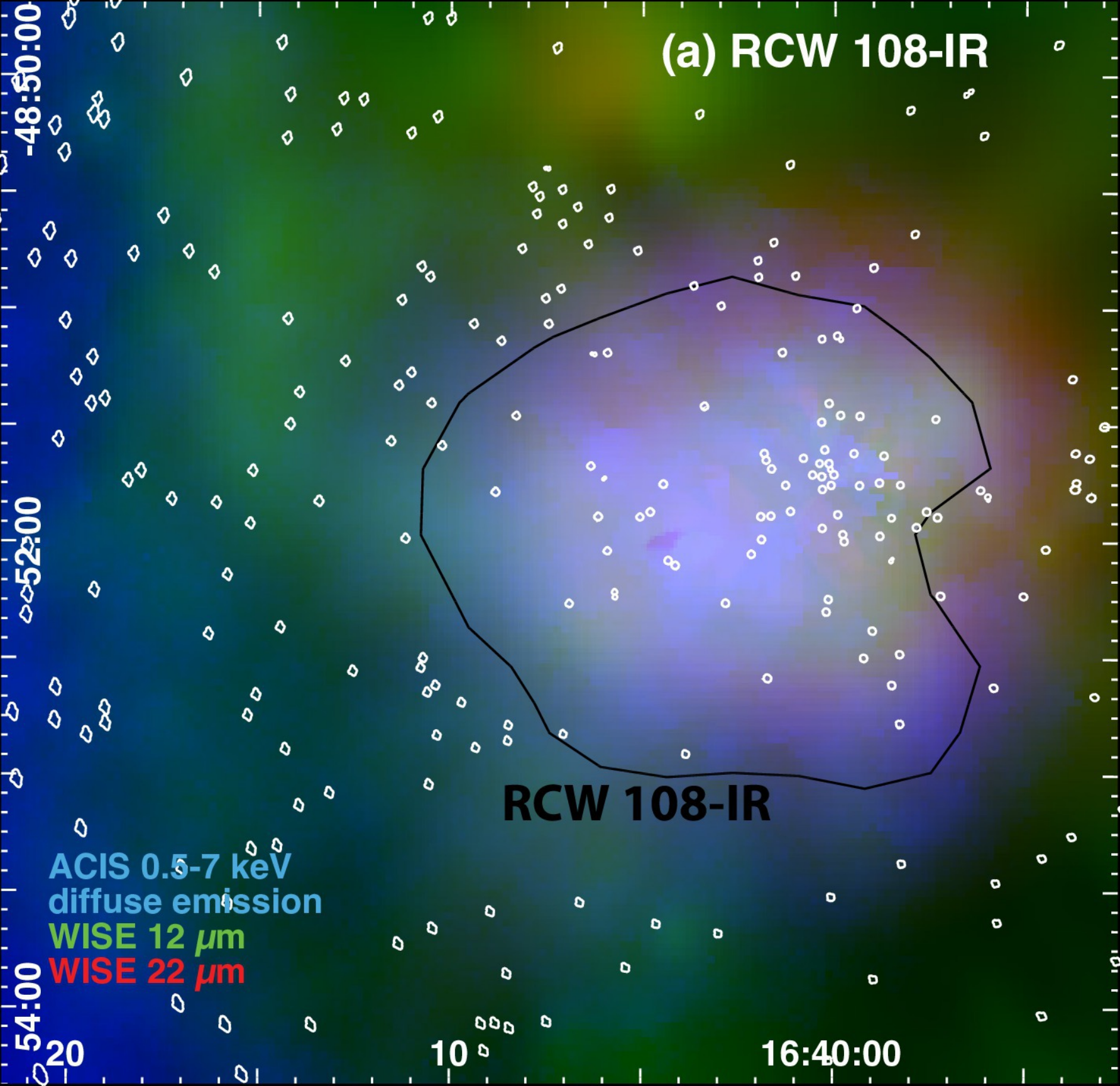}
\includegraphics[width=0.32\textwidth]{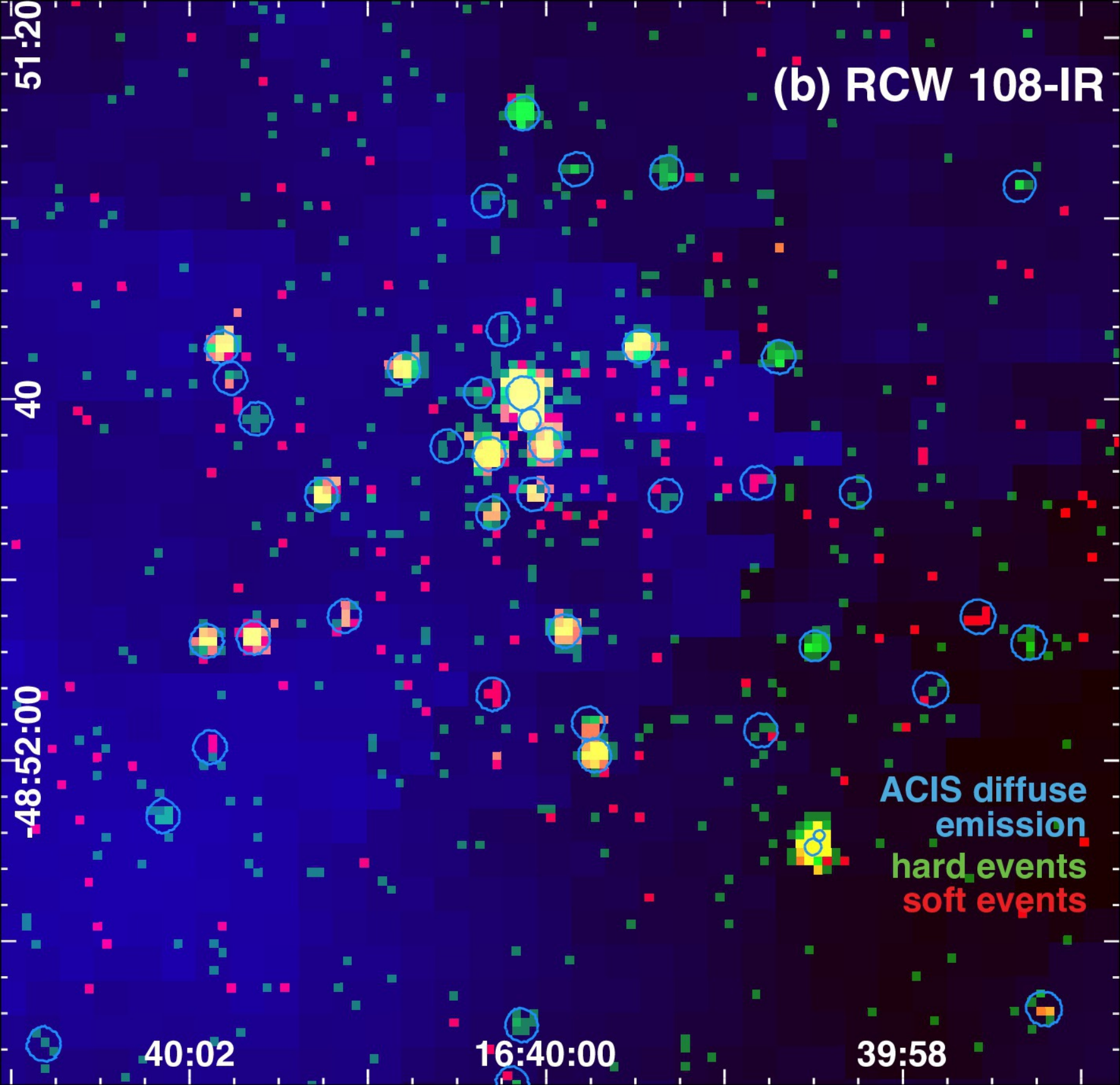}
\includegraphics[width=0.32\textwidth]{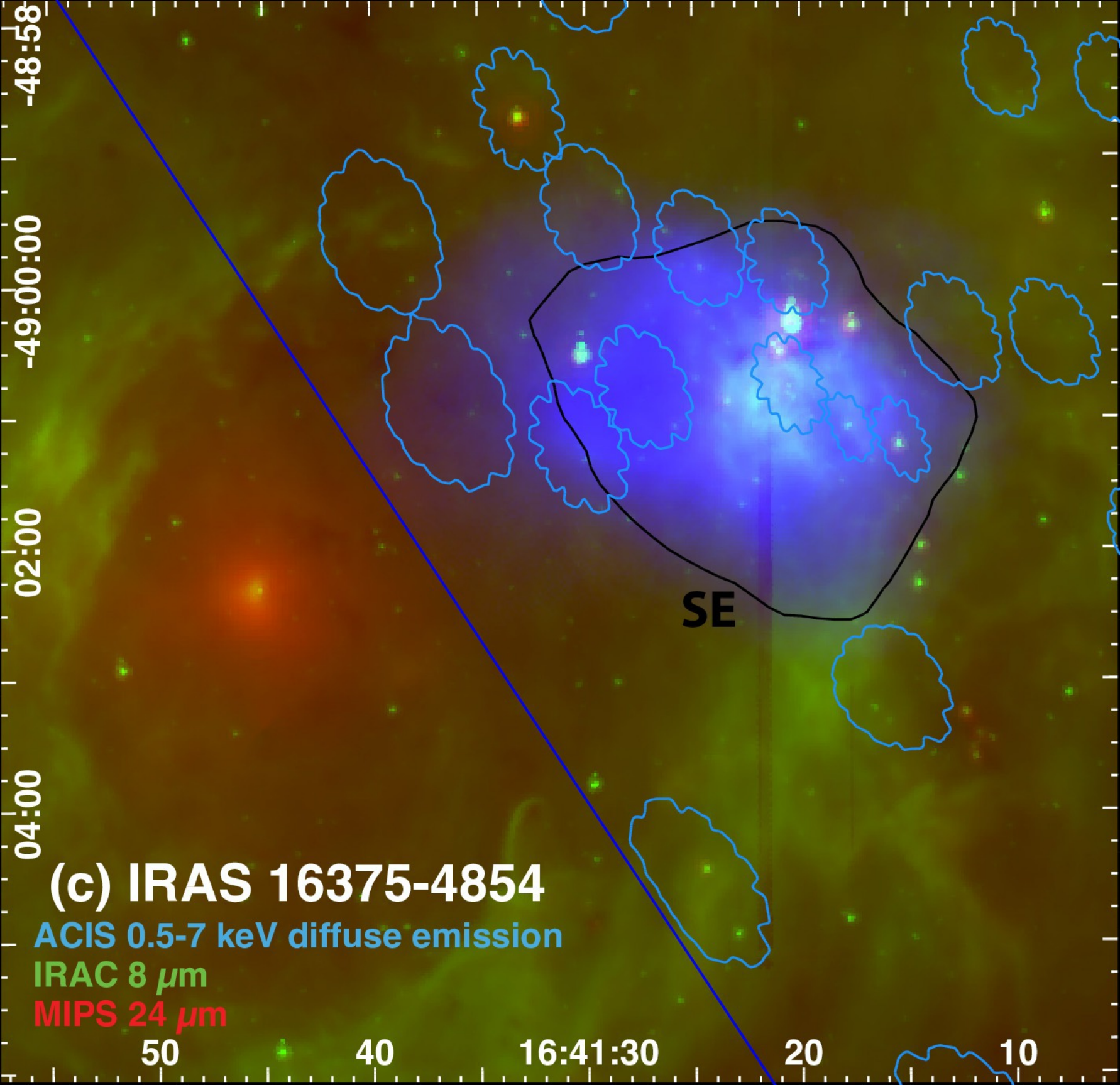}
\caption{
(a) ACIS diffuse emission in the \WISE context, for RCW~108-IR.  The black contour shows our diffuse extraction region.
(b) ACIS event data and diffuse emission for the center of RCW~108-IR.
(c) ACIS diffuse emission in the \Spitzer context, for IRAS~16375-4854.  The dark blue line marks the edge of the ACIS image.  Point source extraction regions are overlaid.  Again the black contour shows our diffuse extraction region.
\label{rcw108.fig}}
\end{figure}

\citet{Wolk08b} present a detailed study of 420 X-ray sources in the 89-ks RCW~108-IR ACIS-I ObsID~4503. 
They report an absence of diffuse X-ray emission in this observation.
We find 989 sources in the same region, plus faint unresolved emission around the embedded cluster.
Figure~\ref{rcw108.fig}(a) and (b) zoom in on the diffuse X-ray emission and X-ray point sources that we find at the center of RCW~108-IR.  The crowded core of the cluster is prominent in X-rays despite the heavy obscuration; Figure~\ref{rcw108.fig}(b) shows 44 (mostly obscured) X-ray sources.

Faint diffuse X-ray emission associated with RCW~108-IR extends east of the embedded cluster (Figure~\ref{rcw108.fig}(a)) but is spatially distinct from the bright diffuse X-ray emission located east of the ionization front (prominent in Figure~\ref{ngc6193.fig}(b) and Figure~\ref{ngc6193gal+spectra.fig} above).  Its spectrum (Figure~\ref{ngc6193gal+spectra.fig} top right) shows a hint of unobscured soft ($kT \sim 0.3$~keV) plasma with very low surface brightness; this must be a foreground plasma that either leaked out of RCW~108-IR or is unrelated to it.  The diffuse spectrum is dominated by unresolved pre-MS stars with absorption consistent with the known obscuring column to the embedded cluster.  This implies that more sensitive X-ray observations would yield more information about the stellar population of this young MSFR.  Any truly diffuse emission from massive star feedback in RCW~108-IR remains undetected in these \Chandra data, due to high obscuration and sensitivity limits.

% From Simbad, IRAS~16375-4854 is at 16 41 16.9 -49 00 33
%              IRAS~16379-4856 is at 16 41 43.6 -49 02 17
{\bf Additional diffuse X-ray emission is present in a large \Spitzer bubble surrounding 
IRAS~16379-4856 and is brightest at the bubble edge, at the location of IRAS~16375-4854 (the region labeled ``SE'' in Figure~\ref{ngc6193gal+spectra.fig} and Figure~\ref{rcw108.fig}(c)).  \citet{Wolk08a} and \citet{Baume11} suggest that this is a site of very recent star formation that may be related to RCW~108-IR.}  This field was captured serendipitously by the off-axis ACIS-S chips in the ACIS-I observation of RCW~108-IR.  We identify a number of X-ray point sources in the region, but the very large off-axis PSF here limits our source-finding capabilities.   

We were unable to fit this ``SE'' diffuse spectrum with the model used for the ``NGC~6193'' diffuse region described above, thus it does not appear to be just the continuation of that plasma emission.  Instead the IRAS~16375-4854 region requires two CIE thermal plasmas with different temperatures:  an unobscured soft component reminiscent of the RCW~108-IR diffuse emission and a heavily obscured, hard component with $kT = 1.2$~keV.  A pre-MS component is also needed; its obscuration was fixed to that of the hard plasma.  The diffuse emission surface brightness is about 50\% larger than that of the unresolved pre-MS stars.  

The hard diffuse component is difficult to explain in terms of wind shocks from massive stars; we will see below that many of MOXC3's diffuse spectra require such high-temperature plasma models (with $kT > 1$~keV).  These hot plasmas might suggest the presence of old, thermalized supernova remnants (SNRs) at the periphery of our MSFRs.  Such phenomena might be expected in large, multi-generational star-forming complexes hosting sequential episodes of massive star formation, feedback, and evolution.

Our overall impression of the NGC~6193/RCW~108-IR field is that much remains to be learned about this multi-generational MSFR.  A modest ACIS-only observation would reveal much more of the rich NGC~6193 cluster population and would almost certainly discover bright, soft diffuse X-ray emission from massive star feedback there.  Given the extent of hard unresolved X-ray emission seen in Figure~\ref{rcw108.fig}, deeper observations of RCW~108-IR also promise interesting results.  A wider mosaic across this rich complex is also justified; IRAS~16375-4854 is just one of several recent star formation sites in Ara OB1a \citep{Wolk08a, Baume11} that should be investigated with dedicated high spatial resolution X-ray observations.

%\clearpage
%-----------------------------------------------------------------------------
\subsection{DR15 \label{sec:dr15}}
% DR15 -- 651 point sources in a single ACIS-I ObsID, no HETG data.
% SF Handbook review:  Reipurth+Schneider08, chapter on Cygnus.  Shown in Fig15 but not mentioned in the text.
% At D = 1.4 kpc (swag from Rygl12), 4*pi*D^2 = 2.3456207e+44 cm^2.
% IRDC: l,b = 79.27 +0.38 -> 20 31 59.42  +40 18 16.5
% No piled up sources in this target.
% Avg A_V is 20 mag, so NH=3.2e22.

\begin{figure}[htb]
\centering
\includegraphics[width=0.49\textwidth]{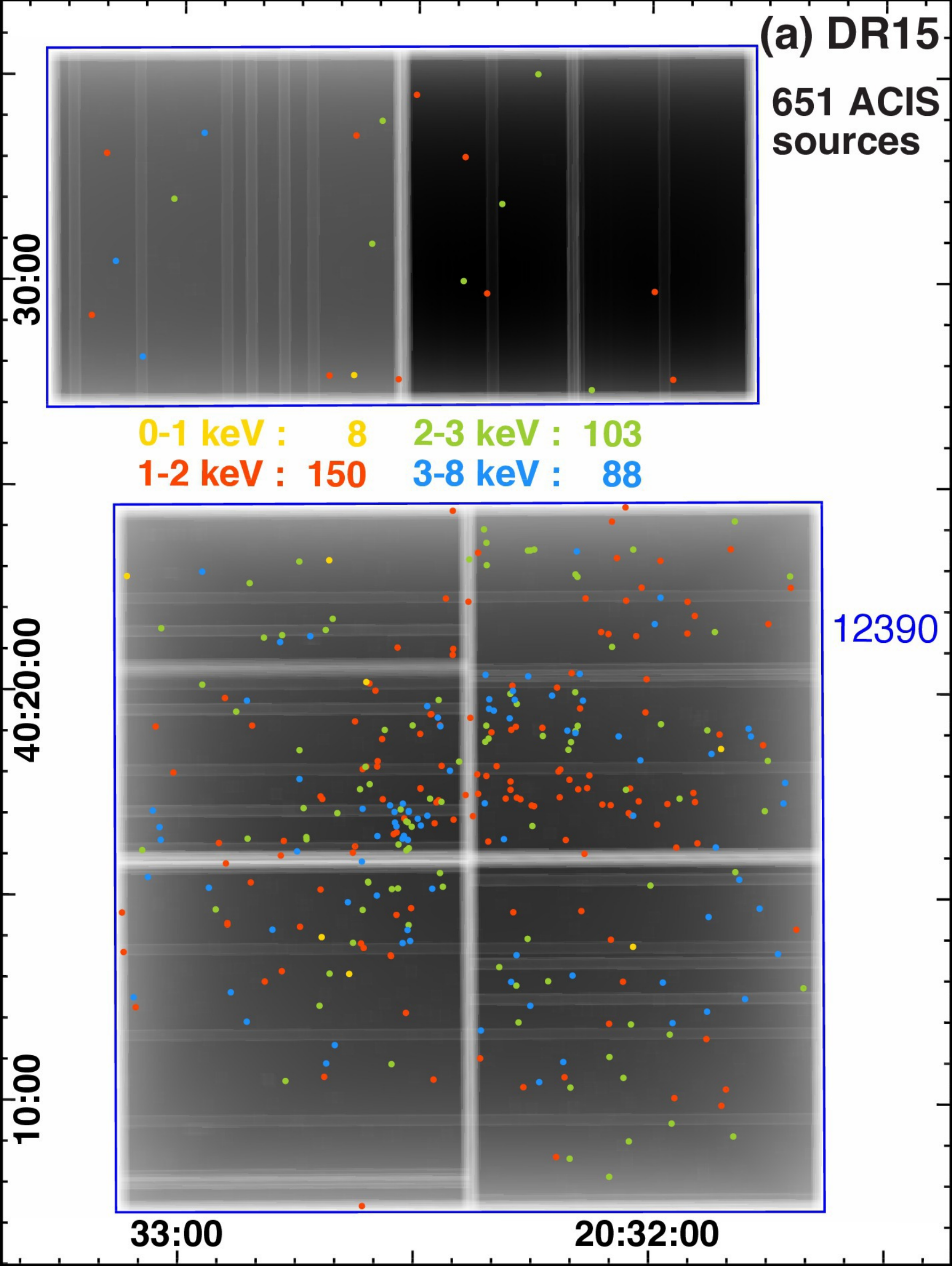}
\includegraphics[width=0.49\textwidth]{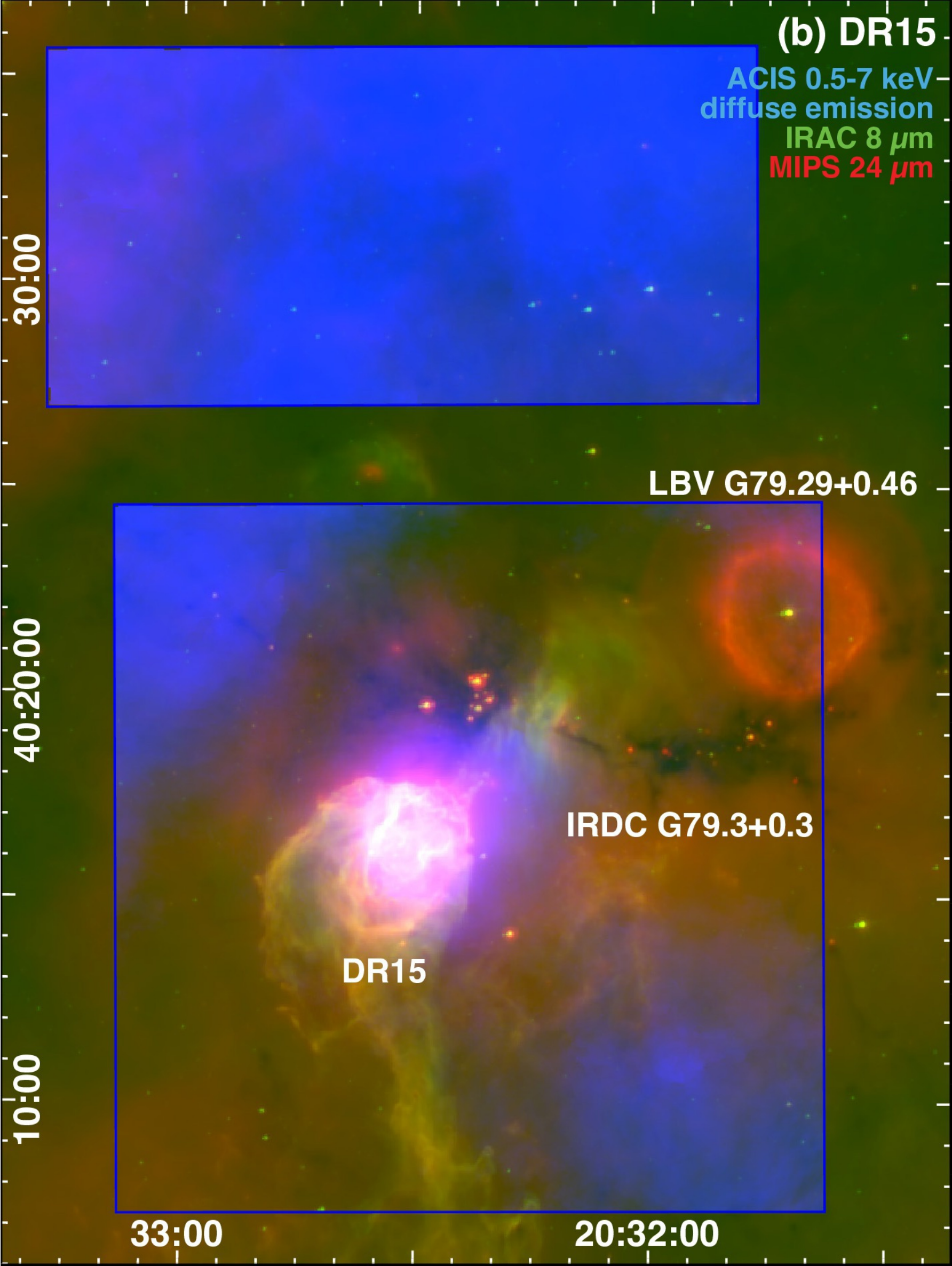}
\caption{DR15.
(a) ACIS exposure map with 349 brighter ($\geq$5 net counts) ACIS point sources overlaid; colors denote median energy for each source.  The ObsID number is given in blue.
(b) ACIS full-band diffuse emission in the \Spitzer context; regions named in the text are labeled.  
\label{dr15.fig}}
\end{figure}

DR15 is one of the many bright radio \hii regions in the vast Cygnus~X complex \citep{Reipurth+Schneider08}, ionized by a pair of B0V stars \citep{Odenwald90,Rivera15}.  It hosts an obscured young cluster and sits at the end of a long, narrow mid-IR pillar; the recent multiwavelength study by \citet{Rivera15} provides the first extensive list of YSOs in the region.  Northwest of DR15 is the prominent IRDC G79.3+0.3 \citep{Oka01,Laws19}; this is captured in the ACIS field (Figure~\ref{dr15.fig}), as is LBV~G79.29+0.46 and its famous ring nebula \citep[e.g.,][]{Agliozzo14,Rizzo14}.

The ACIS I-array data were analyzed by \citet{Rivera15}; they found 131 X-ray sources.  Our analysis includes the off-axis ACIS-S data and pushes to fainter X-ray source limits, revealing almost five times as many point sources and highly structured diffuse X-ray emission that extends onto the ACIS-S CCDs (Figure~\ref{dr15.fig}).  This DR15 study can be compared to the Cygnus~X MSFRs we studied in MOXC2, W75N and IRAS~20126+4104.  All three of these regions feature modest ACIS-I exposures of highly obscured MSFRs ionized by early-B MYSOs, yet all display interesting diffuse X-ray emission as well as large point source populations.

\begin{figure}[htb]
\centering
\includegraphics[width=0.308\textwidth]{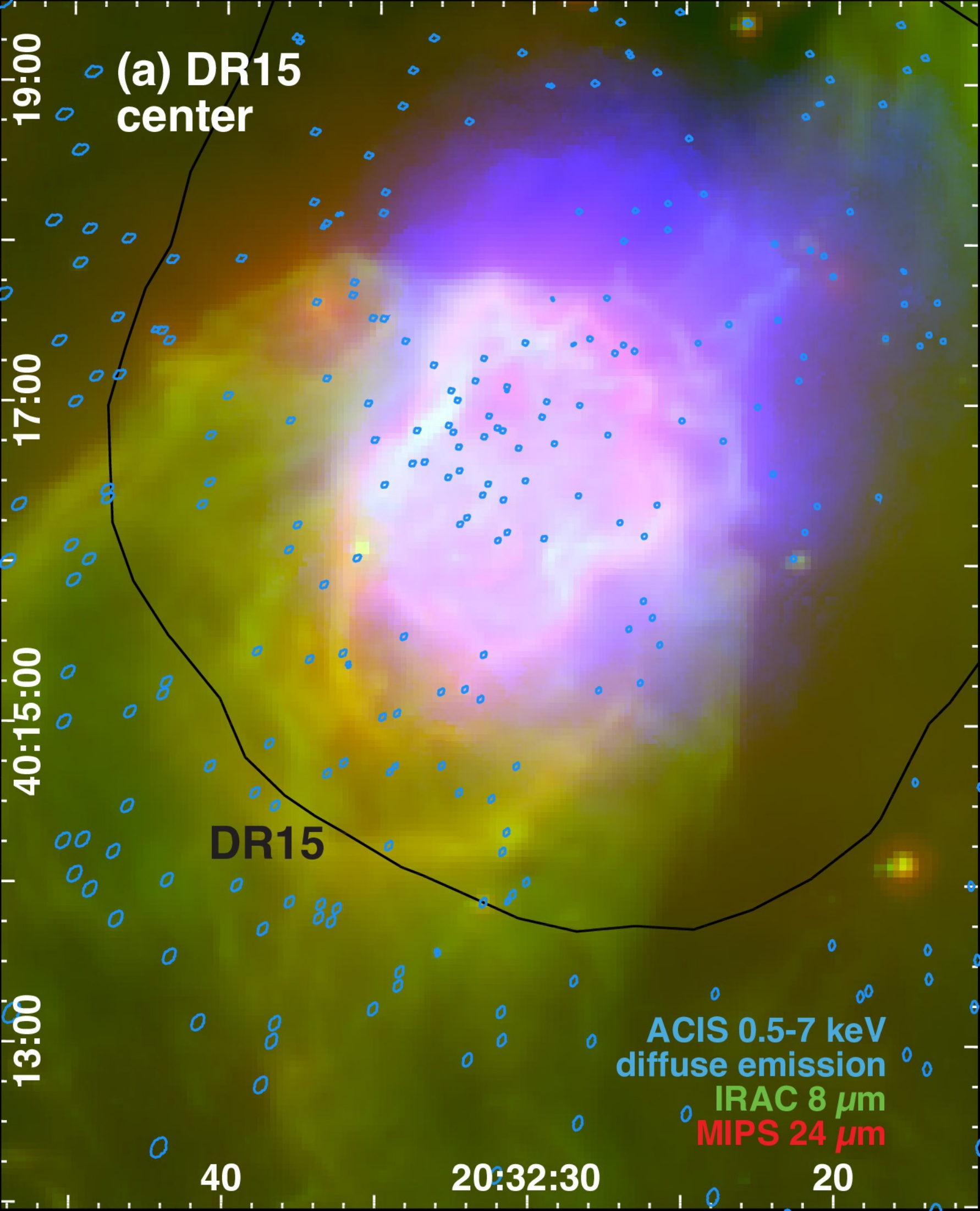}
\includegraphics[width=0.335\textwidth]{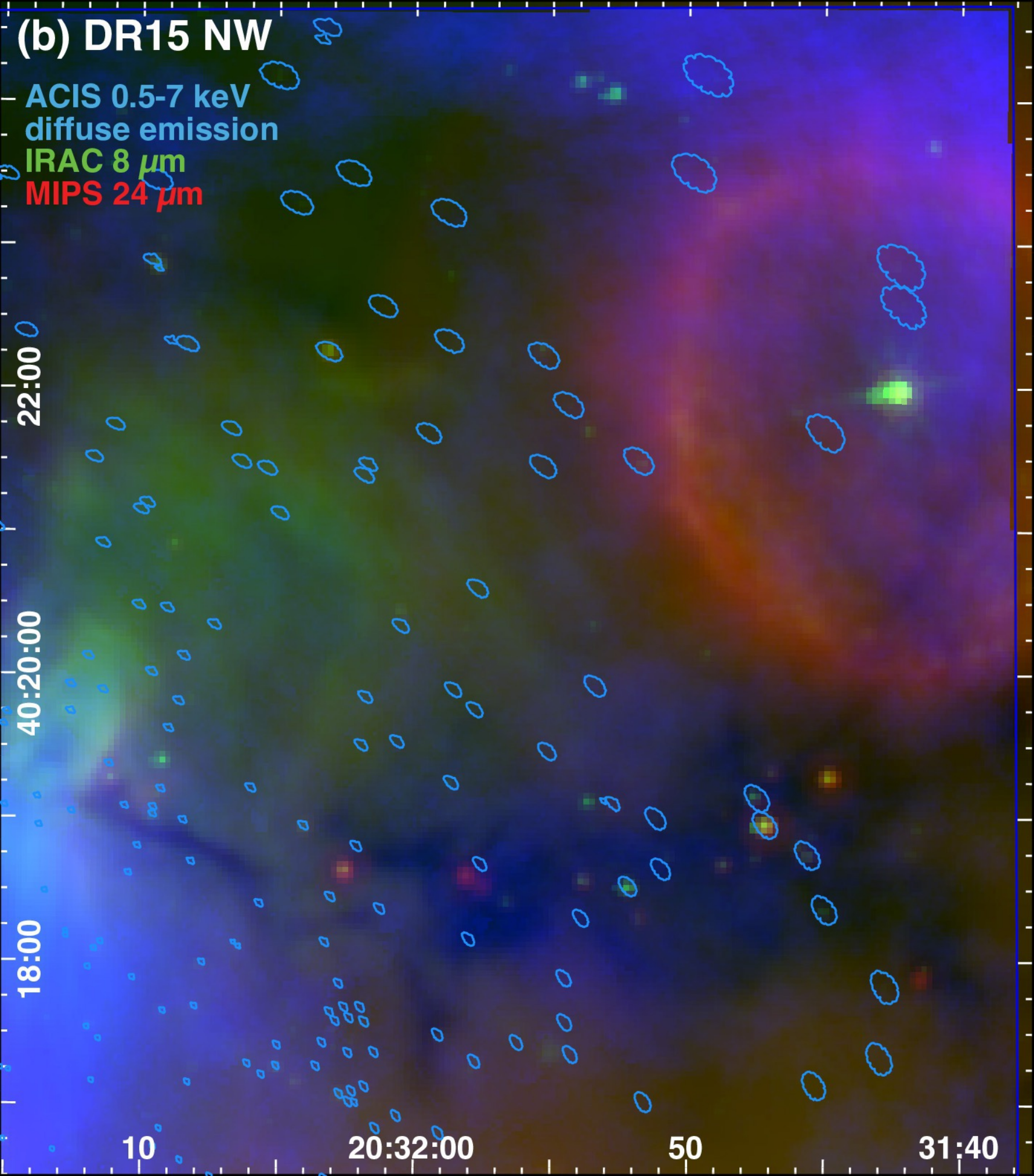}
\includegraphics[width=0.344\textwidth]{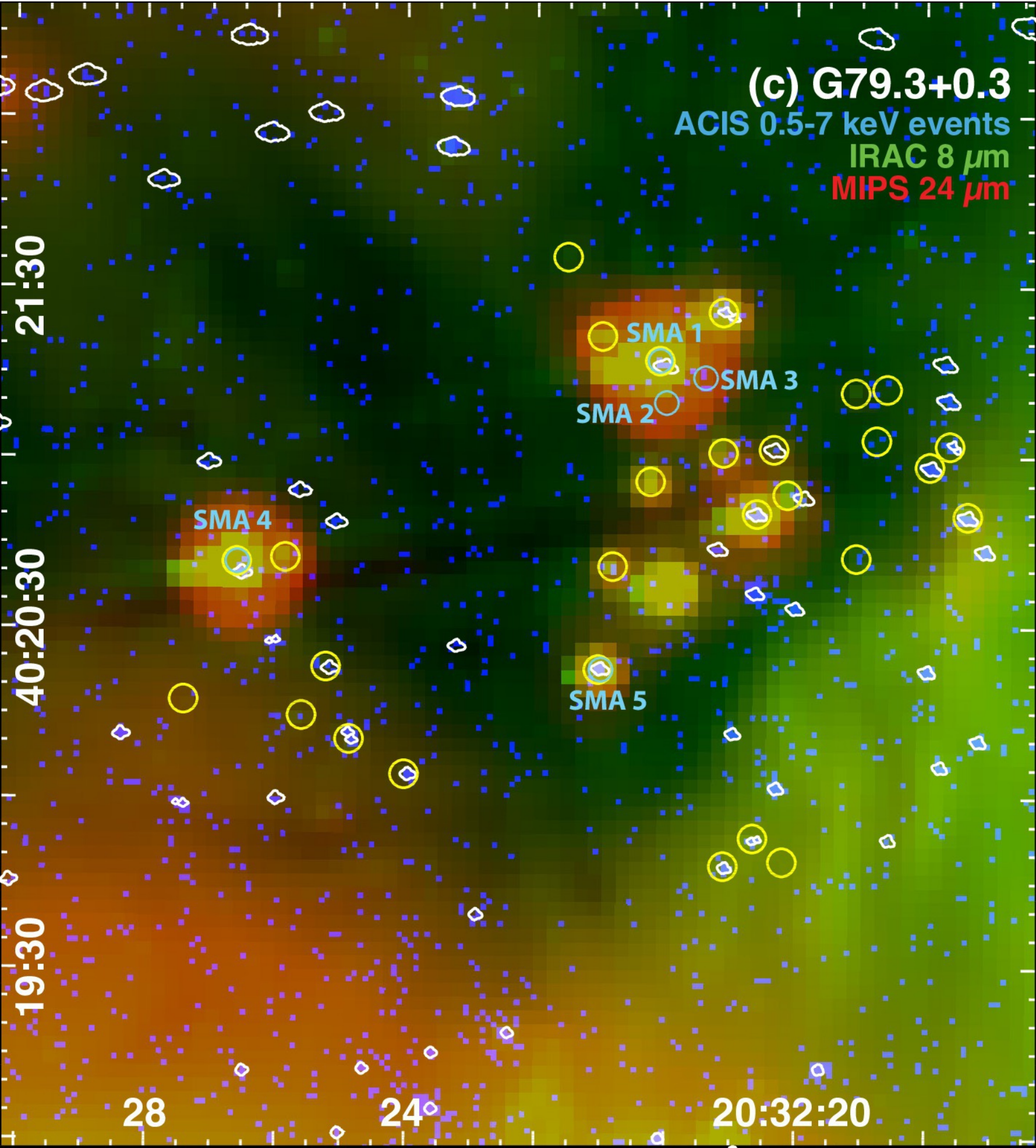}
\caption{DR15 ACIS data in the \Spitzer context.
(a) The DR15 obscured cluster.  ACIS point source extraction regions are overlaid on a diffuse X-ray image.  The black contour shows part of our diffuse extraction region.
(b) The northwest corner of the ACIS image, including the main part of IRDC~G79.3+0.3 and most of the 24~$\mu$m nebula around LBV~G79.29+0.46.
(c) ACIS event data for the eastern part of IRDC~G79.3+0.3, north of the DR15 cluster.  Circles show sources from \citet{Laws19}; see text.
\label{dr15center.fig}}
\end{figure}

The main DR15 cluster is prominent in the ACIS data, containing many obscured X-ray sources (Figure~\ref{dr15center.fig}(a)) with median energy $\sim$3~keV.  The LBV is undetected in X-rays but faint diffuse X-ray emission is superposed on its 24~$\mu$m ring nebula (Figure~\ref{dr15center.fig}(b)).  There is a hint of diffuse X-ray emission across the IRDC at the bottom of this figure; several \Spitzer sources there are detected by ACIS.

We matched our ACIS sources to the \Gaia DR2 catalog \citep{Gaia16,Gaia18} and used 71 matches (assumed to be Cygnus~X members) to estimate a median \Gaia distance of $1.08^{+0.09}_{-0.09}$~kpc.  The parallax distribution has a tail that implies that individual source distances extend to $\sim$1.6~kpc.  Again, the methodology used for this calculation came from \citet{Binder18} and was used in \citet{Povich19}; it is similar to that of \citet{Kuhn19}.  Examining the X-ray properties of these \Gaia matches, we find these sources to be less obscured than those in DR15, with median energy $\sim$1.5~keV.  Thus it appears that \Gaia is seeing the wider Cygnus~X population along this sightline, distributed over several hundred parsecs.  For our purposes in MOXC3, we will continue to assume a distance of 1.4~kpc \citep{Rygl12} for DR15 and the wider ACIS field, with the caveat that individual point sources (and all diffuse X-ray emission) could have substantially different distances.

In the northeast part of IRDC~G79.3+0.3 (Figure~\ref{dr15center.fig}(c)), \citet{Laws19} recently studied the YSO population using \Spitzer (yellow circles) and the Submillimeter Array (SMA; cyan circles).  Of the 28 \Spitzer YSOs they find in this region, we detect 14 of them with ACIS; four of those are resolved into a pair of ACIS sources.  We find X-ray counterparts to SMA objects 1, 4, and 5; \citet{Laws19} determined that SMA objects 2 and 3 are deeply embedded, less evolved objects.  They say that objects 1 and 4 will become massive stars.  Our X-ray detections confirm what we have seen in other ACIS MSFR studies:  MYSOs are hard X-ray emitters, often detected behind many tens of magnitudes of obscuration.  ACIS source CXOU~J203222.02+402016.1 (c696), the counterpart to SMA object 1, is detected with $<$3 net counts.  Such faint but interesting X-ray sources provide additional motivation for pushing X-ray source detection down to faint limits.
% SMA object 1 is ACIS c696, 203222.02+402016.1
% SMA object 4 is ACIS c840, 203228.52+401939.6

\begin{figure}[htb]
\centering
\includegraphics[width=0.99\textwidth]{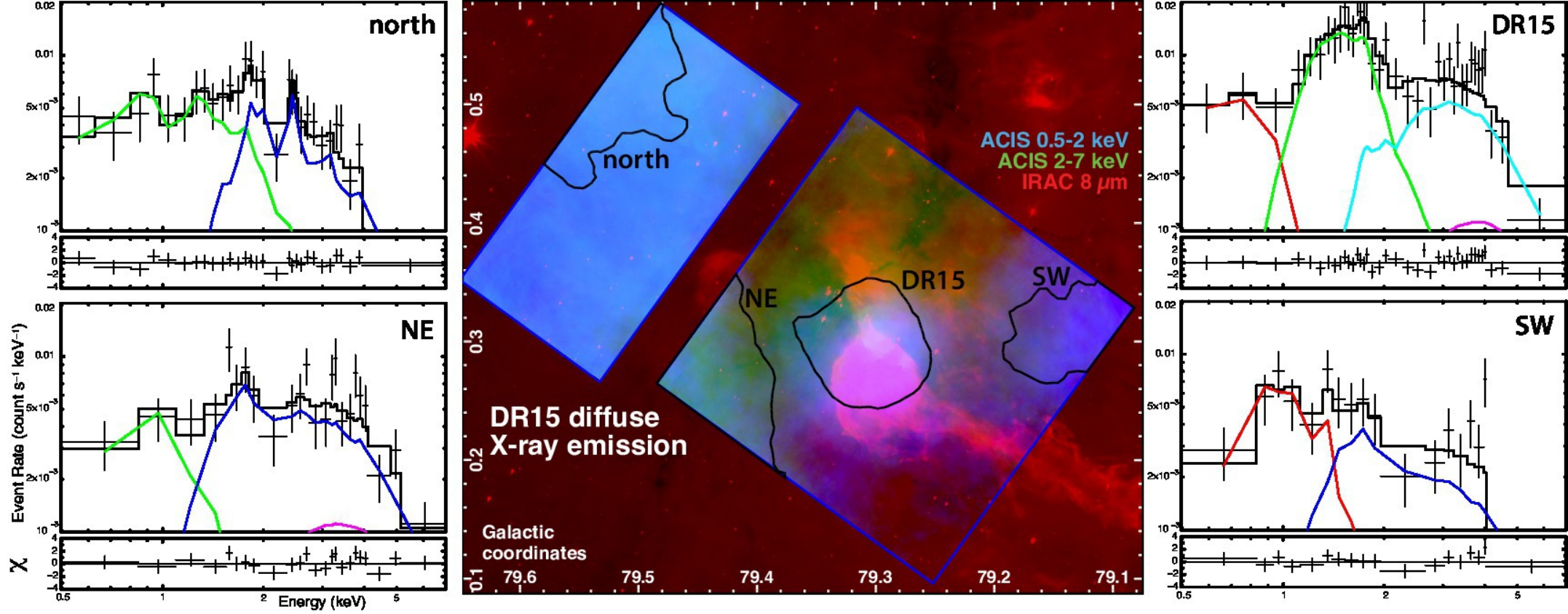}
\caption{Characterizing diffuse X-ray emission in and around DR15.
This image shows ACIS soft-band and hard-band diffuse emission in the \Spitzer context; extraction regions for diffuse spectral fitting are shown in black.  Corresponding spectra flank the image; axis ranges are the same for all spectra.  Table~\ref{tbl:diffuse_spectroscopy_style2} gives fit parameters for the spectral models.
\label{dr15gal+spectra.fig}}
\end{figure}

The diffuse X-ray emission in DR15 is brightest around the central cluster (Figure~\ref{dr15gal+spectra.fig}).  Its spectrum shows a soft, unobscured CIE plasma component with low surface brightness, but the diffuse emission is dominated by an obscured NEI plasma with $kT \sim 0.5$~keV and high surface brightness.  This is similar to the diffuse emission around NGC~2264~IRS2 described above (but five times brighter).  A pre-MS component behind a very high column models the spectrum above 2~keV; its surface brightness is equal to that of the hard plasma.  These results indicate, as usual, that much of DR15's young stellar population remains unresolved by these shallow \Chandra data.  More notable, however, is its NEI plasma; the obscuration to this component is consistent with that of the DR15 cluster, so it probably originates in the cluster.  Such substantial diffuse surface brightness was also seen in the MOXC2 MSFRs W75N and IRAS~20126+4104.  All of these obscured young clusters are ionized by early-B stars; it is remarkable that MSFRs with such modest wind power somehow generate these detectable hot plasmas.

The other three regions of prominent diffuse emission (labeled in Figure~\ref{dr15gal+spectra.fig}) do not contain known young clusters, so we did not include a pre-MS component in their spectral fits.  Regions NE and SW are both modeled by two CIE thermal plasma components:  a soft plasma behind modest obscuration and a very hard plasma behind a high column.  This hard plasma may indicate an underlying pre-MS component, perhaps a distributed population of young stars.  Given the complexity and long history of star formation in the Cygnus~X complex, such a distributed population is plausible.  Patches of soft diffuse emission distributed around the Cygnus~X field are also plausible; only wide-field, high spatial resolution X-ray mapping could determine the extent (and infer the origins) of the responsible plasmas.

Far off-axis, we extracted diffuse emission from the ``north'' region.  This spectrum can be modeled by a single-temperature hard plasma ($kT = 2$~keV) with two different obscuring columns and ionization timescales (see Table~\ref{tbl:diffuse_spectroscopy_style2}).  The more obscured, longer-timescale component accounts for most of the surface brightness, which exceeds $2 \times 10^{32}$~erg~s$^{-1}$~pc$^{-2}$.
% and is the brightest diffuse emission we've encountered so far in MOXC3.  
As we suggested for other MSFRs, this hard recombining plasma suggests an origin in an old thermalized SNR, although a wider mosaic is needed to characterize its spatial extent. 

This relatively short ACIS exposure on the obscured MSFR DR15 significantly advances our understanding of both the X-ray point source population and the diffuse X-ray emission in this corner of the Cygnus~X complex.  As in NGC~2264 IRS2 described above and in other MOXC targets, we find that a young cluster ionized by only early-B stars is generating bright, hot diffuse plasma.  Deeper observations would be justified, to characterize DR15's bright diffuse emission in more detail, to better study the X-ray source population in IRDC~G79.3+0.3, and to quantify the diffuse X-ray emission seen superposed on the IRDC and the LBV nebula (Figure~\ref{dr15center.fig}(b)).

%\clearpage
%-----------------------------------------------------------------------------
\subsection{Berkeley 87 and Onsala 2S \label{sec:berk87}}
% Berkeley 87 -- 879 point sources in a single ACIS-I ObsID, no HETG data.
% SF Handbook review:  mentioned on p. 67 in chapter on Cygnus by Reipurth & Schneider 2008, Vol 1 (north).
% At D = 1.66 kpc, 4*pi*D^2 = 3.2977512e+44 cm^2.  Does this larger distance put Berk87 *behind* Cygnus?
% No piled up sources in this target.
% WR142 is at 20 21 44.34 +37 22 30.5.  It is our source 202144.34+372230.5 ; 'c972', with 41 net counts, median energy 4.3 keV!   WR142 is at 20 21 44.35 +37 22 30.7 according to Sokal10.
% V439 Cyg -- Simbad calls this ``EM* MWC 1015 -- Be Star'', at 20 21 33.58 +37 24 51.7.  No ACIS ctrprt.
% Berk87 at 20 21 43.01 +37 22 13.8 according to Simbad.
% UCHIIR G75.78+0.34 is at 20 21 44.1 +37 26 40 according to Simbad.
% Compact HII region G75.77+0.34 is not in Simbad.
% Avg A_V is 6 mag, so NH=0.96e22.
%
% ON 2S:  At 3.5 kpc, 4*pi*D^2 = 1.466e+45 cm^2.
% Avg A_V is 20 mag, so NH=3.2e22.

Berkeley~87 is a visually revealed MSFR in Cyg~OB1, another one of the many young clusters in the active Cygnus~X star-forming complex \citep{Reipurth+Schneider08}.  Its massive stars have been studied for decades \citep{Turner82, Massey01}; of particular note is the WO2 star WR~142.  Projected on the sky just 3\arcmin-4\arcmin\ north of Berk87 are several prominent small \hii regions in the obscured MSFR Onsala~2 South (ON~2S) \citep{Matthews83,Dent88,Shepherd97}.

WR~142 was studied by both \XMM \citep{Oskinova09} and \Chandra \citep{Sokal10,Skinner19}.  It is a weak, obscured, hard X-ray source; its X-ray emission mechanisms are discussed in these papers.  \citet{Sokal10} discuss other massive stars in Berk87 detected by \Chandra in the same ACIS-I data that we study here.  \citet{Oskinova10} discuss the \XMM massive star detections. 

\begin{figure}[htb]
\centering
\includegraphics[width=0.49\textwidth]{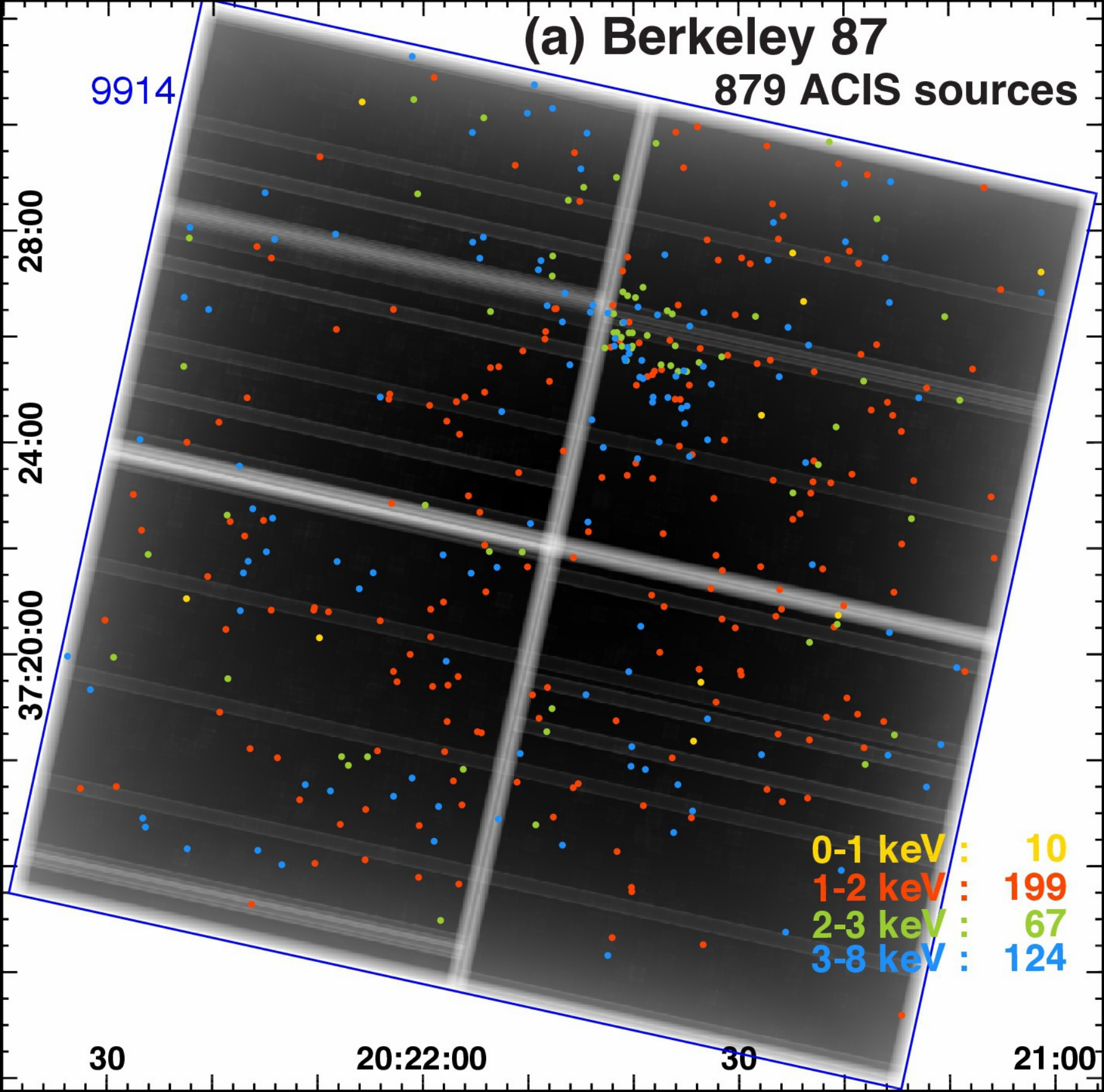}
\includegraphics[width=0.49\textwidth]{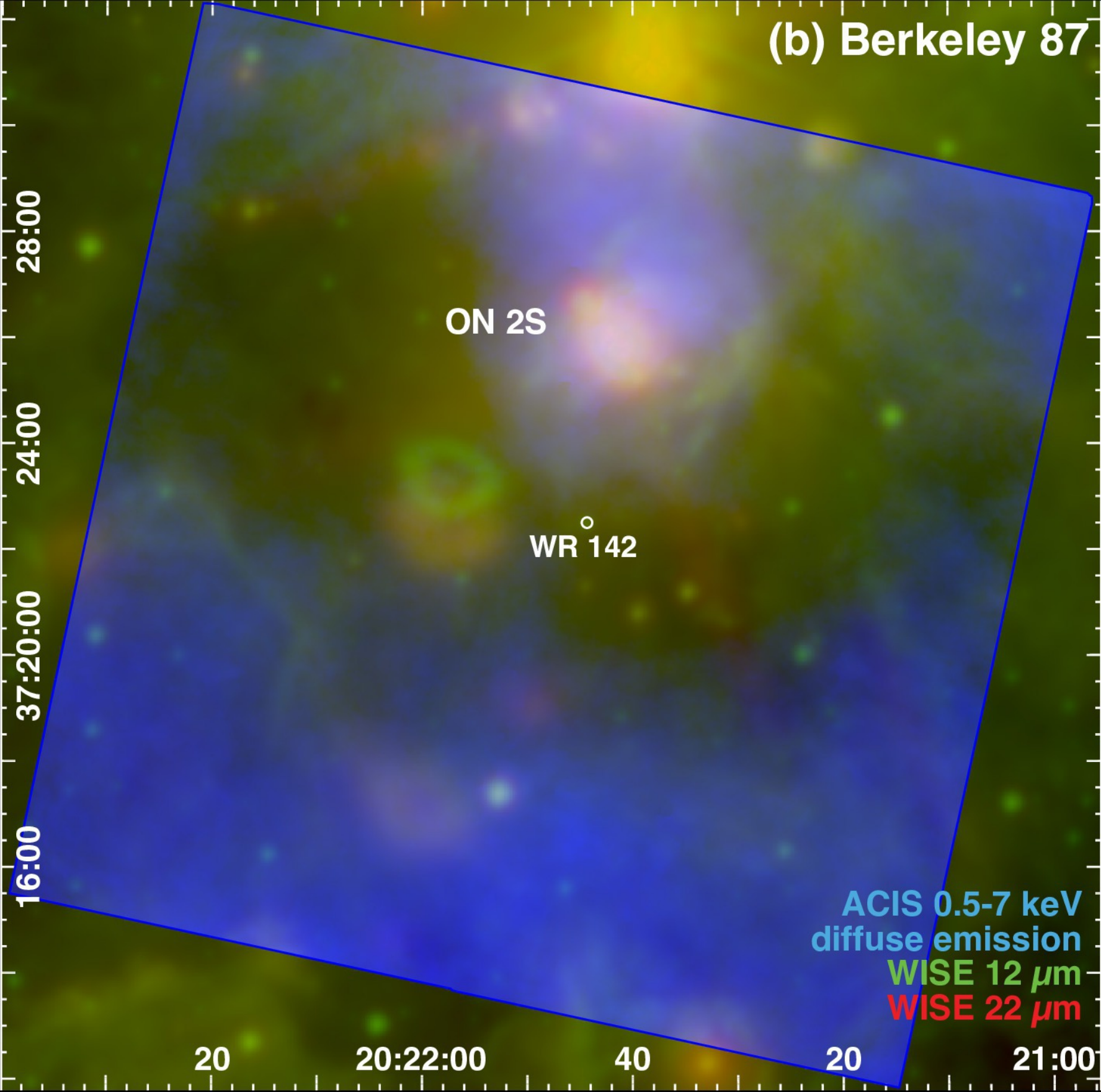}
\caption{Berkeley 87.
(a) ACIS exposure map with 400 brighter ($\geq$5 net counts) ACIS point sources overlaid; colors denote median energy for each source.  The ObsID number is shown in blue.
(b) ACIS diffuse emission in the \WISE context.  The large rings in the \WISE data appear to be artifacts; they are not present in \Spitzer images.
\label{berk87.fig}}
\end{figure}

A second ACIS-I observation of this region was obtained later, centered on ON~2N.  In a recent paper, \citet{Skinner19} analyze these two pointings together (the ON~2N data were not in the public archive when we analyzed the Berk87 ACIS data).  They find 209 point sources in the Berk87 pointing; using a less conservative detection threshold, we find 879 sources in the same data (Figure~\ref{berk87.fig}).  X-ray point sources and diffuse emission in ON~2 are discussed in detail by \citet{Oskinova10} (who assume that Berk87 and ON~2 are at the same distance of 1.23~kpc) and \citet{Skinner19} (who infer from maser parallaxes and Gaia DR2 data that the two MSFRs are at different distances of 1.75~kpc and 3.5~kpc, respectively).  We adopt the \citet{Skinner19} distance for ON~2 and use the Gaia DR2 distance from \citet{Cantat18} for Berk87.

%Our analysis of just the Berk87 ACIS-I pointing shows relatively soft X-ray sources (median energy $<$2~keV) distributed across the field and hard sources concentrated on ON~2S (Figure~\ref{berk87.fig}(a)).  This reflects the spatial distribution expected if the older, less obscured Berk87 cluster is superposed in the foreground on the younger, more obscured, clumpy MSFRs of ON~2S.  

It is notable that the central part of Berk87 (near the aimpoint of the ACIS-I array) does not show an overdensity of X-ray point sources in Figure~\ref{berk87.fig}(a).  The effective area of \Chandra peaks at the ACIS-I aimpoint, so it is hard to avoid a concentration of X-ray sources there when the telescope is pointed at a massive young stellar cluster.  We showed in Table~\ref{targets.tbl} that in these ACIS data, for a cluster with an ONC-like X-ray luminosity function, we should detect at least half of the population all the way down to 0.3~$M_{\sun}$.  Berk87 has many late-O and early-B stars \citep{Massey01}, so we expected to find a robust pre-MS population at the center of the ACIS field; instead the number of X-ray sources is relatively sparse.  We have only seen Berk87's sort of X-ray source distribution when no young cluster was present (e.g., in the AE~Aur data above or in G34.4+0.23 in MOXC2).  

\begin{figure}[htb]
\centering
\includegraphics[width=0.99\textwidth]{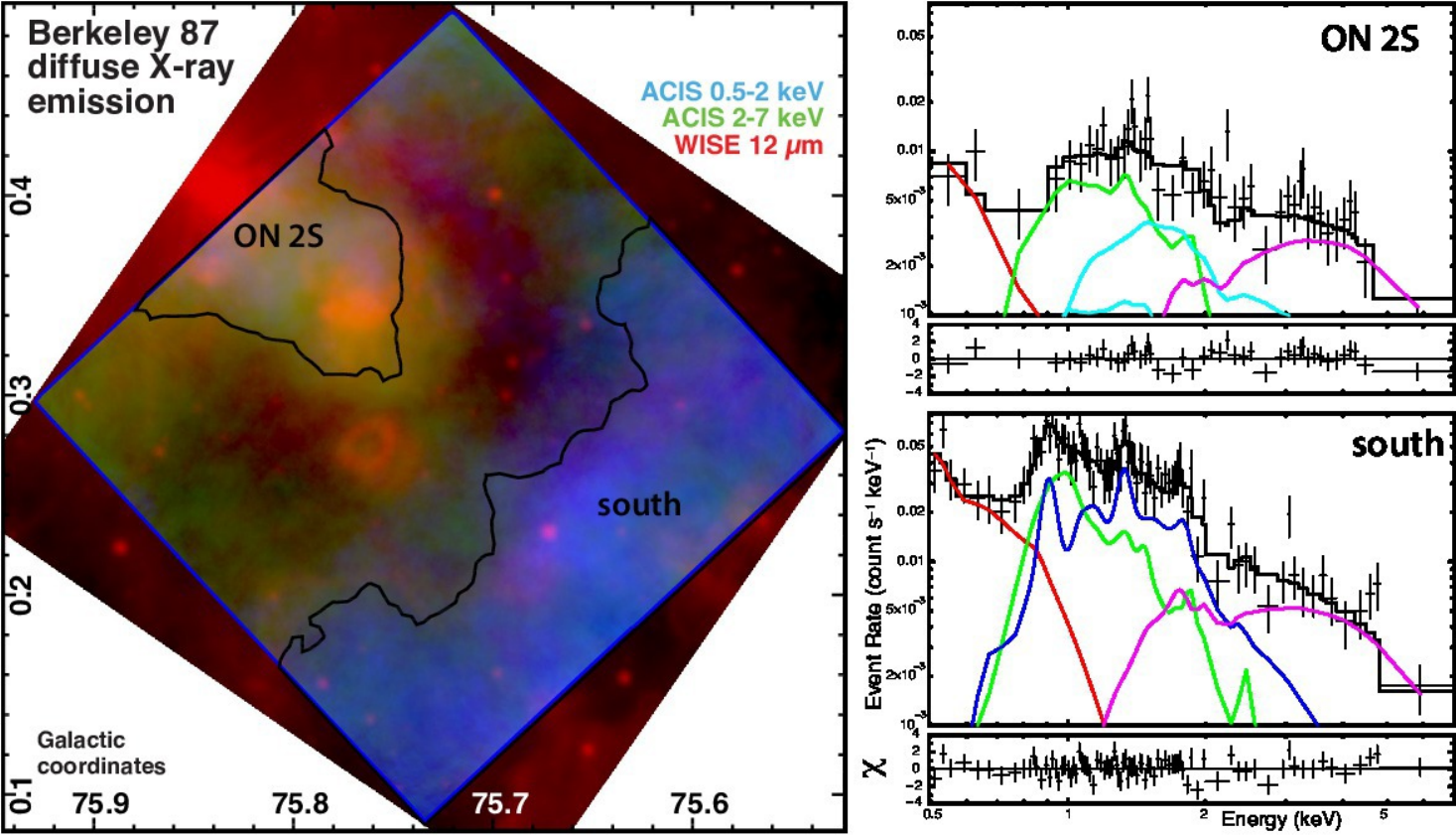}
\caption{Characterizing the Berkeley 87 diffuse X-ray emission.  This image shows ACIS soft-band and hard-band diffuse emission in the \WISE context; extraction regions for diffuse spectral fitting are shown in black.  Corresponding spectra are on the right.  Table~\ref{tbl:diffuse_spectroscopy_style2} gives fit parameters for the spectral models.
\label{berk87gal+spectra.fig}}
\end{figure}

\citet{Skinner19} noted a lack of pre-MS stars in ON~2S and attributed this to insufficient \Chandra sensitivity to this distant, obscured MSFR.  We suggest that this absence of a typical pre-MS population also pertains to Berk87, but our experience from other studies (CCCP, MYStIX, MOXC1, MOXC2) argues that insufficient sensitivity is not the culprit here.  Our images of diffuse X-ray emission (Figure~\ref{berk87.fig}(b), Figure~\ref{berk87gal+spectra.fig}) also show that this region is faint, suggesting that there is little hot plasma contained in Berk87 and no large population of unresolved pre-MS stars just below the detection limit.  Thus it appears that Berk87 lacks both a normal X-ray luminosity function and the hot plasma emission expected from its substantial massive star population.  Further investigation is necessary to explain these observations.

We extracted two diffuse emission regions based on contours of the total-band smoothed diffuse image (Figure~\ref{berk87gal+spectra.fig}).  The first of these includes the ON~2S \hii region complex and extends north of it; we assume that all of this emission is associated with ON~2S and use its distance to calculate the area and luminosity of this diffuse emission in Table~\ref{tbl:diffuse_spectroscopy_style2}.  The spectrum is modeled with an unobscured CIE soft thermal plasma, an obscured harder plasma ($kT = 0.7$~keV) with a timescale tending towards CIE, and a pre-MS component whose absorption was set equal to that of the more obscured {\it pshock} component because it was poorly constrained (Table~\ref{tbl:diffuse_spectroscopy_style2}).  The harder diffuse plasma dominates the diffuse surface brightness, which is 2.5 times brighter than that of the unresolved stars.

The ``south'' region encompasses a large swath of diffuse emission across the southern third of the field.  It samples the complicated Cygnus~X ISM and is modeled by a bright single-temperature ($kT = 0.9$~keV) {\it pshock} plasma transitioning from NEI to CIE (blue and green spectral components), as we saw in the DR15 diffuse region ``north''.  This plasma is significantly softer than that of DR15's ``north'' diffuse region.  A second very soft, unobscured NEI plasma component (shown in red) is also required for a good fit to the Berk87 ``south'' spectrum, although it contributes minimally to the diffuse surface brightness.  

\begin{figure}[htb]
\centering
\includegraphics[width=0.475\textwidth]{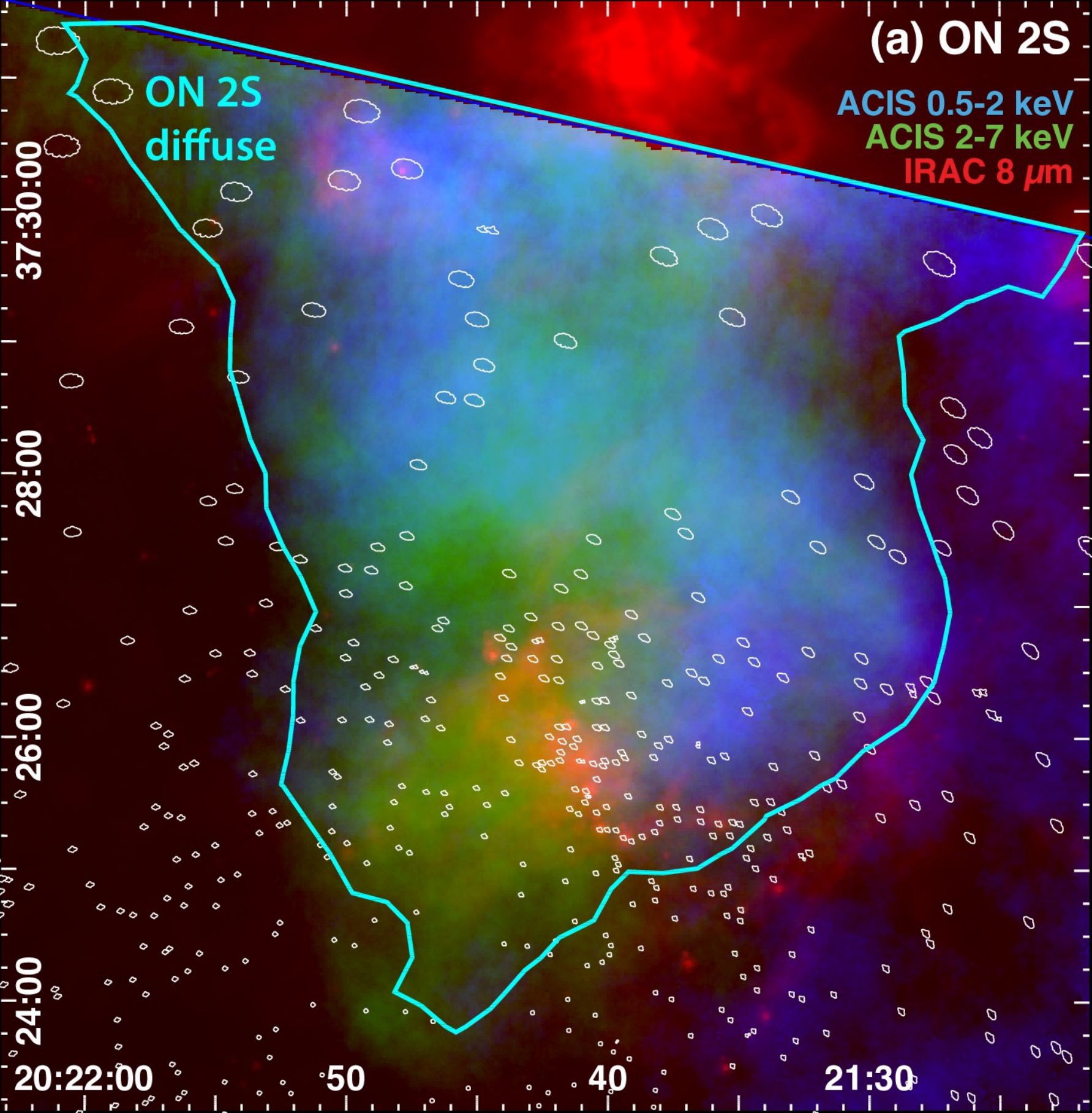}
\includegraphics[width=0.49\textwidth]{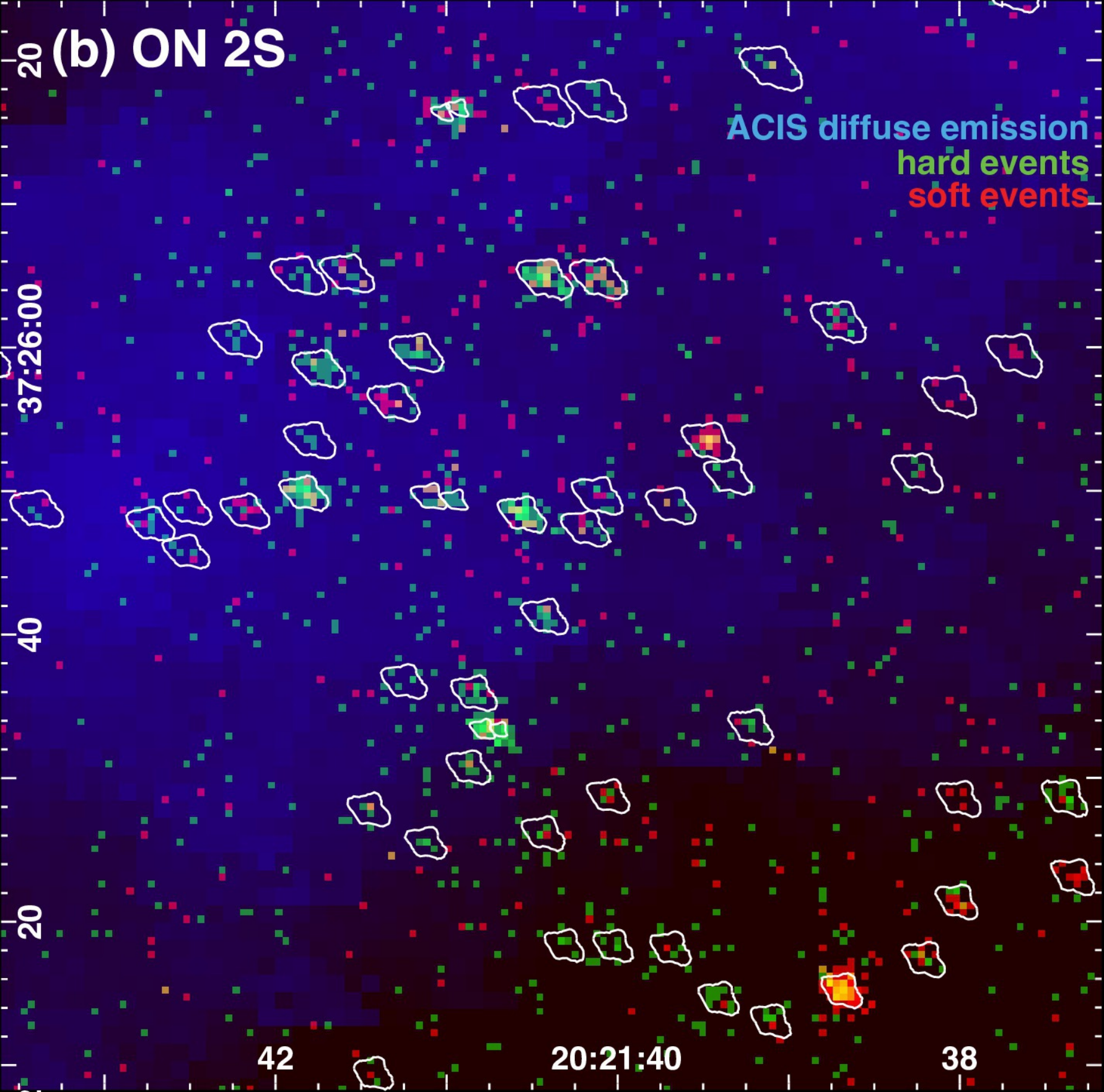}
\caption{ON~2S.
%(a) ACIS diffuse emission in the \WISE and \Spitzer context, for the center of Berkeley~87.
(a) ACIS soft and hard diffuse emission and point source extraction regions in the \Spitzer context, for ON~2S.  The diffuse emission spectral extraction region ``ON2S'' is shown in cyan.
(b) ACIS event data and diffuse emission for the \hii region complex in ON~2S.
\label{on2s.fig}}
\end{figure}

More detail on the ON~2S field is shown in Figure~\ref{on2s.fig}, for comparison with results from \citet{Oskinova10} and \citet{Skinner19} on the same region.  We find the diffuse X-ray emission to be softer and extended over a larger area than described in these earlier studies.  It appears to be more obscured around the ON~2S \hii regions; the soft (blue) component is missing near the bottom of our diffuse extraction region shown in Figure~\ref{on2s.fig}(a).  We also find more X-ray point sources near the ON~2S \hii regions (Figure~\ref{on2s.fig}(b)).  We hope to have an opportunity in the future to include the ACIS data on ON~2N in a re-analysis of this region.  This would allow us to characterize the diffuse X-ray emission across the full extent of ON~2 using smaller diffuse extraction regions and to provide a more complete catalog of its X-ray point source population.  Resolving out more point sources, and modeling the remaining unresolved point source population, is the next step in understanding the truly diffuse X-ray emission in this region.

In summary, Berk87 exhibits a wide distribution of X-ray sources with an atypical lack of central concentration and minimal diffuse X-ray emission.  In contrast, ON~2S shows concentrated, obscured X-ray point sources and a wide swath of bright diffuse X-ray emission encompassing the young \hii regions and extending to the north.  Diffuse X-ray emission south of Berk87 suggests that the wider Cygnus~X ISM is a complicated zone of hot, recombining thermal plasmas, perhaps fueled by a long history of supernova activity.  A faint, unobscured, soft diffuse component is present across the field; it could be associated with massive star feedback from the Berk87 cluster, but again the cluster center lacks such diffuse emission.

%\clearpage
%-----------------------------------------------------------------------------
\subsection{NGC~6231 \label{sec:n6231}}
% NGC 6231 -- 3450 point sources; 2 ObsID's in a single pointing.
% SF Handbook review:  Reipurth08 Vol 2 (south) p. 401.
% At D = 1.71 kpc, 4*pi*D^2 = 3.4994027e+44 cm^2.
% Avg A_V is 3.8 mag, so NH=0.6e22.
% Piled source CD-41 11042 (Simbad) is a.k.a. CPD-41 7742 (Sana03, Sana08)
% WR79 (HD 152270) is a WC+O binary.  Simbad coords are 16 54 19.70 -41 49 11.5.
% The brightest source in the field is 165410.06-414930.1 (c2189, piled up), HD 152248, an eclipsing binary (O7Iabf+O7Ib(f) from GOSSS); Sana04.  NOTE that this is 1 of 5 piled sources in NGC~6231!

NGC~6231, at the heart of Sco~OB1, is a rich monolithic cluster at a modest distance, with many massive stars \citep{Reipurth08}.  As such, it has received substantial attention from both \XMM and {\em Chandra}.  A 180-ks observation to monitor the massive central cluster binary HD~152248 came early in the \XMM mission \citep{Sana04}.  These data also resulted in a catalog of 610 X-ray sources \citep{Sana06a} and studies of the X-ray-emitting massive star \citep{Sana06b} and pre-MS \citep{Sana07} populations.

The \Chandra data on NGC~6231 were obtained as part of the GTO program in 2005 (PI S.~Murray).  \citet{Damiani16} found 1613 ACIS sources in these data; they examined the unusual, very hard ACIS spectrum of the WC+O binary WR~79 (HD~152270) and suggest that it is a colliding-wind binary \citep[CWB,][]{Rauw16}.  \citet{Kuhn17a} found 2411 ACIS sources in these data using our MYStIX-era source-finding machinery; they find 2148 probable cluster members.  They use this X-ray-selected population to study the cluster structure and other properties \citep{Kuhn17b}.

%Damiani18 does a very wide survey of Sco OB1 at many wavelengths to talk more globally about the whole complex.  They present a table of Chandra sources in G345.45+1.50 (IRAS 16562-3959).
 
% WR79 is our source 165419.70-414911.4 (c3496).  It is NOT piled up, according to Table~\ref{tbl:diffuse_spectroscopy_style2} above.  Looks like it might be variable in Obs 6291.  There is some faint diffuse emission south of this source, but nothing right around it.  It is not at the center of the cluster.

\begin{figure}[htb]
\centering
\includegraphics[width=0.49\textwidth]{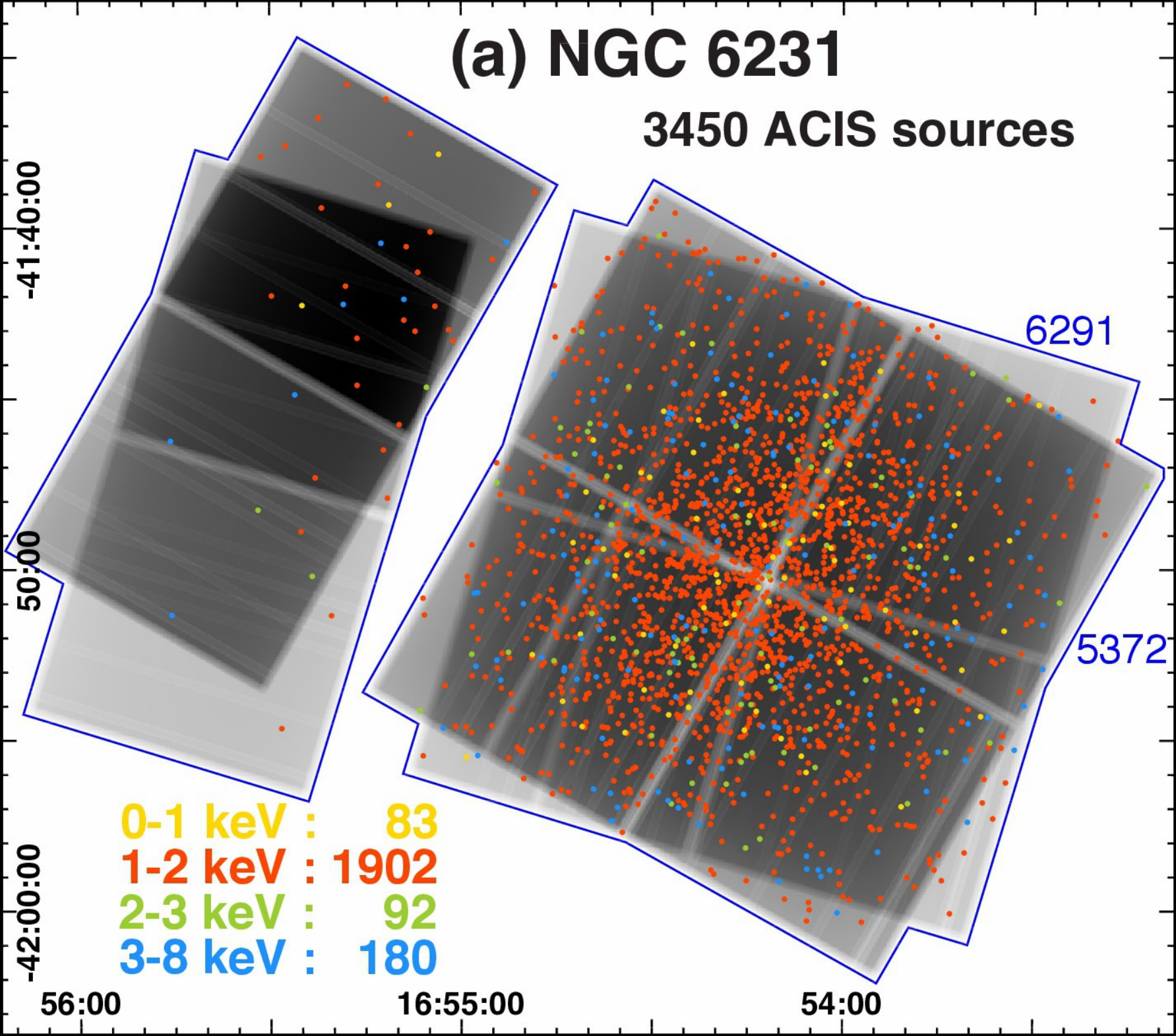}
\includegraphics[width=0.49\textwidth]{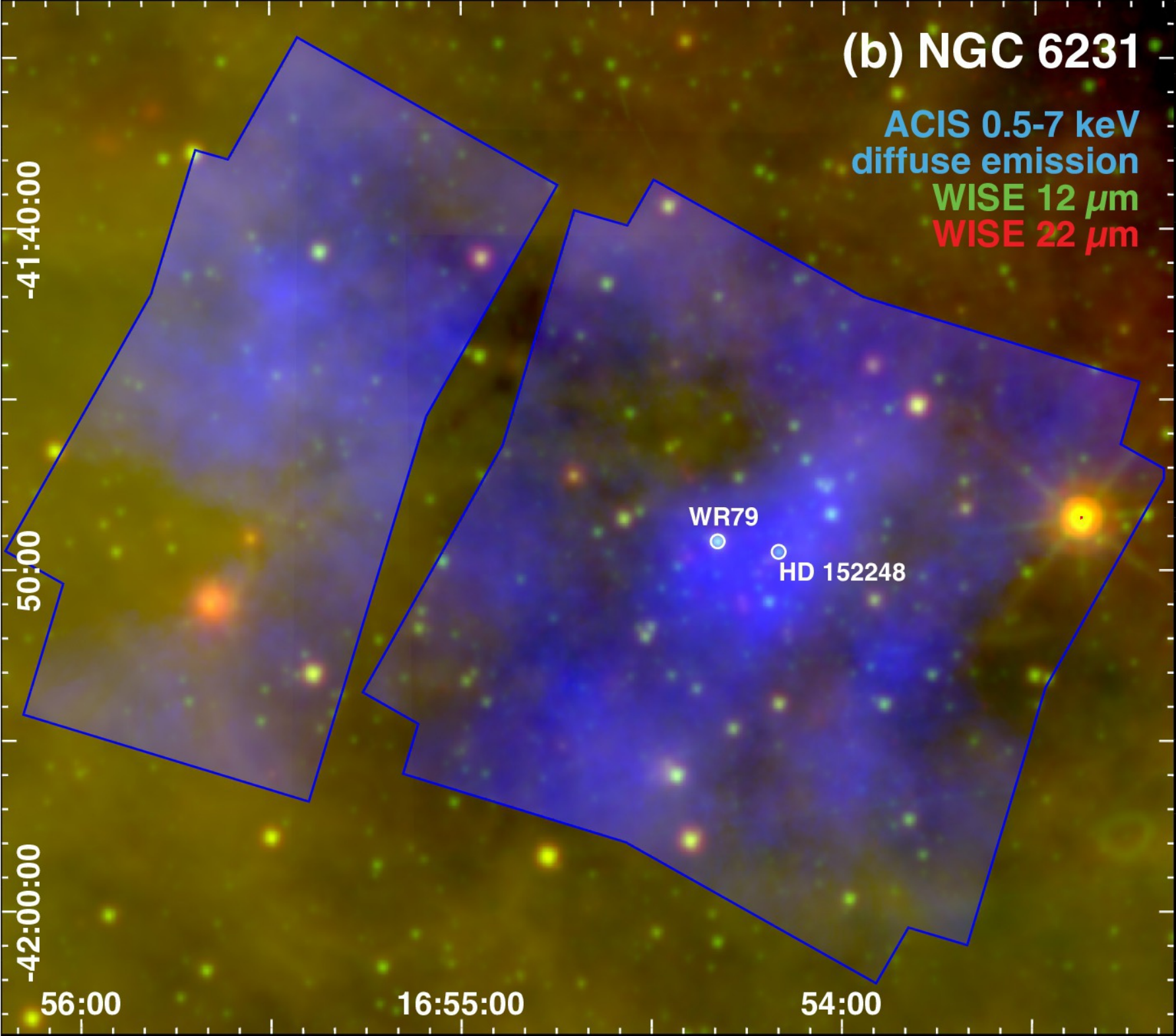}
\caption{NGC~6231.
(a) ACIS exposure map with 2257 brighter ($\geq$5 net counts) ACIS point sources overlaid; colors denote median energy for each source.  ObsID numbers are shown in blue.
(b) ACIS diffuse emission in the \WISE context.  Objects mentioned in the text are marked. 
\label{ngc6231.fig}}
\end{figure}

We re-analyzed the \Chandra data (Figure~\ref{ngc6231.fig}) primarily to search for diffuse emission from massive star feedback.  We find 3450 point sources and highly structured diffuse X-ray emission.  The piled-up massive stars in NGC~6231 (Table~\ref{pile-up_risk.tbl}) are confirmed or likely binaries \citep{Sana08}; they were discussed in studies of the \XMM data \citep{Sana04,Sana06b} so we do not detail them here.  The brightest diffuse emission lies at the cluster center.

%There is a bright source on the S-array devices, 165512.20-414630.7.  Simbad calls this ``[SGR2006] 603 -- X-ray source'' so I should look it up in Sana06.

\begin{figure}[htb]
\centering
\includegraphics[width=0.6\textwidth]{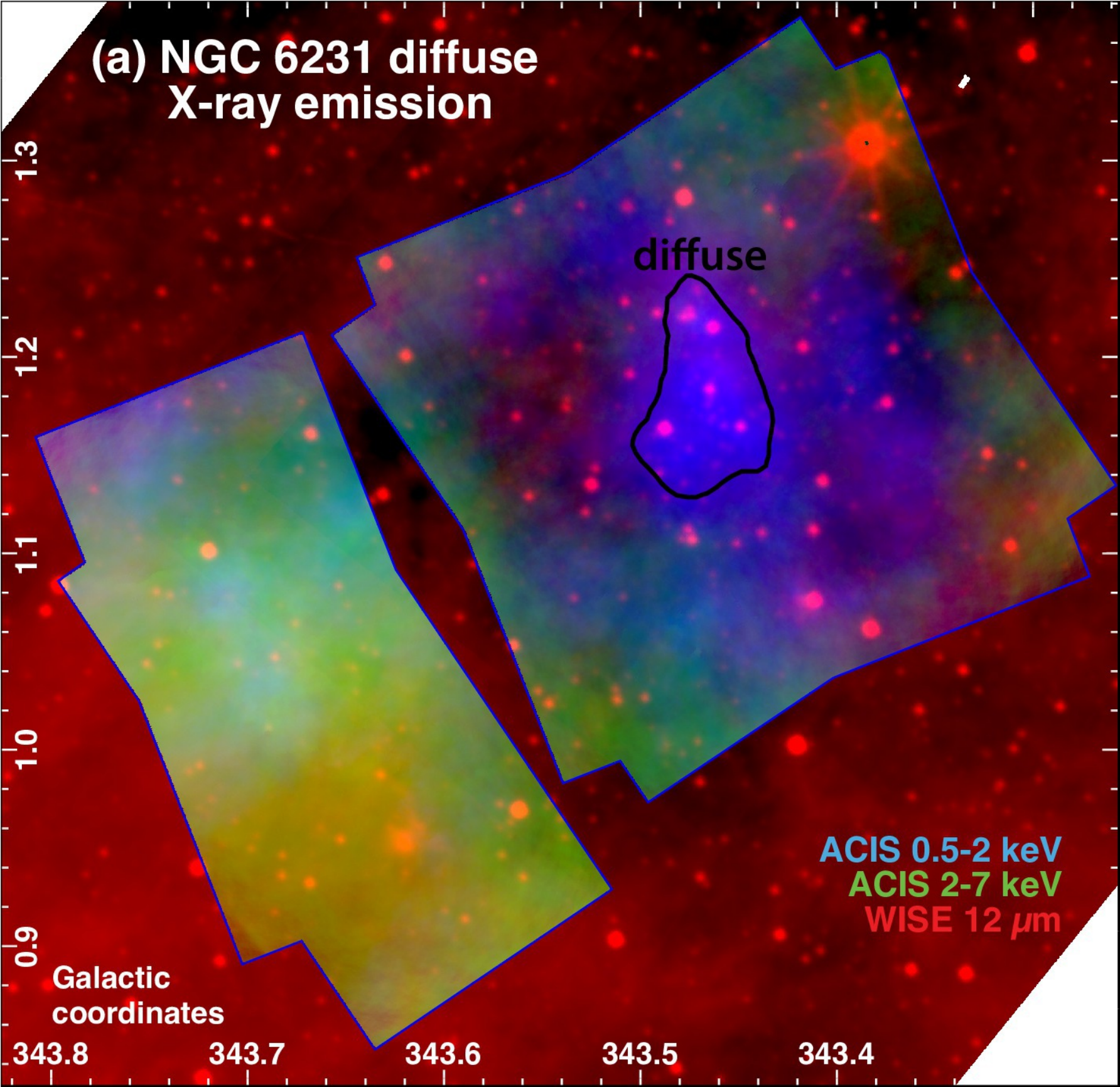}
\parbox[b]{0.39\textwidth}{\hspace{0.15in}
\includegraphics[width=0.34\textwidth]{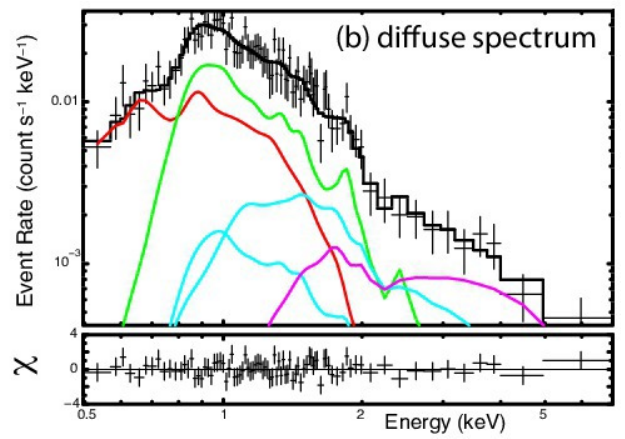}\\
\includegraphics[width=0.39\textwidth]{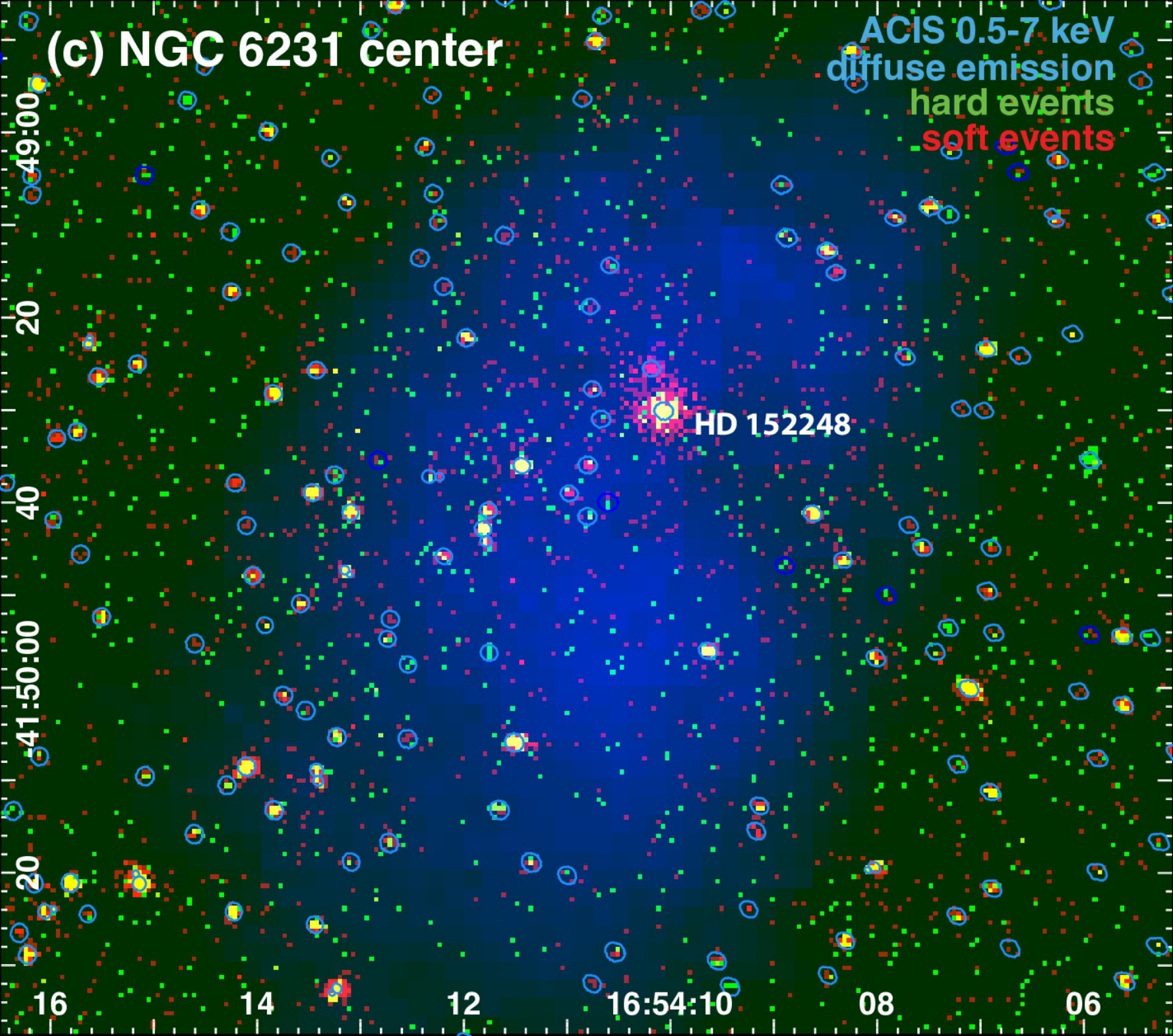}}
\caption{NGC~6231.
(a) ACIS soft-band and hard-band diffuse emission in the \WISE context; the extraction region for diffuse spectral fitting is shown in black.  
(b) The diffuse spectrum from the center of NGC~6231.  Table~\ref{tbl:diffuse_spectroscopy_style2} gives fit parameters for the spectral model. 
(c) ACIS event data and diffuse emission for the central $\sim$2$\arcmin$ of NGC~6231, with 126 point sources.  Dark blue apertures denote ``occasional'' sources.  The diffuse extraction region is much larger than this field.  
\label{ngc6231+spectra.fig}}
\end{figure}

We extracted a diffuse spectrum from the central part of the cluster (Figure~\ref{ngc6231+spectra.fig}) and found substantial hot plasma emission there, with unresolved pre-MS stars contributing $<$7\% of the total diffuse surface brightness.  The emission is dominated by a soft ($kT = 0.3$~keV), short-timescale {\it pshock} plasma behind an obscuring column consistent with the average cluster absorption.  A slightly more absorbed, harder ($kT = 0.8$~keV), long-timescale plasma also contributes to the fit, but with only about a quarter the surface brightness of the soft plasma.  Obscuration for the unresolved pre-MS component was set equal to that of the harder plasma because it was not well-constrained.  This soft plasma at the cluster center is likely due to wind shocks from its massive stars.  Figures~\ref{ngc6231.fig}(b) and \ref{ngc6231+spectra.fig}(a) show that diffuse emission pervades the whole ACIS field and has a complex spatial distribution.   Our analysis splits many of the bright X-ray sources near the cluster center into two or more components (Figure~\ref{ngc6231+spectra.fig}(c)).

% Kuhn suggest that there must have been SN activity based on the runaway massive star, but you can get runaways due to dynamical interactions early in the formation history of a cluster as well.

In summary, the massive dense cluster NGC~6231 offers a wealth of X-ray information on massive stars (especially interacting binaries) and on its large pre-MS population.  Our analysis resolves close pairs of X-ray sources near the cluster center, adds 40\% more X-ray sources, and reveals patchy diffuse X-ray emission across the ACIS field.  More spectral fitting of diffuse structures throughout NGC~6231 should be possible, given their apparent surface brightness, and should prove interesting.

%\clearpage
%-----------------------------------------------------------------------------
\subsection{NGC~6357 \label{sec:n6357}}
% NGC 6357 -- 5269 point sources; many ObsID's in a big mosaic.
% SF Handbook review:  Persi & Tapia 2008 Vol 2 (south) p. 490.
% At D = 1.77 kpc, 4*pi*D^2 = 3.7492832e+44 cm^2.
% Avg A_V is 5 mag, so NH=0.8e22.

We have studied the magnificent southern giant \hii region NGC~6357 \citep{Persi08} with \Chandra since early in the mission.  The original ACIS-I GTO observation of its massive cluster Pismis~24 catalogued $\sim$800 X-ray point sources, both in the concentrated cluster and in a distributed population across the wider field \citep{Wang07}.  We presented a wide mosaic of six ACIS-I pointings in MOXC1, documenting the X-ray sources in two additional rich massive clusters in the complex along with an even more widely dispersed population across NGC~6357's giant molecular cloud (GMC), with a total of $>$3100 X-ray point sources.  MOXC1 also found bright, spatially complex diffuse X-ray emission across the ACIS-I mosaic, filling IR bubbles and threading through cavities in the GMC's highly disrupted ISM.  Several MYStIX science papers provide detailed studies of the three main clusters in NGC~6357; again please see \citet{Feigelson18} for a review.

\begin{figure}[htb]
\centering
\includegraphics[width=0.8\textwidth]{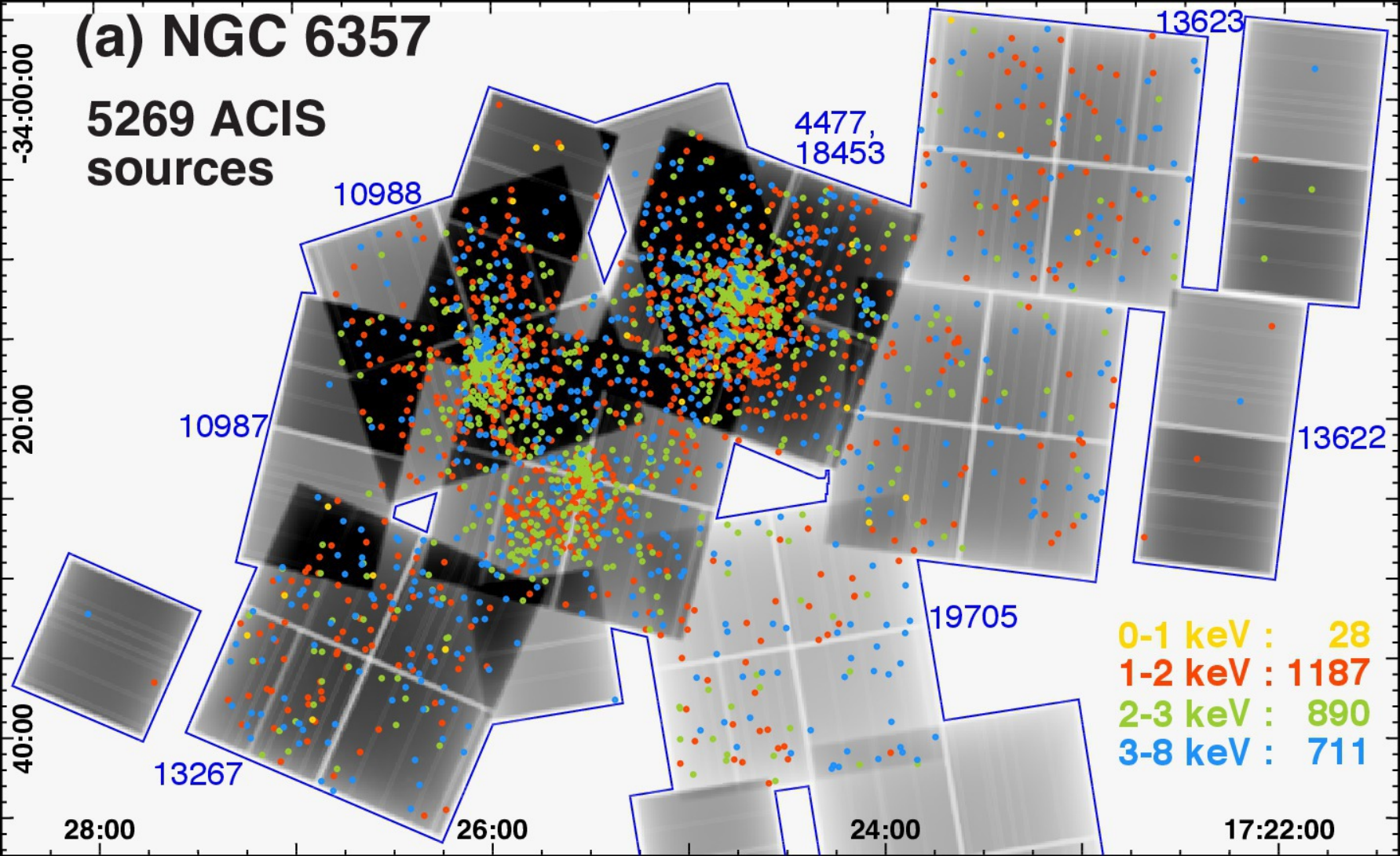}
\includegraphics[width=0.8\textwidth]{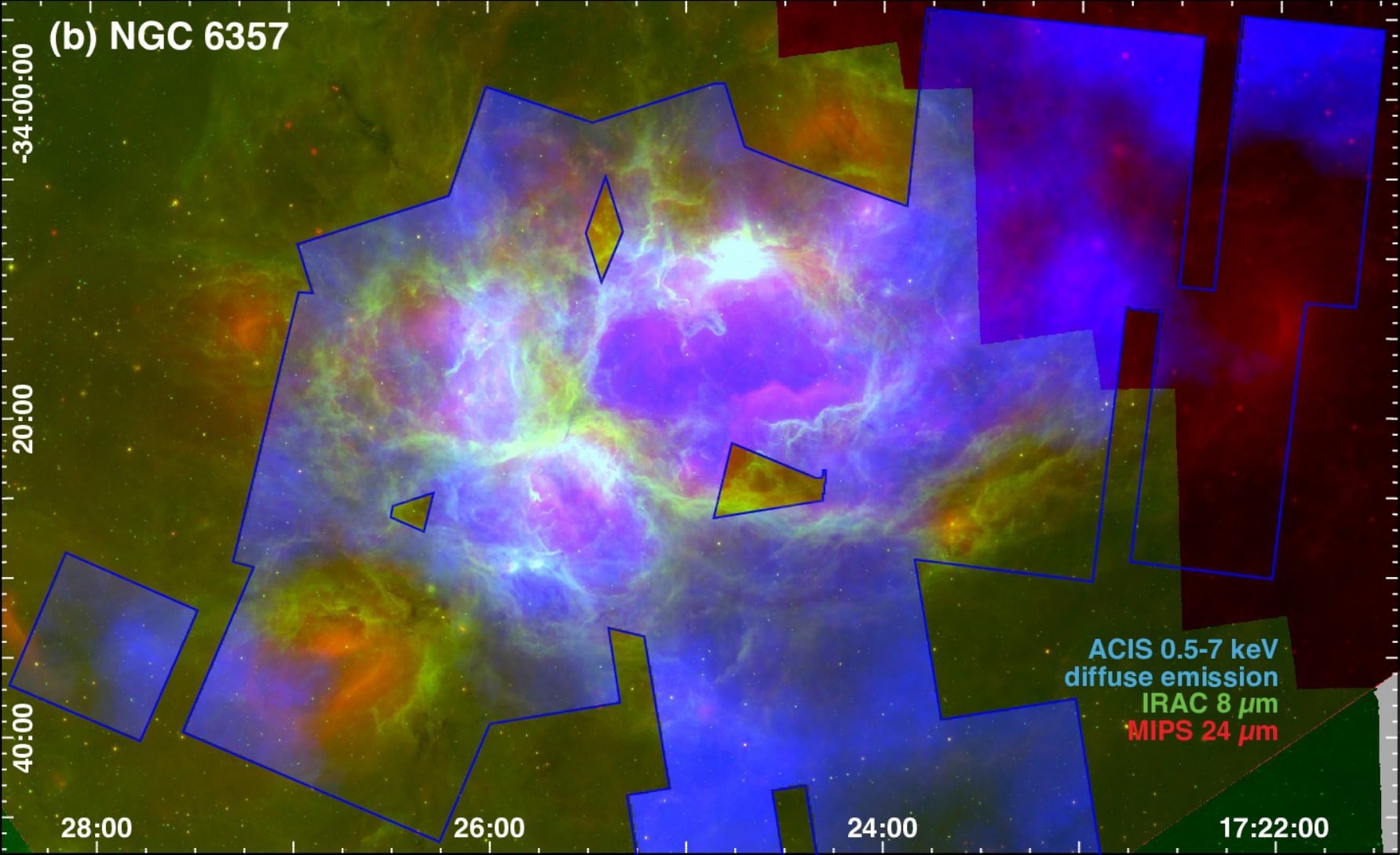}
\caption{NGC~6357.
(a) ACIS exposure map with 2816 brighter ($\geq$5 net counts) ACIS point sources overlaid; colors denote median energy for each source.  ObsID numbers are shown in blue.  Exposure map values for ObsID~19705 are reduced compared to other ObsIDs with the same exposure time due to increasing hydrocarbon contamination on the ACIS optical blocking filters.
(b) ACIS diffuse emission in the \Spitzer context.  The IRAC 8~$\mu$m data do not extend as far west as the ACIS data.
\label{ngc6357.fig}}
\end{figure}

Here we further extend our \Chandra studies of NGC~6357 (Figure~\ref{ngc6357.fig}), adding new ACIS GTO data on Pismis~24 (ObsID~18453) and the interface between NGC~6357 and the G352 giant molecular filament (ObsID~19705).  We include the off-axis ACIS-S CCDs for all seven ACIS-I pointings, to expand our portrait of diffuse X-ray emission in the complex.  In a separate \Chandra Large Project, we will study X-ray emission from the filament connecting NGC~6357 with its twin giant \hii region NGC~6334 \citep{Russeil13}.  The northernmost ACIS-I pointing from that study and its diffuse emission are partially imaged at the bottom of the NGC~6357 mosaics in Figure~\ref{ngc6357.fig}.  X-ray point sources are not shown; they will be included in a separate catalog.  

The CWB WR~93 (HD~157504) shows variability within ObsID~18453 and between ObsIDs.  In addition to the piled-up observations listed in Table~\ref{pile-up_risk.tbl}, it was observed 11.5$\arcmin$ off-axis in ObsID~10988, where it did not pile up; this is a good example of the value added by analyzing the off-axis S-array CCDs in ACIS-I data.  The spectral fit for this ObsID gives $N_{H} = 2.9 \times 10^{22}$~cm$^{-2}$, $kT1 = 0.8$~keV, $kT2 = 2.5$~keV, and $L_{X} = 1.7 \times 10^{33}$~erg~s$^{-1}$ (log $L_{X}$ = 33.24).  Thus the luminosities were the same in ObsIDs 4477 and 10988 but the source was twice as bright in ObsID~18453; the plasma temperatures were similar in all ObsIDs but the absorptions changed.

% *** Could discuss spfits to other piled srcs -- HD 319718 (O3.5If*), [N78] 49 (O5.5IV(f)), 3 2MASS srcs (all 3 have hard spectra and high LX's -- are they IMPS?). ***

The seven-pointing ACIS-I mosaic in Figure~\ref{ngc6357.fig} yields $>$5200 X-ray point sources.  Excising this large point source population and expanding the mosaic with off-axis CCDs shows that the diffuse X-ray emission generated by massive star feedback in NGC~6357 extends far beyond its three massive young clusters (Figure~\ref{ngc6357.fig}(b)).  
 
\begin{figure}[htb]
\centering
\includegraphics[width=0.525\textwidth]{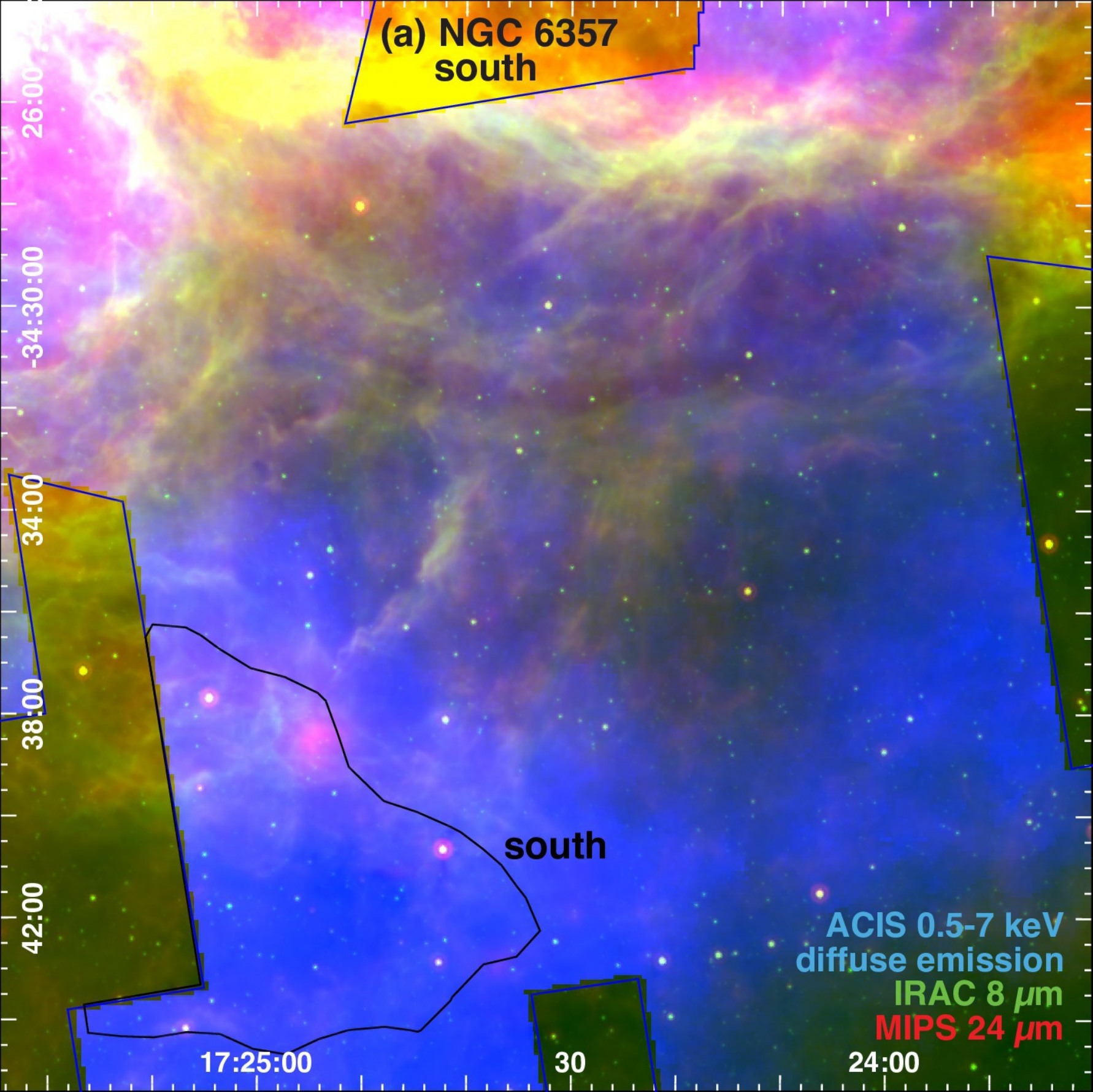}
\includegraphics[width=0.465\textwidth]{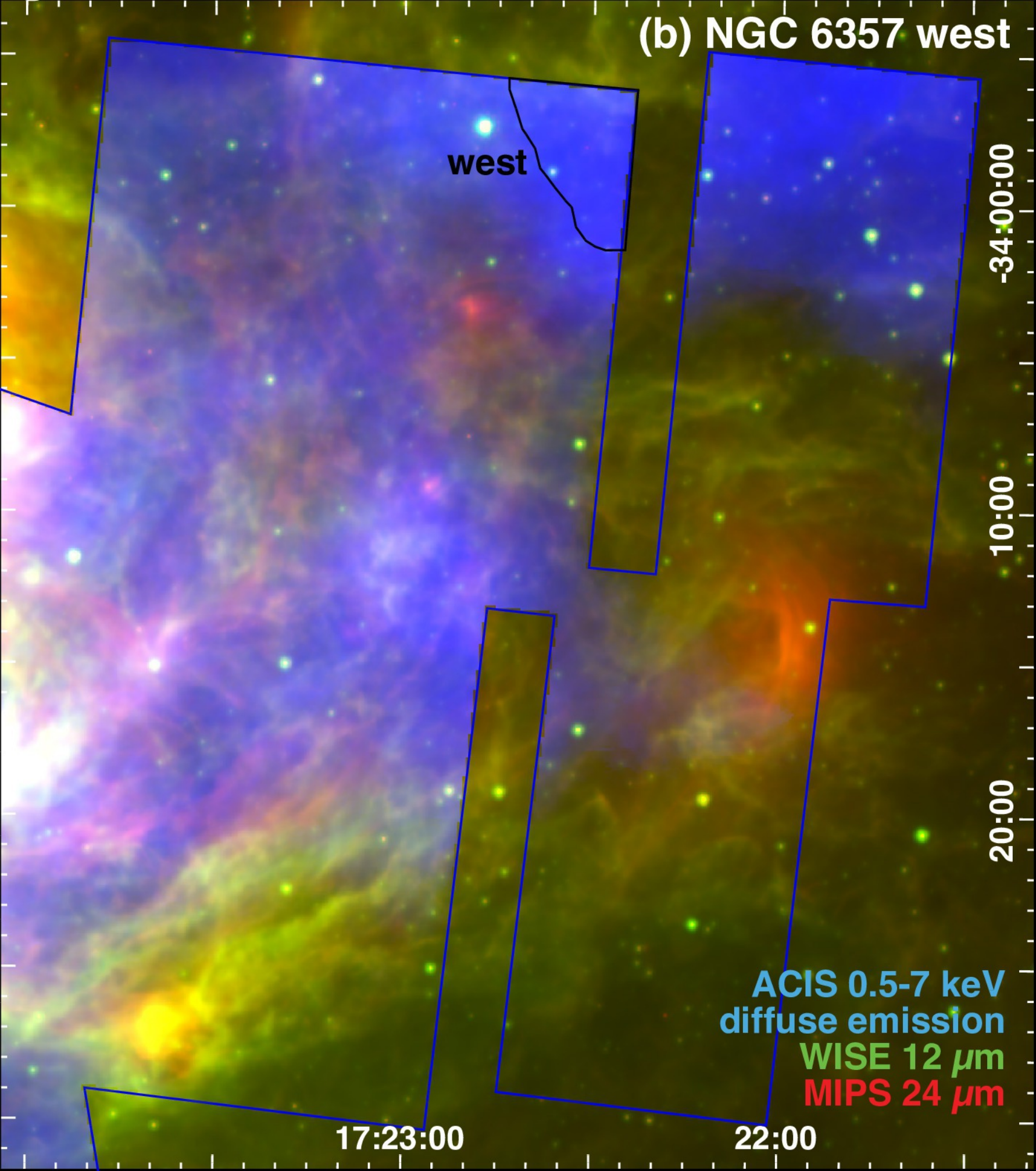}
\caption{Detailed views of NGC~6357.
(a) ACIS diffuse emission in the \Spitzer context, for the new ACIS-I pointing G352.841+0.720 at the southern edge of NGC~6357, where it transitions from a giant \hii region to the more quiescent G352 giant molecular filament.
(b) ACIS diffuse emission in the \WISE and \Spitzer context, for our westernmost ACIS pointings (IRAC data are incomplete for this field).  The off-axis S2 and S3 CCDs are included now, extending the reach of our diffuse X-ray image.
\label{ngc6357outskirts.fig}}
\end{figure}
 
Figure~\ref{ngc6357outskirts.fig}(a) zooms in on the new pointing at the southern edge of NGC~6357.  Rather than fading away, as might be expected at the edge of the complex, the diffuse X-ray emission here is bright.  Similar bright diffuse emission is seen at the northwestern edge of the ACIS mosaic (Figure~\ref{ngc6357outskirts.fig}(b)), where an IR arc opening to the northwest appears to be filled with hot plasma.

\begin{figure}[htb]
\centering
\includegraphics[width=0.85\textwidth]{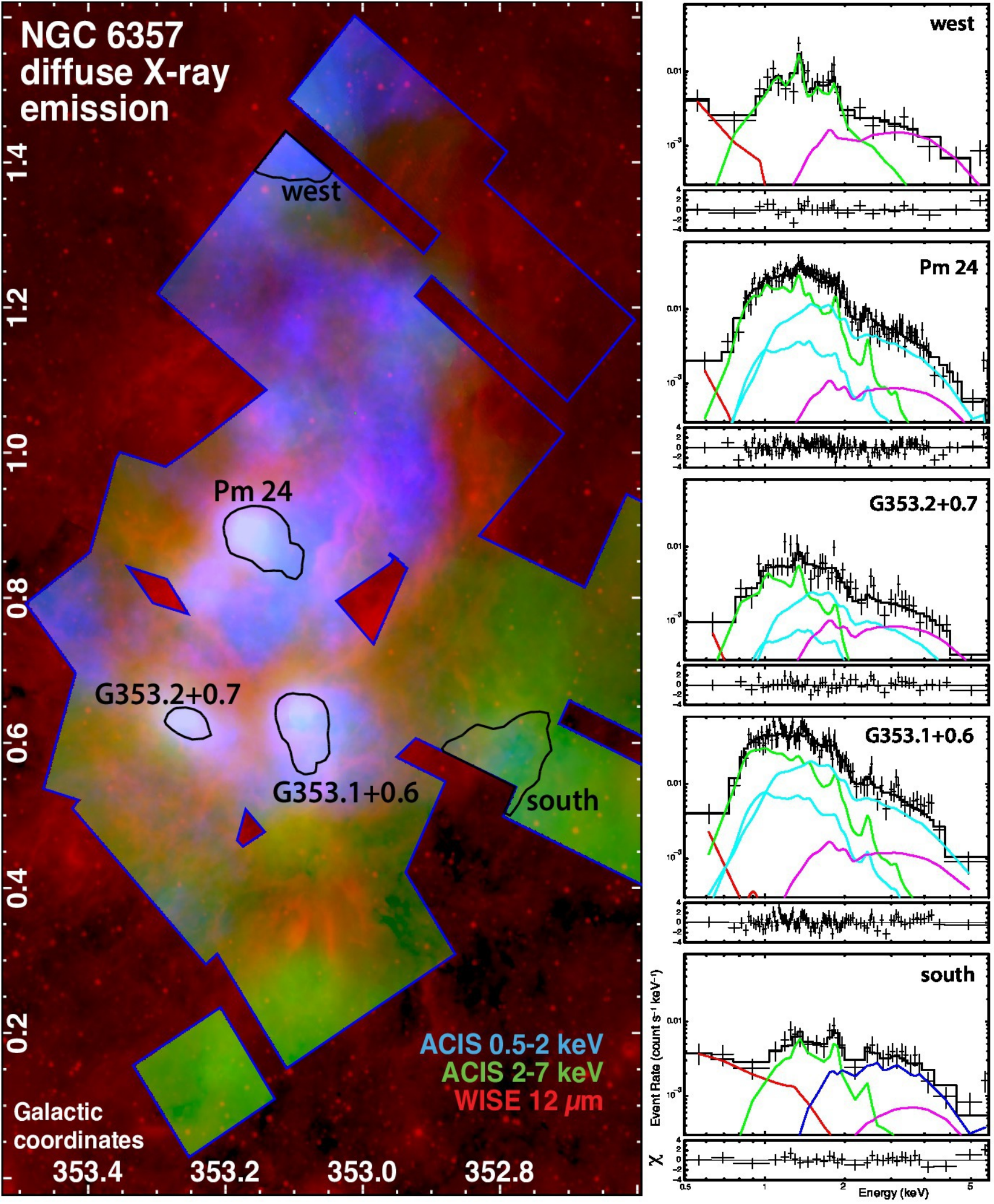}
\caption{Characterizing diffuse X-ray emission across NGC~6357.
This image shows ACIS soft-band and hard-band diffuse emission in the \WISE context; extraction regions for diffuse spectral fitting are shown in black.  Corresponding spectra are on the right; axis ranges are the same for all spectra.  Table~\ref{tbl:diffuse_spectroscopy_style2} gives fit parameters for the spectral models.
\label{ngc6357gal+spectra.fig}}
\end{figure}

Examining the soft-band and hard-band diffuse X-ray emission separately, Figure~\ref{ngc6357gal+spectra.fig}(a) shows a strong gradient in the spectrum of the diffuse X-ray emission as we move down the ACIS mosaic.  The upper part of the field samples soft emission in the wide ``bowl'' structure expanding away from the GMC and the Plane, while the lower part of the field samples the G352 GMC and its increasingly higher obscuration.  

We have selected just a few example diffuse emission regions for spectral fitting (based on their interesting locations and high apparent total-band surface brightnesses); these are outlined in Figure~\ref{ngc6357gal+spectra.fig}(a), with the extracted spectra shown on the right.  Starting from the top, the ``west'' region's spectrum is dominated by a medium-temperature plasma ($kT = 0.8$~keV) with an intermediate ionization timescale.  Its obscuration is about twice the average value for NGC~6357.  No unresolved stellar component is required.  This region has the highest diffuse surface brightness in NGC~6357, and the second highest for all of MOXC3.  

The diffuse emission around Pismis~24 is not particularly well-fit by our spectral model, so its parameters should be interpreted with caution.  With that caveat in mind, it appears that the diffuse emission comes primarily from a {\it pshock} plasma with intermediate temperature ($kT = 0.7$~keV) and ionization timescale.  Its obscuration is about twice the average for NGC~6357.  This is augmented by an unresolved pre-MS component; its obscuring column was set equal to that of the diffuse plasma.  The diffuse plasma's surface brightness is more than twice that of the unresolved stars.  This spectrum appears to show several line-like residuals, including a puzzling high-energy line at $6.80 \pm 0.06$~keV.  Adding a single gaussian at that energy improves the reduced $\chi^{2}$ from 1.33 to 1.24.  
% *** I have no idea what this line is from! ***

The other two young clusters in NGC~6357 also show diffuse X-ray emission.  G353.2+0.7 displays an absorbed CIE soft ($kT = 0.4$~keV) plasma with unresolved pre-MS stars accounting for the hard part of the spectrum; the diffuse plasma has 3.6 times the surface brightness of the unresolved stars.  G353.1+0.6 is similar, but shows a harder CIE plasma ($kT = 0.8$~keV) with only slightly higher surface brightness than the unresolved pre-MS component.  The G353.1+0.6 spectrum also shows line-like residuals.

The ``south'' diffuse spectrum is the most complicated in MOXC3, requiring three {\it pshock} plasmas with different absorptions, temperatures, and ionization timescales.  The first of these is minimally obscured and fairly soft ($kT = 0.3$~keV, shown in red) and tends toward NEI.  The second is substantially obscured and has an intermediate temperature ($kT = 0.7$~keV, shown in green) and ionization timescale.  The third is a hard CIE plasma ($kT = 2.4$~keV, shown in blue) with heavy obscuration.  The intermediate plasma is brightest, but the other two contribute substantially to the total surface brightness in this diffuse region.  As expected, since this region is far from the young clusters, no unresolved pre-MS component is required in the fit.  This diffuse X-ray emission is in the vicinity of the X-ray-bright SNR G352.7-0.1 \citep{Pannuti14} in the G352 GMC, so it may reflect additional supernova activity in the region.

Clearly we have not yet discovered the full extent of the X-ray emission in NGC~6357, or seen the full impact of G352's massive star feedback on the surrounding ISM.  A vast degree-sized ``bowl'' sculpted in heated dust and imaged in the mid-IR continues to the north of our ACIS mosaic; the bright diffuse X-ray emission in our ``west'' region barely samples it.  Future ACIS pointings there would certainly yield a rich example of the distributed point source population and document the tortured paths of the hot gas trying to escape this vast cluster of clusters.  To the south, diffuse X-ray emission is surprisingly bright and hard; supernova activity may be more extensive there than has been appreciated.  Again a wider ACIS mosaic could explore this.

%\clearpage
%-----------------------------------------------------------------------------
\subsection{AFGL 4029 \label{sec:afgl4029}}
% AFGL 4029 -- 833 sources in a single ACIS-I ObsID.
% SF Handbook review:  Megeath et al. article on W3/W4/W5 2008, Vol 1 (north).
% At D = 2.25 kpc, 4*pi*D^2 = 6.0585228e+44 cm^2.  
% No piled up sources in this target.
% Avg A_V is 3 mag, so NH=0.5e22.
% HD 18326 is at 02 59 23.17 +60 33 59.5, from Simbad (l,b = 138.026 +01.500).
% Koenig08 Table 6 lists AGN in W5.
% Deharveng97 star 25 (ionizing IRS1) is at ~ J2000 03 01 31.27 +60 29 13.0
% Deharveng97 star 26 (ionizing IRS2) is at ~ J2000 03 01 34.24 +60 29 13.8  

The embedded MSFR AFGL~4029 sits on the eastern edge of the large \hii region W5-E \citep{Megeath08}, which is ionized by the O6.5V+O9V binary HD~18326 (BD+59~0578) \citep{Sota14}.  It is one of many sites of recent star formation in the vast Outer Galaxy GMC complex W3/W4/W5 \citep{Koenig08}; our \Chandra analysis of W3 and W4 was presented in MOXC1.  \citet{Cantat18} report a Gaia DR2 distance of 2.25~kpc for IC~1848, the OB association powering W5; we assume that AFGL~4029 is at the same distance.  

\begin{figure}[htb]
\centering
\includegraphics[width=0.44\textwidth]{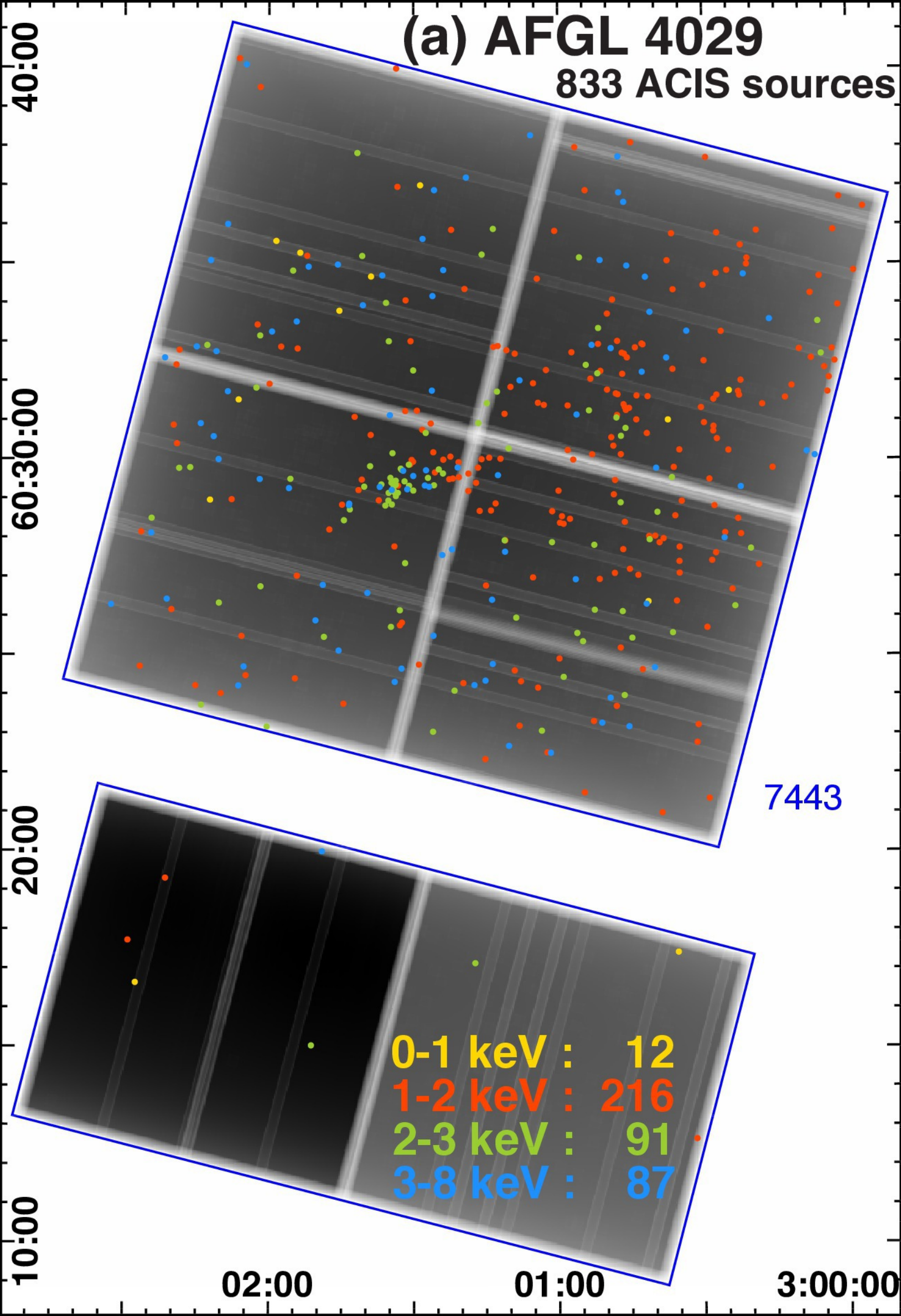}
\includegraphics[width=0.55\textwidth]{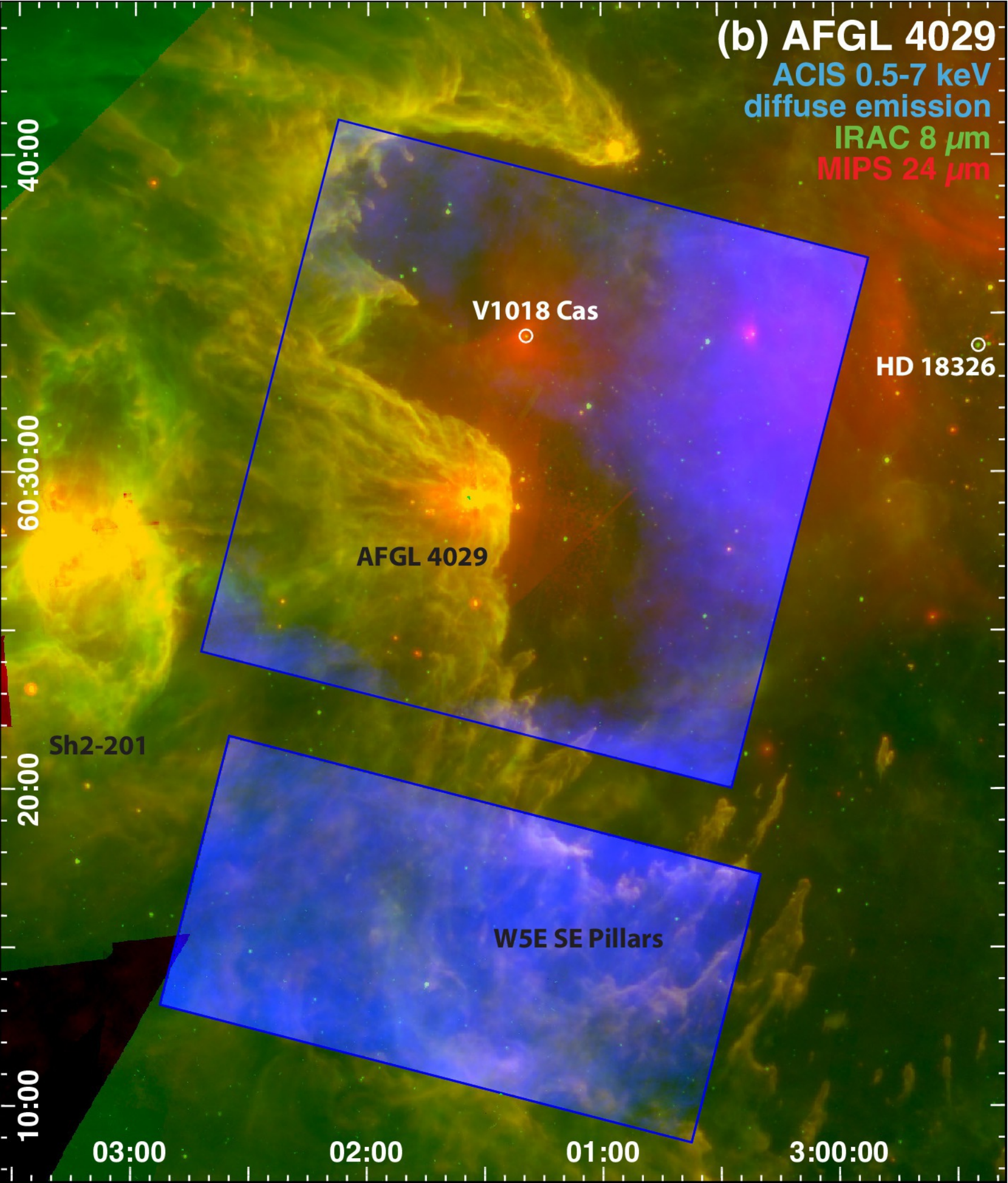}
\caption{AFGL~4029 and W5E.
(a) ACIS exposure map with 406 brighter ($\geq$5 net counts) ACIS point sources overlaid; colors denote median energy for each source.  ObsID numbers are shown in blue.
(b) ACIS diffuse emission in the \Spitzer context.  
\label{afgl4029.fig}}
\end{figure}

This single 82-ks ACIS-I observation yields 833 X-ray point sources in our analysis.  Figure~\ref{afgl4029.fig} shows the brighter half of those sources and diffuse X-ray emission in the wider context of the eastern side of W5-E.  Massive stars are marked in white; the bipolar \hii region Sh2-201 and a large number of pillars pointing towards HD~18326 \citep{Deharveng12} are noted.

Like RCW~108-IR above, AFGL~4029 is a young, obscured cluster at the edge of a large cavity, likely influenced by older star formation and feedback in the vast W5 complex \citep{Megeath08}.  The X-ray sources inside the W5-E cavity (west of the prominent ionization front) are significantly less obscured than those in AFGL~4029 (Figure~\ref{afgl4029.fig}(a)).  Diffuse X-ray emission is found inside the W5-E cavity, but fades near the interface with AFGL~4029's molecular cloud (Figure~\ref{afgl4029.fig}(b)).  This absence of bright diffuse X-ray emission at the cavity edge is unique to W5-E in our experience; such interfaces in other MOXC1, 2, and 3 targets usually show bright diffuse X-rays right up to the bright 8~$\mu$m emission marking the ionization front.  Interestingly, the ACIS-S CCDs in this observation also show bright diffuse emission; this hot plasma must lie in front of the forest of 8~$\mu$m pillars upon which it is superposed, because it is not strongly shadowed by those pillars.

\begin{figure}[htb]
\centering
\includegraphics[width=0.49\textwidth]{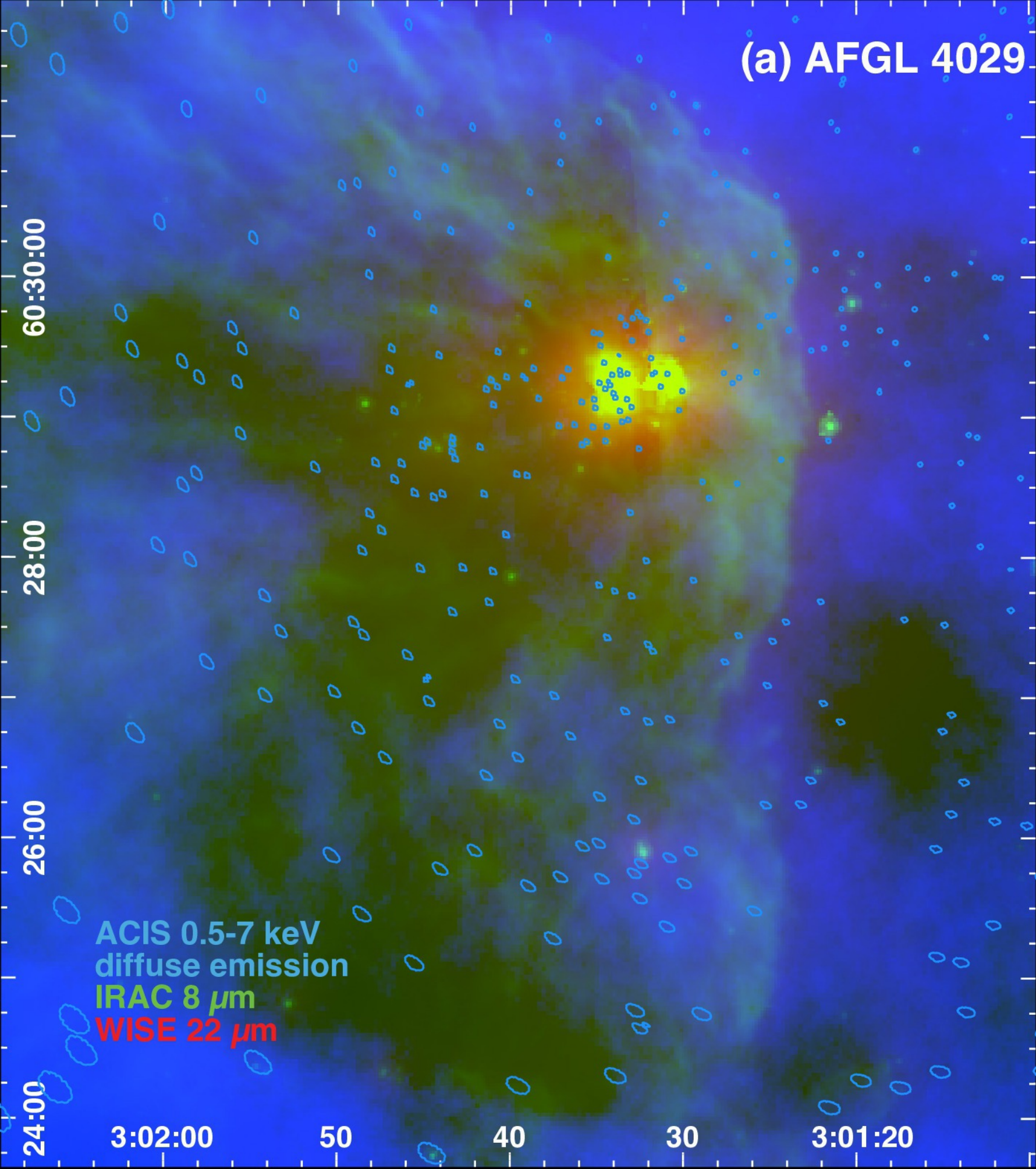}
\includegraphics[width=0.49\textwidth]{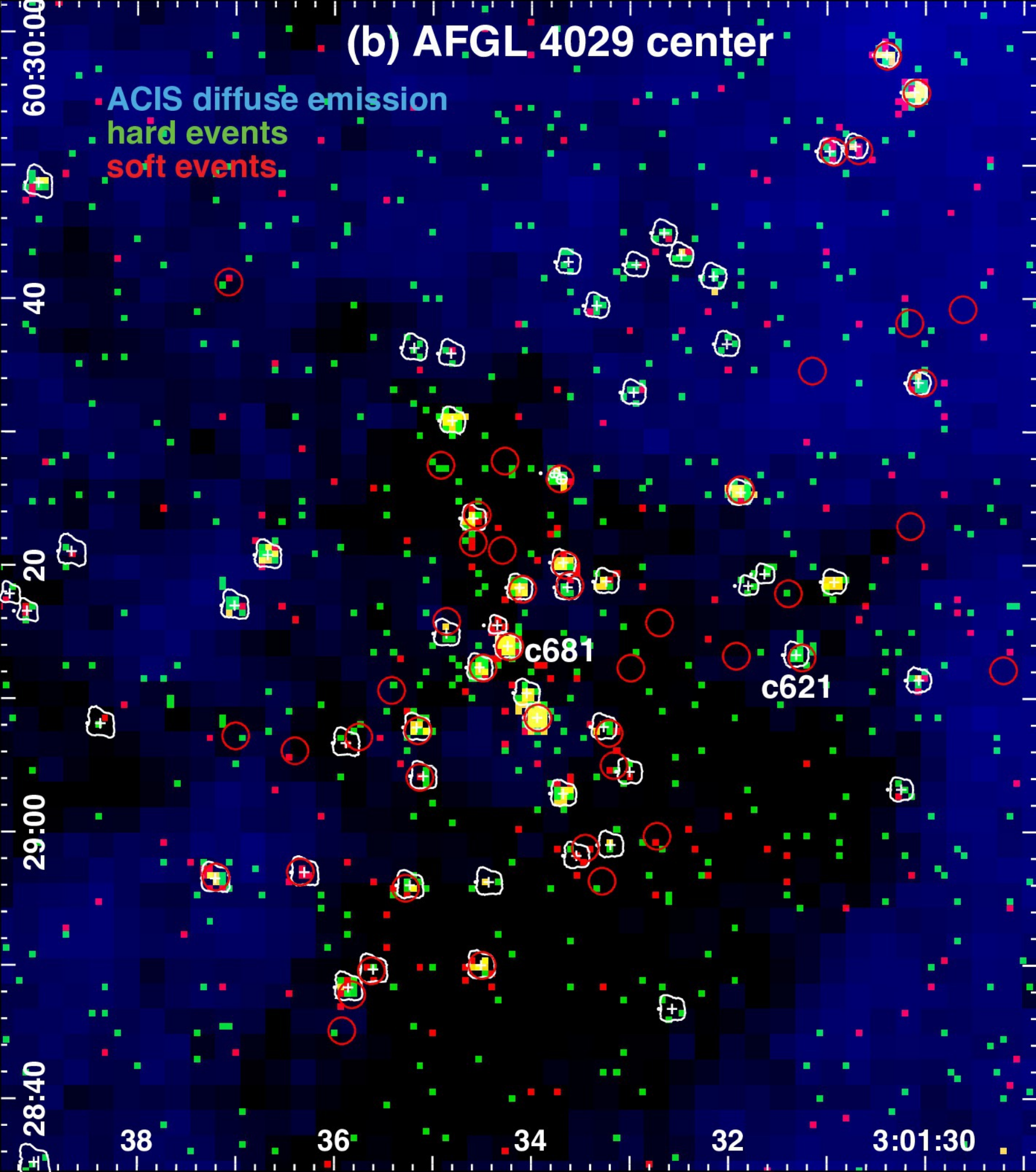}
\caption{Zooming in on AFGL~4029.
(a) ACIS diffuse emission in the \Spitzer and \WISE contexts.  The \Spitzer MIPS data are saturated at the center of AFGL~4029.  We have rescaled the diffuse X-ray emission to show more detail here at the field center.
(b) ACIS event data and diffuse emission for the massive cluster powering AFGL~4029.  Again the diffuse emission has been rescaled to show faint structures around the cluster center.  Red circles mark IR sources from \citet{Deharveng97}; see text.
\label{afgl4029_zoom.fig}}
\end{figure}

Figure~\ref{afgl4029_zoom.fig}(a) zooms in on the embedded cluster; rescaling the faint diffuse X-ray emission there shows that it extends closer to the cluster with a complex, patchy morphology.  The AFGL~4029 obscured cluster sits in a noticeable hole in the diffuse emission.  It appears that hot plasma has yet to escape the confines of the natal cloud that formed AFGL~4029, or it is shadowed by dense material near the cluster and is only detectable a few arcminutes away from the cluster center.  

Continuing to zoom in to the center of AFGL~4029, Figure~\ref{afgl4029_zoom.fig}(b) shows even fainter diffuse emission extending nearly to the heart of the cluster.  It probably includes unresolved pre-MS stars, but it is too faint for spectral fitting.  There are 60 X-ray point sources in this image.  IR sources from \citet{Deharveng97} are shown as red circles; about half have X-ray counterparts.  The brighter X-ray sources have absorptions of $N_{H} \sim 2 \times 10^{22}$~cm$^{-2}$, {\it apec} plasma temperatures of $kT \sim$3--5~keV, and luminosities of $L_{X} \sim 2 \times 10^{31}$~erg~s$^{-1}$.  One of these is CXOU~J030134.24+602913.9 (c681); it is the counterpart of \citet{Deharveng97} ``star 26,'' a B1V star that is the main ionizing source for the compact \hii region IRS2 \citep{DeBuizer17}.  
% It has 94 net counts, a median energy of 2.4~keV and is possibly variable.  A spectral fit gives $N_{H} = 2.1 \times 10^{22}$~cm$^{-2}$, $kT = 2.8$~keV, and $L_{X} = 2.3 \times 10^{31}$~erg~s$^{-1}$.
Our source CXOU~J030131.31+602913.2 (c621) is coincident with \citet{Deharveng97} ``star 25,'' the B0--B3 MYSO ionizing the ultracompact \hii region IRS1.  It has 12 net counts and a median energy of 3.2~keV.  
% Pat tried a CSTAT spectral fit and got NH=24e22, kT=0.8 keV, so HUGE flux correction.

\begin{figure}[htb]
\centering
\includegraphics[width=0.99\textwidth]{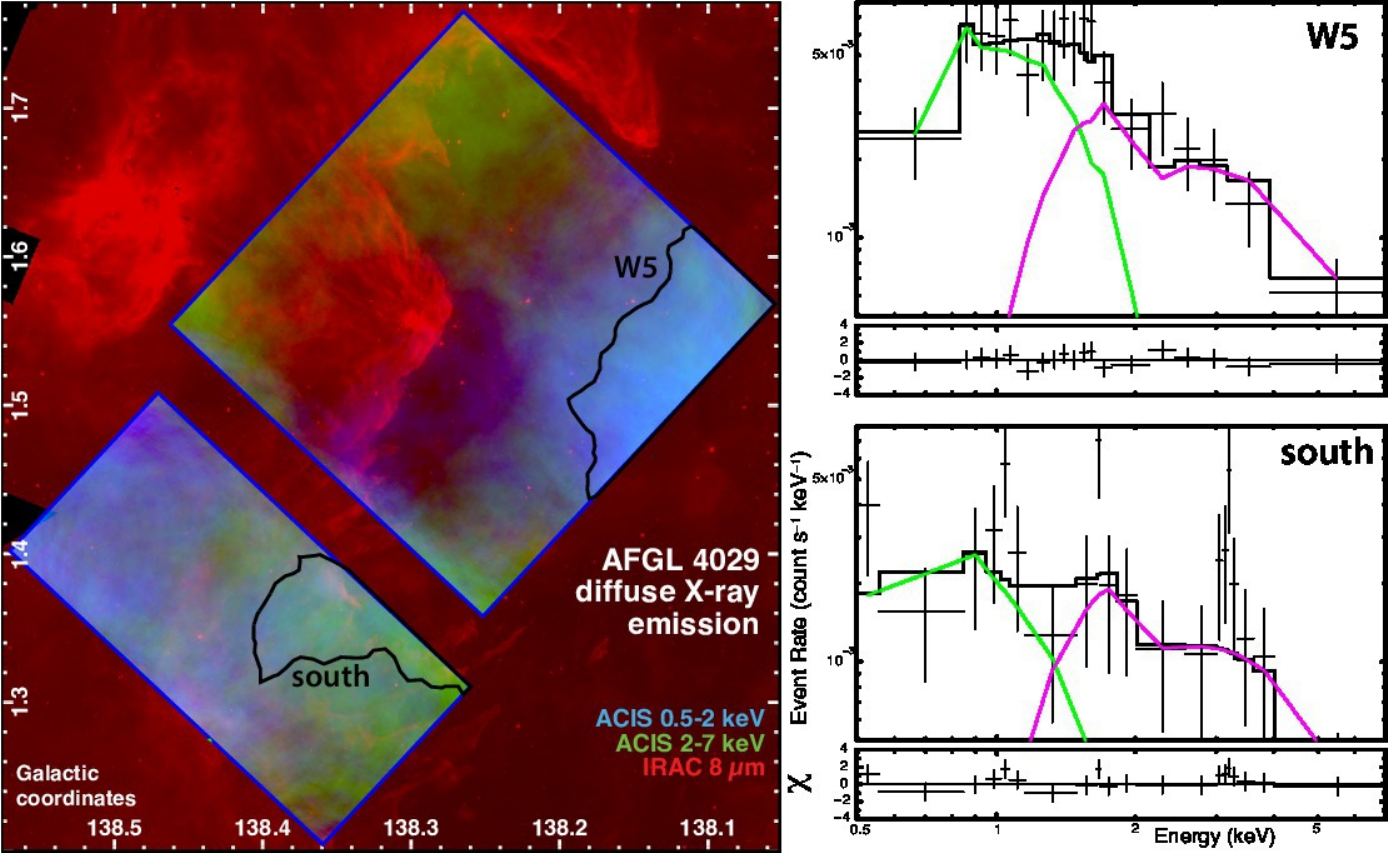}
\caption{Characterizing the AFGL~4029 diffuse X-ray emission.
This image shows ACIS soft-band and hard-band diffuse emission in the \Spitzer context; extraction regions for diffuse spectral fitting are shown in black.  Corresponding spectra are on the right.  Table~\ref{tbl:diffuse_spectroscopy_style2} gives fit parameters for the spectral models.
\label{afgl4029gal+spectra.fig}}
\end{figure}

We extracted two diffuse spectra from this target, one at the edge of the large W5 cavity and one far off-axis in the south, where many pillars are seen in the IRAC data (Figure~\ref{afgl4029gal+spectra.fig}).  Both of these spectra required only a single {\it pshock} component and background; it was not necessary to invoke unresolved pre-MS stars to fit the spectra.  The plasma properties were similar in the two regions:  medium-temperature NEI plasmas with modest to minimal absorption and fairly low surface brightness.  The ``W5'' diffuse region is 2.7 times brighter than the ``south'' region.  All of this diffuse emission may be due to feedback from HD~18326, although the ``south'' spectrum appears to have unmodeled spectral lines (most notably a prominent residual at $\sim$3.2~keV) that may hint at a SNR origin.

This study of a small section of the eastern edge of W5-E is just a minor example of the \Chandra exploration that could be done on W5.  More revealed clusters inhabit the huge W5-E and W5-W cavities and extensive star formation is ongoing all around the periphery of the complex.  All of these regions may well be pervaded by hot plasmas from massive star feedback.  Dense distributions of young stars are found throughout the region \citep{Koenig08}, so \Chandra is needed to separate the tangle of X-ray emission from hot plasmas and young stars.  Using \Herschel data, \citet{Deharveng12} suggest that the W5 complex originated in a massive molecular filament, with MSFRs lining up along the filament axis.  A wide-field ACIS mosaic of W5 would thus make an excellent Perseus Arm counterpoint to CCCP.
\\

%\clearpage
%-----------------------------------------------------------------------------
\subsection{h Per \label{sec:hPer}}
% h Per (NGC 869) -- 3419 sources in a small mosaic.
% SF Handbook review:  none.
% At D = 2.34 kpc, 4*pi*D^2 = 6.5528982e+44 cm^2.  
% The chi Per cluster is NGC 884.  Kosta has 190 ks ACIS-I time on it upcoming in Cycle 20.
% Argiroffi16 says h Per has an age of 13 Myr.
% No piled up sources in this target.
% Avg A_V is 1.9 mag, so NH=0.3e22.

Part of the famous double cluster ``h and $\chi$ Persei,'' h~Per (NGC~869) is the oldest cluster in MOXC3 \citep[13-14~Myr,][]{Currie10,Argiroffi16}.  Although it has lost its most massive stars to supernovae by this age, it retains a large population of B stars \citep{Currie10}.  Like NGC~6231 described above, h~Per is a large, rich cluster with comparatively little obscuration; long \Chandra observations yield a wealth of X-ray-emitting young stars.  

\begin{figure}[htb]
\centering
\includegraphics[width=0.49\textwidth]{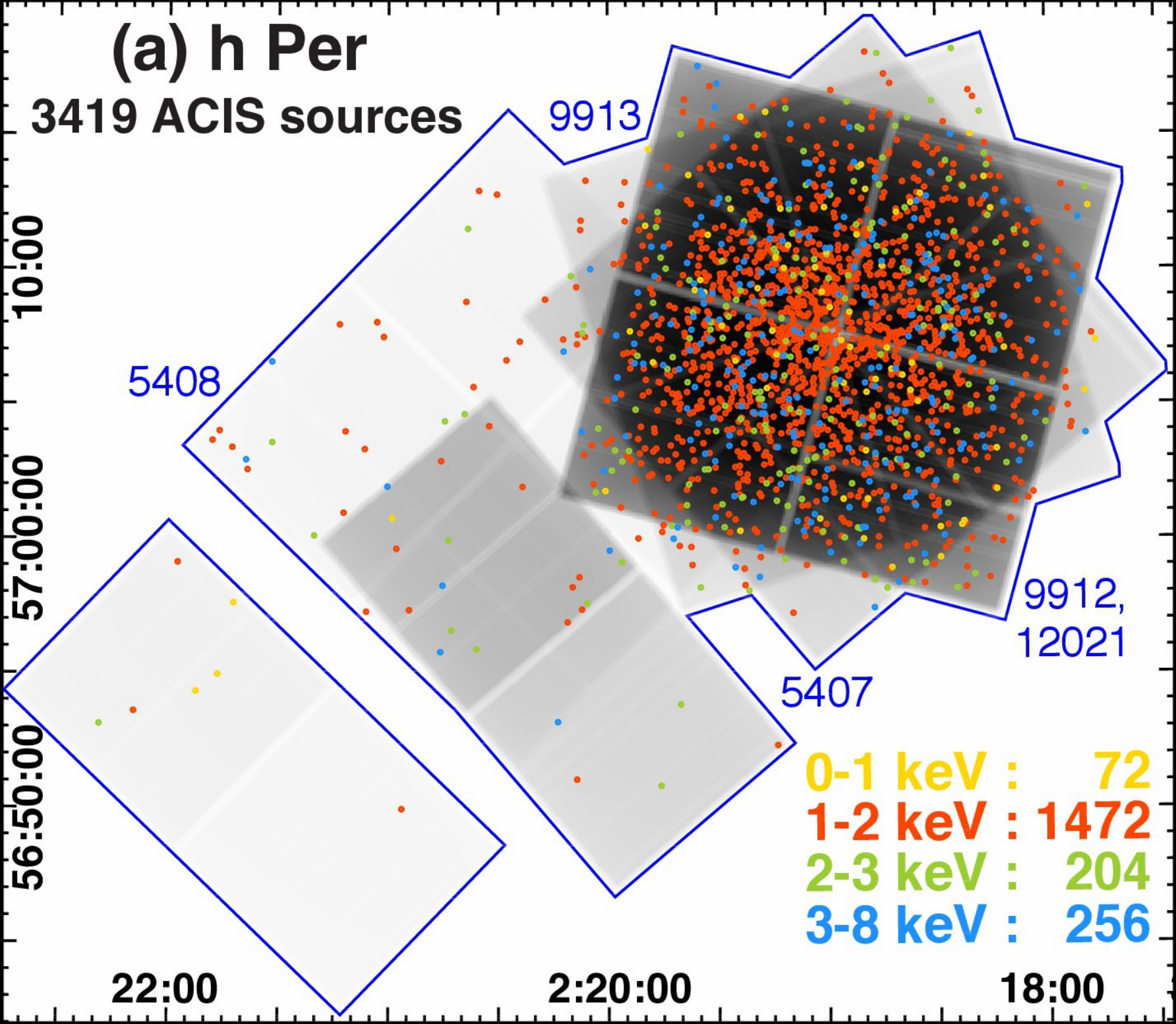}
\includegraphics[width=0.49\textwidth]{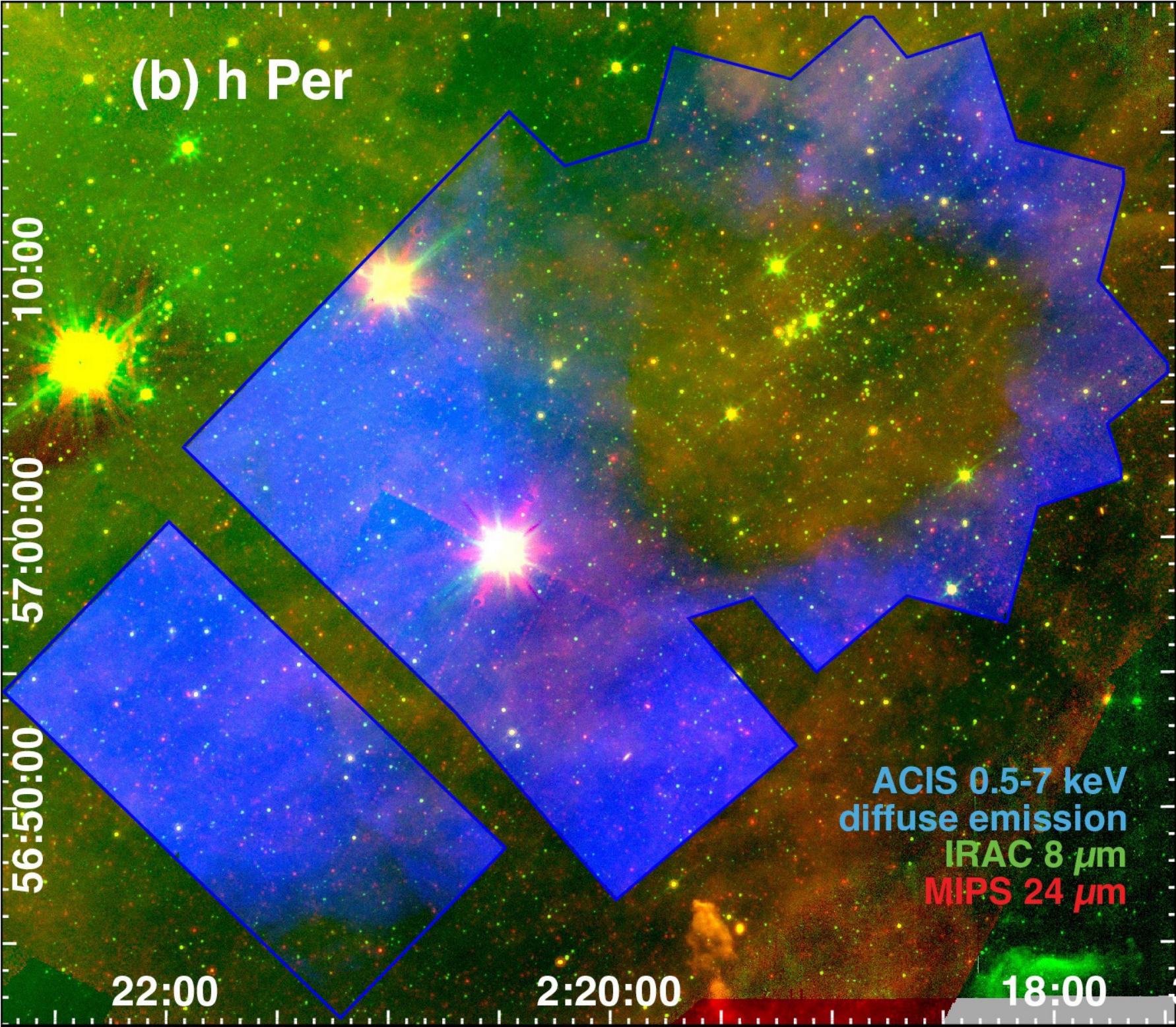}
\caption{h Per.
(a) ACIS exposure map with 2004 brighter ($\geq$5 net counts) ACIS point sources overlaid; colors denote median energy for each source.  ObsID numbers are shown in blue.
(b) ACIS diffuse emission in the \Spitzer context.  
%(c) ACIS event data and the absence of diffuse emission for the center of h~Per.
\label{hPer.fig}}
\end{figure}

\citet{Currie09} analyzed the original 41-ks ACIS-I observation of h~Per (ObsID 5407), finding 330 X-ray sources.  \citet{Argiroffi16} analyzed the three 2009 ACIS-I observations (190~ks, I-array only) and found 1002 X-ray sources.  Our analysis combines all four ACIS-I ObsIDs on h~Per.  With the resulting 231~ks of ACIS-I exposure plus a 10-ks snapshot of the ``Inter-Per'' region between h and $\chi$~Per (ObsID~5408), our analysis recovers $>$3400 X-ray point sources (Figure~\ref{hPer.fig}).

\begin{figure}[htb]
\centering
\includegraphics[width=0.99\textwidth]{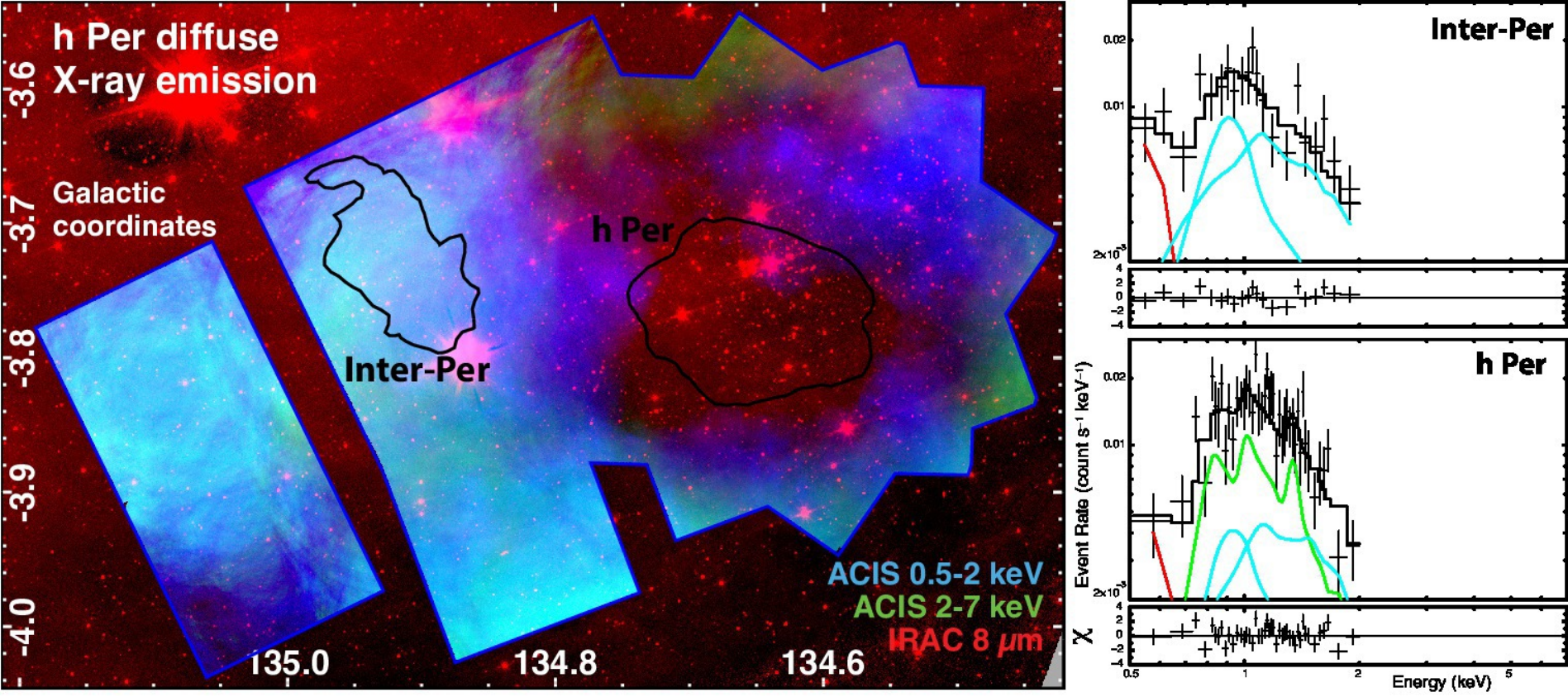}
\caption{Characterizing the h~Per diffuse X-ray emission.
This image shows ACIS soft-band and hard-band diffuse emission in the \Spitzer context; the extraction regions for diffuse spectral fitting are shown in black.  Table~\ref{tbl:diffuse_spectroscopy_style2} gives fit parameters for the spectral models.  
\label{hpergal+spectra.fig}}
\end{figure}

Despite the long ACIS integration centered on h~Per, our diffuse images show only very faint unresolved emission there (Figure~\ref{hpergal+spectra.fig}).  We extracted a diffuse spectrum from the ``Inter-Per'' region between h and $\chi$~Per where the apparent diffuse surface brightness is highest; note that this part of the ACIS mosaic has only a 9.8-ks exposure, so only the soft part of the spectrum had enough counts for fitting.  Its spectrum is dominated by the unresolved pre-MS star component, which has an intrinsic surface brightness that is ten times higher than that of the very soft, unobscured diffuse component invoked to explain the softest channels in the spectrum.

We also extracted a large region of diffuse emission at the center of h~Per (Figure~\ref{hpergal+spectra.fig}).  This samples the region of lowest apparent surface brightness in the field, so it had to be large to get enough counts for a spectrum.  Even with this large region, the spectrum is so noisy above 2~keV that we excluded it from the fit.  We also excluded the softest channel (at $\sim$0.5~keV) because it was extremely high and caused bad fit residuals; this is also probably a manifestation of noise in the spectrum.  For the remaining channels, we achieved a reasonable (not perfect) spectral fit, but the resulting model is not unique, so its fit parameters should be treated with caution.  We froze the absorption on the unresolved pre-MS component at the average value for h~Per rather than linking it to the absorption on the {\it pshock} component, because the latter approach worsened the fit residuals.  In the end, an obscured CIE plasma was needed for this fit, with $kT = 0.5$~keV.  Its intrinsic surface brightness is substantial and consistent with some form of massive star feedback, but its large column suggests that it is not located at the center of this revealed cluster.

In summary, this comparatively old cluster (surprisingly) hosts substantial diffuse X-ray emission behind high absorption.  Only the periphery of our ACIS mosaic shows unobscured diffuse structure; the diffuse spectral fit from the region between h and $\chi$~Per shows that this emission still comes mostly from unresolved stars in this very short ACIS snapshot.  The double clusters have extended halo populations \citep{Currie10,Zhong19}, so we expect most unresolved X-ray emission at the periphery of h~Per to come from this stellar halo.  At the center of the cluster, the obscured plasma emission could be residual cavity supernova signatures from what must have been a robust massive star population, or it could be due to some unrelated structure lying behind the cluster.  In either case, no diffuse component with the average absorption for the cluster is found; perhaps this is not so surprising, for such an old cluster.  \Chandra is scheduled to observe $\chi$~Per (NGC~884) in Cycle~20 (PI K.~Getman); it is also 14~Myr old\citep{Currie10}.  Combined with these h~Per data, the resulting \Chandra mosaic of the double cluster will provide a fine picture (and a large database) of X-ray sources from an ``older'' young stellar population.

%\clearpage
%-----------------------------------------------------------------------------
\subsection{NGC~281 \label{sec:n281}}
% NGC 281 -- 1185 point sources; a single pointing.
% SF Handbook review:  none.
% At D = 2.72 kpc, 4*pi*D^2 = 8.8540000e+44 cm^2.
% No piled up sources in this target.
% Avg A_V is 3 mag, so NH=0.5e22.  Don't know if this is for aimpoint or for IC 1590.

% Main cluster (east of aimpoint) is IC 1590, with center ``Trapezium'' HD 5005ABCD.  AB has sptype O4V((fc))+O9.7II-III from GOSSS.  From Simbad,
%HD 5005A -- O4V((fc)) (from GOSSS), 00 52 49.20 +56 37 39.6
%HD 5005B -- O9.7II-III (from GOSSS), 00 52 49.41 +56 37 39.9
%HD 5005C -- O8.5V(n) (from GOSSS), 00 52 49.56 +56 37 36.8
%HD 5005D -- O9.2V (from GOSSS), 00 52 48.95 +56 37 30.8

NGC~281 offers the third MOXC3 example (along with NGC~6193/RCW~108-IR and W5-E/AFGL~4029 above) of a populous revealed cluster with an adjacent smaller, obscured cluster, with the two clusters separated by a prominent ionization front.  The revealed cluster here is called IC~1590; it sits to the east of the obscured cluster, known as NGC 281 West.  The ACIS-I observation put NGC 281 West at the aimpoint.  These data were analyzed by \citet{Sharma12}, who found 354 X-ray sources, and again by \citet{Hasan18}, who found 446 sources.  

\begin{figure}[htb]
\centering
\includegraphics[width=0.49\textwidth]{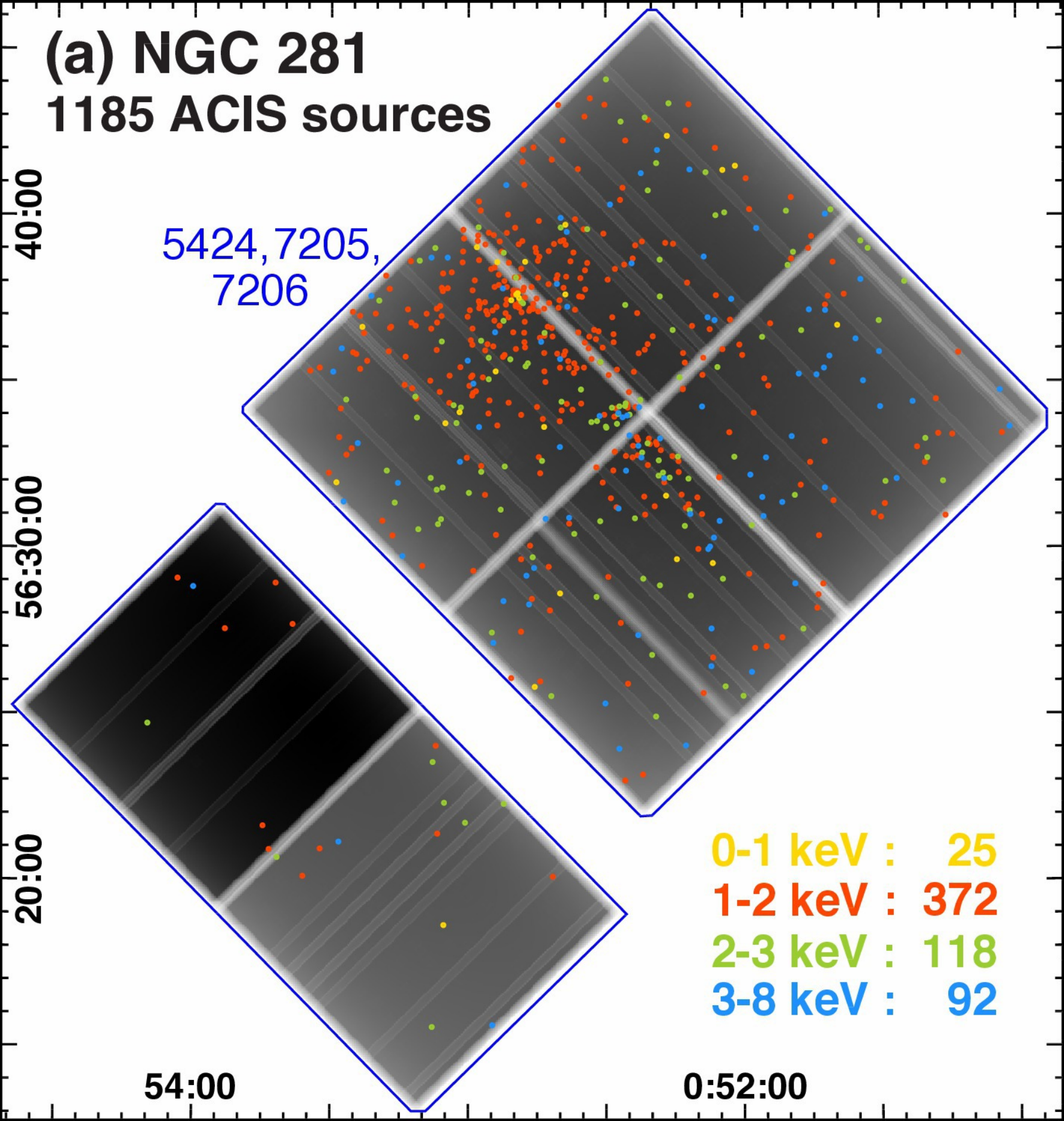}
\includegraphics[width=0.49\textwidth]{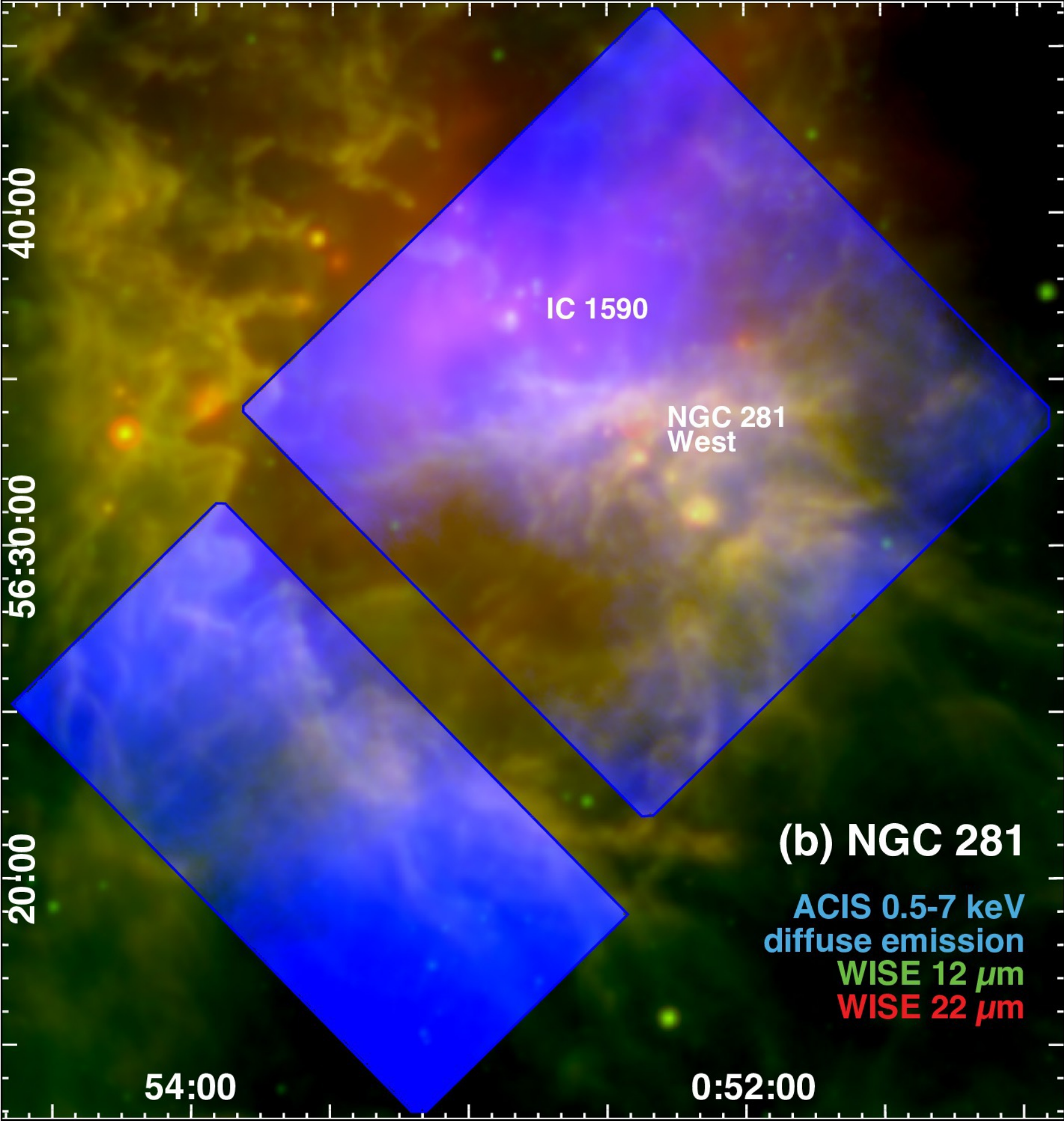}
\caption{NGC~281.
(a) ACIS exposure map with 607 brighter ($\geq$5 net counts) ACIS point sources overlaid; colors denote median energy for each source.  ObsID numbers are shown in blue.
(b) ACIS diffuse emission in the \WISE context.  
\label{ngc281.fig}}
\end{figure}

Our analysis yields 1185 X-ray sources and diffuse X-ray emission on the revealed side of the ionization front that separates the two clusters, as well as far off-axis on the ACIS-S CCDs (Figure~\ref{ngc281.fig}).  Predictably, IC~1590 exhibits softer X-ray sources than the smaller, more obscured NGC 281 West.

\begin{figure}[htb]
\centering
\includegraphics[width=0.325\textwidth]{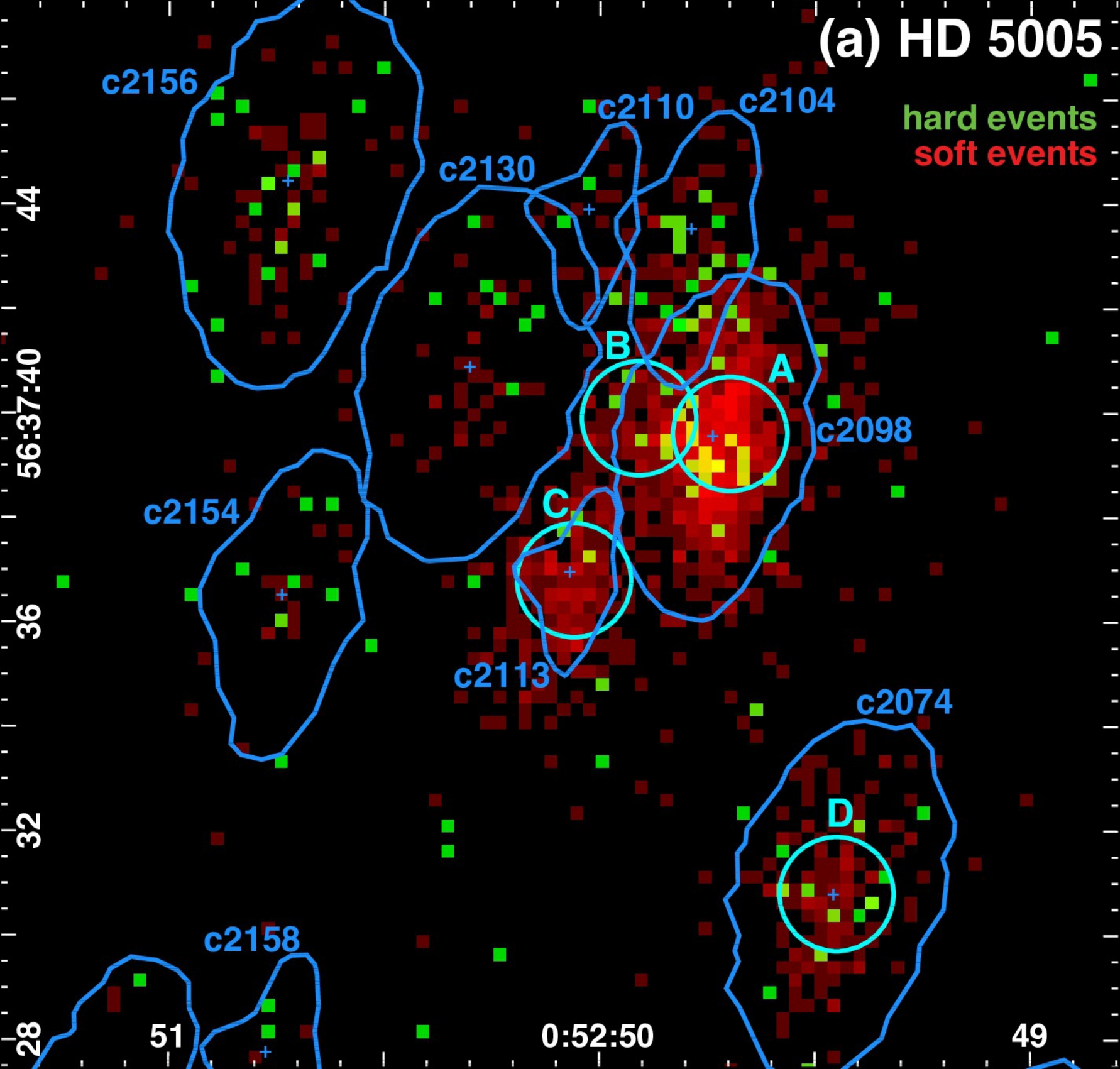}
\includegraphics[width=0.325\textwidth]{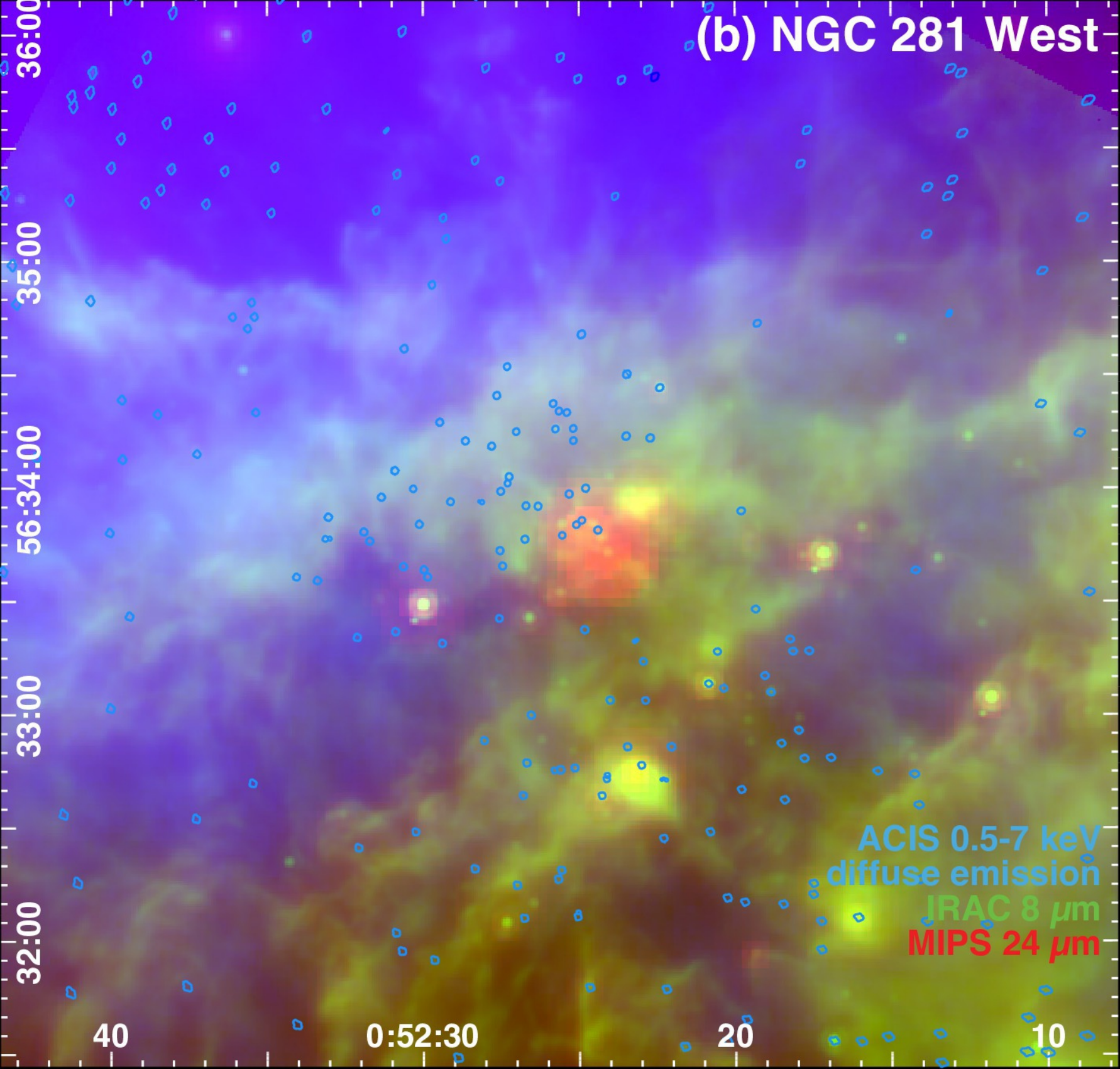}
\includegraphics[width=0.325\textwidth]{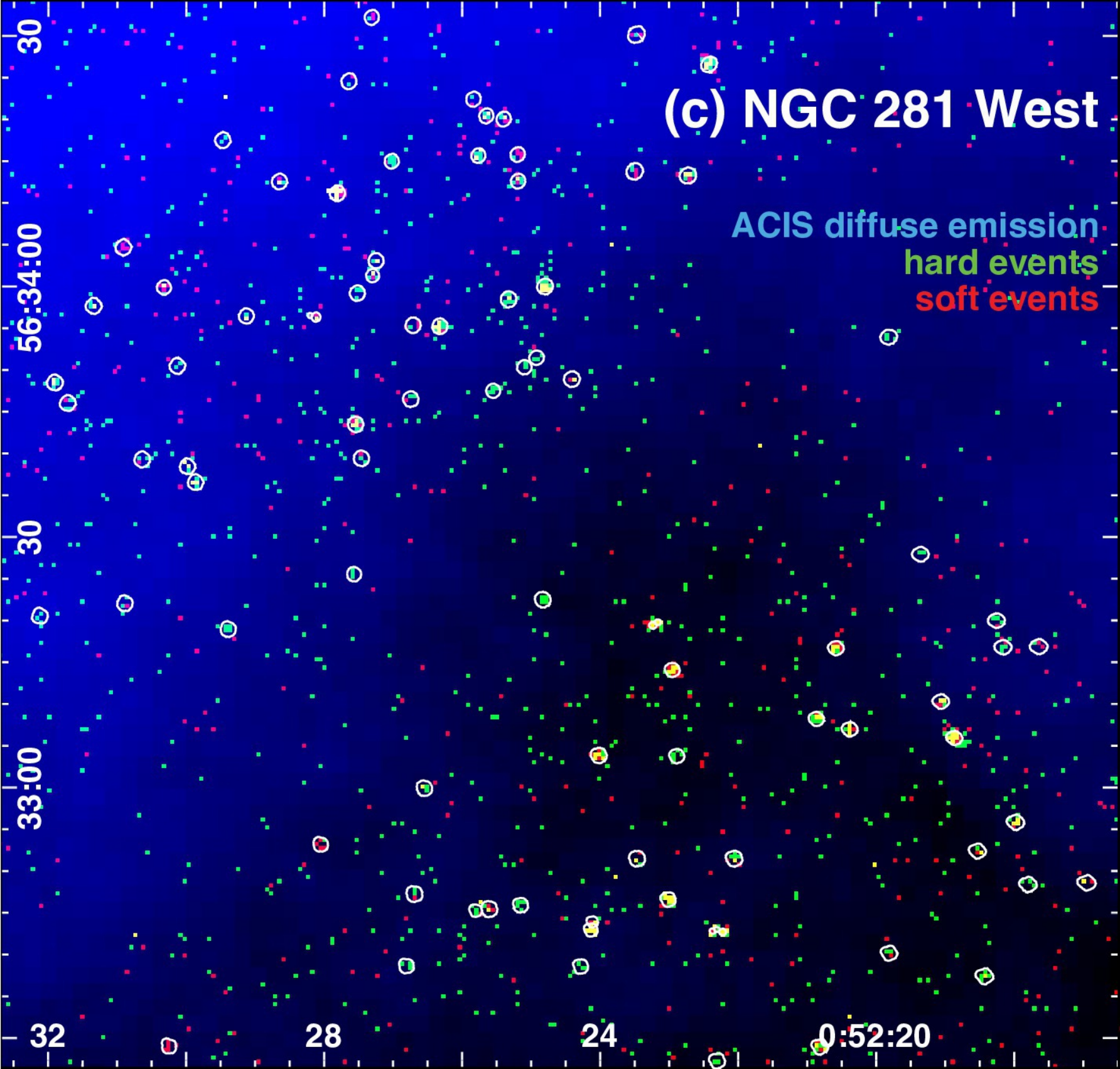}
\caption{
(a) The HD~5005 ``trapezium'' in IC~1590, showing ACIS events binned at 0.5 sky pixels and ACIS extraction regions.  Positions for the massive A--D components of HD~5005 (cyan circles) come from Simbad.
(b) The obscured cluster NGC~281 West, with ACIS point sources outlined.
(c) ACIS event data and diffuse emission in NGC~281 West; $\sim$90 hard X-ray sources are shown.
\label{ngc281_details.fig}}
\end{figure}

The revealed cluster IC~1590 and the field northeast of the ionization front are pervaded by diffuse X-ray emission.  At the center of IC~1590 is the massive Trapezium-like system HD~5005 \citep{Guetter97}; several X-ray sources form a tight clump there (Figure~\ref{ngc281_details.fig}(a)).

We fit the X-ray spectra of the HD 5005 massive components with the usual simple absorbed {\it apec} models.  All of them are soft, constant X-ray sources.  
ACIS source CXOU~J005249.23+563739.5 (c2098) is a blend of the O4V star HD~5005A and the O9.7II-III star HD~5005B \citep{Sota11}.  It is a bright X-ray source that requires a two-temperature {\it apec} fit, with $N_{H} = 0.5 \times 10^{22}$~cm$^{-2}$, $kT1 = 0.2$~keV, $kT2 = 0.5$~keV, and $L_{X} = 5.7 \times 10^{32}$~erg~s$^{-1}$.
ACIS source CXOU~J005249.56+563736.9 (c2113) is the O8.5V HD~5005C \citep{Sota11}, with 155 net counts, $N_{H} = 0.7 \times 10^{22}$~cm$^{-2}$, $kT = 0.2$~keV, and $L_{X} = 2.2 \times 10^{32}$~erg~s$^{-1}$.
ACIS source CXOU~J005248.95+563730.7 (c2074) is the O9.2V HD~5005D \citep{Sota14}, with 161 net counts.  Again a two-temperature {\it apec} fit is needed, with $N_{H} = 0.6 \times 10^{22}$~cm$^{-2}$, $kT1 = 0.1$~keV, $kT2 = 0.7$~keV, and $L_{X} = 1.6 \times 10^{32}$~erg~s$^{-1}$.

Figures~\ref{ngc281_details.fig}(b) and (c) show the obscured cluster NGC~281~West, at the ACIS aimpoint.  This region appears to separate into a series of stellar clumps \citep{Megeath97,Sharma12} that are well-populated with X-ray sources.  The northeast clump of sources \citep[``subcluster a'' in ][]{Sharma12} features diffuse X-ray emission, while ``subcluster b'' to the southwest does not.

\begin{figure}[htb]
\centering
\includegraphics[width=0.99\textwidth]{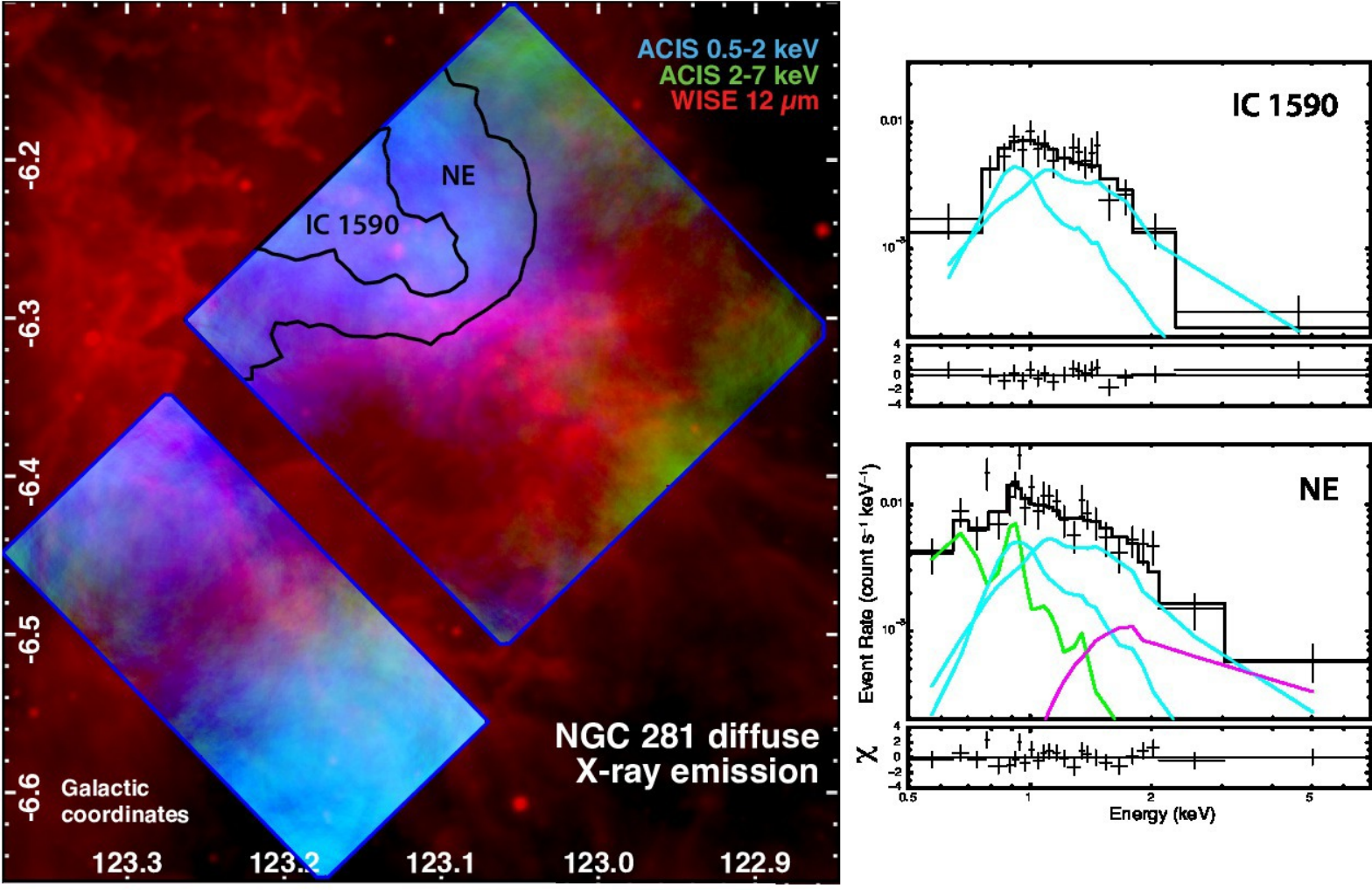}
\caption{Characterizing the NGC~281 diffuse X-ray emission.
This image shows ACIS soft-band and hard-band diffuse emission in the \WISE context; the extraction regions for diffuse spectral fitting are shown in black.  Table~\ref{tbl:diffuse_spectroscopy_style2} gives fit parameters for the spectral models.  
\label{ngc281gal+spectra.fig}}
\end{figure}

We extracted a diffuse spectrum from a region around IC~1590 (Figure~\ref{ngc281gal+spectra.fig}).  A good fit is achieved with just a pre-MS component behind a low absorbing column.  This is the first example in MOXC3 of a diffuse spectrum that appears to consist solely of unresolved stars (others are given below).  

Since soft diffuse emission appears to extend up to the ionization front in Figure~\ref{ngc281.fig}(b), we extracted another diffuse spectrum (``NE'') from a region that surrounds (but excludes) the IC~1590 diffuse region (Figure~\ref{ngc281gal+spectra.fig}).  That spectrum also shows unresolved pre-MS stars, but has additional soft emission from a fairly soft {\it pshock} plasma.  This is probably the expected signature of massive star feedback from IC~1590.

NGC~281 is another example of a massive, revealed cluster (IC~1590) adjacent to an obscured younger cluster (NGC 281 West), with the two separated by a prominent ionization front.  \Chandra adds significantly to the documented stellar populations in both of these clusters.  Three of the four O~stars in the HD~5005 ``trapezium'' in IC~1590 are separately detected; the fourth appears as an excess of X-ray counts east of the brightest component HD~5005A.  The diffuse X-ray emission immediately surrounding IC~1590 is fully explained by unresolved pre-MS stars; any truly diffuse emission associated with massive star feedback in this region is lost in their glow.  This diffuse emission is recovered in the ``NE'' region surrounding IC~1590, which is farther from the cluster center and not so dominated by unresolved stars.

%\clearpage
%-----------------------------------------------------------------------------
\subsection{G305 \label{sec:g305}}
% G305 -- 2184 point sources; HETG data on WR~48a.
% SF Handbook review:  none.
% G305.20+0.21 in Fig7 of Liu19 is out of the ACIS FoV.
% At D = 3.59 kpc, 4*pi*D^2 = 1.5423772e+45 cm^2.
% Avg A_V is 9.5 mag, so NH=1.52e22. 

%Besides WR~48a, there are 4 more WR stars in G305 -- see Mauerhan11. \\ 
%MDM3 = 131209.06-624326.7 ; 'c1790' (isolated bright src several arcmin west of WR~48a) \\
%MDM4 = 131221.32-624012.5 ; 'c2415' and/or 131221.10-624012.6 ; 'c2402' (MDM say 1 src, 131221.2?624012).   There are clearly 2 srcs here; c2415 much brighter than c2402.  MDM's position is in between our 2 srcs.  These sit about 2\arcmin north of Danks~1. \\
%MDM5 = 131225.42-624441.6 ; 'c2791' (isolated faint src sw of WR~48a) \\
%MDM8 = 131228.55-624143.7 ; 'c3225' (isolated bright src in Danks~1) \\

\begin{figure}[htb]
\centering
\includegraphics[width=0.99\textwidth]{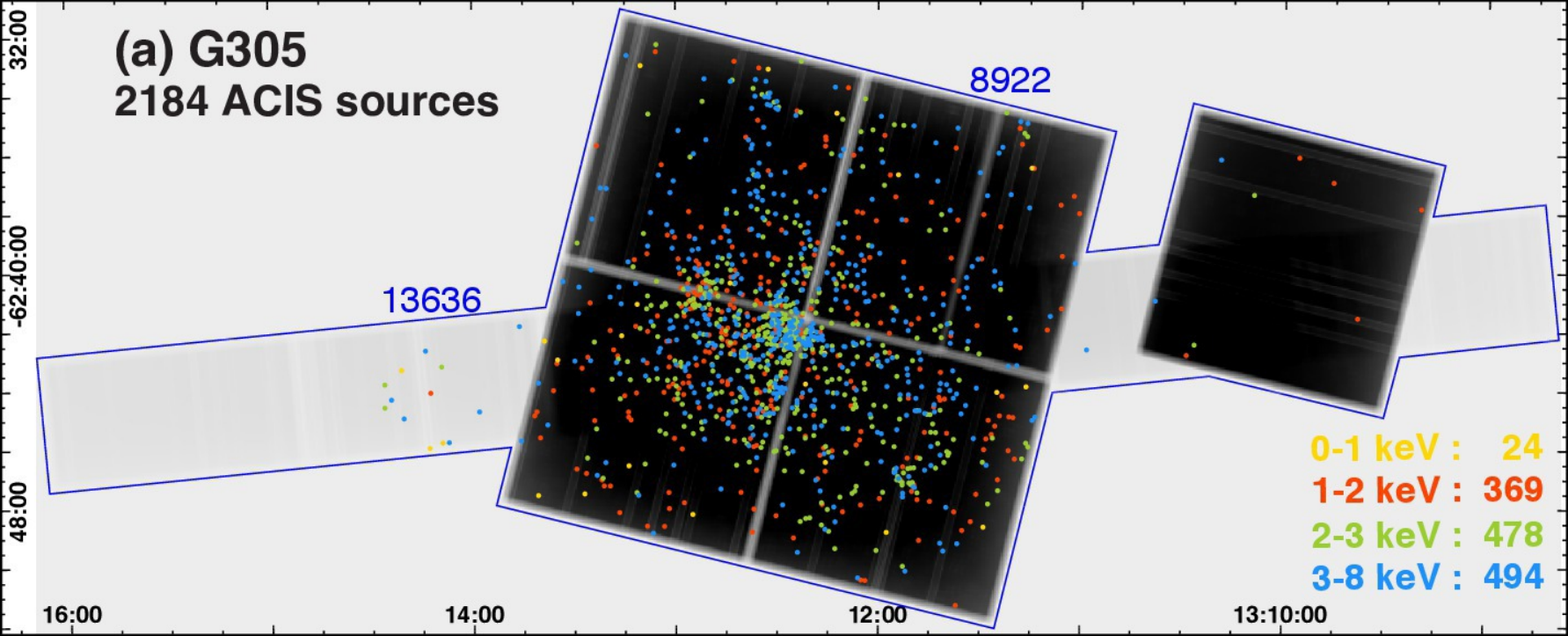}\\[0.05in]
\includegraphics[width=0.7\textwidth]{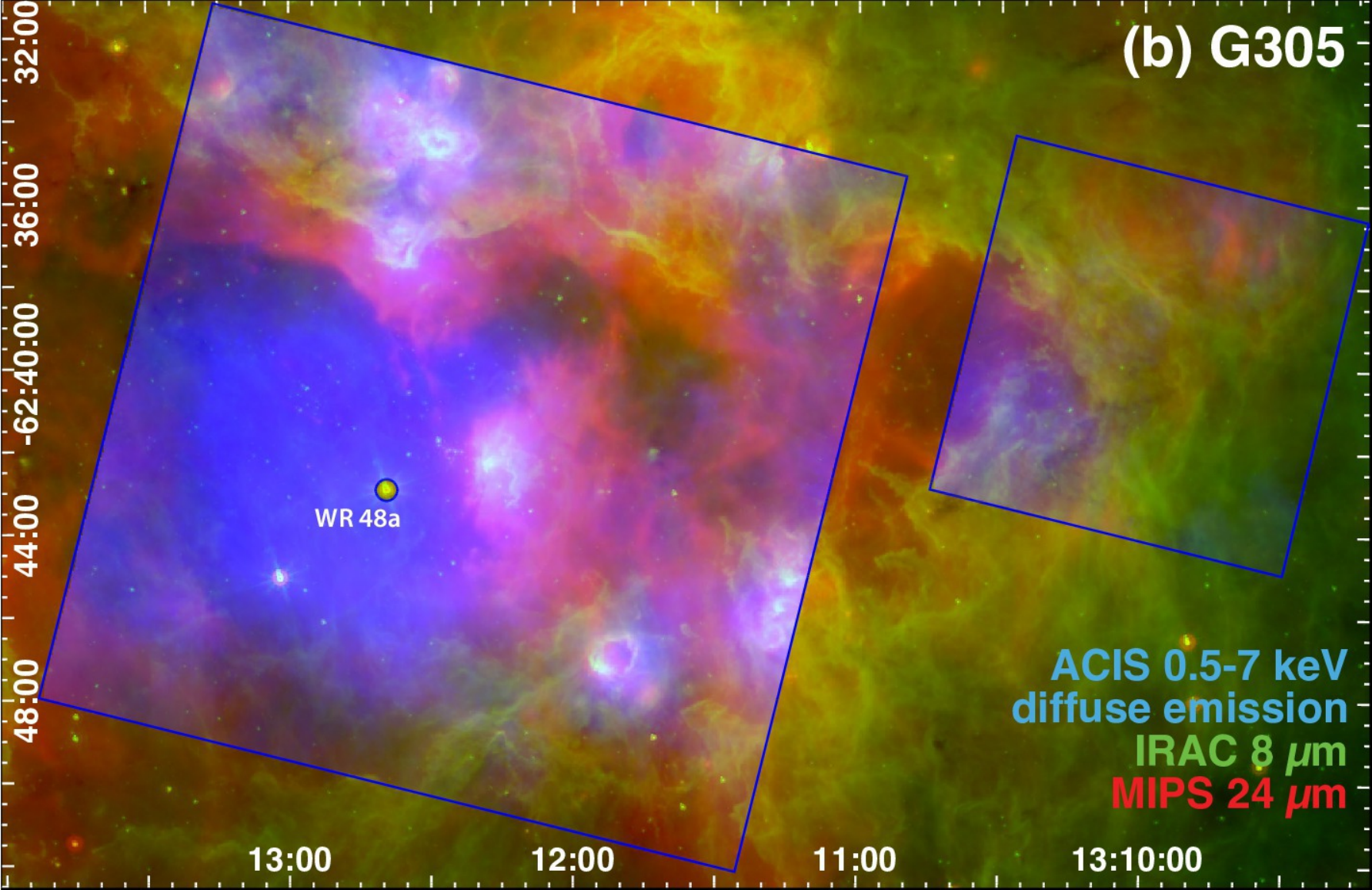} %this uses verysmallmask
\caption{G305.
(a) ACIS exposure map with 1365 brighter ($\geq$5 net counts) ACIS point sources overlaid; colors denote median energy for each source.  ObsID numbers are shown in blue.
(b) ACIS diffuse emission in the \Spitzer context.  The field of view is smaller here than in (a) because HETG data cannot be used to map diffuse X-ray emission.  A small mask has been applied around the piled-up source WR~48a to reduce its contamination to the diffuse emission.
\label{g305.fig}}
\end{figure}

G305 is a vast cluster of clusters complex \citep{Baume09} spanning more than two square degrees  \citep[e.g.][]{Danks84, Clark04, Hindson13}.  It hosts a large number of massive stars \citep{Mauerhan11, Davies12}, including the X-ray-bright CWB WR~48a \citep{Zhekov11, Zhekov14a, Zhekov14b}.  The combined bolometric luminosity of all known massive stars in G305 is only half its total IR luminosity \citep{Binder18}, so its massive star census is still incomplete.  Most of its clusters are quite young; they ionize a wide range of \hii regions and many still host star-forming clumps and cores \citep{Clark04,Faimali12}.  \Chandra has only observed the center of the complex (Figure~\ref{g305.fig}), capturing the massive clusters Danks~1 and Danks~2 \citep{Danks84} with an ACIS-I imaging observation (PI M.~Gagn{\'e}) and WR~48a with the HETG \citep{Zhekov14a}.  Many of G305's known massive stars are detected in the ACIS-I data; \citet{Zhekov17} suggests that many of its WR stars are CWBs based on their high X-ray luminosities.

%\clearpage

\begin{figure}[htb]
\centering
\includegraphics[width=0.47\textwidth]{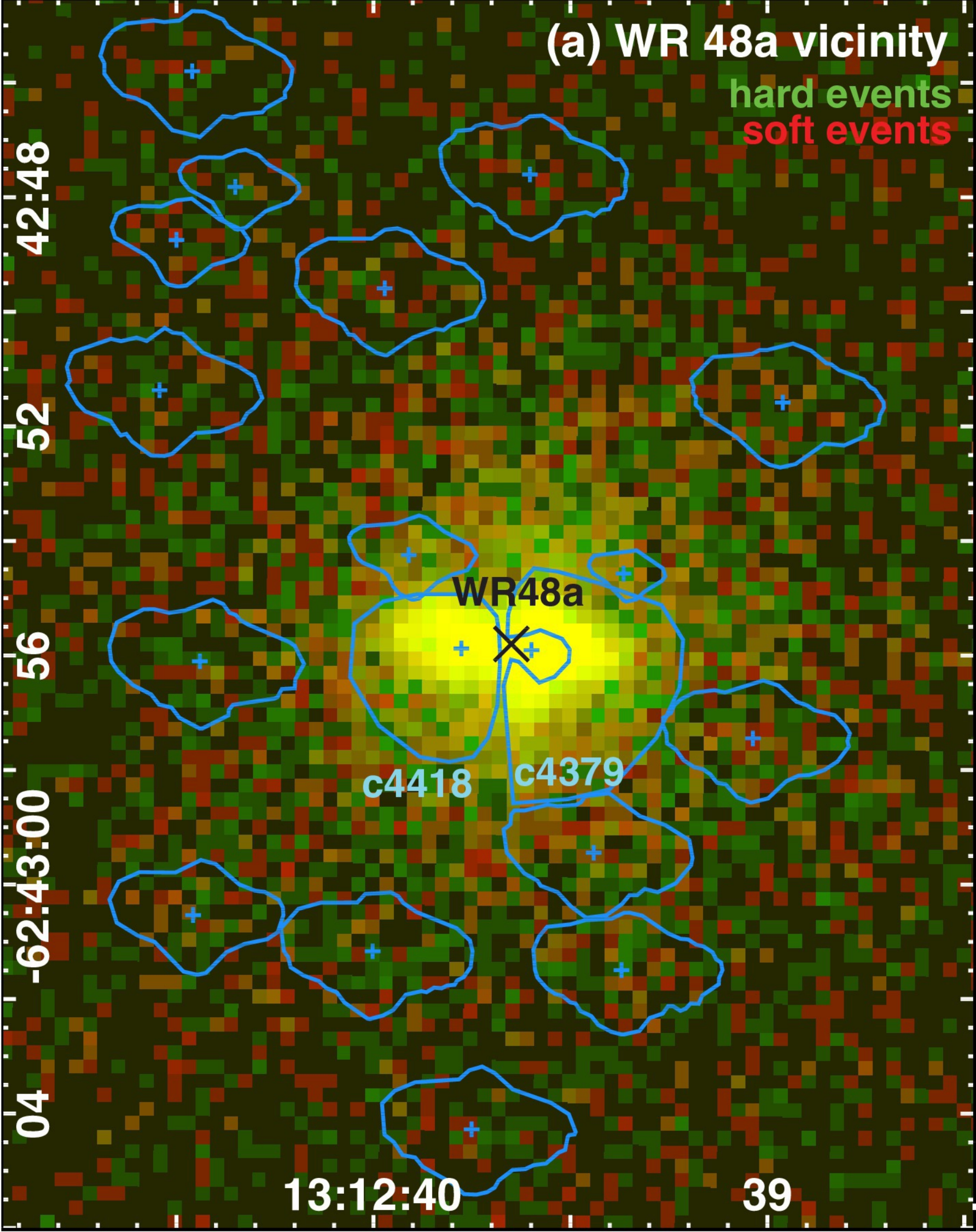}
\parbox[b]{0.52\textwidth}{
\includegraphics[width=0.52\textwidth]{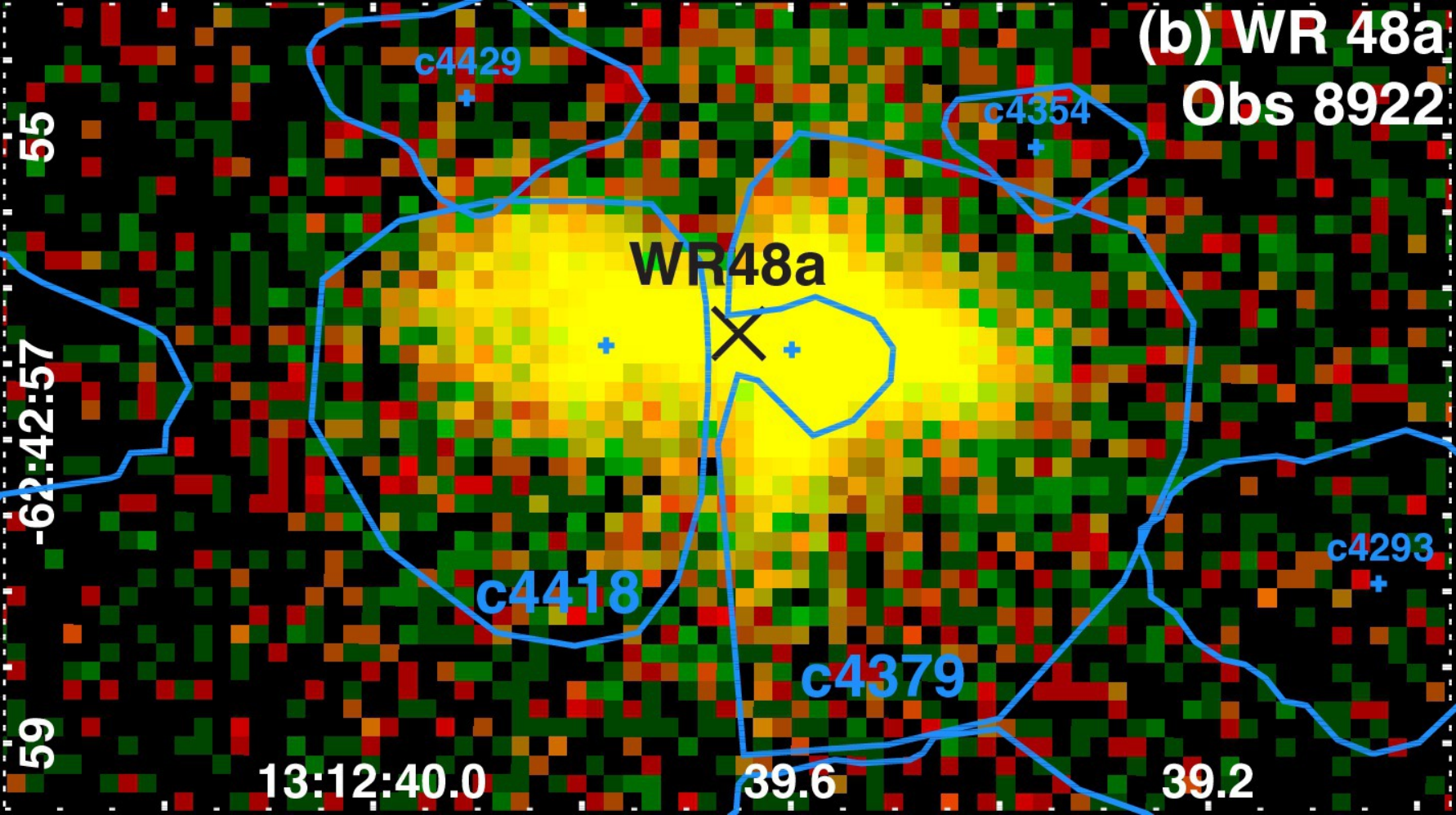}\\ [0.05in]
\includegraphics[width=0.52\textwidth]{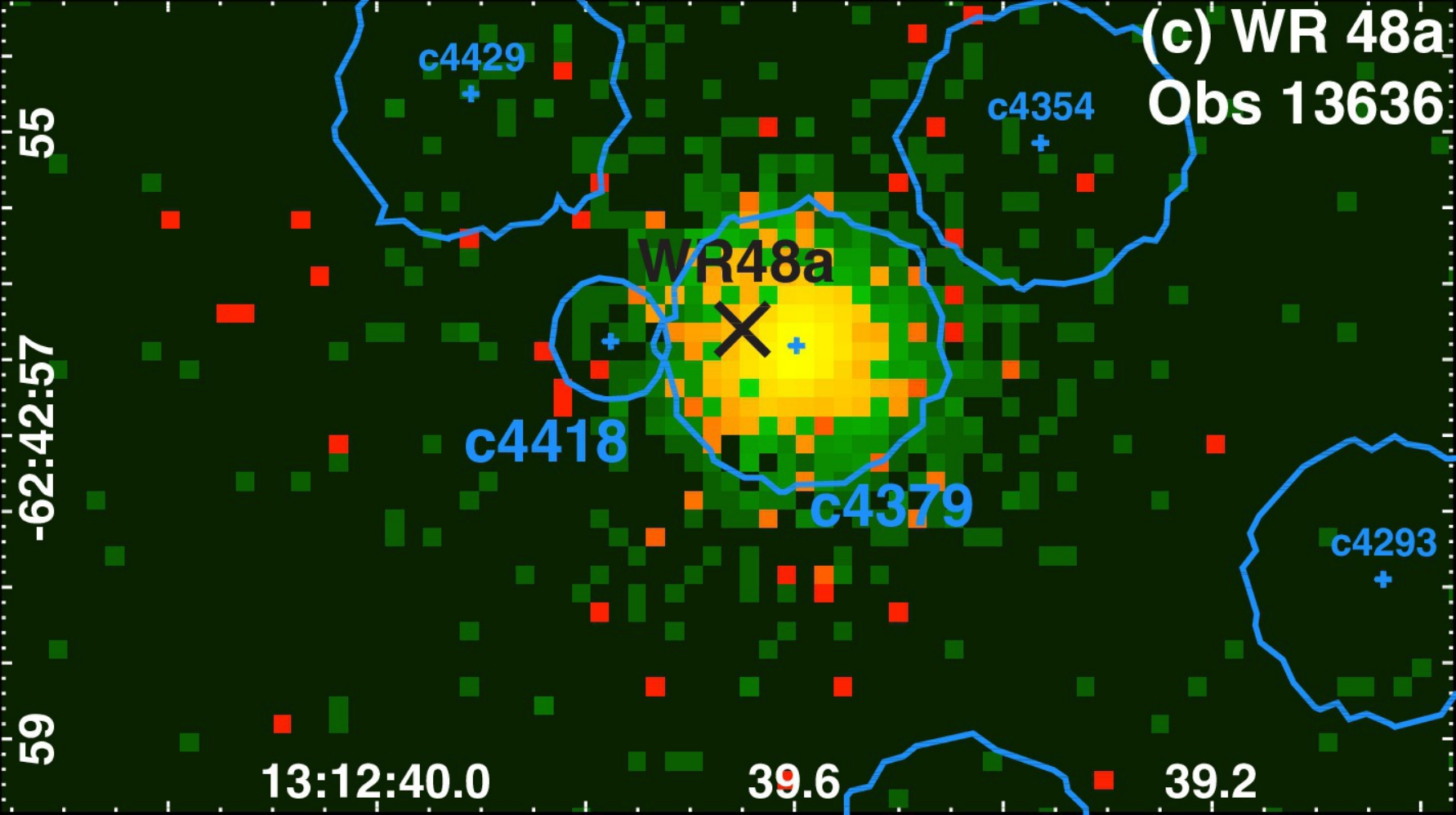}
}
\caption{The WR~48a region; the black cross shows the \citet{Davies12} position of WR~48a.  ACIS source positions come from image reconstruction.
(a) A clump of X-ray sources around WR~48a.  % Diffuse X-ray emission is omitted for clarity; it is bright and centered on WR~48a. 
(b) A zoom of (a) for just the ACIS-I ObsID 8922, now shown with 0.125$\arcsec$ pixels.  We find two X-ray sources coincident with WR~48a.  This field is 3$\arcmin$ off-axis. 
(c) The same field as (b), now for just the HETG ObsID 13636, where WR~48a is imaged on-axis.  Source c4418 has disappeared.
\label{WR48a.fig}}
\end{figure}

%\subsubsection{WR~48a \label{sec:WR~48a}}
\citet{Zhekov14b} find WR~48a to be a WC8+WN8h CWB and a bright, variable X-ray source.  Our analysis suggests that this important CWB may host a subcluster of fainter X-ray sources (Figure~\ref{WR48a.fig}(a)).  Although not obvious as separate sources in this image, these sources appear as peaks in the maximum likelihood image reconstruction that we use for source-finding; they persisted through the entire \AEacro\ source validation process.  They may be spurious peaks from over-reconstruction of the bright PSF wings of WR~48a, but we have no quantitative evidence of this.  Thus we have retained them as possible X-ray sources; it will be difficult to confirm their existence in other wavelengths because of the overwhelming brightness of WR~48a.

%Pat notes that when we say a source is piled, we really mean that the extraction aperture is piled.  Thus the close neighbor to WR~48a may not be piled itself, but it sits on a piled background from WR~48a.  We need to do a spectral fit to its reconstructed spectrum, get the luminosity, then use PIMMS to estimate the count rate -- from that, we can tell whether the source itself is piled.  Using our spectral fit params after pile-up correction, PIMMS shows count rate was 2.13e-2; their pile-up estimation is 3\%. 

Of particular importance is CXOU~J131239.77-624255.8 (c4418), the close companion to WR~48a (ACIS source CXOU~J131239.59-624255.9, c4379) that we find in the ACIS-only ObsID~8922.  This source is too far away from WR~48a to be the massive companion discussed in \citet{Williams12}.  It appears prominently in ObsID~8922 (Figure~\ref{WR48a.fig}(b)) with $L_{X} = 1.2 \times 10^{34}$~erg~s$^{-1}$, but has disappeared in ObsID~13636 (Figure~\ref{WR48a.fig}(c)).  We calculate an upper limit of 6.9 counts for c4418 in ObsID~13636; using the spectral fit parameters for this source from Table~\ref{pile-up_risk.tbl} in the \anchorfoot{http://asc.harvard.edu/toolkit/pimms.jsp}{{\em PIMMS}} count rate simulator, we estimate that its apparent flux decreased by two orders of magnitude between these two observations. % See e-mail from 31 May 2019 G305 folder for details.

Custom extraction apertures were drawn for sources c4418 and c4379 in ObsID~8922, to isolate the events from each source and, for c4379, to avoid its heavily piled-up core.  These two sources were modeled simultaneously for pile-up, since c4379 generates a large (and piled-up) background in the extraction aperture of c4418.  This background makes c4418 appear to be more piled up than it is.  Quantities in Table~3 pertain to the extraction apertures for the sources shown there; for c4418, most of the pile-up in its ObsID~8922 extraction aperture comes from the wings of c4379.  Similarly, the Table~\ref{pile-up_risk.tbl} pile-up ratio for c4379 in ObsID~8922 is the highest in MOXC3, but it is far smaller than it would have been without the custom extraction aperture that avoided the PSF core.  Such extreme measures were not needed for this source in ObsID~13636; there, a normal 90\% extraction aperture centered on c4379 showed only modest pile-up.  %This was due to two factors:  the sensitivity was reduced by the HETG, and the source itself produced less X-ray emission.

We fit the pile-up reconstructed spectra for WR~48a (c4379) (Table~\ref{pile-up_risk.tbl}).  We find that the total X-ray luminosity of this source was the same in its two observations, but the luminosities of the two spectral components changed; the 1.5~keV plasma got brighter in the second observation, while the hard plasma got fainter.  The absorption increased by a factor of two between these observations, as already reported by \citet{Zhekov14a}.  We do not mark this source as variable in Table~\ref{pile-up_risk.tbl} because its $L_{tc}$ did not change between the two observations, but the character of the emission changed substantially.

% Pat has made reconstructed lightcurves for each source, but they still have backgrounds, so c4418's (neighbor) lightcurve retains an imprint of c4379's (WR~48a). 

%\clearpage
%\subsubsection{The Danks~1 and Danks~2 Clusters \label{sec:danks}}

\begin{figure}[htb]
\centering
\includegraphics[width=0.49\textwidth]{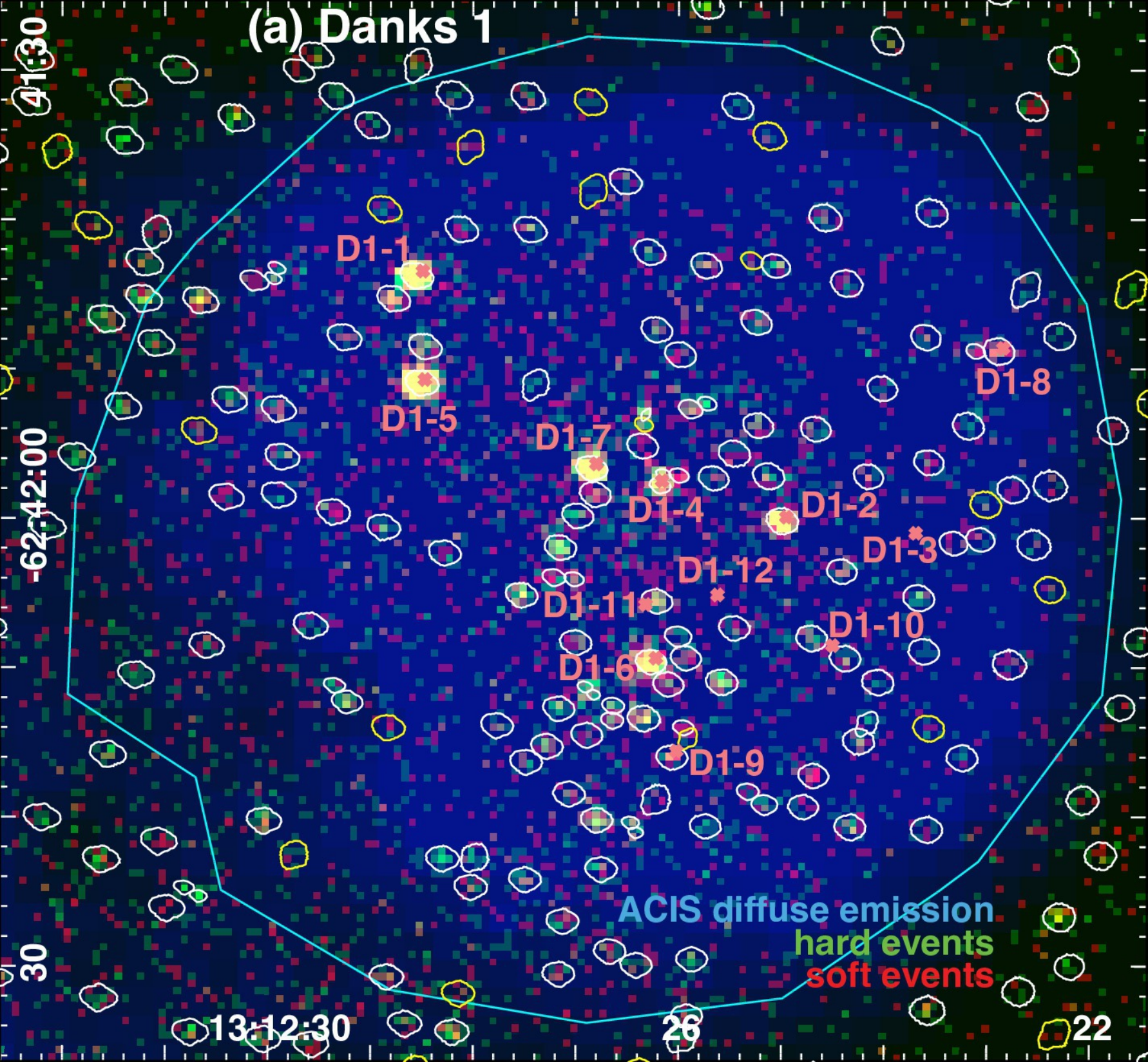}
\includegraphics[width=0.49\textwidth]{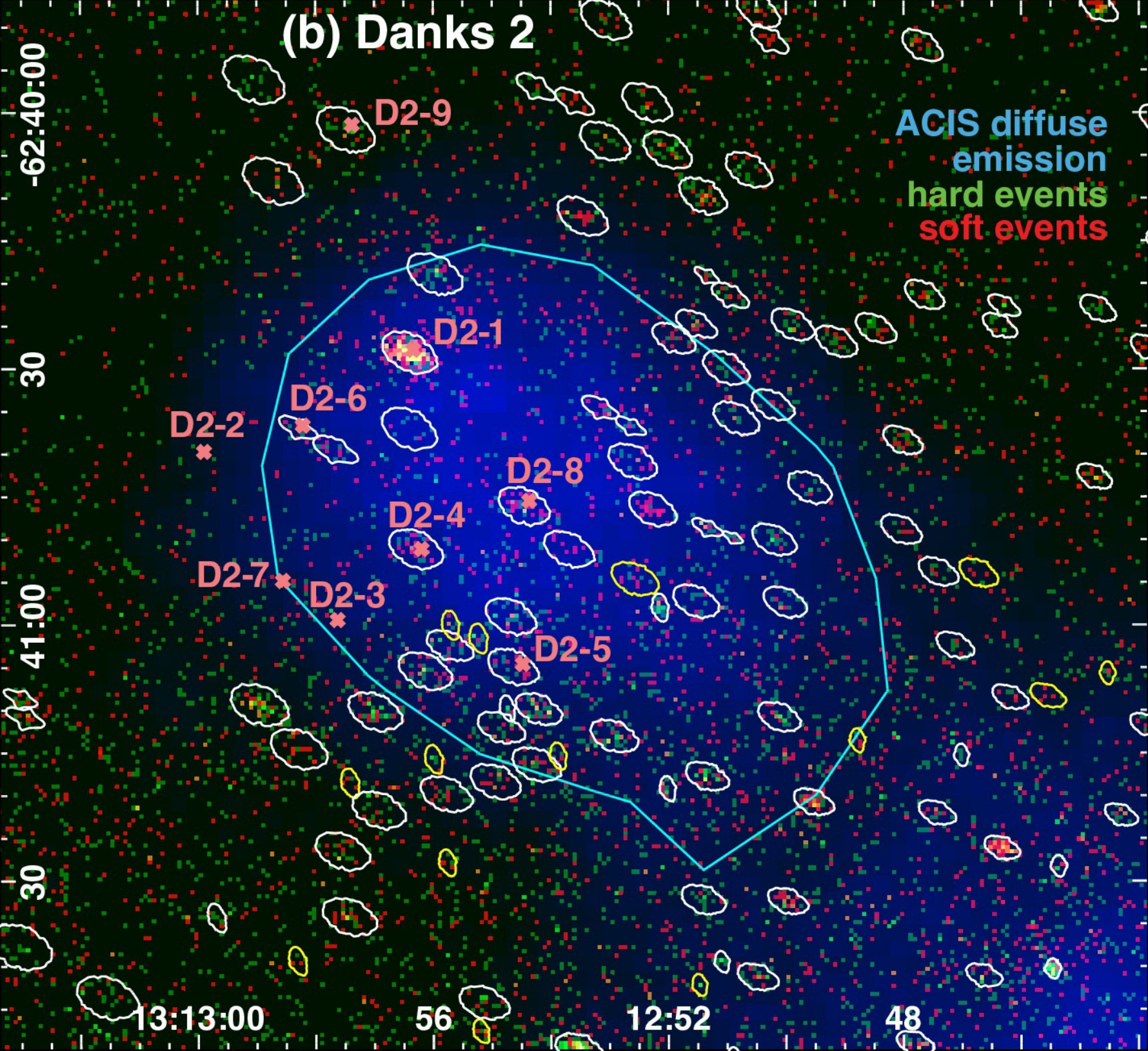}
\caption{ACIS event data and diffuse emission for the main clusters in G305.  Yellow apertures denote ``occasional'' sources.  ``D1'' and ``D2'' sources are from \citet{Davies12}.  Diffuse extraction regions are shown in cyan.
(a) Danks 1. 
(b) Danks 2.  The diffuse emission in the southwest corner is centered on WR~48a.
\label{danks.fig}}
\end{figure}

The \Chandra data for the massive clusters Danks~1 and Danks~2 \citep{Danks84} are shown in Figure~\ref{danks.fig}.  Massive stars from \citet{Davies12} are marked.  Many have X-ray counterparts; for some, the massive star position lies between two X-ray sources.  \citet{Zhekov17} gives X-ray luminosities for the WR stars in these clusters, from ACIS ObsID 8922.

% WNLh stars:
% D1-1 is 131228.55-624143.7 ; 'c3225' -- inter-ObsID variability suggested; need separate spectra from each ObsID to get fluxes -- fit to regular spectrum gives 4.6e22, 0.8 + 2.2 keV, 0.8e33
% D1-2 is 131224.99-624200.1 ; 'c2731' -- not variable; fit gives 3.3e22, 0.6 + 3.4 keV, 0.7e33
% D1-5 is 131228.50-624150.9 ; 'c3219' -- inter-ObsID variability suggested; need separate spectra from each ObsID to get fluxes -- fit to regular spectrum gives 3.0e22, 1.0 keV, 0.8e33

% D1-3 (O8-B3I) undetected
% D1-4 (O6-8If) sits between 131226.15-624157.0 ; 'c2894' (18 netcts, medE=3.3) and 131226.18-624157.8 ; 'c2896'  (55 netcts, medE=1.8)
% D1-6 (O6-8If) is 131226.26-624209.6 ; 'c2913' -- 328 netcts, medE=1.9
% D1-7 (O4-6) is 131226.84-624156.6 ; 'c3023' -- 426 netcts, medE=2.0, not variable.
% D1-8 (O4-6) is 131222.87-624148.8 ; 'c2543' -- 24 netcts, medE=1.8
% D1-9 (O4-6) is 131226.06-624216.0 ; 'c2880' -- 17 netcts, medE=1.6
% D1-10 (O6-8) probably undetected, sits between c2700 and c2653
% D1-11 (O8-B3) probably undetected, sits at the edge of c2903
% D1-12 (O8-B3) undetected

In Danks~1, the three WNLh stars (D1-1, D1-2, and D1-5) are all bright X-ray sources \citep{Zhekov17}.  Our X-ray counterparts to D1-1 (CXOU~J131228.55-624143.7, c3225) and D1-5 (CXOU~J131228.50-624150.9, c3219) are variable.  D1-1 dropped an order of magnitude in X-ray luminosity between the two ACIS ObsIDs; D1-5 dropped by a factor of almost 5.

The brightest O star is the O4-6V star D1-7, which is ACIS source CXOU~J131226.84-624156.6 (c3023).  Its spectral fit gives $N_{H} = 3.2 \times 10^{22}$~cm$^{-2}$, $kT1 = 0.9$~keV, $kT2 = 4.3$~keV, and $L_{X} = 0.6 \times 10^{33}$~erg~s$^{-1}$.  The hard spectral component suggests binarity, although the source is not variable in these ACIS observations.
  
In Danks~2, the brightest X-ray source is the O8-B3 supergiant D2-1, which is ACIS source CXOU~J131256.41-624028.1 (c5413).  Its spectral fit gives $N_{H} = 3.7 \times 10^{22}$~cm$^{-2}$, $kT = 0.5$~keV, and $L_{X} = 3.0 \times 10^{33}$~erg~s$^{-1}$.  In contrast, D1-3, with the same spectral type, is undetected by ACIS.

The O6.5V+O6.5V eclipsing binary 2MASS J13130841-6239275 \citep{Kourniotis15} sits $\sim$2$\arcmin$ northeast of Danks~2.  The ACIS counterpart is CXOU~J131308.43-623927.2 (c5723); it has 51 net counts, a median energy of 1.8~keV, and is not variable.  A spectral fit yields $N_{H} = 3.6 \times 10^{22}$~cm$^{-2}$, $kT = 0.5$~keV, and $L_{X} = 3.0 \times 10^{32}$~erg~s$^{-1}$.  This large absorbing column is nearly twice the extinction reported by \citet{Kourniotis15}.

\begin{figure}[htb]
\centering
\includegraphics[width=0.99\textwidth]{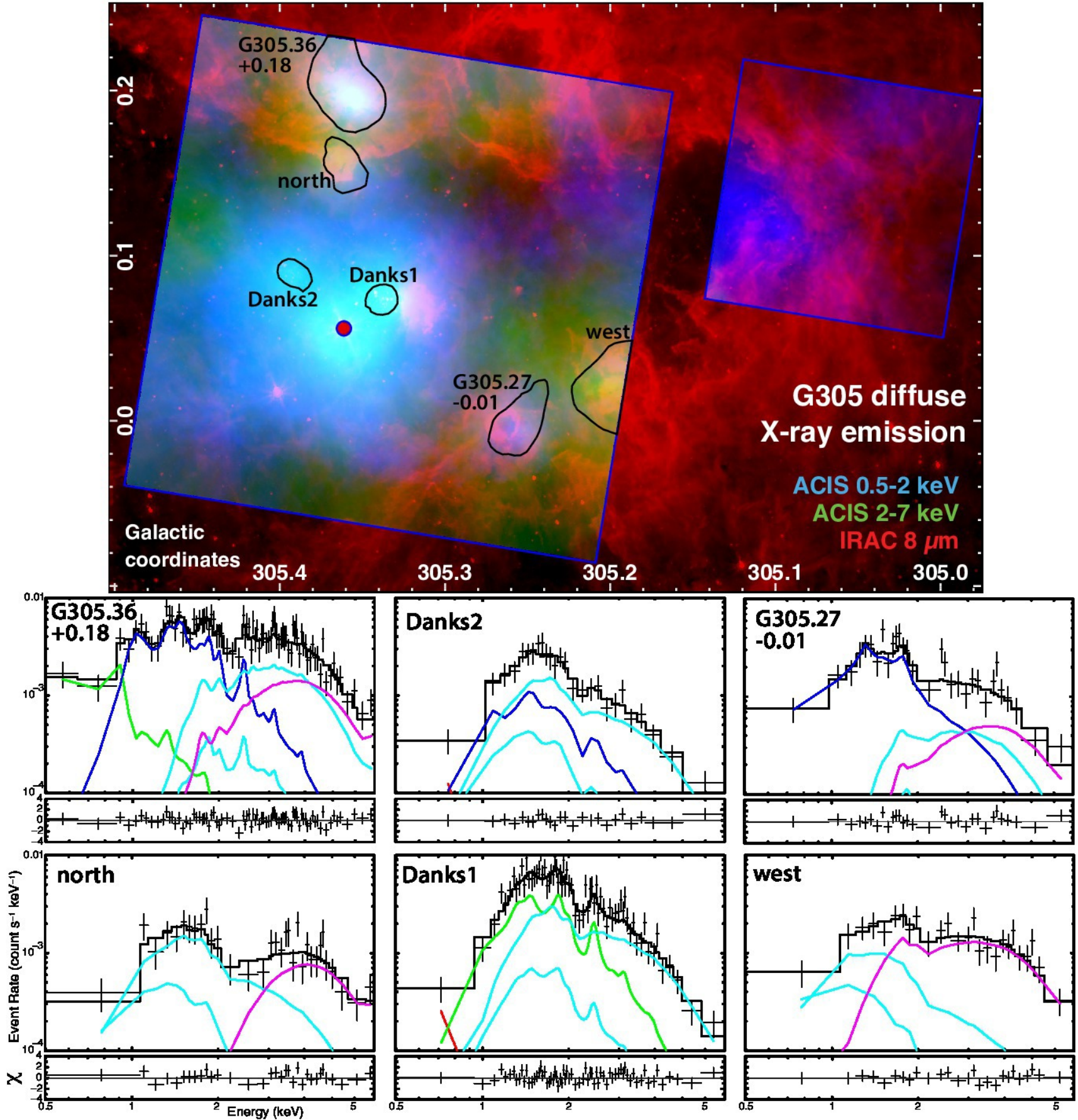} %this uses verysmallmask, WR~48a is masked before smoothing, and sqrt(smoothed ACIS images) are shown.
\caption{Characterizing the G305 diffuse X-ray emission.
This image shows ACIS soft-band and hard-band diffuse emission in the \Spitzer context; extraction regions for diffuse spectral fitting are shown in black.  Square root versions of the ACIS images are employed here, to accommodate the large dynamic range of the diffuse X-ray emission across the G305 field.  Corresponding spectra are shown below the image; axis ranges are the same for all spectra.  Table~\ref{tbl:diffuse_spectroscopy_style2} gives fit parameters for the spectral models.
\label{g305gal+spectra.fig}}
\end{figure}

We extracted the diffuse X-ray emission in regions immediately surrounding Danks~1 and Danks~2 (Figure~\ref{danks.fig}).  The resulting spectra are similar (Figure~\ref{g305gal+spectra.fig}); both show bright, obscured CIE {\it pshock} plasmas with high temperatures ($kT = 0.9$~keV for Danks~1, $kT = 1.3$~keV for Danks~2) and unresolved pre-MS components with high surface brightnesses.  Absorptions for the pre-MS components were not independently constrained in the spectral fits, so they were set equal to that of their respective hard {\it pshock} components.  Surface brightnesses for both the hot plasma and unresolved pre-MS components were substantially brighter for Danks~1 than for Danks~2.  This is perhaps to be expected, since Danks~1 is denser and younger than Danks~2 \citep{Davies12}.

%\subsubsection{Other SFRs in G305 \label{sec:g305other}}
% Baume09 Table~\ref{tbl:diffuse_spectroscopy_style2}:  
% DBS2003 130 has AV = 10.4, located at 13:11:54.0 -62:47:02.0.  Simbad:  13 11 54.0 -62 47 00, or l,b = 305.27 -00.00.
% DBS2003 131 has AV = 11.7, located at 13:11:39.4 -62:33:11.5 -- this is off the ACIS FoV!!!  Simbad:  13 11 39.400 -62 33 11.55, or l,b =  305.26 +00.23.
% DBS2003 132, from Simbad:  13 12 18.0 -62 42 18, or l,b = 305.32 +00.07.  This sits just west of Danks~1.

From \citet{Baume09}, the \hii region G305.3+0.2 contains cluster [DBS2003]~131 \citep{Dutra03}; this is just outside the ACIS field of view.  These authors also say that cluster [DBS2003]~132 \citep{Dutra03}, which sits just west of Danks~1, is associated with the \hii region G305.3+0.1, but it is too close to the bright foreground G0 star HD~114515 for them to study.  We detect HD~114515 with 69 net counts (CXOU~J131215.80-624250.4, c2094) and find $N_{H} = 0.2 \times 10^{22}$~cm$^{-2}$, $kT = 0.8$~keV, and $L_{X} = 1.3 \times 10^{29}$~erg~s$^{-1}$ using a distance of 434~pc from \citet{Bailer18}.  There are a number of X-ray sources around this star, but no obvious cluster centered on the \hii region and no concentration of diffuse X-ray emission around it.

%HD 114515 is a foreground G0 star at 13 12 15.79 -62 42 50.47, d=434pc from Bailer-Jones18.  Amazingly, we detect it:  131215.80-624250.4 ; 'c2094', 69 net counts, median energy 0.99 keV. 
%A spectral fit to c2094 yields $N_{H} = 0.2 \times 10^{22}$~cm$^{-2}$, $kT = 0.8$~keV, and $L_{X} = 1.3 \times 10^{29}$~erg~s$^{-1}$ using d=434pc (4*pi*D^2 = 2.254e+43 cm^2).
%*** Add a figure panel to Fig~\ref{g305_clumps.fig} showing region around HD 114515?  No -- not interesting.  No concentration of diffuse emission, stars look like an extension of Danks~1. ***

\begin{figure}[htb]
\centering
\includegraphics[width=0.482\textwidth]{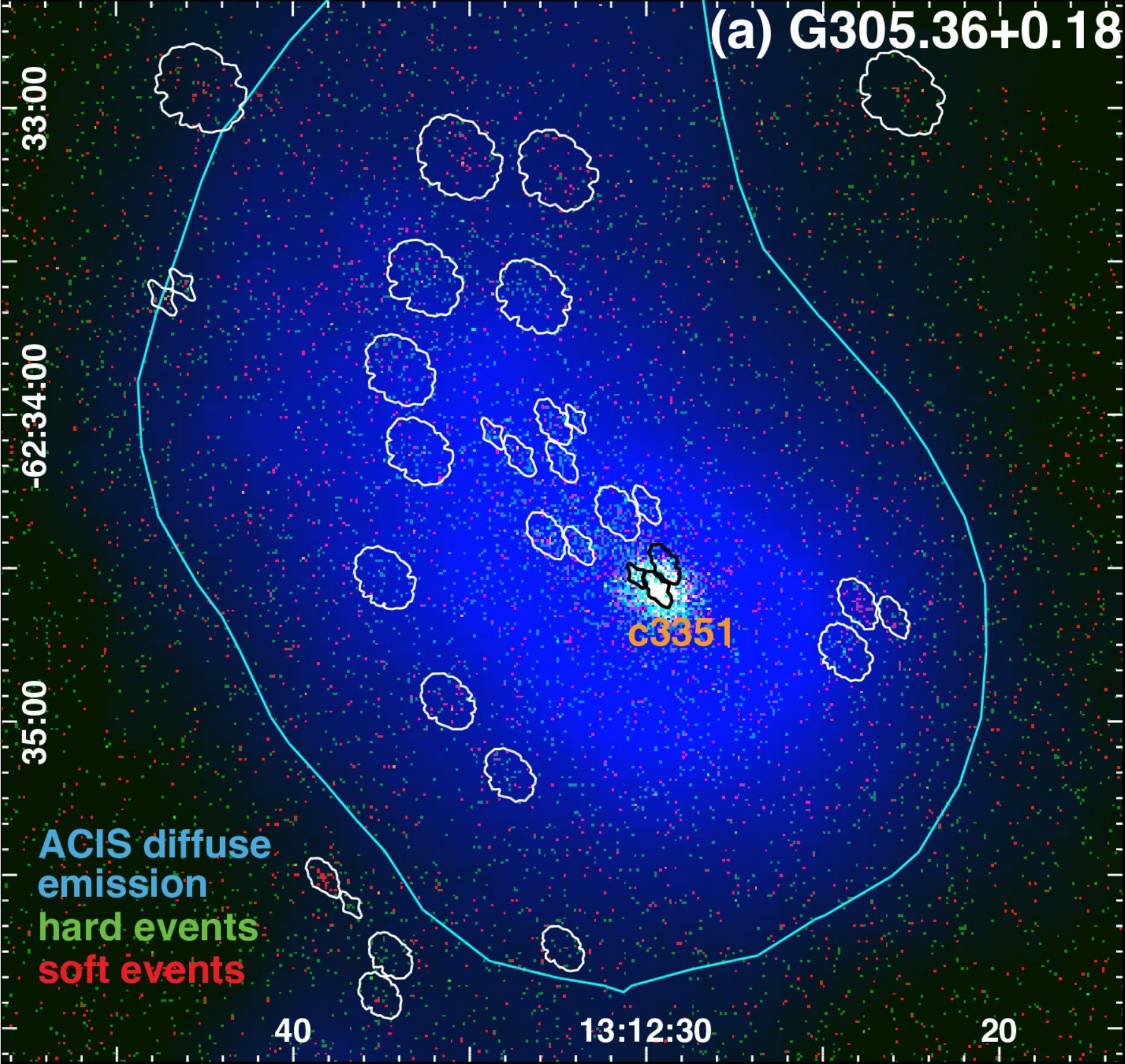}
\includegraphics[width=0.50\textwidth]{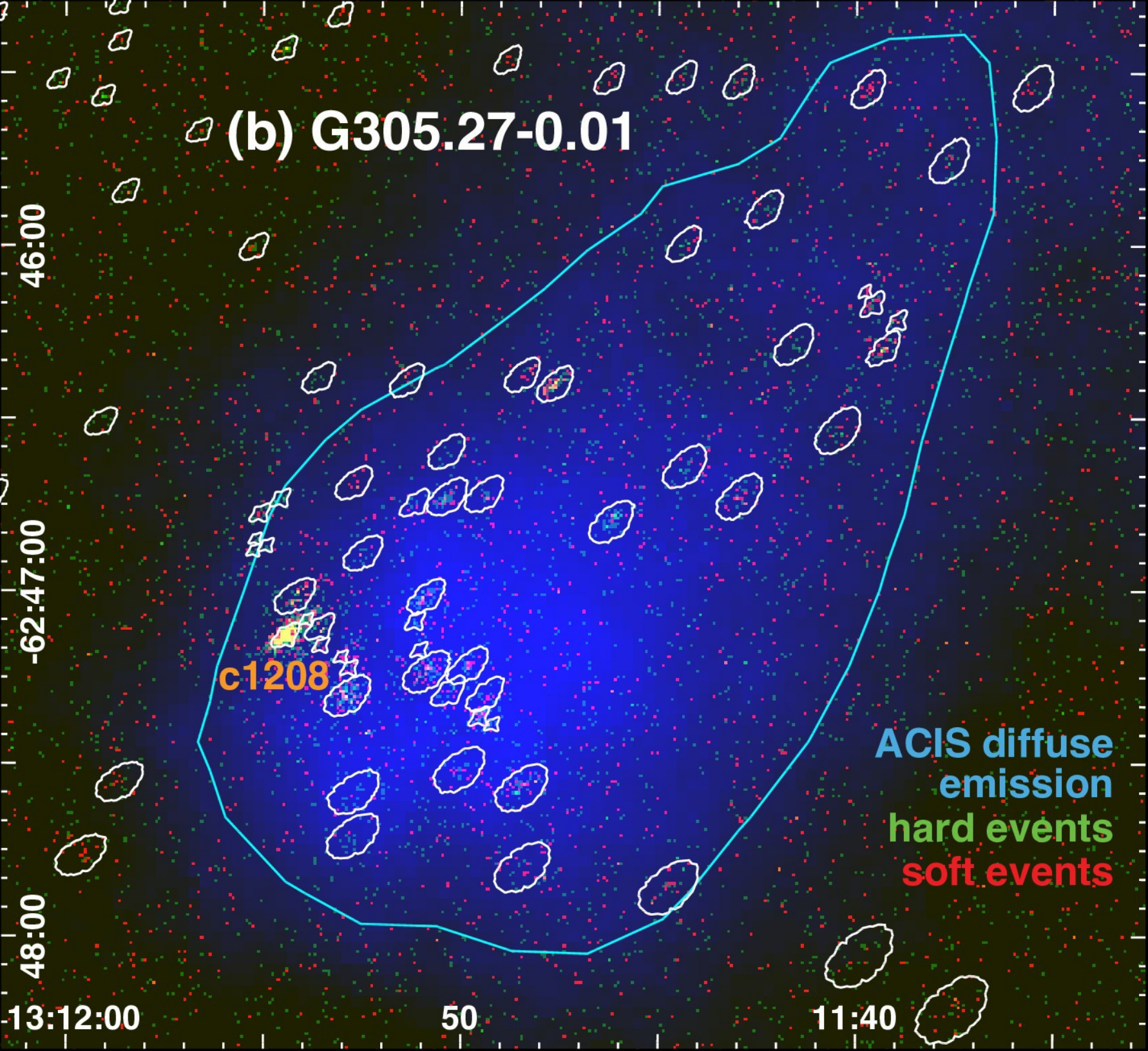}
\caption{ACIS event data and diffuse emission for other clumps of X-ray sources in G305.  The brightest ACIS source in each field is labeled.  Diffuse extraction regions are shown in cyan.
(a) G305.36+0.18 (cluster VVV~CL022).
(b) G305.27-0.01 (cluster [DBS2003]~130). 
\label{g305_clumps.fig}}
\end{figure}

The \hii region G305.36+0.18 hosts the cluster VVV~CL022 \citep{Borissova11}; a concentration of X-ray sources is found here (Figure~\ref{g305_clumps.fig}(a)).  There is a bright ACIS source in this field, CXOU~J131229.67-623433.9 (c3351, with close neighbors c3310 and c3468).  A spectral fit to c3351 yields $N_{H} = 6.4 \times 10^{22}$~cm$^{-2}$, $kT = 1.9$~keV, and $L_{X} = 2.9 \times 10^{33}$~erg~s$^{-1}$.  This high luminosity suggests a massive star; the hard spectrum suggests more than one component in an interacting system.

%Simbad id's ``Spitzer Dark Cloud'' SDC G305.361+0.180 from Peretto \& Fuller 2009 (A\&A 505, 405); also HII region GAL 305.36+00.18, mm source G305.362+0.185 from Hill05 (MNRAS 363, 405) and others; VVV CL022 -- Cluster of Stars is at 13 12 36.0 -62 37 18, from Borissova11; Spitzer bubble S153 \citep{Churchwell06}.  There is a bright ACIS source in this field, 131229.67-623433.9 ; 'c3351' (with close neighbors c3310 and c3468).  No Simbad counterpart, but lots in VizieR:  Bailer-Jones18 gives a Gaia DR2 distance of about 3.24 kpc.  There is a 2MASS (and other IR) counterpart; src is in the CSC.

The \hii region G305.27-0.01 contains the cluster [DBS2003]~130 \citep{Dutra03}, which is also seen in X-rays (Figure~\ref{g305_clumps.fig}(b)).  It also includes a bright ACIS source, CXOU~J131154.42-624708.0 (c1208, with close neighbor c1191).  \citet{Borissova16} assign a spectral type of B0Ve to this source (which they call DBS130 Object 2).   A spectral fit to c1208 yields $N_{H} = 4.3 \times 10^{22}$~cm$^{-2}$, $kT1 = 0.8$~keV, $kT2 = 4.0$~keV, and $L_{X} = 1.2 \times 10^{33}$~erg~s$^{-1}$.  Again this high luminosity and very hard spectrum suggest a massive, interacting binary.
%Again lots of VizieR counterparts:  Bailer-Jones18 gives a Gaia DR2 distance of about 3.47 kpc.    There is a 2MASS (and other IR) counterpart; src is in the CSC and XMM serendipitous source catalogs.  Also in RMS survey (Mottram10), MSX G305.2694-00.0072.  Simbad calls this ``[BRA2016] DBS130 Obj2 -- Star'' from Borissova16.

We find diffuse X-ray emission concentrated on both G305.36+0.18 and G305.27-0.01 (Figures~\ref{g305gal+spectra.fig} and \ref{g305_clumps.fig}).  The diffuse extraction region for G305.36+0.18 is large compared to other G305 diffuse X-ray regions; it captures a hard ($kT = 1.1$~keV) plasma transitioning from NEI to CIE.  The two {\it pshock} components are seen behind very different absorbing columns.  This spectrum has a prominent hard component that is modeled by highly absorbed unresolved pre-MS stars; in this case, the surface brightness for this component is almost twice as bright as that for the diffuse hot plasma.  The diffuse emission around G305.27-0.01 is dominated by a hard ($kT = 1.4$~keV), short-timescale {\it pshock} component with the average absorption for G305.  The hard part of the spectrum is modeled by a highly absorbed unresolved pre-MS component with much lower surface brightness.

Additional diffuse X-ray regions (``north'' and ``west'' in Figure~\ref{g305gal+spectra.fig}) can be modeled simply as moderately obscured unresolved pre-MS populations and background.  The absorbing column for the ``north'' region is consistent with the average G305 column; that for the ``west'' region is lower than average.  The surface brightness of unresolved stars in the ``north'' region is comparable to that for G305.27-0.01; in the ``west'' region, it is almost four times fainter.  We confirm earlier claims \citep[e.g.,][]{Willis15} that these regions may be additional sites of recent star formation in the G305 complex.

%\subsubsection{G305 Summary}

G305 appears to be a Carina-like cluster of clusters, with widespread star formation and extensive, morphologically and spectrally complex diffuse X-ray emission.  The existing ACIS data sample just part of the G305 complex; substantial star formation is ongoing outside the ACIS field of view \citep{Baume09,Willis15}.  Detailed study of the brighter X-ray sources may yield candidate massive stars to add to the current (incomplete) census.  Understanding the extent of diffuse X-ray emission in G305 would require tiling the full G305 complex with ACIS observations; ACIS is necessary to separate out the large number of pre-MS stars found across the field.  A second HETG observation of WR~48a is planned for 2019 November (PI S.~Zhekov); we hope that its close neighbor CXOU~J131239.77-624255.8 can also be studied with these new data.

%\clearpage
%-----------------------------------------------------------------------------
\subsection{RCW 49 \label{sec:rcw49}}
% RCW 49 -- 3139 point sources; wide-field mosaic includes MSFR snapshot on G284.0-0.9 (3649 srcs total), but I've separated that western part of the mosaic out for the MSFR snapshots paper (510 srcs).
% SF Handbook review:  none.
% At D = 4.21 kpc, 4*pi*D^2 = 2.1211232e+45 cm^2.
% Avg A_V is 6.5 mag, so NH=1.04e22. 
% MSP = Moffat, A. F. J., Shara, M. M., & Potter, M. 1991, AJ, 102, 642.  
% WR~20a: 10 23 58.00 -57 45 49.0
% WR~20b: 10 24 18.40 -57 48 29.8   
% WR 20aa:  l,b = 284.331634, -0.583555 from Drew18; 10:23:23.49 -58:00:20.8 from Roman-Lopes11.  This is our source 102323.35-580021.2 ; 'c1856', barely captured far off-axis!  It has 365 net counts in 45 ks total exposure, median energy 1.7 keV.  No variability.
% WR 20c:  l,b = 284.175755, +0.077817 from Drew18; 10:25:02.60 -57:21:47.3 from Roman-Lopes11.  Not in ACIS FoV.
% HESS J1023?575:  10 23 00.0 -57 30 00 from Simbad -- coords not very precise.
% MSP18:  10:24:02.44 -57:44:36.1, from Hur15.
% WR~21a:  10 25 56.50 -57 48 43.5, from Simbad.

RCW~49 is a spectacular southern MSFR made famous by the \Spitzer GLIMPSE survey \citep{Whitney04,Churchwell04}.  At its heart is the dense, massive young cluster Westerlund~2 (Wd2), containing many O stars and two CWBs, WR~20a and WR~20b \citep{Rauw04,Bonanos04,Rauw05,Rauw07,Naze08,Rauw11,Montes13,Vargas13,Hur15}.  A third CWB, WR~21a \citep{Benaglia05,Niemela08,Tramper16,Gosset16}, lies $\sim$15$\arcmin$ to the east of Wd2; its membership in the cluster is uncertain \citep{Roman-Lopes11,Carraro13}.  Two more WR stars, WR20c and WR~20aa \citep{Roman-Lopes11,Drew18} are found in the outskirts of Wd2 and may be an ejected pair from dynamical interactions in Wd2.

\begin{figure}[htb]
\centering
\includegraphics[width=0.49\textwidth]{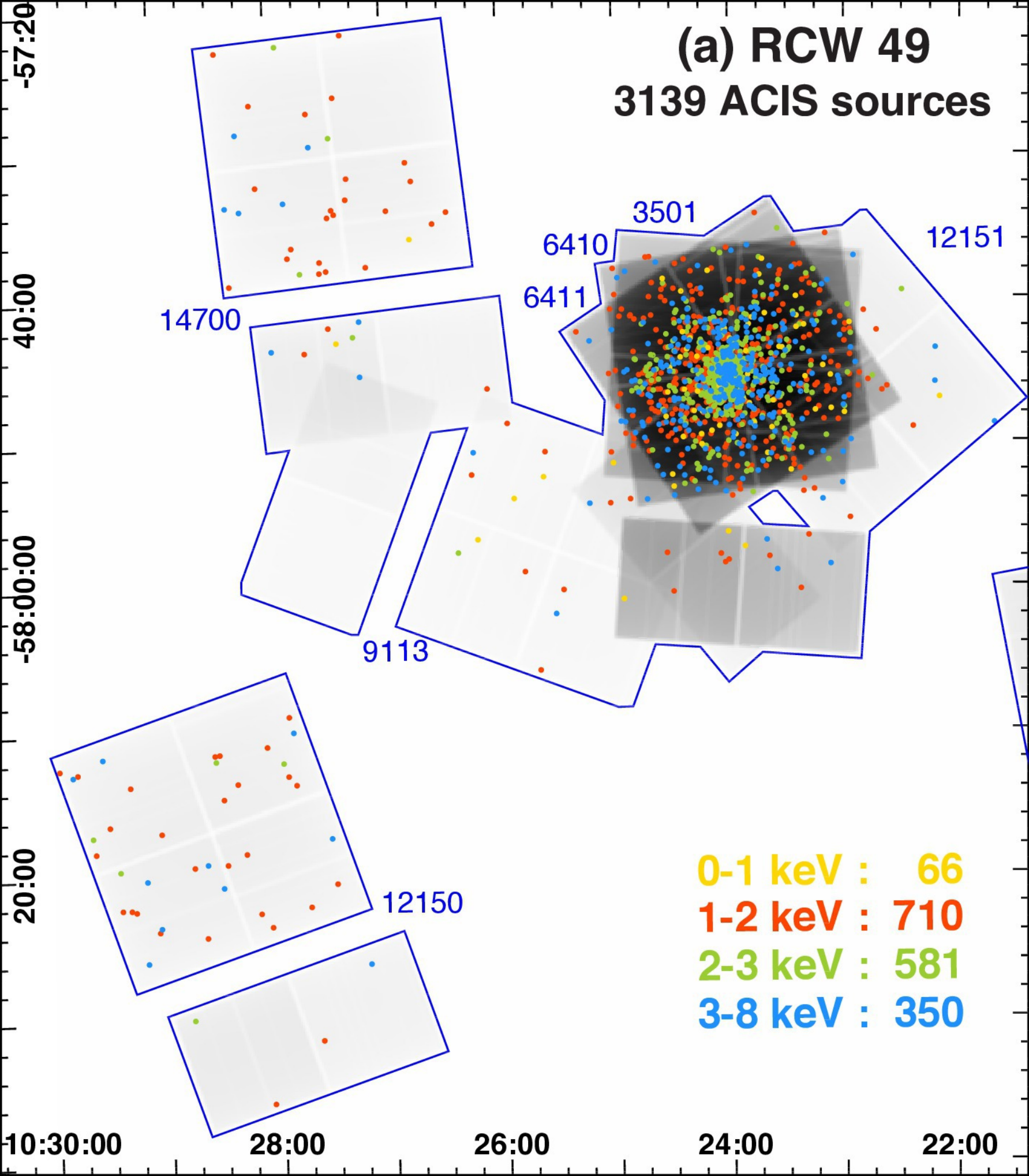}
\includegraphics[width=0.49\textwidth]{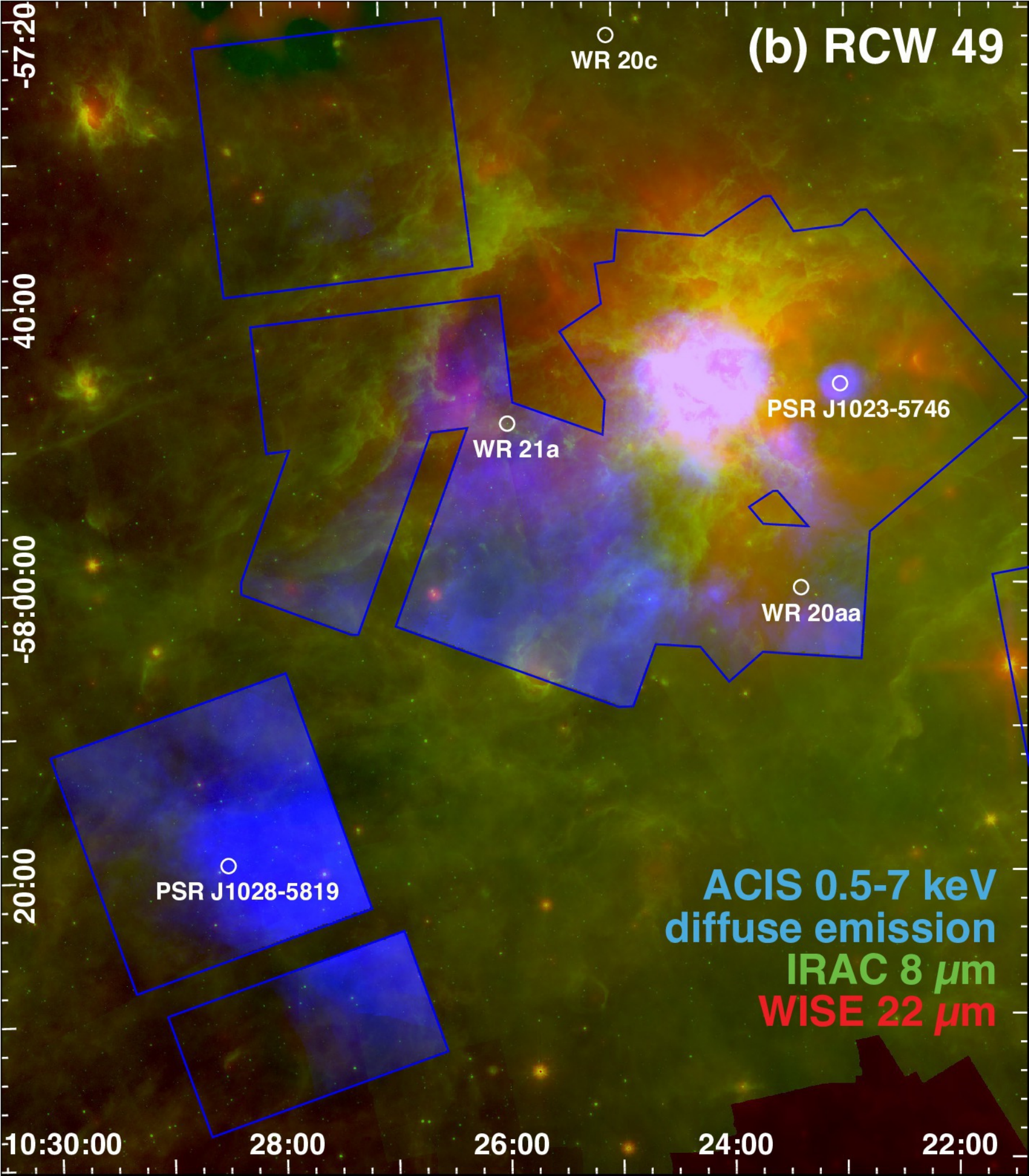}
\caption{The wide-field ACIS mosaic around RCW~49.
(a) ACIS exposure map with 1707 brighter ($\geq$5 net counts) ACIS point sources overlaid; colors denote median energy for each source.  ObsID numbers are shown in blue.
(b) ACIS diffuse emission in the \Spitzer and \WISE contexts.  Sources discussed in the text are marked.
\label{rcw49.fig}}
\end{figure}

The original 35-ks ACIS GTO observation of RCW~49 was described by \citet{Tsujimoto07}; they found and characterized 468 X-ray point sources.  An additional 97~ks of ACIS-I data followed, to study the properties of WR~20a.  \citet{Naze08} analyzed the combined 132-ks ACIS dataset, concentrating on the WR stars and many other massive stars in Wd2, plus some flaring X-ray sources.  We re-analyzed the 132-ks RCW~49 dataset and expanded it to include several surrounding short, archival ACIS observations (Figure~\ref{rcw49.fig}).  We find $>$3000 X-ray point sources, most centered on Wd2.  Spatially and spectrally complex diffuse X-ray emission is detected throughout our wide ACIS mosaic.  The partial outline of another ACIS observation is shown at the western edge of the field, to indicate that the full mosaic we analyzed is even larger.  This western field (which includes the MSFR G284.0-0.9 and WR~18) will be described in a future paper.

\begin{figure}[htb]
\centering
\includegraphics[width=0.49\textwidth]{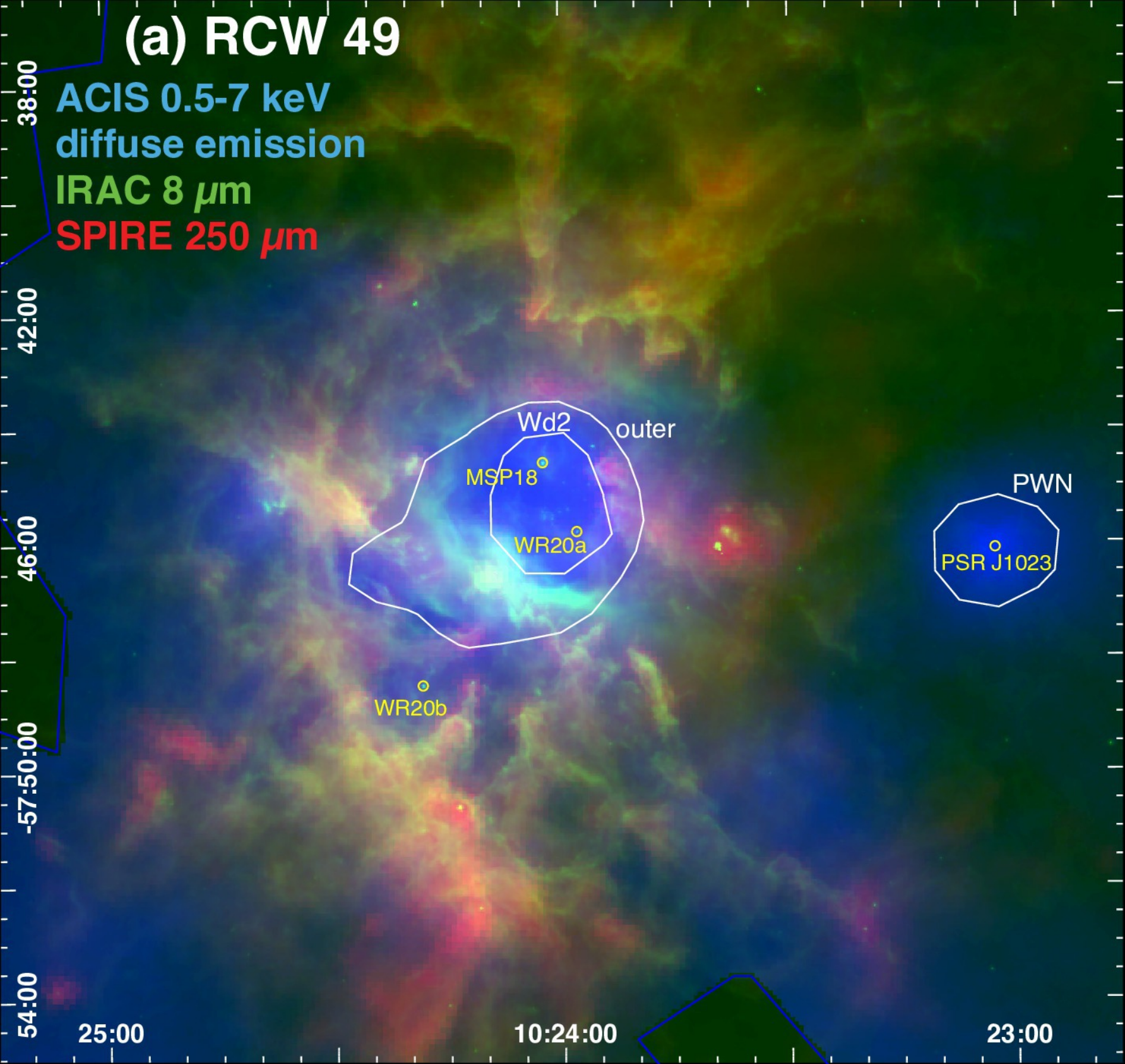}
\includegraphics[width=0.49\textwidth]{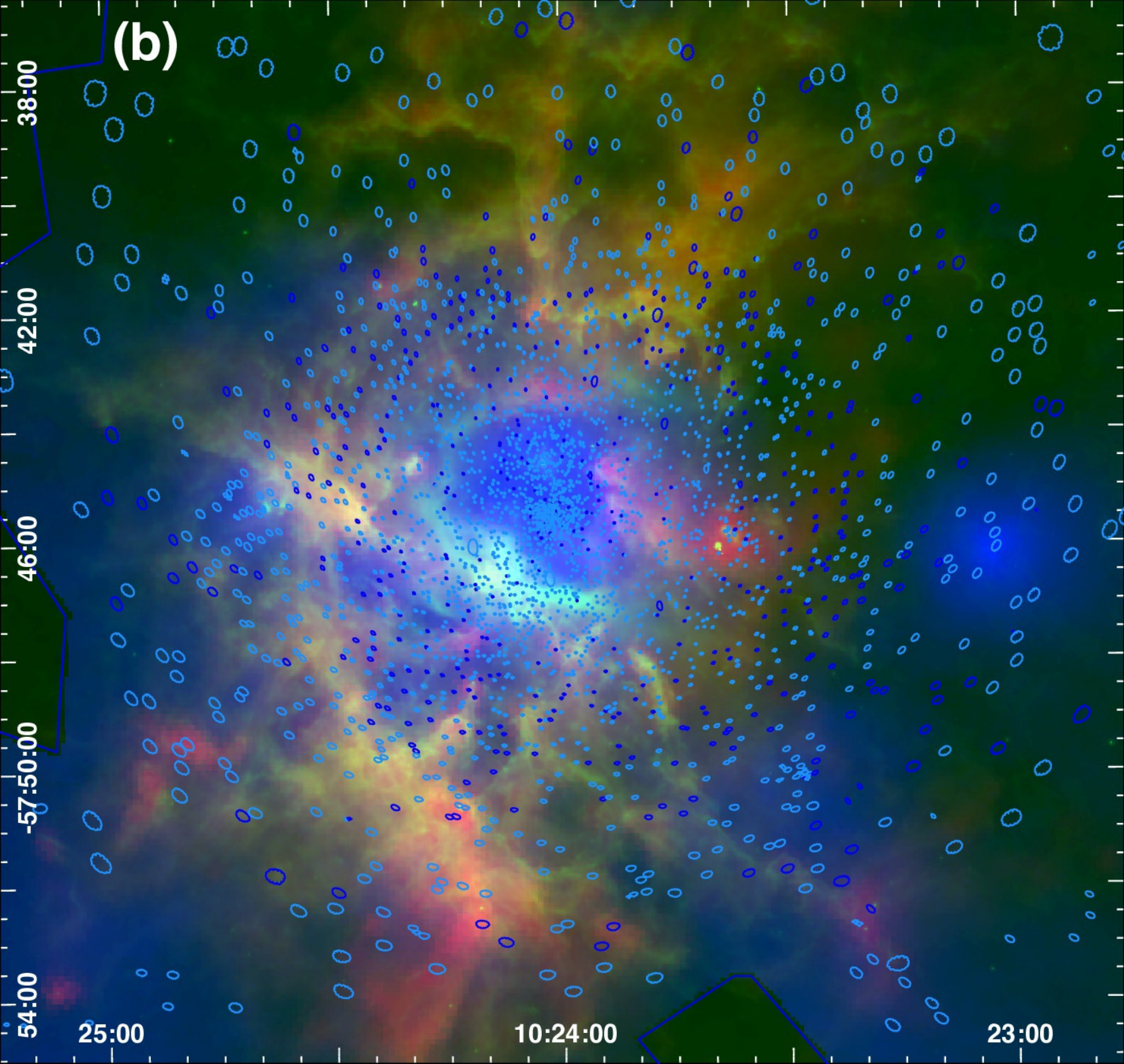}
\caption{RCW 49 itself.
(a) ACIS diffuse emission in the \Spitzer and \Herschel contexts.  The \WISE 22~$\mu$m data are saturated at the center of RCW~49, so the {\em Herschel}/SPIRE data are shown instead.  
(b) The same image as (a), now with X-ray point source extraction apertures overlaid.  Dark blue apertures denote ``occasional'' sources.
\label{rcw49_zoom.fig}}
\end{figure}

Figure~\ref{rcw49_zoom.fig} zooms in on RCW~49, showing the diffuse X-ray emission and thousands of X-ray point sources found in this long ACIS-I observation.  Some point sources and diffuse regions discussed in the text below are marked in panel (a).

WR~21a was captured serendipitously at the edge of 4.7-ks ObsID~9113 (Figure~\ref{rcw49.fig}(b)); its scattered light spills over onto the far off-axis corner of ObsID~14700; we recognized this because of a similar situation with WR~22 in the Carina Nebula \citep{Townsley11a}.  This scattered light was masked to avoid corrupting our images of diffuse X-ray emission.  The ACIS lightcurve shows measurable brightening during this very short observation.  We find no strong concentration of diffuse X-ray emission centered on this star.  WR~21a is slightly piled up despite being imaged 9$\arcmin$ off-axis.  Our fit to the pile-up reconstructed spectrum (Table~\ref{pile-up_risk.tbl}) shows it to be one of the brightest X-ray sources in MOXC3, with $L_{X} = 1.4 \times 10^{34}$~erg~s$^{-1}$ (assuming the RCW~49 distance).  \citet{Gosset16} provide a detailed study of the X-ray characteristics of this source.

The O2~If*/WN6 star WR~20aa, a possible runaway from Wd2 \citep{Roman-Lopes11,Drew18}, is found far off-axis on the ACIS-S devices (Figure~\ref{rcw49.fig}(b)), as MOXC3 source CXOU~J102323.35-580021.2 (c1856).  It has 365 net counts in 45~ks total exposure; a good spectral fit is obtained with a simple {\em tbabs*apec} model, with $N_{H} = 2.4 \times 10^{22}$~cm$^{-2}$, $kT = 0.9$~keV, and 0.5--8~keV corrected flux of $1.15 \times 10^{-12}$~erg~cm$^{-2}$~s$^{-1}$.  At 4.21~kpc, this gives $L_{X} = 2.4 \times 10^{33}$~erg~s$^{-1}$.  Its partner WR~20c is outside the ACIS field of view.

\begin{figure}[htb]
\centering
\includegraphics[width=0.49\textwidth]{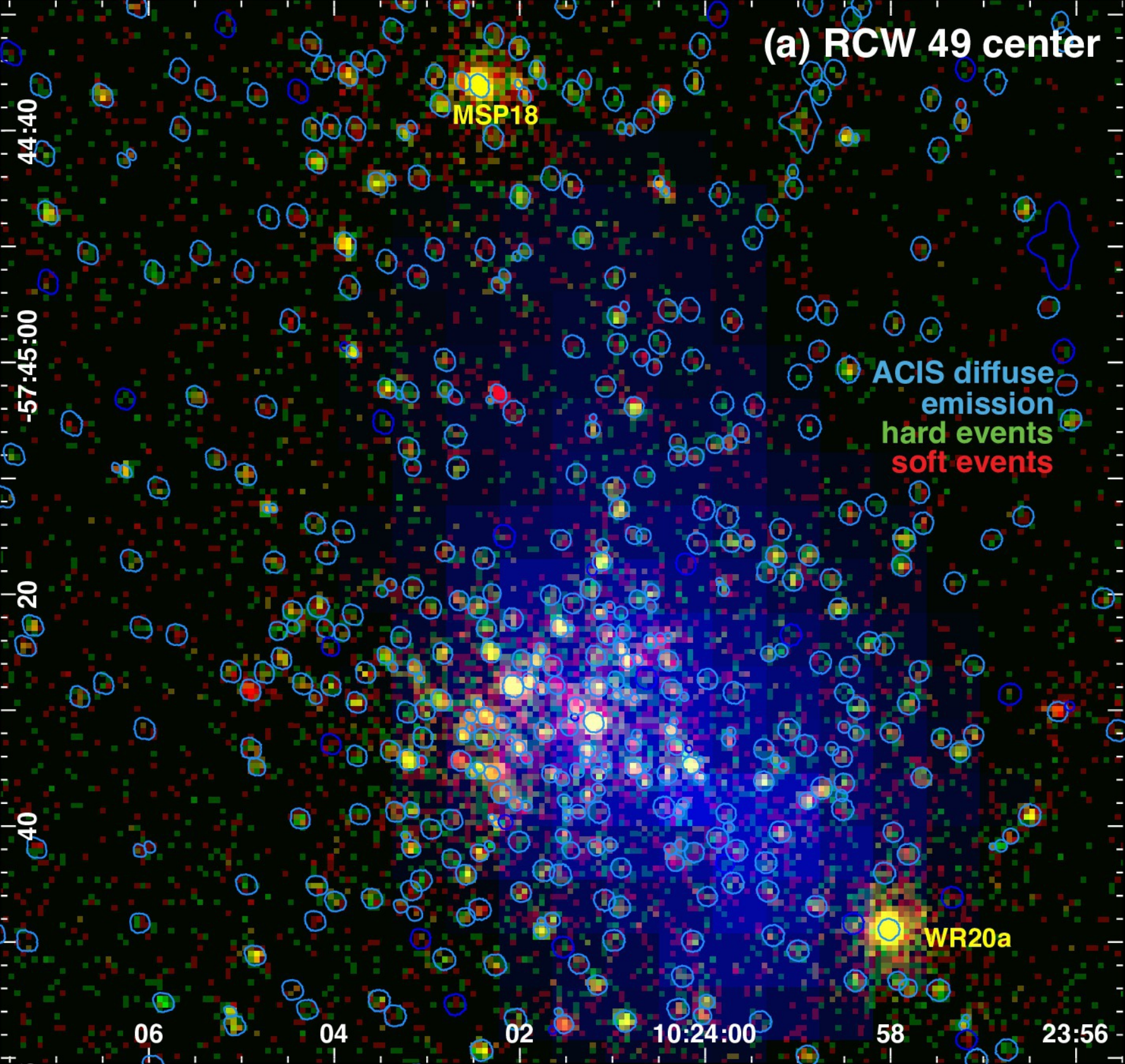} 
\includegraphics[width=0.49\textwidth]{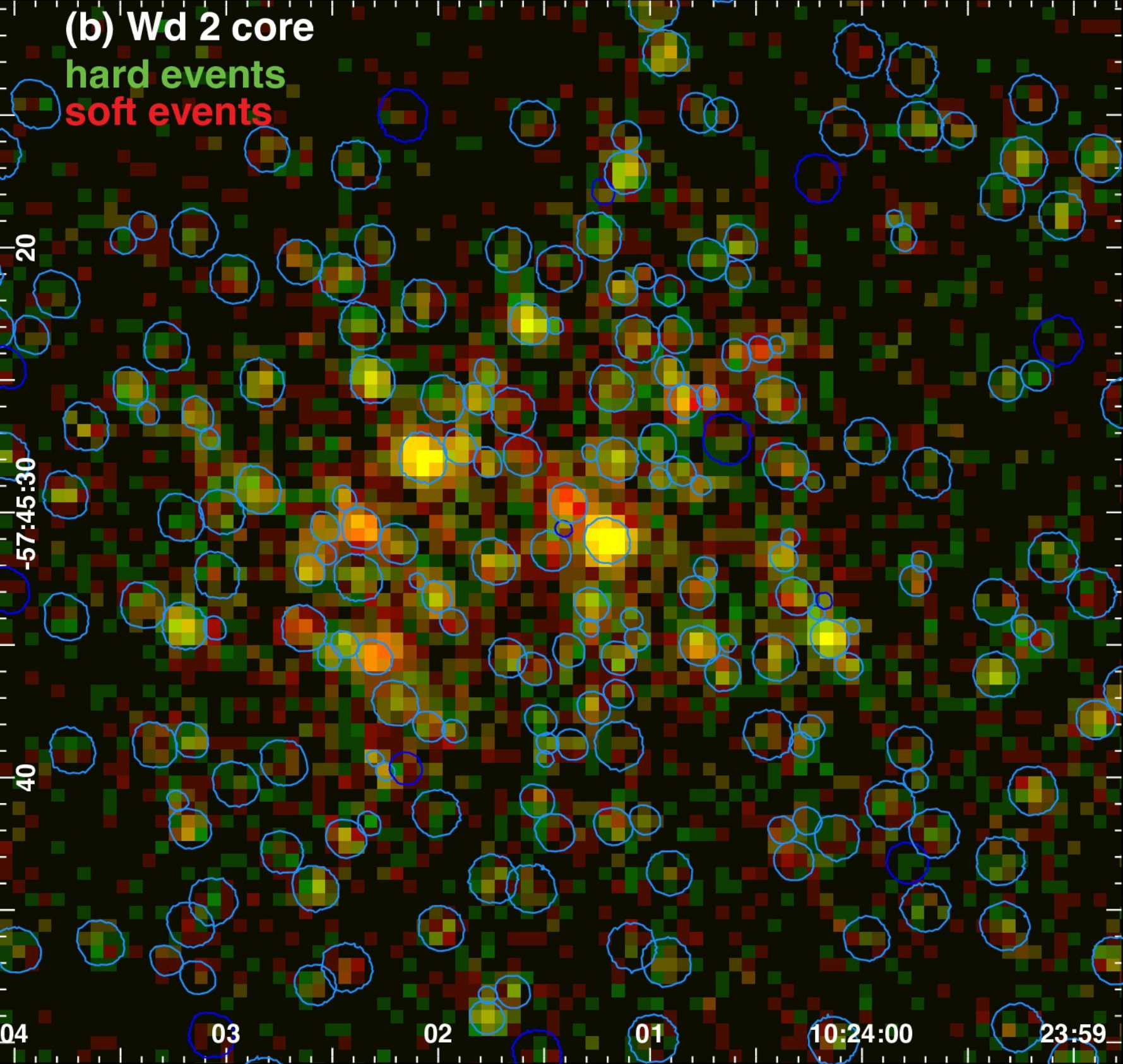}
\caption{Westerlund 2. 
(a)  ACIS event data and diffuse emission at the center of RCW~49; 394 ACIS sources are shown.
(b)  Zoomed version of (a) showing the crowded cluster core; 144 ACIS sources are shown.  The diffuse X-ray image is omitted for clarity. 
\label{Wd2.fig}}
\end{figure}

Figure~\ref{Wd2.fig} focuses in on the ACIS data at the center of RCW~49 and into the core of Wd2.  Panel (a) shows the central $\sim$1.5$\arcmin$ of the field; piled-up sources WR~20a and the O4V star MSP18 \citep{Hur15} are not at the center of the cluster.  Only the central $\sim$20$\arcsec$ of the cluster is crowded.  The ``Wd2'' diffuse extraction region (Figure~\ref{rcw49_zoom.fig}(a)) is larger than the field displayed here.  Panel (b) features ACIS events in the central $\sim$40$\arcsec$ of Wd2.  In the cluster core, many source extraction apertures are reduced to minimize contamination from close neighbors.  Few ACIS observations of MSFRs are this confusion-limited; a similarly dramatic example is NGC~3603 from MOXC1.

We performed pile-up correction on MSP18 and WR~20a (Table~\ref{pile-up_risk.tbl}).  Our simple spectral fits to the pile-up reconstructed spectra confirm the findings of \citet{Naze08} that MSP18 is a hard ($kT > 3$~keV) X-ray source with $L_{X} \sim 1.2 \times 10^{33}$~erg~s$^{-1}$ and no evidence of variability.  \citet{Rauw11} searched for evidence of binarity in MSP18 but found none.  

WR~20a showed a variable lightcurve in ObsID~6411 and between ObsIDs.  It was piled up in three of its four ACIS observations; our spectral fits after pile-up correction show that its X-ray luminosity changed by more than a factor of two, reaching the high value of $L_{X} = 1.3 \times 10^{34}$~erg~s$^{-1}$ in ObsID~6411.  It was observed 7.6$\arcmin$ off-axis in ObsID~12151 and did not pile up there.  A spectral fit to the 139 net counts in those data gives $N_{H} = 4.3 \times 10^{22}$~cm$^{-2}$, $kT = 1.2$~keV, and $L_{X} = 9.1 \times 10^{33}$~erg~s$^{-1}$.  Time-resolved pile-up correction and spectroscopy are warranted for this important CWB.

\begin{figure}[htb]
\centering
\includegraphics[width=0.99\textwidth]{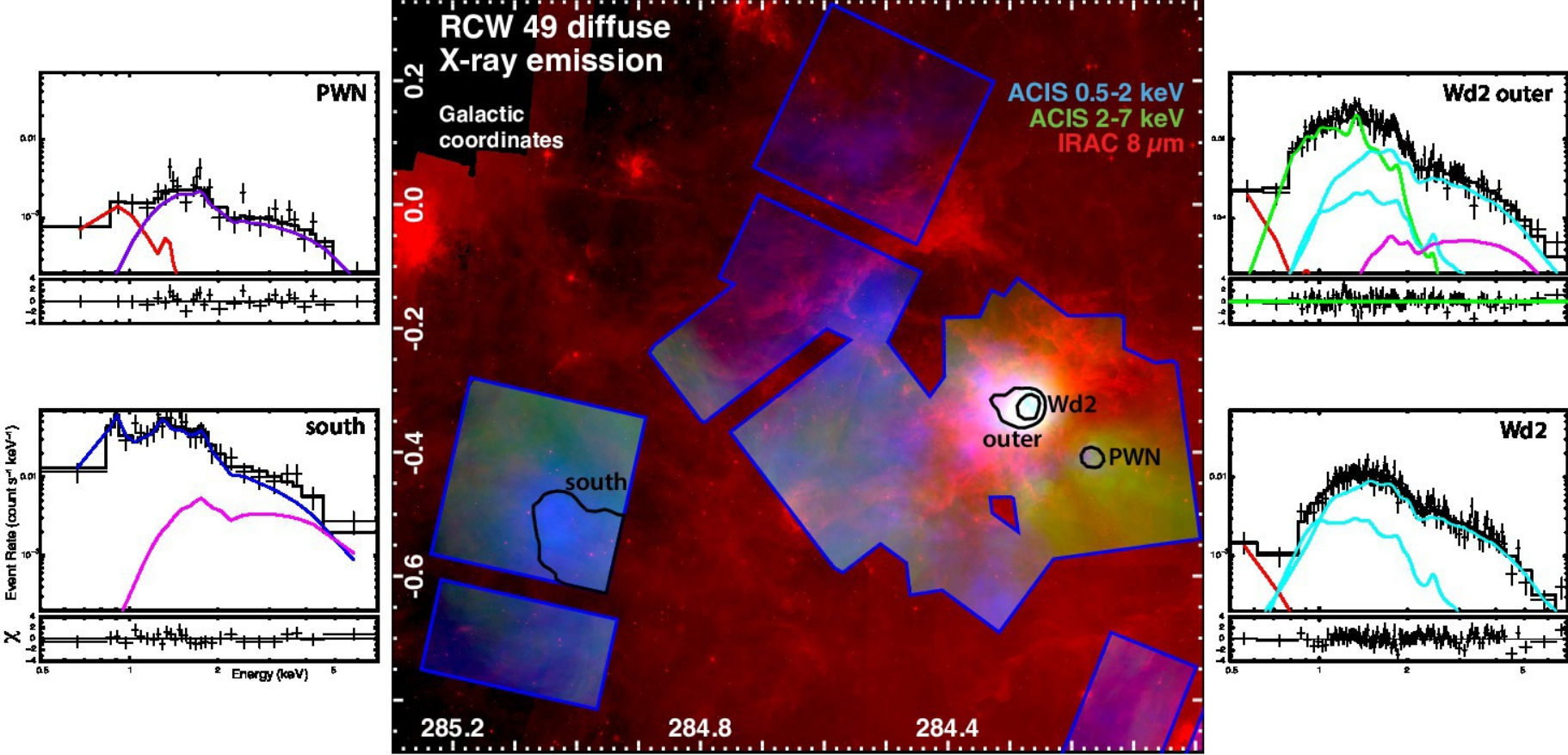}
\caption{Characterizing diffuse X-ray emission across the RCW~49 mosaic.
This image shows ACIS soft-band and hard-band diffuse emission in the \Spitzer context; extraction regions for diffuse spectral fitting are shown in black.  Corresponding spectra are shown below the image; axis ranges are the same for all spectra.  Table~\ref{tbl:diffuse_spectroscopy_style2} gives fit parameters for the spectral models.
\label{rcw49gal+spectra.fig}}
\end{figure}

Although our diffuse X-ray images suggest that truly diffuse X-ray emission pervades RCW~49, the spectrum from our small diffuse extraction region at the center of Wd2 (Figures~\ref{rcw49_zoom.fig}(a) and \ref{rcw49gal+spectra.fig}) is dominated by unresolved pre-MS stars.  This component has very high surface brightness, comparable to that seen in our diffuse spectra of Danks~1 and 2 in G305.  These are all massive clusters that will resolve out into even larger stellar populations with deeper \Chandra observations.  Such efforts are necessary in order to characterize the hot plasmas from massive star feedback that likely pervade these dense cluster cores.

We searched harder for Wd2's hot plasma by extracting another diffuse spectrum from a region surrounding Wd2 but excluding the ``Wd2'' diffuse region described above.  This ``outer'' region (Figures~\ref{rcw49_zoom.fig}(a) and \ref{rcw49gal+spectra.fig}) should be less dominated by unresolved pre-MS stars, and this is in fact what we find.  A soft, absorbed CIE {\it pshock} plasma dominates the soft part of the spectrum.  Absorption for the unresolved pre-MS component was not well-constrained, so it was set equal to that of the hot plasma.  The surface brightness of this plasma is nine times brighter than that of the unresolved stars; it is the second brightest diffuse plasma in MOXC3 (after Danks~1 in G305).  

In retrospect, a good fit to the spectrum of the smaller central Wd2 diffuse region was achieved with this ``outer'' region model, but it reverted to the earlier fit dominated by unresolved pre-MS stars when the parameters were perturbed slightly for error calculations.  Thus it is likely that the core of Wd2 is suffused by soft diffuse emission, but its spectral signature is being overwhelmed by unresolved stars.

%\subsubsection{The Wider Field}
% See sources_of_interest.reg.
The wide-field ACIS mosaic around RCW~49 (Figure~\ref{rcw49.fig}) features four short ACIS-I observations of evolved objects in addition to the deep observation of RCW~49 itself.  As we have come to expect in MOXC studies, these ACIS snapshots and the off-axis ACIS-S CCDs included in the RCW~49 mosaic add substantially to our understanding of the field.  All of these regions reveal interesting diffuse X-ray emission; the most extensive and spatially complex diffuse emission is found in ObsID 9113, with an exposure time of just 4.7~ks.  
% ***Describe spectral characteristics -- if soft, this could be hot plasma leaking from RCW~49!***

The target of the 10-ks ObsID 14700 was the {\em Fermi} source 2FGL~J1027.4-5730c; we find no X-ray sources within 1.4$\arcmin$ of its position.   
% (10 27 27.4 -57 30 40 from Simbad), an ``unclassified Fermi source'' (PI Pavlov); they were looking for a pulsar.  
This observation does capture diffuse X-ray emission in its far southwestern corner, hinting that the large bubble east of RCW~49 may be filled with hot plasma leaking from the MSFR.  There is also faint diffuse emission on the I-array (Figure~\ref{rcw49gal+spectra.fig}). 

% PSR~J1023-5746:  10 23 02.388 -57 46 09.86 from Simbad
ObsID~12151 was part of the \Chandra Pulsar Survey \citep[ChaPS][]{Kargaltsev12}; it obtained a 10-ks ACIS-I snapshot of PSR~J1023-5746, west of RCW~49.  The ChaPS distance for this pulsar is uncertain and consistent with the distance to RCW~49 used here.  We extracted a spectrum for PSR~J1023-5746 (MOXC3 source CXOU~J102302.86-574606.4, c1430) from all four RCW~49 ObsIDs that included it, recovering 365 net counts in 142~ks of observations.  A simple \XSPEC~model ({\em tbabs*pow}) gives a good fit, with $N_{H} = 1.3 \times 10^{22}$~cm$^{-2}$, photon index 0.9, and 0.5--8~keV corrected flux of $8.3 \times 10^{-14}$~erg~cm$^{-2}$~s$^{-1}$.  At 4.21~kpc, this gives $L_{X} = 1.8 \times 10^{32}$~erg~s$^{-1}$.  These results are similar to those found in ChaPS.

PSR~J1023-5746 is surrounded by a $>$1$\arcmin$ patch of diffuse emission (Figures~\ref{rcw49.fig}(b) and \ref{rcw49_zoom.fig}(a)), confirming the suggestion by ChaPS of a PWN.  We fit the PWN spectrum with a moderately-absorbed CIE diffuse plasma ($kT = 0.3$~keV) with fairly low surface brightness, plus a power law with photon index 2.0.  The power law dominates the spectrum, with $L_{X} = 3 \times 10^{32}$~erg~s$^{-1}$ within our diffuse extraction region.   

% HESS J1023-575 -- See Yang17, Aharonian18.  I don't want to get into this CR/gamma ray business.  

Surrounding PSR~J1023-5746 and its PWN, hard diffuse X-ray emission (Figure~\ref{rcw49gal+spectra.fig}) fills the large radio shell to the southwest of RCW~49 described by \citet{Whiteoak97} (found at roughly 284.1, -0.5 in Figure~\ref{rcw49gal+spectra.fig}).  This may trace a cavity SNR, suggesting that the radio shell is the result of an earlier generation of massive star formation associated with the RCW~49 complex.

% PSR~J1028-5819:  10 28 27.888 -58 19 06.14 from Simbad
ObsID~12150 was also part of ChaPS; it is a 10-ks ACIS-I snapshot of PSR~J1028-5819 (MOXC3 source CXOU~J102827.90-581906.3, c7406).  ChaPS gives a distance of 2.76~kpc to this pulsar.
% Simbad doesn't list a SNR within a 2$\arcmin$ cone around 10:28:10.774 -58:19:00.986, the peak in the diffuse X-ray emission.
A region of bright diffuse emission (labeled ``south'' in Figure~\ref{rcw49gal+spectra.fig}) is centered $\sim$$2\arcmin$ west of PSR~J1028-5819 (Figure~\ref{rcw49.fig}(b)).  We find a good fit to its spectrum using a single {\it pshock} component with a short ionization timescale and a hard thermal plasma ($kT = 2.4$~keV).  The surface brightness and X-ray luminosity (assuming the same distance as PSR~J1028-5819) of this emission are high compared to many other MOXC3 diffuse regions, but comparable to the ``south'' region in Berkeley~87.  This could be the SNR associated with PSR~J1028-5819.

%\subsubsection{RCW~49 Summary}
In summary, soft diffuse emission surrounds Wd2 and documents its massive star feedback.  Pulsars surrounded by their own diffuse X-ray emission with hard spectra suggest SNRs associated with RCW~49 and the foreground PSR~J1028-5819.  Our wide mosaic around RCW~49 hints at interesting morphology for more extended diffuse X-ray structures, although the observations are too short to provide much detail.  

Wd2 has been re-observed with ACIS (2018 September, 265~ks ACIS-I, PI L.~Lopez).  The combined ACIS dataset of nearly 400~ks duration will reach 0.5~dex deeper into the X-ray luminosity function of the pre-MS population and will provide a long baseline for variability studies.  It should facilitate more detailed characterization of diffuse X-ray emission both east of Wd2 and in the southwest radio shell; we look forward to the results of these deep \Chandra observations.

%\clearpage
%=============================================================================
\section{Summary and Conclusions \label{sec:summary}}

As in MOXC1 and MOXC2, we find a menagerie of X-ray emission in MOXC3.  \Chandra resolves nearly 28,000 point sources in MOXC3's 14 MSFRs.  Diffuse X-ray emission pervades and surrounds these MSFRs, offering ghostly evidence of massive star feedback at work throughout the lifetimes of these massive stars, from MYSOs to SNRs.  X-ray point sources---even bright ones---come and go in the ever-varying X-ray sky (e.g., WR~48a's neighbor in G305).  

Giant \hii regions sport multiple young clusters and cavities filled with hot plasma (e.g., NGC~6357, G305).  Throughout MOXC3, distributed populations of X-ray sources add to the evidence that star formation is multi-generational, rarely restricted to a single impulsive event.  Obscured clumps of X-ray sources indicate recent or ongoing star formation; X-ray emission mechanisms must commence early in the star formation process in order for us to detect these sources (e.g., NGC~2264~IRS1, DR15, AFGL~4029, ON~2S).  

As often illustrated by the far off-axis ACIS-S CCDs included in our analysis, MOXC3's regions of ample star birth are never far from sites of massive star death, which leave behind hard diffuse X-ray emission at MSFR peripheries (e.g., RCW~49).  These faint SNR candidates probably fill voids in the disturbed ISM around today's MSFRs; wider ACIS mosaics are needed to map their extent and to separate their diffuse emission from the distributed populations of young stars that also tend to inhabit multi-generational MSFR complexes.

Three MOXC3 targets (NGC~2264, NGC~6193, and G305) include HETG data.  Although these data add complexity to our analysis, they are useful for accessing parts of the sky that lack regular ACIS imaging data and for providing extra counts and time sampling for fields that do have ACIS data.  Although the bright massive stars that were the subject of the HETG observations are still piled up in the zeroth-order images that we analyzed, they are much less affected than in ACIS-only data, making their lightcurves and CCD-resolution spectra easier to correct for pile-up distortion.  

Three MOXC3 targets (NGC~6193 + RCW~108-IR, AFGL~4029, and NGC~281) share the same relatively simple geometry:  a revealed massive stellar cluster adjacent to a molecular cloud faced by a prominent ionization front seen in the mid-IR, with a younger obscured cluster on the far side of that ionization front still emerging from the molecular cloud.  We find evidence of massive star feedback in the form of bright diffuse X-ray emission in the cavity created by the older cluster, sharply cut off at or near the ionization front.  \citet{Dale15} warn against assuming that this geometry constitutes a triggering scenario, but it seems possible (at least) that feedback from the older cluster is influencing star formation near its interface with the nearby molecular cloud. 

Two MOXC3 targets (AE~Aur and Berk87) have sightlines that also include more distant MSFRs (Aur~OB1 and ON~2S, respectively).  Such fortunate (or unfortunate) geometries are common in Galactic Plane studies, making it difficult to assign distances to all the X-ray sources found in ACIS data; this is especially true for patches of diffuse emission.

Many of our predictions from the last CCCP paper \citep{Townsley11c}, which compared diffuse X-ray emission in several famous MSFRs to that in the Carina Nebula, have proven true here.  We said that unresolved young stars could overwhelm the signature of hot plasma emission at the centers of massive clusters in spectra from short ACIS observations; that was seen here in IC~1590 (in NGC~281) and in Wd2 (in RCW49).  The plasma signature was recovered, in those examples, in a spectrum extracted from a region immediately surrounding the cluster cores.  We said that multi-generational MSFRs left a complicated mix of diffuse X-ray emission from cavity SNRs and wind shocks superposed on both concentrated and distributed populations of young stars.  This is certainly the case with all MOXC targets; much more could be learned with wider and deeper \Chandra observations.

The plasma temperatures and surface brightnesses of the soft diffuse emission seen often in MOXC3 (Table~\ref{tbl:diffuse_spectroscopy_style2}) are comparable to those found in Carina \citep{Townsley11b} and other famous MSFRs \citep{Townsley11c}.  Such faint, soft X-ray plasmas may be ubiquitous around mid- to late-O stars and young early-B stars, but their emission is easily absorbed by surrounding natal clouds or intervening ISM material, making this signature of massive star feedback often difficult to detect.  

As seen in MOXC1 and MOXC2, again MOXC3 shows that early-B MYSOs ionizing ultracompact \hii regions and young embedded clusters often exhibit harder diffuse X-ray emission seen behind heavy obscuration; brighter, softer plasmas may also exist in these regions but remain undetected because their X-ray emission is completely absorbed by the surrounding dense material.  Scenarios for the early evolution of massive stars (and their accompanying clusters, disks, and planets) should not ignore these high-energy radiation fields; we find them too often to dismiss them as anomalous.  In particular, NGC~2264~IRS1 has very hard, variable X-ray emission that mimics pre-MS star flares; this early-B MYSO strongly suggests that it formed in the same way as lower-mass stars. 

We continue our analysis efforts on archival {\em Chandra}/ACIS observations of MSFRs.  Future MOXC studies will concentrate on more distant targets, including some MSFRs in the Galactic Center and the Magellanic Clouds, then will return to relatively nearby complexes, emphasizing MSFRs hosting early- and mid-O stars within 2~kpc.  Once again, we encourage the star formation community to use our Zenodo archives and to incorporate X-ray data products into a multiwavelength approach to MSFR science analysis.  Such a broad foundation is necessary for a sophisticated understanding of star formation, evolution, and feedback across the Galaxy and beyond.

%\clearpage
% =============================================================================
\acknowledgements Acknowledgments:  
We appreciate the time and effort donated by our anonymous referee to comment on this paper.  
We are grateful to Breanna Binder for making us aware of the \Chandra observations of AE~Aurigae and to Michael Kuhn for a summary of Gaia DR2 systematics.
This work was supported by the {\em Chandra X-ray Observatory} archive grant
AR7-18004X   % AO17 (co-I's Broos, Povich)
(PI:  L.\ Townsley)
and by the Penn State ACIS Instrument Team Contract SV4-74018.  Both of these were issued by the \Chandra X-ray Center, which is operated by the Smithsonian Astrophysical Observatory for and on behalf of NASA under contract NAS8-03060.
The ACIS Guaranteed Time Observations included here were selected by the ACIS Instrument Principal Investigator, Gordon P.\ Garmire, of the Huntingdon Institute for X-ray Astronomy, LLC, which is under contract to the Smithsonian Astrophysical Observatory; Contract SV2-82024.
M.S.P.\ is supported by the National Science Foundation through grant CAREER-1454333.
This research used data products from the \Chandra Data Archive, software provided by the \Chandra X-ray Center in the application package {\it CIAO}, and {\it SAOImage DS9} software developed by the Smithsonian Astrophysical Observatory.  This research also used data products from the {\em Spitzer Space Telescope}, operated by the Jet Propulsion Laboratory (California Institute of Technology) (JPL/CalTech) under a contract with NASA, data products from the {\em Wide-field Infrared Survey Explorer} ({\em WISE}), which is a joint project of the University of California, Los Angeles and JPL/CalTech, funded by NASA, and data products from the European Space Agency (ESA) {\em Herschel Space Observatory}, with science instruments provided by European-led Principal Investigator consortia and with important participation from NASA.  This work also used data from the ESA mission {\it Gaia} (\url{https://www.cosmos.esa.int/gaia}), processed by the {\it Gaia} Data Processing and Analysis Consortium (DPAC,
\url{https://www.cosmos.esa.int/web/gaia/dpac/consortium}); funding for the DPAC has been provided by national institutions, in particular the institutions participating in the {\it Gaia} Multilateral Agreement.  This research used NASA's Astrophysics Data System Bibliographic Services, and the SIMBAD database and VizieR catalog access tool provided by CDS, Strasbourg, France.

% See http://journals.aas.org/authors/aastex/facility.html
\vspace{5mm}
\facilities{CXO (ACIS), Gaia, Herschel (SPIRE), Spitzer (IRAC, MIPS), WISE}.

\software{
	{\em ACIS Extract} \citep{Broos10,AE12,AE16},
	\CIAO\ \citep{Fruscione06},
	\DSnine\ \citep{Joye03},
	\MARX\ \citep{Davis12},
	\TOPCAT\ \citep{Taylor05},
	\XSPEC\ \citep{Arnaud96}.
}

% =============================================================================
% APPENDICES
%\appendix
% =============================================================================

% =============================================================================
% Bibliography
\input{bib.tex}

\end{document}

%% file: target_table.tex
\noindent
\setlength{\tabcolsep}{0.05in}
\renewcommand{\arraystretch}{0.9}  % to change row spacing in table

\begin{deluxetable}{lrcDDDrccrl}
\tablecaption{MOXC3 Targets \label{targets.tbl}}
\tabletypesize{\tiny}

\tablehead{
\colhead{MOXC3 \rule{0mm}{3ex}} & 
\colhead{Galactic} &  
\colhead{Celestial J2000} & 
\twocolhead{Distance} & 
\twocolhead{Scale} & 
\twocolhead{$\langle A_V \rangle$} & 
\colhead{Nominal} & 
\colhead{$\log L_{tc}$} & 
\colhead{$M_{50\%}$}   & 
\colhead{X-ray Srcs} &
\colhead{Distance Reference} \\
\colhead{Target}  & 
\colhead{($l,b$)}  &                                                                         
\colhead{($RA,Dec$)} & 
\twocolhead{(kpc)}    & 
\twocolhead{($\arcmin$/pc)} & 
\twocolhead{(mag)} & 
\colhead{Exp (ks)} &
\colhead{(erg/s)} & 
\colhead{($M_{\sun}$)} & 
\colhead{(\#)} &
\colhead{} 
}
\decimalcolnumbers
\startdata
\\
% NOTE -- cols Distance, Scale, and <A_V> can't be blank, because of "D" formats above!
%         Target           Galactic coords           Celestial coords        Distance     Scale    <A_V>      Exp    lim L_tc    M_50%    X-ray srcs     Dist ref
 \object{AE Aur     }&   172.08  $-$2.26    &    05 16 18.2  $+$34 18 44    &  0.402   &  8.55  &   1.8    &  141   & \nodata & \nodata &    945     &   \citet{Bailer18}         \\ % Gaia DR2 182071570715713024 (AE Aur) has a B-J18 distance of 401.55 pc.  
 \object{NGC 2264   }&   203.23  $+$2.05    &    06 40 58.7  $+$09 34 14    &  0.738   &  4.66  &   0.5    &  293   & 28.04   &    0.1  &   3373     &   \citet{Kuhn19}           \\ % Cantat18 gets D = 723pc.  Coords and exposure are for Micela's LP.  
 \object{NGC 6193	}&	 336.71  $-$1.57 	& 	 16 41 20.4	 $-$48 45 47 	&  1.19    &  2.89  &   2.6    & $\sim$42\tablenotemark{a} & 29.23   &    0.2  &   2596     &   \citet{Cantat18}         \\ % Coords, 358 ks HETG exposure, and A_V (Wolk08 AJ paper Table 9 (east region)) are for NGC 6193.  Cantat18 gets D = 1190pc for NGC6193.  The emap shows 47.5% as much exposure at aimpoint here as for RCW 108 -- so effectively 42ks of bare ACIS-I time.  
 \object{RCW 108-IR }&   336.49  $-$1.48    &    16 40 00.1  $-$48 51 45    &  1.19    &  2.89  &  14      &   89   & 29.56   &    0.3  &see NGC 6193&   \citet{Cantat18}         \\ % Coords, exposure, and A_V (Wolk08 AJ paper Table 9) are for RCW 108.  We assume same distance as NGC6193.
 \object{DR15       }&    79.29  $+$0.31    &    20 32 20.3  $+$40 16 33    &  1.4     &  2.46  &  20      &   40   & 30.17   &    0.7  &    651     &   \citet{Rygl12}           \\ % Ionized by B0 stars.  Captures part of IRDC G79.3+0.3 in ACIS-I FoV of ObsID 12390.  Distance is a swag.
 \object{Aur OB1    }&   172.08  $-$2.26    &    05 16 18.2  $+$34 18 44    &  1.63    &  2.11  &   1.5    &  141   & 29.23   &    0.2  & see AE Aur &   this paper               \\ % Aur OB1 distance is from Matt's Gaia calc.																																											 % 
 \object{Berkeley 87}&    75.73  $+$0.30    &    20 21 44.3  $+$37 22 31    &  1.66    &  2.07  &   6      &   70   & 29.68   &    0.3  &    879     &   \citet{Cantat18}         \\ % Cantat18 gets D = 1661pc.  Binder18 gets D = 1.74 kpc.
 \object{NGC 6231   }&   343.46  $+$1.16    &    16 54 14.3  $-$41 50 37    &  1.71    &  2.01  &   3.8    &  120   & 29.37   &    0.2  &   3450     &   \citet{Kuhn19}           \\ % Cantat18 gets D = 1617pc.
 \object{NGC 6357   }&   353.17  $+$0.90    &    17 24 43.4  $-$34 11 56    &  1.77    &  1.94  &   5      &   40   & 29.96   &    0.5  &   5269     &   \citet{Kuhn19}           \\ % Cantat18 gets D = 1678pc.  Binder18 gets D = 1.78 kpc.  Coords are for Pismis 24.  Total #srcs includes the GTO ptg at the MSFR/IRDC interface but not the 3 IRDC ptgs.  <A_V>=5 mag came from MOXC1.
 \object{AFGL 4029  }&   138.26  $+$1.57    &    03 01 19.4  $+$60 30 49    &  2.25    &  1.53  &   3      &   82   & 29.70   &    0.3  &    833     &   \citet{Cantat18}         \\ % Assuming same distance as IC1848, Cantat18 gets D = 2251pc.
 \object{h Per      }&   134.64  $-$3.75    &    02 19 02.2  $+$57 07 12    &  2.34    &  1.47  &   1.9    &  231   & 29.23   &    0.2  &   3419     &   \citet{Cantat18}         \\ % NGC 869.  Cantat18 gets D = 2336pc.
 \object{NGC 281    }&   123.07  $-$6.31    &    00 52 27.1  $+$56 33 54    &  2.72    &  1.26  &   3      &   98   & 29.81   &    0.4  &   1185     &   \citet{Cantat18}         \\ % Cluster is called IC 1590.  Sato08 got D = 2.82 kpc.  Cantat18 gets D = 2719pc.
 \object{Onsala 2S  }&    75.78  $+$0.34    &    20 21 44.1  $+$37 26 40    &  3.5     &  0.98  &  20      &   70   & 30.70   &    1.5  & see Berk87 &   \citet{Skinner19,Xu13}   \\ % Coords are for UCHIIR G75.78+0.34, from Simbad.  Skinner's distance to ON 2S comes from Xu13's maser parallaxes.
 \object{G305       }&   305.32  $+$0.07    &    13 12 17.5  $-$62 42 20    &  3.59    &  0.96  &   9.5    &  119   & 30.23   &    0.8  &   2184     &   \citet{Binder18}         \\ % Params only for Obs8922.  Second ptg HETG 98ks on WR48A adds an effective ~11.5 ks.  Cantat18 gets D = 5324pc to Danks1, 6324pc to Danks2.  These seem crazy!
 \object{RCW 49     }&   284.27  $-$0.33    &    10 24 02.5  $-$57 45 23    &  4.21    &  0.82  &   6.5    &  132   & 30.25   &    0.8  &   3139     &   \citet{Cantat18}         \\ % Cantat18 gets D = 4208pc to Wd2.  
% 																																													 % Full mosaic (3649 total srcs) includes western bit on G284.0-0.9, WR18, and a planetary nebula.  I've separated that out into the MSFR snapshots paper.  It has 510 srcs. 
\enddata

\tablecomments{
Col.\ (1):  Name we will use for the MSFR in MOXC3 (except for the first entry, which is the star AE~Aurigae).
\\Col.\ (2):  Galactic coordinates for the MSFR, given to facilitate comparison with the literature.
\\Col.\ (3):  Coordinates for the primary {\em Chandra}/ACIS observation of the MSFR, similar to those shown in the \Chandra X-ray Center's {\em Chaser} observation search tool.  % Coords requested by the observer (derived from event file header keywords RA_TARG, DEC_TARG).  
\\Col.\ (4):  Distance from the literature or estimated in this paper; citations are given in Col.\ (11).  These are often derived from \Gaia DR2 data \citep{Gaia16,Gaia18}; exceptions are DR15 and Onsala~2S.  Papers deriving \Gaia distances make different assumptions about DR2 systematics; we do not attempt to resolve these differences.
\\Col.\ (5):  Image scale (in arcminutes per parsec) assuming the distance given in Col.\ (4).
\\Col.\ (6):  Approximate average absorption to the target, estimated from a variety of literature sources.  Most MSFRs have highly variable and spatially complex obscuration, so this value should be used only as a rough indicator.
\\Col.\ (7):  Typical exposure time for the main MSFR.  Most mosaics have a wide range of exposures; detailed exposure maps are shown in Section~\ref{sec:targets}.
\\Col.\ (8):  Rough limiting luminosity where the brighter half of the X-ray population is sampled, using {\em PIMMS} simulations for a 5-count pre-MS star on-axis, with an {\it apec} plasma ($kT = 2.7$~keV) and 0.4*Z$_\odot$ abundances \citep{Preibisch05}.  The subscripts ``tc'' mean total band (0.5--8~keV), corrected for extinction.
\\Col.\ (9):  Corresponding limiting mass, where the brighter half of the X-ray population is captured \citep[][Figure 3]{Preibisch05}.  % For shallow observations, this limit is higher than pre-MS masses, so only ``bright'' sources (some massive stars and IMPS) are expected.
\\Col.\ (10):  Total number of X-ray point sources found across the entire mosaic.
}
\tablenotetext{a}{
This time is shown as approximate because we are using the zeroth-order image from an HETG observation, so the effective ACIS-only integration time is energy-dependent (1~keV here).   
}

\end{deluxetable} 

%% file: observing_log.tex
%  2019 May 23 08:50
%  acis_extract, version 5288  2018-06-23; ae_flatten_collation, version 5300  2018-08-08; hmsfr_tables, version 5422  2019-05-13

% COLUMN SPACING
\setlength{\tabcolsep}{0.5mm}

% ROW SPACING
% The traditional method for controlling row spacing, e.g.
%   \renewcommand{\arraystretch}{1.5} 
% does not work with AASTeX tables.  The AASTeX manual says you must edit the class file:
%
%    "Authors can control the space between data lines in tables by making a simple modification to the classfile. The line \def\arraystretch{1.0} is set to the default value of 1. To increase the space between data lines by 10% change the argument to 1.1. Likewise, to decrease the space between data lines by 10% to produce a tighter table use 0.9."

%\begin{longrotatetable}
\startlongtable
\begin{deluxetable}{crrrccrlllllC}
\centering 
\tabletypesize{\tiny} \tablewidth{0pt}
%\tablecolumns{8}

\tablecaption{ Log of {\em Chandra} Observations 
 \label{tbl:obslog}}

\tablehead{
\colhead{Target} & 
\colhead{ObsID} & 
\colhead{Start Date} & 
\colhead{Exp} & 
\multicolumn{2}{c}{On-axis Position} & 
\colhead{Roll} & 
%\multicolumn{2}{c}{} & 
\colhead{Config} & 
\colhead{Mode} &
\colhead{PI}  &
\colhead{TGAIN} &
\colhead{OBF} &
\colhead{Shift}  \\
\cline{5-6}
\colhead{} & 
\colhead{} & 
\colhead{(UT)} & 
\colhead{(s)} & 
\colhead{$\alpha_{\rm J2000}$} & 
\colhead{$\delta_{\rm J2000}$} & 
\colhead{(\arcdeg)} & 
\colhead{} &
\colhead{} &
\colhead{} &
\colhead{} &
\colhead{} &
\colhead{(SKY pixel)} 
}
%\colnumbers
\decimalcolnumbers
\startdata
\multicolumn{3}{l}{\bf AE Aur; Aur OB1 } \\
                   Ae Aur & \dataset[ 19943]{  ADS/Sa.CXO#obs/19943} &       2016-11-16 &   14884 & 05:16:20.95 & +34:17:58.1 &  122 &     I (0123) &  TE-VF & B Rangelov & { 2016-11-01N6 } & N12 & (-0.113, +2.336) \\ % 0\% exposure discarded for high background.
                   Ae Aur & \dataset[ 19445]{  ADS/Sa.CXO#obs/19445} &       2016-11-17 &   44491 & 05:16:21.05 & +34:17:57.3 &  122 &     I (0123) &  TE-VF & B Rangelov & { 2016-11-01N6 } & N12 & (-0.250, +2.153) \\ % 0\% exposure discarded for high background.
                   Ae Aur & \dataset[ 19979]{  ADS/Sa.CXO#obs/19979} &       2017-01-03 &   26725 & 05:16:08.27 & +34:18:45.8 &  250 &     I (0123) &  TE-VF & B Rangelov & { 2016-11-01N6 } & N12 & (+0.548, -1.059) \\ % 0\% exposure discarded for high background.
                   Ae Aur & \dataset[ 19941]{  ADS/Sa.CXO#obs/19941} &       2017-01-04 &   26725 & 05:16:08.27 & +34:18:45.4 &  250 &     I (0123) &  TE-VF & B Rangelov & { 2016-11-01N6 } & N12 & (+0.288, -0.730) \\ % 0\% exposure discarded for high background.
                   Ae Aur & \dataset[ 19951]{  ADS/Sa.CXO#obs/19951} &       2017-01-06 &   27716 & 05:16:08.28 & +34:18:43.9 &  250 &     I (0123) &  TE-VF & B Rangelov & { 2016-11-01N6 } & N12 & (+0.348, -0.758) \\ % 0\% exposure discarded for high background.
\hline
\multicolumn{3}{l}{\bf NGC 2264   } \\
                 NGC 2264 & \dataset[  2550]{  ADS/Sa.CXO#obs/02550} &       2002-02-09 &   48135 & 06:40:47.77 & +09:50:43.5 &  281 &   I (012367) &   TE-F & J Stauffer & { 2002-02-01N6 } & N11 & (+0.038, -0.213) \\ % 0\% exposure discarded for high background.
                 NGC 2264 & \dataset[  2540]{  ADS/Sa.CXO#obs/02540} &       2002-10-28 &   94156 & 06:40:58.73 & +09:34:14.2 &   78 &   I (012367) &   TE-F & S Sciortino & { 2002-08-01N6 } & N11 & (-0.093, +0.287) \\ % 1\% exposure discarded for high background.
                   15 Mon & \dataset[  5401]{  ADS/Sa.CXO#obs/05401} &       2005-11-20 &   54751 & 06:40:59.18 & +09:53:47.1 &   71 &   H (456789) &   TE-F & W Waldron & { 2005-11-01N6 } & N11 & (+0.307, -0.032) \\ % 0\% exposure discarded for high background.
                   15 Mon & \dataset[  6248]{  ADS/Sa.CXO#obs/06248} &       2005-12-13 &    7614 & 06:40:59.19 & +09:53:44.6 &   52 &   H (456789) &   TE-F & W Waldron & { 2005-11-01N6 } & N11 & (+0.078, -0.245) \\ % 0\% exposure discarded for high background.
                   15 Mon & \dataset[  6247]{  ADS/Sa.CXO#obs/06247} &       2006-01-31 &   37471 & 06:40:58.40 & +09:53:38.6 &  287 &   H (456789) &   TE-F & W Waldron & { 2005-11-01N6 } & N11 & (-0.221, +0.006) \\ % 0\% exposure discarded for high background.
         NGC 2264 field 1 & \dataset[  9768]{  ADS/Sa.CXO#obs/09768} &       2008-03-12 &   27786 & 06:41:12.59 & +09:29:32.3 &  270 &     I (0123) &   TE-F & G Micela & { 2008-02-01N6 } & N11 & (+0.426, +0.507) \\ % 0\% exposure discarded for high background.
         NGC 2264 field 1 & \dataset[  9769]{  ADS/Sa.CXO#obs/09769} &       2008-03-28 &   29757 & 06:41:12.47 & +09:29:31.7 &  266 &     I (0123) &   TE-F & G Micela & { 2008-02-01N6 } & N11 & (+0.373, +0.321) \\ % 0\% exposure discarded for high background.
                 NGC 2264 & \dataset[ 14368]{  ADS/Sa.CXO#obs/14368} &       2011-12-03 &   73520 & 06:40:58.99 & +09:34:42.7 &   63 &    I (01237) &  TE-VF & G Micela & { 2011-11-01N6 } & N11 & (-0.941, -0.040) \\ % 1\% exposure discarded for high background.
                 NGC 2264 & \dataset[ 13610]{  ADS/Sa.CXO#obs/13610} &       2011-12-05 &   92537 & 06:40:58.99 & +09:34:42.8 &   63 &    I (01237) &  TE-VF & G Micela & { 2011-11-01N6 } & N11 & (-0.876, -0.262) \\ % 0\% exposure discarded for high background.
                 NGC 2264 & \dataset[ 13611]{  ADS/Sa.CXO#obs/13611} &       2011-12-07 &   60235 & 06:40:59.00 & +09:34:42.7 &   63 &    I (01237) &  TE-VF & G Micela & { 2011-11-01N6 } & N11 & (-0.816, -0.184) \\ % 0\% exposure discarded for high background.
                 NGC 2264 & \dataset[ 14369]{  ADS/Sa.CXO#obs/14369} &       2011-12-08 &   66157 & 06:40:59.00 & +09:34:42.7 &   63 &    I (01237) &  TE-VF & G Micela & { 2011-11-01N6 } & N11 & (-0.853, -0.488) \\ % 0\% exposure discarded for high background.
\hline
\multicolumn{3}{l}{\bf NGC 6193; RCW 108-IR} \\
                HD 150136 & \dataset[  2569]{  ADS/Sa.CXO#obs/02569} &       2002-06-27 &   90335 & 16:41:18.89 & -48:45:39.3 &  312 &   H (456789) &   TE-F & S Skinner & { 2002-05-01N6 } & N11 & (+0.085, +0.194) \\ % 0\% exposure discarded for high background.
                  RCW 108 & \dataset[  4503]{  ADS/Sa.CXO#obs/04503} &       2004-10-25 &   88812 & 16:39:58.68 & -48:51:54.2 &  236 &   I (012367) &  TE-VF & S Wolk & { 2004-08-01N6 } & N11 & (+0.387, +0.533) \\ % 0\% exposure discarded for high background.
                HD 150136 & \dataset[ 14598]{  ADS/Sa.CXO#obs/14598} &       2013-06-06 &  148104 & 16:41:21.12 & -48:45:38.5 &  351 &    H (56789) &   TE-F & J-C Leyder & { 2013-05-01N6 } & N11 & (+0.925, -0.129) \\ % 0\% exposure discarded for high background.
                HD 150136 & \dataset[ 14599]{  ADS/Sa.CXO#obs/14599} &       2013-10-09 &  119677 & 16:41:21.02 & -48:45:55.6 &  247 &   H (456789) &   TE-F & J-C Leyder & { 2013-08-01N6 } & N11 & (-0.385, -0.180) \\ % 0\% exposure discarded for high background.
\hline
\multicolumn{3}{l}{\bf DR15       } \\
                     DR15 & \dataset[ 12390]{  ADS/Sa.CXO#obs/12390} &       2011-01-25 &   39876 & 20:32:22.68 & +40:16:41.3 &  359 &   I (012367) &  TE-VF & N Wright & { 2010-11-01N6 } & N10 & (+0.421, -0.019) \\ % 0\% exposure discarded for high background.
\hline
\multicolumn{3}{l}{\bf Berkeley 87; Onsala 2S} \\
              Berkeley 87 & \dataset[  9914]{  ADS/Sa.CXO#obs/09914} &       2009-02-02 &   70148 & 20:21:46.42 & +37:22:44.9 &   12 &     I (0123) &   TE-F & S Skinner & { 2009-02-01N6 } & N10 & (+0.054, -0.175) \\ % 0\% exposure discarded for high background.
\hline
\multicolumn{3}{l}{\bf NGC 6231   } \\
                 NGC 6231 & \dataset[  5372]{  ADS/Sa.CXO#obs/05372} &       2005-07-03 &   76187 & 16:54:14.37 & -41:49:53.1 &  299 &   I (012367) &   TE-F & S Murray & { 2005-05-01N6 } & N10 & (+0.145, +0.018) \\ % 0\% exposure discarded for high background.
                 NGC 6231 & \dataset[  6291]{  ADS/Sa.CXO#obs/06291} &       2005-07-16 &   44384 & 16:54:14.04 & -41:49:53.1 &  286 &   I (012367) &   TE-F & S Murray & { 2005-05-01N6 } & N10 & (+0.113, +0.070) \\ % 0\% exposure discarded for high background.
\hline
\multicolumn{3}{l}{\bf NGC 6357   } \\
         NGC 6357 Field I & \dataset[  4477]{  ADS/Sa.CXO#obs/04477} &       2004-07-09 &   37689 & 17:24:43.30 & -34:12:12.7 &  288 &   I (012367) &  TE-VF & G Garmire & { 2004-05-01N6 } & N11 & (+0.023, -0.434) \\ % 0\% exposure discarded for high background.
             G353.2$+$0.7 & \dataset[ 10988]{  ADS/Sa.CXO#obs/10988} &       2010-05-07 &   39651 & 17:26:01.71 & -34:14:46.3 &   72 &   I (012367) &  TE-VF & L Townsley & { 2010-05-01N6\_revA } & N11 & (-0.106, -1.111) \\ % 0\% exposure discarded for high background.
             G353.1$+$0.6 & \dataset[ 10987]{  ADS/Sa.CXO#obs/10987} &       2010-07-17 &   40527 & 17:25:35.41 & -34:23:36.5 &  283 &   I (012367) &  TE-VF & L Townsley & { 2010-05-01N6\_revA } & N11 & (+0.228, -1.266) \\ % 0\% exposure discarded for high background.
           G353.08$+$0.36 & \dataset[ 13267]{  ADS/Sa.CXO#obs/13267} &       2012-07-06 &   56211 & 17:26:40.03 & -34:34:45.8 &  293 &    I (01236) &  TE-VF & G Garmire & { 2012-05-01N6 } & N11 & (+1.059, -0.720) \\ % 0\% exposure discarded for high background.
           G352.90$+$1.02 & \dataset[ 13622]{  ADS/Sa.CXO#obs/13622} &       2013-01-29 &   39459 & 17:23:29.74 & -34:20:43.9 &   96 &   I (012367) &  TE-VF & L Townsley & { 2012-11-01N6 } & N11 & (-0.905, -0.134) \\ % 0\% exposure discarded for high background.
           G353.08$+$1.24 & \dataset[ 13623]{  ADS/Sa.CXO#obs/13623} &       2013-01-31 &   39458 & 17:23:05.99 & -34:04:09.7 &   95 &   I (012367) &  TE-VF & L Townsley & { 2012-11-01N6 } & N11 & (-0.910, -1.036) \\ % 0\% exposure discarded for high background.
            NGC6357\_CS61 & \dataset[ 18453]{  ADS/Sa.CXO#obs/18453} &       2016-07-05 &   41526 & 17:24:44.87 & -34:12:14.2 &  292 &     I (0123) &  TE-VF & G Garmire & { 2016-05-02N6 } & N11 & (+1.351, -1.204) \\ % 0\% exposure discarded for high background.
%        G352.279$+$0.808 & \dataset[ 18909]{  ADS/Sa.CXO#obs/18909} &       2017-07-05 &   39553 & 17:22:40.02 & -34:59:12.2 &  293 &     I (0123) &  TE-VF & L Townsley & { 2017-05-02N6 } & N11 & (+0.538, -0.566) \\ % 0\% exposure discarded for high background.
%        G352.015$+$0.705 & \dataset[ 18908]{  ADS/Sa.CXO#obs/18908} &       2017-10-11 &   36979 & 17:22:19.69 & -35:15:54.9 &  260 &   I (012367) &  TE-VF & L Townsley & { 2017-05-02N6 } & N11 & (+1.069, +0.690) \\ % 0\% exposure discarded for high background.
%        G352.557$+$0.744 & \dataset[ 18910]{  ADS/Sa.CXO#obs/18910} &       2017-10-13 &   39446 & 17:23:40.47 & -34:47:49.1 &  260 &   I (012367) &  TE-VF & L Townsley & { 2017-05-02N6 } & N11 & (-0.013, -0.200) \\ % 0\% exposure discarded for high background.
         G352.841$+$0.720 & \dataset[ 19705]{  ADS/Sa.CXO#obs/19705} &       2017-10-14 &   39532 & 17:24:32.87 & -34:34:34.0 &  260 &    I (01237) &  TE-VF & G Garmire & { 2017-05-02N6 } & N11 & (+0.377, -0.141) \\ % 0\% exposure discarded for high background.
\hline
\multicolumn{3}{l}{\bf AFGL 4029  } \\
                AFGL 4029 & \dataset[  7443]{  ADS/Sa.CXO#obs/07443} &       2007-11-20 &   81707 & 03:01:19.25 & +60:29:58.5 &  194 &   I (012367) &  TE-VF & M Gagn{\'e} & { 2007-11-01N6 } & N10 & (+0.153, +0.265) \\ % 0\% exposure discarded for high background.
\hline
\multicolumn{3}{l}{\bf h Per      } \\
              Inter$-$Per & \dataset[  5408]{  ADS/Sa.CXO#obs/05408} &       2004-11-28 &    9826 & 02:20:30.09 & +57:03:22.0 &  224 &   I (012367) &  TE-VF & N Evans & { 2004-11-01N6 } & N10 & (+0.121, +0.127) \\ % 0\% exposure discarded for high background.
                    h Per & \dataset[  5407]{  ADS/Sa.CXO#obs/05407} &       2004-12-02 &   41105 & 02:19:00.36 & +57:07:04.4 &  229 &   I (012367) &  TE-VF & N Evans & { 2004-11-01N6 } & N10 & (+0.170, -0.021) \\ % 0\% exposure discarded for high background.
                    h Per & \dataset[  9913]{  ADS/Sa.CXO#obs/09913} &       2009-10-16 &   36677 & 02:18:58.62 & +57:07:12.2 &  162 &     I (0123) &   TE-F & G Micela & { 2009-08-01N6 } & N10 & (+0.253, +0.397) \\ % 0\% exposure discarded for high background.
                    h Per & \dataset[ 12021]{  ADS/Sa.CXO#obs/12021} &       2009-11-08 &   51448 & 02:18:59.18 & +57:06:56.3 &  195 &     I (0123) &   TE-F & G Micela & { 2009-11-01N6 } & N10 & (+0.425, +0.191) \\ % 0\% exposure discarded for high background.
                    h Per & \dataset[  9912]{  ADS/Sa.CXO#obs/09912} &       2009-11-11 &  101706 & 02:18:59.18 & +57:06:56.3 &  195 &     I (0123) &   TE-F & G Micela & { 2009-11-01N6 } & N10 & (+0.538, +0.137) \\ % 0\% exposure discarded for high background.
\hline
\multicolumn{3}{l}{\bf NGC 281    } \\
                  NGC 281 & \dataset[  7206]{  ADS/Sa.CXO#obs/07206} &       2005-11-08 &   23197 & 00:52:25.22 & +56:33:47.5 &  225 &   I (012367) &  TE-VF & S Wolk & { 2005-11-01N6 } & N11 & (+0.136, +0.298) \\ % 0\% exposure discarded for high background.
                  NGC 281 & \dataset[  5424]{  ADS/Sa.CXO#obs/05424} &       2005-11-10 &   61866 & 00:52:25.23 & +56:33:47.5 &  225 &   I (012367) &  TE-VF & S Wolk & { 2005-11-01N6 } & N11 & (+0.138, +0.127) \\ % 0\% exposure discarded for high background.
                  NGC 281 & \dataset[  7205]{  ADS/Sa.CXO#obs/07205} &       2005-11-12 &   12995 & 00:52:25.24 & +56:33:47.4 &  225 &   I (012367) &  TE-VF & S Wolk & { 2005-11-01N6 } & N11 & (+0.169, +0.124) \\ % 0\% exposure discarded for high background.
\hline
\multicolumn{3}{l}{\bf G305       } \\
                     G305 & \dataset[  8922]{  ADS/Sa.CXO#obs/08922} &       2008-12-13 &  119449 & 13:12:15.31 & -62:41:55.0 &  103 &    I (01236) &  TE-VF & M Gagn{\'e} & { 2008-11-01N6 } & N11 & (-0.274, -0.251) \\ % 0\% exposure discarded for high background.
                   WR 48a & \dataset[ 13636]{  ADS/Sa.CXO#obs/13636} &       2012-10-12 &   98603 & 13:12:37.41 & -62:43:03.3 &  174 &   H (456789) &   TE-F & M Gagn{\'e} & { 2012-08-01N6 } & N10 & (-0.098, +0.162) \\ % 0\% exposure discarded for high background.
\hline
\multicolumn{3}{l}{\bf RCW 49} \\
                   RCW 49 & \dataset[  3501]{  ADS/Sa.CXO#obs/03501} &       2003-08-23 &   35308 & 10:24:00.52 & -57:45:17.6 &  184 &   I (012367) &  TE-VF & G Garmire & { 2003-08-01N6 } & N11 & (+0.108, +0.272) \\ % 2\% exposure discarded for high background.
%            G284.0$-$0.9 & \dataset[  6577]{  ADS/Sa.CXO#obs/06577} &       2006-05-28 &    9632 & 10:19:09.82 & -58:02:34.7 &  258 &   I (012367) &   TE-F & G Garmire & { 2006-05-01N6 } & N11 & (+0.202, -0.316) \\ % 0\% exposure discarded for high background.
                    WR20a & \dataset[  6410]{  ADS/Sa.CXO#obs/06410} &       2006-09-05 &   47761 & 10:23:56.21 & -57:45:40.1 &  171 &     I (0123) &   TE-F & G Rauw & { 2006-08-01N6 } & N11 & (-0.114, -0.022) \\ % 3\% exposure discarded for high background.
                    WR20a & \dataset[  6411]{  ADS/Sa.CXO#obs/06411} &       2006-09-28 &   49377 & 10:23:56.84 & -57:45:35.0 &  146 &     I (0123) &   TE-F & G Rauw & { 2006-08-01N6 } & N11 & (-0.325, +0.067) \\ % 0\% exposure discarded for high background.
%            CPD$-$572874 & \dataset[  7423]{  ADS/Sa.CXO#obs/07423} &       2007-09-15 &   19698 & 10:15:18.26 & -57:51:40.7 &  158 &   I (012367) &   TE-F & Joel Kastner & { 2007-08-01N6 } & N11 & (-0.225, +0.058) \\ % 0\% exposure discarded for high background.
        AX J1025.6$-$5757 & \dataset[  9113]{  ADS/Sa.CXO#obs/09113} &       2008-04-27 &    4695 & 10:25:36.37 & -57:57:19.1 &  290 &   I (012367) &  TE-VF & M Roberts & { 2008-02-01N6 } & N11 & (+0.373, +0.730) \\ % 0\% exposure discarded for high background.
%                   WR 18 & \dataset[  8910]{  ADS/Sa.CXO#obs/08910} &       2008-10-29 &   19685 & 10:17:00.60 & -57:54:35.8 &  108 &    S (23678) &   TE-F & S Skinner & { 2008-08-01N6 } & N11 & (-0.375, -0.380) \\ % 0\% exposure discarded for high background.
         PSR J1022$-$5746 & \dataset[ 12151]{  ADS/Sa.CXO#obs/12151} &       2011-07-01 &    9930 & 10:23:01.56 & -57:46:33.7 &  229 &   I (012367) &   TE-F & G Garmire & { 2011-05-01N6 } & N11 & (-0.537, +0.463) \\ % 0\% exposure discarded for high background.
         PSR J1028$-$5819 & \dataset[ 12150]{  ADS/Sa.CXO#obs/12150} &       2011-09-16 &    9929 & 10:28:36.30 & -58:17:30.4 &  161 &   I (012367) &  TE-VF & G Garmire & { 2011-08-01N6 } & N11 & (-0.764, +0.444) \\ % 0\% exposure discarded for high background.
     2FGL J1027.4$-$5730c & \dataset[ 14700]{  ADS/Sa.CXO#obs/14700} &       2013-09-03 &    9845 & 10:27:23.87 & -57:30:45.7 &  173 &   I (012367) &  TE-VF & G Pavlov & { 2013-08-01N6 } & N11 & (+0.478, -0.358) \\ % 0\% exposure discarded for high background.
\enddata
\tablecomments{The table is divided by target; the name of each target from Table~\ref{targets.tbl} is shown in bold.
\\Col.\ (1): Name of the target in the \anchorparen{https://cda.harvard.edu/chaser/mainEntry.do}{\Chandra Data Archive search and retrieval tool {\em ChaSeR}}.
\\Col.\ (2): \Chandra Observation Identification (ObsID) number.
\\Col.\ (3): Calendar date when the observation began.
\\Col.\ (4): Exposure times are the net usable times after various filtering steps are applied in the data reduction process. 
For the following ObsIDs, we discarded exposure time as noted to remove periods of high instrumental background: 
3501 (2\%),
6410 (3\%).
The time variability of the ACIS background is discussed in \S6.16.3 of the \anchorparen{http://asc.harvard.edu/proposer/POG/}{\Chandra Proposers' Observatory Guide} 
and in the ACIS Background Memos at \url{http://asc.harvard.edu/cal/Acis/Cal_prods/bkgrnd/current/}.
\\Col.\ (5) \& (6): The On-axis Position is the \anchorparen{http://cxc.harvard.edu/ciao/faq/nomtargpnt.html}{time-averaged location of the optical axis (CIAO parameters RA\_PNT,DEC\_PNT)}.
Units of right ascension ($\alpha$) are hours, minutes, and seconds; units of declination ($\delta$) are degrees, arcminutes, and arcseconds.
\\Col.\ (7): Observatory roll angle.
\\Col.\ (8): ACIS observations can take on three \Chandra configurations: I (optical axis on ACIS-I, no grating), S (optical axis on ACIS-S, no grating), or H (optical axis on ACIS-S, HETG in place).
The subset of the ACIS CCD detectors enabled for the observation is listed in parentheses; the layout of the ten detectors (numbered 0--9 here) in the ACIS focal plane is shown in \S6.1 
of the \anchorparen{http://asc.harvard.edu/proposer/POG/}{\Chandra Proposers' Observatory Guide}.
\\Col.\ (9): ACIS observing modes are described in \S6.12 of the \anchorparen{http://asc.harvard.edu/proposer/POG/}{\Chandra Proposers' Observatory Guide}.
\\Col.\ (10): Principal Investigator of the observation.
\\Col.\ (11): Abbreviated name of the ACIS Time-Dependent Gain file used for calibration of event energies, e.g., ``2002-02-01N6'' = ``acisD2002-02-01t\_gainN0006.fits''.
\\Col.\ (12): The version of the Optical Blocking Filter model used for calibration of Ancillary Response Files and exposure maps.
\\Col.\ (13): The shift (in RA and Dec) applied to the ObsID's aspect file (via the {\em CIAO} tool {\em wcs\_update}) to achieve astrometric alignment, 
expressed as (dx,dy) in the \anchorparen{http://asc.harvard.edu/ciao/ahelp/coords.html}{\Chandra ``SKY'' coordinate system}; 1 SKY pixel = 0.492$\arcsec$.
}
\end{deluxetable}
%\end{longrotatetable} 

%% file: xray_column_labels.tex
% COLUMN SPACING
%\setlength{\tabcolsep}{1mm}

% ROW SPACING
% The traditional method for controlling row spacing, e.g.
%   \renewcommand{\arraystretch}{1.5} 
% does not work with AASTeX tables.  The AASTeX manual says you must edit the class file:
%
%    "Authors can control the space between data lines in tables by making a simple modification to the classfile. The line \def\arraystretch{1.0} is set to the default value of 1. To increase the space between data lines by 10% change the argument to 1.1. Likewise, to decrease the space between data lines by 10% to produce a tighter table use 0.9."

%\todo{I believe that the chi-square variability index is adequately described by table note ``c''.  }

%\begin{longrotatetable}
\startlongtable
\begin{deluxetable*}{lll}
\tablecaption{MOXC3 X-ray Sources and Properties \label{xray_properties.tbl}
}
\tablewidth{7in}
\tabletypesize{\scriptsize}
\tablecolumns{3}
%\tablenum{Table 2}

\tablehead{   
  \colhead{Column Label} & 
  \colhead{Units} & 
  \colhead{Description} 
}
\colnumbers
\startdata
\cutinhead{Name and position, derived from the ObsIDs that minimize the position uncertainty \citep[][\S6.2 and 7.1]{Broos10}}
 RegionName                 & \nodata              & name of the MSFR \\ 
 Name                       & \nodata              & \parbox[t]{4.0in}{X-ray source name in IAU format; prefix is CXOU~J} \\
 Label\tnm{a}               & \nodata              & X-ray source name used within the project \\
 RAdeg\tnm{b}               & deg                  & right ascension (ICRS) \\
 DEdeg\tnm{b}               & deg                  & declination (ICRS) \\
 PosErr                     & arcsec               & 1-$\sigma$ error circle around (RAdeg,DEdeg) \\
 PosType                    & \nodata              & algorithm used to estimate position \citep[][\S7.1]{Broos10}  \\
\\[0.10em]
\cutinhead{Validity metrics\tnm{c}, derived from a pre-defined set of ObsID combinations \citep[][\S2.2.3]{Townsley18}}
 ProbNoSrc\_MostValid       & \nodata              & smallest of ProbNoSrc\_t, ProbNoSrc\_s, ProbNoSrc\_h, ProbNoSrc\_v \\
%Merge\_MostValid           & \nodata              & merge that produced ProbNoSrc\_MostValid \\
%Band\_MostValid            & \nodata              & energy band that produced ProbNoSrc\_MostValid \\
%SrcCounts\_MostValid       & count                & observed counts used in calculaton of ProbNoSrc\_MostValid \\
 ProbNoSrc\_t               & \nodata              & \parbox[t]{4.0in}{smallest {\em p}-value under the no-source null hypothesis \citep[][\S4.3]{Broos10} among validation merges}\\
 ProbNoSrc\_s               & \nodata              & \parbox[t]{4.0in}{smallest {\em p}-value under the no-source null hypothesis among validation merges}\\
 ProbNoSrc\_h               & \nodata              & \parbox[t]{4.0in}{smallest {\em p}-value under the no-source null hypothesis among validation merges}\\
%ProbNoSrc\_v               & \nodata              & \parbox[t]{4.0in}{smallest {\em p}-value under the no-source null hypothesis among validation merges}\\
%NumValidObsIDs             & \nodata              & number of individual ObsIDs in which the source was valid \\
%ValidObsIDList             & \nodata              & list of the individual ObsIDs in which the source was valid\\
IsOccasional                & boolean              & \parbox[t]{4.0in}{flag indicating that source validation failed in all multi-ObsID merges; source validation comes from a single ObsID}\\
\\[0.10em]
\cutinhead{Variability indices\tnm{d}, derived from all ObsIDs}
 ProbKS\_single             & \nodata              & \parbox[t]{4.0in}{smallest {\em p}-value under the null hypothesis (no variability within each single ObsID) for the Kolmogorov--Smirnov test on the timestamps of each ObsID's event list } \\
 ProbKS\_merge\tnm{e}       & \nodata              & \parbox[t]{4.0in}{{\em p}-value under the null hypothesis (no variability) for the Kolmogorov--Smirnov test on the timestamps of the multi-ObsID event list } \\
 ProbChisq\_PhotonFlux\tnm{e} & \nodata              & \parbox[t]{4.0in}{{\em p}-value under the null hypothesis (no variability) for the $\chi^2$ test on the single-ObsID measurements of PhotonFlux\_t } \\
\\[0.10em]
\cutinhead{Observation details and photometric quantities, derived from the set of ObsIDs that optimizes photometry \citep[][\S6.2 and 7]{Broos10}}
 ExposureTimeNominal        & s                    & total exposure time in merged ObsIDs \\
 ExposureFraction\tnm{f}    & \nodata              & fraction of ExposureTimeNominal that source was observed \\
 RateIn3x3Cell\tnm{g}       &  count/frame         & 0.5:8 keV, in 3$\times$3 CCD pixel cell \\
 NumObsIDs                  & \nodata              & total number of ObsIDs extracted \\
 NumMerged                  & \nodata              & \parbox[t]{4.0in}{number of ObsIDs merged to estimate photometry properties} \\
 MergeBias                  & \nodata              & fraction of exposure discarded in merge \\
\\[0.10em]
 Theta\_Lo                  & arcmin               & smallest off-axis angle for merged ObsIDs \\
 Theta                      & arcmin               &  average off-axis angle for merged ObsIDs \\
 Theta\_Hi                  & arcmin               &  largest off-axis angle for merged ObsIDs \\
\\[0.10em]
 PsfFraction                & \nodata              & average PSF fraction (at 1.5 keV) for merged ObsIDs \\
 SrcArea                    & (0.492 arcsec)$^2$   & average aperture area for merged ObsIDs \\
 AfterglowFraction\tnm{h}   & \nodata              & suspected afterglow fraction\\
% Correction\_t        & \nodata              & pile-up correction to SrcCounts\_t, NetCounts\_t \\
\\[0.10em]
 SrcCounts\_t               & count                & observed counts in merged apertures \\
 SrcCounts\_s               & count                & observed counts in merged apertures \\
 SrcCounts\_h               & count                & observed counts in merged apertures \\
\\[0.10em]
 BkgScaling                 & \nodata              & scaling of the background extraction \citep[][\S5.4]{Broos10} \\
\\[0.10em]
 BkgCounts\_t               & count                & observed counts in merged background regions \\
 BkgCounts\_s               & count                & observed counts in merged background regions \\
 BkgCounts\_h               & count                & observed counts in merged background regions \\
\\[0.10em]
 NetCounts\_t               & count                & net counts in merged apertures \\
 NetCounts\_s               & count                & net counts in merged apertures \\
 NetCounts\_h               & count                & net counts in merged apertures \\
\\[0.10em]
 NetCounts\_Lo\_t\tnm{i}    & count                & 1-$\sigma$ lower bound on NetCounts\_t \\
 NetCounts\_Hi\_t           & count                & 1-$\sigma$ upper bound on NetCounts\_t \\
%\\[0.10em]
 NetCounts\_Lo\_s           & count                & 1-$\sigma$ lower bound on NetCounts\_s \\
 NetCounts\_Hi\_s           & count                & 1-$\sigma$ upper bound on NetCounts\_s \\
%\\[0.10em]
 NetCounts\_Lo\_h           & count                & 1-$\sigma$ lower bound on NetCounts\_h \\
 NetCounts\_Hi\_h           & count                & 1-$\sigma$ upper bound on NetCounts\_h \\  
\\[0.10em]
 MeanEffectiveArea\_t\tnm{j}& cm$^2$~count~photon$^{-1}$ & mean ARF value \\
 MeanEffectiveArea\_s       & cm$^2$~count~photon$^{-1}$ & mean ARF value \\
 MeanEffectiveArea\_h       & cm$^2$~count~photon$^{-1}$ & mean ARF value \\
\\[0.10em]
 MedianEnergy\_t\tnm{k}   & keV                  & median energy, observed spectrum \\
 MedianEnergy\_s            & keV                  & median energy, observed spectrum \\
 MedianEnergy\_h            & keV                  & median energy, observed spectrum \\
\\[0.10em]
 PhotonFlux\_t\tnm{l}     & photon~cm$^{-2}$~s$^{-1}$     & apparent photon flux \\
 PhotonFlux\_s              & photon~cm$^{-2}$~s$^{-1}$     & apparent photon flux \\
 PhotonFlux\_h              & photon~cm$^{-2}$~s$^{-1}$     & apparent photon flux \\
\\[0.10em]
 EnergyFlux\_t              & erg~cm$^{-2}$~s$^{-1}$ & max(EnergyFlux\_s,0) + max(EnergyFlux\_h,0) \\
 EnergyFlux\_s\tnm{l}       & erg~cm$^{-2}$~s$^{-1}$ & apparent energy flux \\
 EnergyFlux\_h              & erg~cm$^{-2}$~s$^{-1}$ & apparent energy flux 
\enddata
%\tablecomments{Rows are sorted by R.A.}
                                                                                                                                              
\tablecomments{
These X-ray columns are produced by the
{\em ACIS Extract} \citep[\AEacro,][]{Broos10,AE12,AE16} software package.  Similar column labels were previously published by the CCCP \citep{Broos11}, MOXC1, and MOXC2.
\AEacro\ and its User's Guide are publicly available from the Astrophysics Source Code Library, from Zenodo, and at \href{http://personal.psu.edu/psb6/TARA/ae_users_guide.html}{\url{http://personal.psu.edu/psb6/TARA/ae_users_guide.html}}.
}
\tablecomments{
The suffixes ``\_t'', ``\_s'', and ``\_h'' on names of photometric quantities designate the {\em total} (0.5--8~keV), {\em soft} (0.5--2~keV), and {\em hard} (2--8~keV) energy bands. 
The suffixes ``\_t'', ``\_s'', and ``\_h'' on source validation {\em p}-values (``ProbNoSrc\_*'') designate the 0.5--7~keV, 0.5--2~keV, and 2--7~keV energy bands. 
}
\tablecomments{
Source name and position quantities (Name, RAdeg, DEdeg, PosErr, PosType) are computed using a subset of each source's extractions chosen to minimize the position uncertainty \citep[][\S6.2 and 7.1]{Broos10}.   
Source significance quantities (``ProbNoSrc\_*'') are computed using a pre-defined set of ObsID combinations which do not depend on the data observed \citep[][\S2.2.3]{Townsley18}.
Variability indices (ProbKS\_single, ProbKS\_merge, ProbChisq\_PhotonFlux ) are computed using all ObsIDs.
All other quantities are computed using a subset of ObsIDs, chosen independently for each source, to balance the conflicting goals of minimizing photometric uncertainty and of avoiding photometric bias \citep[][\S6.2 and 7]{Broos10}. 
}
\tablecomments{
As with many \ACIS\ studies, source extractions in MOXC3 do not share identical calibrations.
Front-illuminated and back-illuminated \ACIS\ detectors have moderately different Effective Area curves.
ACIS-only and zeroth-order HETG configurations have very different Effective Area curves.
Even observations taken on the same detector in the same configuration suffer from time-dependent Effective Area curves due to time-varying contamination on the \ACIS\ Optical Blocking Filters.
These calibration variations among source extractions introduce variable biases into certain uncalibrated source properties (e.g., MedianEnergy---see note {\em k}) and introduce source-dependent calibration errors into other source properties (e.g., PhotonFlux, EnergyFlux, ProbKS\_merge, ProbChisq\_PhotonFlux---see notes {\em l,e}).
Spectral fitting is needed to account for these calibration differences and to establish actual flux changes in sources that might be variable.
}

\tablenotetext{a}{Source ``labels'' identify each source during data analysis, as the source position (and thus the Name) is subject to change.}

\tablenotetext{b}{\ACIS\ ObsIDs are shifted to align with our astrometric reference catalog, the Two Micron All Sky Survey \citep[2MASS,][]{Skrutskie06}, which uses the ICRS coordinate system (\url{https://old.ipac.caltech.edu/2mass/releases/allsky/doc/sec4_6.html}).}

\tablenotetext{c}{In statistical hypothesis testing, the {\em p}-value is the probability of obtaining a test statistic at least as extreme as the one that was actually observed, when the null hypothesis is true.
The {\em p}-value of the observed extraction under the no-source hypothesis is calculated by the method described by \citet[][Appendix~A2]{Weisskopf07}, which is derived under the assumption that X-ray extractions follow Poisson distributions.}

\tablenotetext{d}{See \citet[][\S7.6]{Broos10} for a description of the ProbKS\_single and ProbKS\_merge variability indices, and caveats regarding possible spurious indications of variability using the ProbKS\_merge index.   The ProbChisq\_PhotonFlux variability index is the {\em p}-value under the null hypothesis (no variability) for the standard $\chi^2$ test on the single-ObsID measurements of PhotonFlux\_t.  }

\tablenotetext{e}{The accuracy of the inter-ObsID variability metrics (ProbKS\_merge and ProbChisq\_PhotonFlux) depends on the consistency of the Effective Area curves for the extractions of the source.
Sources observed with both the ACIS-only and zeroth-order HETG configurations will often produce spurious indications of  inter-ObsID variability (because those configurations have very different Effective Area curves).
Inter-ObsID variability metrics are proper {\em p}-values only when all observations of the source have identical Effective Area curves.
}

\tablenotetext{f}{Due to dithering over inactive portions of the focal plane, a \Chandra source often is not observed during some fraction of the nominal exposure time (\url{http://cxc.harvard.edu/ciao/why/dither.html}).  We report here the \CIAO\ quantity ``FRACEXPO'' produced by the tool {\em mkarf}.}

\tablenotetext{g}{ACIS suffers from a non-linearity at high count rates known as {\em photon pile-up}, which is described in MOXC1, MOXC2, and references therein. 
Source properties in this table are {\em not} corrected for pile-up effects.
Column RateIn3x3Cell is an estimate of the observed count rate falling on an event detection cell of size 3$\times$3 \ACIS\ pixels, centered on the source position.
When RateIn3x3Cell $>0.05$ (count/frame), the reported source properties may be biased by pile-up effects.
See Table~\ref{pile-up_risk.tbl} for a list of source extractions confirmed to have significant pile-up.
}

\tablenotetext{h}{Some background events arising from an instrumental effect known as ``afterglow'' (\url{http://cxc.harvard.edu/ciao/why/afterglow.html}) may contaminate source extractions, despite careful procedures to identify and remove them during data preparation \citep[][\S3]{Broos10}.
After extraction, we attempt to identify afterglow events using the \AEacro\ tool {\em ae\_afterglow\_report}, and report the fraction of extracted events attributed to afterglow; see the \AEacro\ manual.}

\tablenotetext{i}{Confidence intervals (68\%) for NetCounts quantities are estimated by the \CIAO\ tool {\em aprates} (\url{http://asc.harvard.edu/ciao/ahelp/aprates.html}).}

\tablenotetext{j}{The Ancillary Response File (ARF) in \ACIS\ data analysis represents both the effective area of the Observatory and the fraction of the observation time for which data were actually collected for the source (column ExposureFraction).}

\tablenotetext{k}{MedianEnergy is the median energy of extracted events, corrected for background \citep[][\S7.3]{Broos10}. 
As an uncalibrated property of the observed events, MedianEnergy depends on the shape of the Effective Area curves for the extractions of the source.
} % \AEacro\ quantity ENERG\_PCT50\_OBSERVED

\tablenotetext{l}{PhotonFlux = (NetCounts / MeanEffectiveArea / ExposureTimeNominal) \citep[][\S7.4]{Broos10}. 
EnergyFlux = $1.602 \times 10^{-9} {\rm (erg/keV)} \times$ PhotonFlux $\times$ MedianEnergy \citep[][\S2.2]{Getman10}. 
Because MeanEffectiveArea depends on the shape of the Effective Area curves for the extractions of the source, PhotonFlux and EnergyFlux exhibit source-dependent calibration errors.
}

\end{deluxetable*}
%\end{longrotatetable}

%% file: piled_table.tex
\renewcommand{\arraystretch}{1.0}  % to change row spacing in table

\startlongtable
\begin{deluxetable}{@{\hspace{1em}}llcl|cccchhrh|ccc@{\hspace{2em}}ccc}
\tablecaption{Sources Exhibiting Photon Pile-up \label{pile-up_risk.tbl}}
\tabletypesize{\tiny} \tablewidth{0pt}

\tablehead{
\\[-7pt]
\multicolumn{4}{c}{\hrulefill\ Source\ \hrulefill\ } & 
\multicolumn{8}{c}{\ \hrulefill\ Extractions\ \hrulefill\ } & 
\multicolumn{6}{c}{\ \hrulefill\ Spectral Fitting\ \hrulefill} \\[-7pt] 
\colhead{Name} & \colhead{Label} & \colhead{Var} & \colhead{Identifier} & \colhead{ObsID} &  \colhead{$\theta$} & \colhead{PSF-Frac} & \colhead{Ratio} &
\nocolhead{label} & \nocolhead{ObsID} & \colhead{Counts} & \nocolhead{SNR} & 
\colhead{$N_{H}$} &  \colhead{$kT_1$} & \colhead{$kT_2$} & \colhead{$L_{1tc}$} & \colhead{$L_{2tc}$} & \colhead{$L_{tc}$} \\[-7pt] % \colhead{SRC_CNTS?}  & \colhead{(count frame$^{-1}$)}
\colhead{(CXOU J)} &&&&& \colhead{(\arcmin)} &&& &&&& \colhead{($10^{22}$cm$^{-2}$)} & \multicolumn{2}{c}{(keV)} &  \multicolumn{3}{c}{(log erg s$^{-1}$)}   
}
%\colnumbers
\decimalcolnumbers
\startdata
%                     label MVar                         ObsID                      hidden     hidden  NetCounts_t  SNR      NH                  KT1                KT2                   L1tc      L2tc      Ltc
\multicolumn{5}{l}{{\bf AEAur} (D=0.402 kpc)}\\
051618.17+341845.3 & c1412 && AE Aur                  & 19943 & 1.0 & 0.91 & 1.025 &c1412    &  19943 &   270 &    15.5 & 0.54    & 0.28      & \nodata           & \nodata & \nodata & 31.55\tablenotemark{a} \\ % 267 (ct),  0.064 (ct/frame)
                   &       && AE Aur                  & 19445 & 1.0 & 0.91 & 1.030 &c1412    &  19445 &   870 &    28.5 & "       & "         & \nodata           & \nodata & \nodata & "     \\ % 846 (ct),  0.067 (ct/frame)
                   &       && AE Aur                  & 19979 & 2.0 & 0.91 & 1.043 &c1412    &  19979 &   500 &    21.4 & "       & "         & \nodata           & \nodata & \nodata & "     \\ % 482 (ct),  0.059 (ct/frame)
                   &       && AE Aur                  & 19941 & 2.0 & 0.90 & 1.027 &c1412    &  19941 &   560 &    22.6 & "       & "         & \nodata           & \nodata & \nodata & "     \\ % 542 (ct),  0.066 (ct/frame)
                   &       && AE Aur                  & 19951 & 2.0 & 0.90 & 1.030 &c1412    &  19951 &   560 &    22.6 & "       & "         & \nodata           & \nodata & \nodata & "     \\ % 545 (ct),  0.065 (ct/frame)
%\multicolumn{3}{l}{D=0.402 kpc}\\                                                                                                                                                                                                                                           
\hline
\multicolumn{5}{l}{{\bf NGC2264} (D=0.738 kpc)}\\
064040.44+095050.4 & c1564 &v& V339 Mon                &  2550     & 1.8 & 0.90 & 1.081  & c1564    &   2550 &  1900 &    42.8 &  0.78  &  0.30   &  2.9              &   31.65 &   31.50 & 31.88  \\ %  1779 (ct),  0.137 (ct/frame) Pre-MS star (no Gaia DR2 parallax in SIMBAD).  The SIMBAD terminology "V*" means "variable star."
064046.08+094917.3 & c2471 &v& V642 Mon                &  2550     & 1.5 & 0.62 & 1.192  & c2471    &   2550 &  1400 &    37.0 &  0.016 &  0.84   &  1.6              &   31.34 &   31.22 & 31.59  \\ %  1212 (ct),  0.254 (ct/frame) Pre-MS star at 220 pc (Gaia DR2).  The SIMBAD terminology "V*" means "variable star."
064048.61+093557.3 & c2927 &v& 2MASS J06404862+0935578 & 13610     & 2.8 & 0.90 & 1.054  & c2927    &  13610 &  1200 &    34.2 &  3.4   &  9.5    & \nodata           & \nodata & \nodata & 31.81  \\ %  1195 (ct),  0.062 (ct/frame) Emission-line star at 750 pc (Gaia DR2); see Rapson14, ApJ 794, 124.  SIMBAD calls this source "[RPG2014] 33477" but I prefer the 2MASS name.  This source was not listed in the pile-up table (Table 3) in Kuhn13.
064056.50+095410.4 & c4507 &v& NGC 2264 121            &  2550     & 4.1 & 0.90 & 1.053  & c4507    &   2550 &  2300 &    47.3 &  0.063 &  0.90   &  3.9              &   30.72 &   31.46 & 31.53  \\ %  2216 (ct),  0.097 (ct/frame) Emission-line star at 718 pc (Gaia DR2); CXORRS J064056.5+095410 (Ramirez04, AJ 128, 787).  This is 2MASS J06405650+0954104.
064058.51+093331.7 & c5011 &v& HD 261902               &  2540     & 0.7 & 0.91 & 1.076  & c5011    &   2540 &  3000 &    53.8 &  0.019 &  3.2    & \nodata           & \nodata & \nodata & 31.30  \\ %  2800 (ct),  0.109 (ct/frame) Pre-MS B8 star at 985 pc (Gaia DR2); CXOANC J064058.5+093331 (I think the ANC is for ANCHORS).
%
% PROB_KS shows no intra-ObsID variability for 15 Mon.
% The 4-ObsID lightcurve for 15 Mon (c5052) shows a factor-of-two difference in AE Photon Flux between the bare ACIS ObsID and the 3 HETG ObsIDs.
% However, AE Photon Flux is affected by the radically different shape of the bare ACIS and HETG ARFs.
% XSPEC apparent fluxes are pretty similar: FH8 =1.650E-12, 1.382E-12, 1.333E-12, 1.443E-12 (bare ACIS is ~15% higher than HETGs).
% Considering ACIS/HETG calibration errors, and pile-up correction errors, I'm not inclined to claim variability. (PSB)
064058.66+095344.7 & c5052 & & 15 Mon                   &  2550     & 4.0 & 0.90 & 1.143  & c5052    &   2550 &  6800 &    81.2 &0.038,0.26& 0.038 & 0.23 & 31.43 & 32.35 & 32.40\tablenotemark{a}  \\ %  5904 (ct),  0.263 (ct/frame)  Be Star -- HETG target, at 282 pc from Hipparcos I think; binary, O7V + B1.5V.  Separate SIMBAD entry for "* 15 Mon A -- Double or multiple star" of type O7V((f))zvar C from GOSSS (2014).
                   &       & & 15 Mon                   &  5401 (H) & 0.1 & 0.90 & 1.050  & c5052    &   5401 &  1000 &    30.6 & "        & "     & "    & "     & "     & "      \\ %   946 (ct),  0.061 (ct/frame) HETG obs -- still piled in zeroth order
                   &       & & 15 Mon                   &  6248 (H) & 0.1 & 0.90 & 1.039  & c5052    &   6248 &   160 &    11.4 & "        & "     & "    & "     & "     & "      \\ %   148 (ct),  0.069 (ct/frame) HETG obs -- still piled in zeroth order
                   &       & & 15 Mon                   &  6247 (H) & 0.1 & 0.90 & 1.057  & c5052    &   6247 &   750 &    26.4 & "        & "     & "    & "     & "     & "      \\ %   710 (ct),  0.067 (ct/frame) HETG obs -- still piled in zeroth order
064059.36+093333.3 & c5292 &v& 2MASS J06405936+0933333 & 13610      & 1.2 & 0.91 & 1.040  & c5292    &  13610 &   670 &    24.7 &  1.1   &  3.6    & \nodata           & \nodata & \nodata & 31.41  \\ %   643 (ct),  0.058 (ct/frame) YSO at 1140 pc (Gaia DR2)
064059.54+093510.8 & c5354 &v& V421 Mon                & 14368      & 0.5 & 0.91 & 1.031  & c5354    &  14368 &  1200 &    33.4 &  0.49  &  3.0    & \nodata           & \nodata & \nodata & 31.23  \\ %  1158 (ct),  0.056 (ct/frame) Pre-MS star at 495 pc (Gaia DR2).  The SIMBAD terminology "V*" means "variable star."
064102.53+093455.8 & c6104 &v& V781 Mon                & 13610      & 0.9 & 0.91 & \nodata& c6104                                                                                                   \\ %   486 (ct),  0.063 (ct/frame) ERROR: extracted events must come from only one CCD!  Pre-MS star at 725 pc (Gaia DR2).  The SIMBAD terminology "V*" means "variable star."
064103.50+093118.4 & c6395 &v& NGC 2264 154            &  2540      & 3.2 & 0.89 & 1.027  & c6395    &   2540 &  1800 &    40.9 &  0.033 &  0.75   &  1.6              &   30.43 &   30.85 & 30.99  \\ %  1718 (ct),  0.052 (ct/frame) Pre-MS star at 715 pc (Gaia DR2)
064105.36+093313.5 & c6993 &v& V812 Mon                &  2540      & 1.9 & 0.90 & 1.066  & c6993    &   2540 &  2700 &    51.2 &  0.20  &  0.39   &  2.3              &   30.62 &   31.24 & 31.33  \\ %  2565 (ct),  0.100 (ct/frame) Pre-MS star at 823 pc (Gaia DR2)  Separate SIMBAD entry for CXOANC J064105.4+093313 (ANCHORS).  The SIMBAD terminology "V*" means "variable star."
064105.54+093140.6 & c7048 &v& 2MASS J06410554+0931405 &  9768      & 2.8 & 0.90 & 1.110  & c7048    &   9768 &  1800 &    41.6 &  0.15  &  3.2    & \nodata           & \nodata & \nodata & 31.71  \\ %  1643 (ct),  0.178 (ct/frame) Pre-MS star at 805 pc (Gaia DR2)
064105.74+093101.3 & c7114 &v& 2MASS J06410574+0931012 &  9769      & 2.2 & 0.90 & 1.056  & c7114    &   9769 &   960 &    29.9 &  0.15  &  0.32   &  2.8              &   30.61 &   31.32 & 31.39  \\ %   912 (ct),  0.102 (ct/frame) Pre-MS star, no Gaia DR2 parallax given in SIMBAD
064106.19+093622.9 & c7248 &v& LkHA 53                 &  2540      & 2.8 & 0.89 & 1.027  & c7248    &   2540 &  1800 &    41.6 &  0.057 &  1.2    &  4.2              &   30.64 &   30.91 & 31.10  \\ %  1769 (ct),  0.055 (ct/frame) Pre-MS star at 701 pc (Gaia DR2).  SIMBAD's "EM*" means "emission line star," "LkHA" means Lick Observatory H-alpha emission-line star."
064106.83+092732.3 & c7418 &v& LkHA 54                 &  9768      & 2.5 & 0.90 & 1.059  & c7418    &   9768 &   910 &    29.1 &  0.19  &  0.36   &  2.7              &   30.82 &   31.31 & 31.43  \\ %   862 (ct),  0.097 (ct/frame) Pre-MS star at 585 pc (Gaia DR2).  SIMBAD's "EM*" means "emission line star," "LkHA" means Lick Observatory H-alpha emission-line star."
                   &       & & LkHA 54                 &  9769      & 2.4 & 0.90 & 1.035  & c7418    &   9769 &   550 &    22.5 &  0.050 &  0.32   &  2.1              &   30.21 &   31.02 & 31.08  \\ %   534 (ct),  0.056 (ct/frame)
064108.59+092933.8 & c7888 &v& MPCM 064108.59+092933.7 &  9769      & 1.0 & 0.90 & 1.065  & c7888    &   9769 &   400 &    19.0 &  7.8   &  3.9    & \nodata           & \nodata & \nodata & 32.18  \\ %   383 (ct),  0.103 (ct/frame) ***No SIMBAD object within 2" search radius!***  VizieR gives lots of Chandra and MYStIX IDs and IR counterparts, but no famous source and, oddly, no 2MASS counterpart.  I'll give MPCM ID from Broos13.  Povich13 calls this MIRES G203.3139+02.0509.  It is also "SST 13241" in Sung09, "A Spitzer View of NGC 2264," but only [3.6] and [4.5] mags given.
064109.54+092925.3 & c8098 &v& 2MASS J06410954+0929250 & 14369      & 5.9 & 0.89 & 1.027  & c8098    &  14369 &  4100 &    63.4 &  3.4   &  6.8    & \nodata           & \nodata & \nodata & 32.27  \\ %  4085 (ct),  0.055 (ct/frame) Pre-MS star, no Gaia parallax in SIMBAD
064110.00+092746.1 & c8223 &v& 2MASS J06410999+0927460 &  9768      & 1.9 & 0.90 & \nodata& c8223        9768                                                                                       \\ %   247 (ct),  0.068 (ct/frame) ERROR: extracted events must come from only one CCD!  Pre-MS star at 527 pc (Gaia DR2)
                   &       & & 2MASS J06410999+0927460 &  9769      & 1.9 & 0.90 & 1.076  & c8223    &   9769 &   850 &    28.2 &  0.26  &  0.38   &  2.1              &   31.09 &   31.41 & 31.58  \\ %   796 (ct),  0.145 (ct/frame)
064113.04+092732.0 & c8854 &v& 2MASS J06411303+0927319 &  9769      & 2.0 & 0.89 & 1.045  & c8854    &   9769 &   600 &    23.5 &  0.31  &  0.33   &  1.8              &   30.95 &   31.04 & 31.30  \\ %   574 (ct),  0.068 (ct/frame) Pre-MS star at 824 pc (Gaia DR2).  A second SIMBAD entry in this 2" search is 0.3" away, "[RRA2004] NGC 2264 MM VLA 8", a radio source from Reipurth04, AJ 127, 1736, looks like the driving source for HH 108.
%\multicolumn{3}{l}{D=0.738 kpc}\\                                                                                                                                                                                                                                           
\hline
\multicolumn{5}{l}{{\bf NGC6193} (D=1.19 kpc)}\\
164120.42-484546.7 & c4062 &v& HD 150136                &  2569 (H) & 0.3 & 0.98 & 1.153 & c4062    &   2569 &  8700 &    90.8 &  0.84  &  0.13  &  0.61             &   33.96 &   33.43 & 34.08 \\ %  7572 (ct),  0.214 (ct/frame)
                   &       & & HD 150136                & 14598 (H) & 0.2 & 0.98 & 1.221 & c4062    &  14598 & 15500 &   120.8 &  0.86  &  0.15  &  0.60             &   34.06 &   33.58 & 34.19 \\ % 12812 (ct),  0.273 (ct/frame) Spectroscopic binary, the aimpoint of these HETG observations, O3.5III(f*) + O6IV from Pourbaix04, A&A 424, 727; at 1060 pc from Gaia DR2
                   &       & & HD 150136                & 14599 (H) & 0.2 & 0.98 & 1.206 & c4062    &  14599 & 11100 &   102.4 &  0.90  &  0.12  &  0.58             &   34.27 &   33.59 & 34.36 \\ %  9345 (ct),  0.255 (ct/frame)
%\multicolumn{3}{l}{D=1.19 kpc}\\                                                                                                                                                                                                                                           
\hline
% No piled sources in any of the following targets:
%TARGET: DR15
%TARGET: Sh2-187
%
\multicolumn{5}{l}{{\bf NGC6231} (D=1.71 kpc)}\\
% The dates of the two ObsIDs in NGC6231 are identical, so if extractions are all FI (or BI) then MERGE_KS and MERG_CHI are valid.
165401.84-414823.0 & c1235 &v& HD 152234                & 5372 & 2.8 & 0.64 & 1.072   & c1235    &   5372 &  2900 &    52.8 &  0.74  &  0.28  &  1.3              &   32.81 &   31.74 & 32.85\\ % 1945 (ct)   0.111 (ct/frame) Blue supergiant at 794 pc (from Hipparcos), sptype B0.5Ia (from Bidelman 1954, PASP 66, 249)
                   &       & & HD 152234                & 6291 & 2.7 & 0.68 & 1.077   & c1235    &   6291 &  2100 &    44.9 &  0.71  &  0.25  &  0.89             &   32.86 &   32.18 & 32.94\\ % 1530 (ct)   0.144 (ct/frame)
165403.59-414729.8 & c1464 & & HD 152233                & 5372 & 3.1 & 0.90 & \nodata & c1464        5372                                                                                    \\ % 1186 (ct)   0.101 (ct/frame) ERROR: extracted events must come from only one CCD!  Spectroscopic binary, sptype O6II(f) from GOSSS; at 1670 pc from Gaia DR1 (maybe there's a better estimate in DR2?  SIMBAD doesn't list it.)
                   &       & & HD 152233                & 6291 & 3.1 & 0.90 & 1.050   & c1464    &   6291 &  1700 &    40.7 &  0.85  &  0.22  &  0.80             &   33.01 &   32.29 & 33.09\\ % 1656 (ct)   0.111 (ct/frame)
165410.06-414930.1 & c2189 &v& HD 152248                & 5372 & 0.9 & 0.91 & 1.371   & c2189    &   5372 &  5400 &    72.4 &  0.90  &  0.17  &  0.47             &   33.74 &   33.25 & 33.86\\ % 3922 (ct)   0.422 (ct/frame) "Eclipsing binary of beta Lyr type" according to SIMBAD.  At 1700 pc from Gaia DR2, sptype O7Iabf+O7Ib(f) from GOSSS
                   &       & & HD 152248                & 6291 & 0.8 & 0.91 & \nodata & c2189        6291                                                                                    \\ % 1829 (ct)   0.339 (ct/frame) ERROR: extracted events must come from only one CCD!
165411.63-415057.3 & c2407 & & HD 152249                & 5372 & 1.2 & 0.61 & 1.148   & c2407    &   5372 &   890 &    27.9 &  0.84  &  0.24  &  0.50             &   32.93 &   28.32 & 32.93\tablenotemark{a}\\ % 1094 (ct)   0.078 (ct/frame) Blue supergiant, sptype OC9Iab from GOSSS, at 2067 pc from Gaia DR2
                   &       & & HD 152249                & 6291 & 1.2 & 0.56 & 1.130   & c2407    &   6291 &   270 &    15.5 &  "     &  "     & "                 &   "     &   "     & "    \\ %  276 (ct)   0.080 (ct/frame)
165419.83-415009.3 & c3512 &v& CD-41 11042              & 5372 & 1.1 & 0.55 & 1.146   & c3512    &   5372 &   690 &    25.2 &  0.95  &  0.62  & \nodata           & \nodata & \nodata & 32.40\\ %  750 (ct)   0.060 (ct/frame) "Eclipsing binary of Algol type" from SIMBAD.  Sptype O9.5IV from GOSSS, at 1659 pc from Gaia DR2
%\multicolumn{3}{l}{D=1.71 kpc}\\                                                                                                                                                                                                                                           
\hline
%
% No piled sources in the following target:
%TARGET: Berk87
%
%\tablebreak
\multicolumn{5}{l}{{\bf NGC6357} (D=1.77 kpc)}\\
 172259.77-340439.5 & p6\_135  & & 2MASS J17225977-3404395 & 13623 & 1.4 & 0.90 & 1.048 & p6\_135 &  13623 &   870 &    28.4 &  0.22  &  2.0                & \nodata           & \nodata & \nodata & 32.00  \\ %  828 (ct)   0.078 (ct/frame)  This is the name we used in MOXC1.  ***No SIMBAD object within 2" search radius!***  VizieR has lots of visual and IR counterparts.  Might want to note its MPCM name from Broos13. 
 172438.73-341202.9 & p1\_337  &v& 2MASS J17243873-3412028 & 18453 & 1.3 & 0.90 & 1.116 & p1\_337 &  18453 &  2100 &    45.2 &  1.8   &  4.8                & \nodata           & \nodata & \nodata & 32.80  \\ % 1932 (ct)   0.166 (ct/frame)  SIMBAD just calls this an "infrared source" and gives no Gaia parallax.  It is Wang07 src #188.  Might want to note its MPCM name from Broos13.  Ah, in MOXC1 it was not piled!
 172443.49-341156.9 & p1\_591  & & HD 319718               &  4477 & 0.3 & 0.90 & 1.118 & p1\_591 &   4477 &  2100 &    44.7 &  2.2   &  0.69               & \nodata           & \nodata & \nodata & 33.43\tablenotemark{a}  \\ % 1872 (ct)   0.183 (ct/frame)
                    &          & & HD 319718               & 18453 & 0.4 & 0.91 & 1.094 & p1\_591 &  18453 &  1900 &    42.0 &  "     &  "                  & \nodata           & \nodata & \nodata & "      \\ % 1700 (ct)   0.145 (ct/frame) HD 319718AB is a spectroscopic binary.  Also known as "Cl Pismis 24 1"; component A aka Cl Pismis 24-1SW in Maiz-Apellaniz07, ApJ 660, 1480.  A & B are separated by 0.4".  This position is closer to A (at 17 24 43.50 -34 11 56.8) but B is not far off (17 24 43.48 -34 11 57.2).  Sptype O3.5If* from GOSSS.  No Gaia DR2 parallax.  In MOXC1 we called this "Pismis 24-1 SW + NEab".
 172443.95-341145.6 & p1\_644  &v& 2MASS J17244396-3411458 &  4477 & 0.5 & 0.89 & 1.060 & p1\_644 &   4477 &   800 &    27.3 &  1.7   &  5.5                & \nodata           & \nodata & \nodata & 32.37  \\ %  769 (ct)   0.076 (ct/frame) This is Wang07's [WTF2007] 379 (that's the SIMBAD name).  Oddly, SIMBAD doesn't give the CXOU name, but notes other names:  2MASS J17244396-3411458 (which we use here because we used it in MOXC1), Gaia DR2 5976057078081524352 (but no Gaia parallax given).  Sptype K4 according to Fang12, A&A 539, 119.
%172444.73-341202.6 & p1\_713  & & Pismis 24 17            &  4477 & 0.3 & 0.91 & 0.995 \\ %  552 (ct)   0.054 (ct/frame) This is the name we used in MOXC1.  Sptype O3.5III(f*) from GOSSS.  [WTF2007] 420; 2MASS J17244473-3412027.  No Gaia parallax in SIMBAD (which calls this "Cl Pismis 24 17").  ***ASK PAT -- is this commented out because it's not really piled?  Why is "correction" so much different for all 6357 sources than it was in MOXC1???***
%
% The XSPEC apparent fluxes for WR 93 tell the same story as AE's photon flux lightcurves---the BI observation had a much lower apparent flux.
 172508.86-341112.4 & p1\_1372 &v& WR 93                    &  4477 & 5.4 & 0.90 & 1.048 & p1\_1372 &   4477 &  4200 &    63.5 &  1.6  &  1.0                &  2.7              &   33.03 &   32.75 & 33.21\\ % 4008 (ct)   0.119 (ct/frame)
%                   &          & & WR 93                    & 10988 &11.5 & 0.41 &not piled& p1\_1372 &10988 &   800 &         &  2.9  &  0.80               &  2.5              & LKT   & LKT   & LKT \\ % F1cH8 =3.646E-12  F2cH8 =1.007E-12  FcH8 =4.653E-12
                    &          & & WR 93                    & 18453 & 5.1 & 0.90 & 1.061 & p1\_1372 &  18453 &  4100 &    62.6 &  2.0  &  0.78               &  2.4              &   33.44 &   32.91 & 33.55\\ % 3860 (ct)   0.131 (ct/frame) This is the name we used in MOXC1 was "WR 93 (HD 157504)".  WR star, sptype WC7+O7-9 from van der Hucht 1988, A&A 199, 217.  At 1867 pc from Gaia DR2.  Part of the cluster VVV CL167 according to Lima14, A&A 568, 16.
 172534.23-342311.7 & p3\_718  & & [N78] 49                 & 10987 & 0.5 & 0.91 & 1.041 &  p3\_718 &  10987 &   870 &    28.5 &  2.5  &  0.32               &  2.1              &   33.44 &   32.02 & 33.45\\ %  841 (ct)   0.076 (ct/frame) This is the name we used in MOXC1.  SIMBAD just calls this "Star" and gives a sptype of O5, with no citation.  This is 2MASS J17253421-3423116.  No Gaia DR2 parallax.  
%\multicolumn{3}{l}{D=1.77 kpc}\\                                                                                                                                                                                                                                           
\hline
% No piled sources in any of the following targets:
%TARGET: AFGL4029
%TARGET: IRAS06548-0815
%TARGET: hPer
%TARGET: BDSB2003+81
%TARGET: NGC281
%
%\tablebreak
\multicolumn{5}{l}{{\bf G305} (D=3.59 kpc)}\\       
% WR48 shows no single-ObsID variability.                      
% Although the two extractions have somewhat inconsistent AE Photon Flux, one is bare FI and one is gratings BI so variability is not certain.
% XSPEC apparent flux and Ltc are:
%  EPOCH_8922_FI : FH8 =6.974E-12    34.55
%  EPOCH_13636_BI: FH8 =2.718E-12    34.40
% I'm not inclined to claim variability. (PSB)
131239.59-624255.9 & c4379 &   & WR 48a &   8922     & 3.0 & 0.41 & 1.502 & c4379    &   8922 & 14200 &   141.7 &  4.0  &  1.5  & 9.5 & 34.37 & 34.07 & 34.55 \\ %  grp10.0_2T:Solar,Solar_noerr 13614 (ct)   0.709 (ct/frame)
                   &       &   & WR 48a &  13636 (H) & 0.3 & 0.90 & 1.072 & c4379    &  13636 &  4000 &    63.8 &  8.7  &  1.5  & 9.5 & 34.49 & 33.44 & 34.52 \\ %   grp5.0_2T:Solar,Solar_noerr  3922 (ct)   0.075 (ct/frame) WR star, sptype WC6 from Davies12, but of course the Zhekov papers note that this is a CWB.
%
% The neighbor (c4418) shows no single-ObsID variability; it's aperture has a lot of power from WR48a (which is not variable).
% XSPEC apparent flux in the crowded and pile-up corrected bare ACIS extraction (8922_FI) is FH8 =6.517E-13, ~10% of WR 48a.
% In the HETG extraction (13636_BI) NET_CNTS is negative, and 1-sigma upper limit (NET_CNTS_SIGMA_UP) = 6.9.
% Converting the upper limit of 6.9 net counts to an upper limit on FH8 in XSPEC would be complicated. 
% This feels like significant inter-ObsID variability (PSB).
% LKT -- see e-mail from 27 May 2019 -- I did a PIMMS calculation of the upper limit on FH8 for Obs13636 and got 3.5e-15, 2 orders of magnitude fainter than Obs8922, so definitely variable.
131239.77-624255.8 & c4418 &v& none   &  8922     & 3.0 & 0.79 & 1.333 & c4418    &   8922 &  9000 &    60.3 &8.2,1.7& 1.1   &  9.5    & 34.05   &  32.89   & 34.08 \\ %  8818 (ct)   0.236 (ct/frame)  This is the companion to WR 48a, just to its east.  Is this source actually piled, or is it just sitting on the piled wing of WR 48a?  Neither SIMBAD nor VizieR shows anything spatially close enough to be the counterpart to this source -- everything there is really just WR 48a.  With K=5 mag, it is so bright that it makes a giant zone of avoidance.
%\multicolumn{3}{l}{D=3.59 kpc}\\                                                                                                                                                                                                                                           
\hline
%TARGET: Sh2-271
%
%G284.0-0.9 & 101940.09-580547.0 & c1069 & HD 89725 & 6577 & 5.1 & 0.90 & 1.027 \\ %   542 (ct)   0.074 (ct/frame)  SIMBAD calls this a "Star" with sptype F0/3IV/V from Houk & Cowley 1975, https://ui.adsabs.harvard.edu/?#abs/1975MSS...C01....0H .  At 375 pc, from Gaia DR2, so very much a foreground source.  This is on Obs 6577 (MSFR snapshot of G284.0-0.9), far west of Wd2/RCW49.  This came out of the MOXC3 pile-up table when we separated these western pointings off into the MSFR snapshots paper.
%
%
%
\multicolumn{5}{l}{{\bf RCW49} (D=4.21 kpc)}\\
102358.01-574548.9 & c3658 &v& WR 20a & 3501 & 0.6 & 0.91 & 1.130 & c3658    &   3501 &   770 &    26.8 &  3.4   &  1.5    & \nodata           & \nodata & \nodata & 33.75 \\ %   687 (ct)   0.134 (ct/frame)  WR star.  SIMBAD calls this "V* V712 Car" aka MSP240, but I've called it by its more usual name.  No parallax listed in SIMBAD.  WR 20b, presumably not piled, is at 10 24 18.397 -57 48 29.78 and has a Gaia DR2 distance of 3571 pc.  Our source c5840 (102418.39-574829.7) is its counterpart.
                   &       & & WR 20a & 6410 & 0.3 & 0.91 & 1.097 & c3658    &   6410 &  1900 &    42.5 &  3.4   &  0.96   &  2.7              &   33.86 &   33.16 & 33.94 \\ %  1740 (ct)   0.130 (ct/frame)
                   &       & & WR 20a & 6411 & 0.3 & 0.91 & 1.097 & c3658    &   6411 &  2300 &    46.8 &  3.8   &  0.92   &  2.2              &   34.07 &   33.24 & 34.13 \\ %  2094 (ct)   0.150 (ct/frame)
102402.43-574436.1 & c4397 & & MSP18  & 3501 & 0.7 & 0.91 & 1.101 & c4397    &   3501 &   400 &    19.1 &  1.2   &  3.1    & \nodata           & \nodata & \nodata & 33.08\tablenotemark{a} \\ %   368 (ct)   0.090 (ct/frame) SIMBAD calls this "Cl Westerlund 2 5, Star in Cluster," sptype O5/5.5V/III(f) from Rauw11, A&A 535, 40.  I use this more common name from the literature.
                   &       & & MSP18  & 6410 & 1.4 & 0.89 & 1.027 & c4397    &   6410 &  1000 &    30.3 &  "     &  "      & \nodata           & \nodata & \nodata & "    \\ %   967 (ct)   0.078 (ct/frame)
                   &       & & MSP18  & 6411 & 1.2 & 0.89 & 1.075 & c4397    &   6411 &  1200 &    34.0 &  "     &  "      & \nodata           & \nodata & \nodata & "    \\ %  1148 (ct)   0.090 (ct/frame)
102556.50-574842.9 & c6948 &v& WR 21a & 9113 & 9.0 & 0.91 & 1.029 & c6948    &   9113 &   900 &    29.0 &  1.8   &  1.9    & \nodata           & \nodata & \nodata & 34.16 \\ %   890 (ct)   0.060 (ct/frame)  WR star, aka Wack 2134, at 5500 pc according to Gaia DR2 parallax -- that distance puts it a kpc behind Wd2, assuming a distance of 4.4 kpc from Binder&Povich18 (but note their +/- 1 kpc error on that distance).  This src was barely captured by the 4.7-ks Obs 9113 (see Gosset16), far off-axis at the very corner of the I-array -- amazing that it's still piled that far off-axis.  Its scattered light spills over onto Obs14700, way off at the SW corner of the S2 chip.
%\multicolumn{3}{l}{D=4.21 kpc}
\enddata
\tablecomments{The table is divided by target; the name of each target from Table~\ref{targets.tbl} is shown in bold, followed by the target distance from Table~\ref{targets.tbl}.  For each source, only piled-up ObsIDs are listed, i.e., ObsIDs in which the source was observed but did not pile up are not included in this table.  Pile-up correction could not be performed in some instances, because the source dithered across multiple CCDs.  Such instances have no results shown in this table; entries appear nonetheless, to document the existence of pile-up in the listed ObsID.  
\\
\XSPEC\ fits were performed on spectra corrected for pile-up.  For massive stars, the model form was {\em tbabs*apec$^1$}, {\em tbabs(apec$^1$ + apec$^2$)}, or {\em tbabs$^1$*apec$^1$ + tbabs$^2$*apec$^2$} with solar abundances, using the Tuebingen-Boulder ISM absorption model.  For pre-MS stars in NGC~2264, the model form was changed to accommodate pre-MS star abundances:  {\em tbabs*vapec$^1$} or {\em tbabs(vapec$^1$ + vapec$^2$)} with abundances frozen at the values adopted by the XEST study \citep{Gudel07}, relative to \citet{Anders89}, scaled to \citet{Wilms00}.
\\Col.\ (1): X-ray source name in IAU format.
\\Col.\ (2): X-ray source label used within the project.
\\Col.\ (3): ``v'' indicates that the source is variable, either exhibiting ``definite'' single-ObsID variability (ProbKS\_single $<0.005$ for some ObsID) or obvious inter-ObsID variability.
\\Col.\ (4): Source counterpart from SIMBAD, VizieR, or the literature.
\\Col.\ (5): \Chandra ObsID number.  The ``(H)'' denotes an HETG observation.
\\Col.\ (6): Off-axis angle for the source in the specified ObsID.
\\Col.\ (7): Fraction of the PSF (at 1.497 keV) enclosed within the extraction region for the specified ObsID.  A reduced PSF fraction (significantly below 90\%) indicates that the source is in a crowded region. 
\\Col.\ (8): The level of pile-up, characterized by the estimated ratio of pile-up-free to observed (piled-up) count rates in the 0.5--8~keV energy band.  Note that the core of any source is more piled up than the wings, so this ratio depends on the chosen aperture.   % No quantities in Table~\ref{xray_properties.tbl} have been corrected for pile-up effects.
\\Col.\ (9): Approximate number of X-ray events extracted from the chosen aperture.
\\Col.\ (10): Absorption column density from the \XSPEC\ model {\em tbabs}; the lower limit is fixed at $0.016 \times 10^{22}$cm$^{-2}$ \citep[$A_V$ = 0.1 mag,][]{Vuong03}.  When two values are listed, they are $N_{H}^1$ and $N_{H}^2$ from the model {\em tbabs$^1$*apec$^1$ + tbabs$^2$*apec$^2$}.
\\Col.\ (11,12): Plasma temperature from the \XSPEC\ {\em apec} or {\em vapec} models; the upper limit on plasma temperature is fixed at 9.5~keV.
\\Col.\ (13,14,15): Luminosities calculated from the spectral fit, in the total (t) band (0.5--8 keV), corrected (c) for absorption, assuming the distance shown in the table.
}
\tablenotetext{a}{
The model shown provided adequate fits to all pile-up reconstructed spectra.  We find no evidence that the spectral shape or flux varied among the observations. 
}

\end{deluxetable}

%Position matches are very good for all of these -- generally <0.2".

%% file: diffuse_spectroscopy_style2.tex
%  2019 Apr 16 12:11
%  acis_extract, version 5386  2019-02-18; ae_flatten_collation, version 5411  2019-04-14; hmsfr_tables, version 5411  2019-04-14
%\setlength{\tabcolsep}{0.03in}
%\renewcommand{\arraystretch}{2.0}  % to get double spacing in table

%  Use the aastex \nocolhead command and "h" column type to hide unwanted abundance columns!
\startlongtable
\begin{longrotatetable}
\movetabledown=0.9in
\begin{deluxetable}{lrrhrhhhhhh|llllc|llllc|llllc|llcchhhh|cchh@{}}   % 0 abundances
% \begin{deluxetable}{@{\hspace{1em}}lrrhrhhhhhh||llllc|llllc|llllc|llcchhhh||ccrr@{}}   % 0 abundances
%\begin{deluxetable}{@{\hspace{1em}}lrrhr|lllllh||llllc|llllc|llllc|llcc|lhlc||ccrr@{}}   % 0 abundances
%\begin{deluxetable}{@{\hspace{1em}}lrrhr|llllll||llllc|llllc|llllc|llcc|lhlc||ccrr@{}}   % 0 abundances
 
%\tablewidth{0pt}
%\tabletypesize{\scriptsize} 
 \tabletypesize{\tiny}
%\tablecolumns{13}

\tablecaption{Spectral Fits for Diffuse Regions \label{tbl:diffuse_spectroscopy_style2}}

%{\ \hrulefill\ Extractions\ \hrulefill\ }

\tablehead{
 \multicolumn{4}{c}{\ \hrulefill\ Diffuse Region\tablenotemark{a}\ \hrulefill\ } &
\multicolumn{32}{c}{\ \hrulefill\ Spectral Fit Parameters\tablenotemark{b}\ \hrulefill\ }      % 6 abundances                                                            
%& \colhead{Notes\tablenotemark{d}} 
\\ [-7pt]
%
% \multicolumn{4}{c}{\hrulefill} &
%\multicolumn{32}{c}{\hrulefill}       % 6 abundances
\\[-7pt]
\multicolumn{4}{c}{} &
\colhead{Quality} &
\multicolumn{21}{c}{\ \hrulefill\ Diffuse Components\ \hrulefill\ } &
\multicolumn{4}{c}{\ \hrulefill\ Unresolved Pre-MS Stars\ \hrulefill\ } &
\multicolumn{4}{c}{} & %Background}  
\multicolumn{2}{c}{Total Diffuse}
\\[-3pt]
%
%\multicolumn{4}{c}{} &
%\multicolumn{1}{c}{} &
% \multicolumn{6}{c}{} &                  % 0 abundances
%\multicolumn{15}{c}{\hrulefill} &
%\multicolumn{3}{c}{} &                                                 
%\multicolumn{4}{c}{} \\[-3pt]
%
\multicolumn{4}{c}{} &
\multicolumn{1}{c}{} &
%\multicolumn{6}{c}{Abundances} &                  % 6 abundances
 \multicolumn{6}{c}{}           &                  % 0 abundances
\multicolumn{5}{c}{\ \hrulefill\ pshock$^1$\ \hrulefill\ } &
\multicolumn{5}{c}{\ \hrulefill\ pshock$^2$\ \hrulefill\ } &
\multicolumn{5}{c}{\ \hrulefill\ pshock$^3$\ \hrulefill\ } &
\multicolumn{4}{c}{\ \hrulefill\ apec$^4$+apec$^5$\ \hrulefill\ } &
\multicolumn{4}{c}{} & %apec$^6$} &
\multicolumn{2}{c}{\ \hrulefill\ (1+2+3)\ \hrulefill\ } \\[-5pt]
% 
%\multicolumn{4}{c}{\hrulefill} &
%\multicolumn{1}{c}{\hrulefill} &
%%\multicolumn{6}{c}{Abundances} &                  % 6 abundances
% \multicolumn{6}{c}{}           &                  % 0 abundances
%\multicolumn{5}{c}{\hrulefill} &
%\multicolumn{5}{c}{\hrulefill} &
%\multicolumn{5}{c}{\hrulefill} &
%\multicolumn{4}{c}{\hrulefill} &
%\multicolumn{4}{c}{} & %\hrulefill} &
%\multicolumn{2}{c}{\hrulefill}   \\[-5pt]
%
%
\colhead{Name} & \colhead{NetCt} & \colhead{Area} & \nocolhead{} & 
\colhead{$\chi^2 \!/\! \rm{DOF}$} &
%
%  \colhead{O} &   \colhead{Ne} &   \colhead{Mg} &   \colhead{Si} &   \colhead{S} &   \colhead{Fe} & % 6 abundances
 \nocolhead{O} & \nocolhead{Ne} & \nocolhead{Mg} & \nocolhead{Si} & \nocolhead{S} & \nocolhead{Fe} & % 0 abundances
\colhead{$N_{H}^{1}$} & \colhead{$kT^1$} & \colhead{${\tau}^1$} & \colhead{$SEM^1$} & \colhead{$SB^{1}_{tc}$} &
\colhead{$N_{H}^{2}$} & \colhead{$kT^2$} & \colhead{${\tau}^2$} & \colhead{$SEM^2$} & \colhead{$SB^{2}_{tc}$} &
\colhead{$N_{H}^{3}$} & \colhead{$kT^3$} & \colhead{${\tau}^3$} & \colhead{$SEM^3$} & \colhead{$SB^{3}_{tc}$} &
\colhead{$N_{H}^{\bigstar}$} &                                       \colhead{$SEM^5$} & \colhead{$SB^{\bigstar}_{tc}$} & \colhead{$L^{\bigstar}_{tc}$} &
\nocolhead{$N_{H}^{6}$} & \nocolhead{$kT^6$} &                      \nocolhead{$SEM^6$} & \nocolhead{$SB^{6}_{tc}$} &
\colhead{$SB^{}_{tc}$} & \colhead{$L^{}_{tc}$}
%
%\colhead{SB4$_{tc}$} & \colhead{SB5$_{tc}$}
%
\\[-5pt]
\colhead{} & \colhead{(cts)} & \colhead{(pc$^{2}$)} & \nocolhead{($\arcmin^{2}$)} &
\colhead{()} & 
%\multicolumn{6}{c}{(solar units)} &        % 6 abundances
 \multicolumn{6}{c}{}              &        % 0 abundances
\multicolumn{23}{c}{(\dotfill\ $N_H$: $10^{22}$cm$^{-2}$\ \dotfill\ $kT$:  keV\ \dotfill\ $\tau$: log cm$^{-3}$ s\ \dotfill\ $SEM$: log cm$^{-3}$ pc$^{-2}$\ \dotfill $SB$: log erg s$^{-1}$ pc$^{-2}$ \dotfill $L$: log erg s$^{-1}$ \dotfill)} 
\\[-10pt]
\colhead{(1)} & \colhead{(2)} & \colhead{(3)} & \nocolhead{} & \colhead{(4)} & 
\nocolhead{} & \nocolhead{} & \nocolhead{} & \nocolhead{} & \nocolhead{} & \nocolhead{} & 
\colhead{(5)} & \colhead{(6)} & \colhead{(7)} & \colhead{(8)} & \colhead{(9)} & 
\colhead{(10)} & \colhead{(11)} & \colhead{(12)} & \colhead{(13)} & \colhead{(14)} & 
\colhead{(15)} & \colhead{(16)} & \colhead{(17)} & \colhead{(18)} & \colhead{(19)} & 
\colhead{(20)} & \colhead{(21)} & \colhead{(22)} & \colhead{(23)} & 
\nocolhead{(24)} & \nocolhead{()} & \nocolhead{(25)} & \nocolhead{(26)} & 
\colhead{(27)} & \colhead{(28)} 
}
%
% NOTE:  the \phantom calls in this table are used to make the numbers line
% up properly.
%
% You can NOT simply put pairs of '$' characters around each field in the format string below to get math mode for all the columns because, unfortunately, the AASTEX /nodata command fails in math mode.
% The table generator will substitute (via a 'sed' process) /nodata for each NaN printed by the format string below.
% We COULD carefully craft these format strings to produce '$/nodata$' in such cases and then have the table generator's sed look for that exact string.
%
%\decimalcolnumbers   % This command does not work, so we write column numbers explicitly above.
\startdata
\multicolumn{20}{l}{{\bf AEAur} D=0.402 kpc; Aur OB1 D=1.63 kpc (used for Unresolved Pre-MS Star luminosities in both regions)}\\
AE Aur (0.402 kpc)  &    382 &     0.4 &    28.3 &    14/ 14 & & & & & &  & $0.0\phd\!*$  & $0.27\phd_{\cdots}^{+0.4}$ & $12.0\phd_{\cdots}^{\cdots}$ & $ 53.5\phd_{\cdots}^{+0.4}$ &   30.70& \nodata   & \nodata & \nodata & \nodata & \nodata& \nodata  & \nodata & \nodata & \nodata & \nodata& \nodata                    & \nodata                     & \nodata & \nodata& \nodata & \nodata          & \nodata                   & \nodata                    &   30.70 &   30.29& {grp1.8_diffuse_3pshock} &            HandFit \\
\enddata
\end{deluxetable}
\end{longrotatetable}

%% file: bib.tex
% =============================================================================
% BIBLIOGRAPHY
% =============================================================================